\documentclass[a4paper,preprintnumbers,floatfix,twocolumn,aps,prb,unsortedaddress,superscriptaddress]{revtex4-2}

\usepackage[para]{threeparttable} 
\usepackage{amsmath}
\usepackage{graphicx} 
\usepackage{booktabs}
\usepackage{caption}
\usepackage{subcaption}

\usepackage{url}
\usepackage{hyperref}
\usepackage{miller}
\usepackage{bm}
\usepackage{enumitem}
\usepackage{gensymb}
\usepackage{color}
\usepackage{arydshln}
\usepackage{tabu}
\usepackage{xcolor}
\usepackage{tabularx}

\newcommand{\etal}{{\it et al. }}
\usepackage[paperwidth=210mm,paperheight=297mm,centering,hmargin=1.7cm,vmargin=2.5cm]{geometry}

\newcommand{\dashrule}[1][black]{%
  \color{#1}\rule[\dimexpr.5ex-.2pt]{4pt}{.4pt}\xleaders\hbox{\rule{4pt}{0pt}\rule[\dimexpr.5ex-.2pt]{4pt}{.4pt}}\hfill\kern0pt%
}
\newcommand{\rulecolor}[1]{%
  \def\CT@arc@{\color{#1}}%
}

\usepackage[utf8]{inputenc}

\renewcommand{\selectlanguage}[1]{} 

\begin{document}

\title{Comprehensive study of massively overlapping cascades in common elemental metals}

\author{A. Fellman}
\thanks{Corresponding author}
\email{aslak.fellman@helsinki.fi}
\affiliation{Department of Physics, P.O. Box 43, FI-00014 University of Helsinki, Finland}
\author{J. Byggmästar}
\affiliation{Department of Physics, P.O. Box 43, FI-00014 University of Helsinki, Finland}
\author{F. Granberg}
\affiliation{Department of Physics, P.O. Box 43, FI-00014 University of Helsinki, Finland}

\date{\today}

\begin{abstract}
Massively overlapping cascades simulations were carried out in 21 elemental metals: Be, Al, Ti, V, Cr, Fe, Co, Ni, Cu, Zr, Nb, Mo, Rh, Pd, Ag, Hf, Ta, W, Pt, Au and Pb. These elements have simple FCC, BCC and HCP lattice structures. For each element, 2000 cumulative 5 keV cascades were simulated using molecular dynamics. The overlapping cascades were simulated using various classical analytical interatomic potentials as well as with machine-learning interatomic potentials for many of the elements. The general conclusion is that results from massively overlapping cascades simulations are very sensitive to the choice of interatomic potential across the periodic table. Furthermore, we see that FCC metals all form stacking fault tetrahedra due to irradiation, and also typically include interstitial type Shockley partial and Frank type dislocations. In BCC materials typically form interstitial $1/2$\hkl<111> dislocations, but vacancy type $1/2$\hkl<111> dislocations were also observed in V and Nb. HCP materials tend to form complex dislocation structures mainly consisting of $a$-type dislocations that can be of both vacancy or interstitial type. Correlations between saturated defect concentrations and fundamental underlying material properties are explored.
\end{abstract}

\maketitle

\section{Introduction}
\label{sec:intro}

Radiation alters materials in various manners via radiation-induced damage. The damage can for instance be energetic particles hitting space crafts, neutrons and heavier ions interacting with the nuclear power plant vessel or fusion reactor walls, or decay of nuclear waste affecting the storage containers. The defects formed in the lattice will affect many different aspects of the materials, such as loss of ductility/embrittlement, change in electric and thermal properties, and swelling of the material~\cite{JENKINS199316,ZINKLE1995123}. All these effects have been studied in one way or another in the last century, however, they are not yet fully understood. Especially as the response of the material usually is not linear with dose, and there can be onset doses required for certain material property changes. All of these changes are based on modifications to the material at the atomistic scale, leading to lack of deep knowledge, as this scale is very hard to study. 

In elemental metals, a substantial amount of radiation damage studies have been conducted. For instance, in FCC metals it has been observed that dislocation loops and stacking fault tetrahedra (SFTs) are formed~\cite{HASHIMOTO2006295,schaublin2005irradiation}. Swelling has also been observed to be significant at higher doses in these materials, albeit very temperature dependent~\cite{zinkle1989void}. BCC metals mainly form $1/2\hkl<111>$ dislocation loops, and in some cases \hkl<100> loops, and the loops are observed to be both of interstitial- and vacancy-type, however mainly interstitial-type~\cite{eyre1973damage,JENKINS199316,Eyre01071965}. Also in HCP materials several dislocation loops with different Burgers vectors are observed, both vacancy- and interstitial-type. These loops are found to be either $c$- or $a$-type (or $c+a$), depending on their habit plane, and this has been seen to be determined by the c/a ratio of the material~\cite{griffiths1993evolution}. In all materials, it is known that exact type, Burgers vector and habit plane of dislocation loops will affect the properties of the material. Additionally, all other defect clusters formed will also affect both the dislocation movement and the material itself, and are, therefore, crucial to understand. 

In order to investigate the defect build-up and the nature of the dislocations, several experimental techniques can be utilized. One of the most common techniques is transmission electron microscopy (TEM), which can both see the size and the nature of the dislocation loops and other large defect clusters, usually with a resolution limitation of a nanometer or two~\cite{zhou2007diffraction,yi_direct_2015,yi_-situ_2016-1}. TEM is the go-to tool for investigating dislocation loops in metals, and have shown their evolution as a function of irradiation dose. Another method to indirectly investigate defects is Rutherford backscattering spectroscopy in channeling mode (RBS/c), which can detect the disorder due to defects in the crystalline material~\cite{zhang_influence_2015}. With this method the defect evolution (mainly the ones of interstitial type) can be determined, although usually not the exact nature of the defects, if not combined with simulations~\cite{bruncrona2024understanding}. Vacancies, especially single and small clusters can be studied with positron annihilation spectroscopy (PAS), and their concentrations and evolution can be determined~\cite{tuomisto2026positron}. Other techniques to determine and understand the defect structures are resistivity measurements to see changes in electric conductivity due to defects and resistivity recovery (RR) that can determine the migration barriers of defects found in the material~\cite{vajda1977anisotropy}. Additionally, techniques such as transient grating spectroscopy (TGS) and calorimetry studies can with some assumptions be used to complement other techniques to understand the underlying defect structure~\cite{reza_thermal_2020,hirst2022revealing}. All of these techniques have been successfully used to understand metallic systems under irradiation, however, a single one is usually not enough due to their limitations. Most of them also needs to be carried out ex-situ, limiting the number of data points, and full evolutions cannot be obtained. Crucially, none of these experimental methods can give the full atomistic resolution of the defect microstructure.

Molecular dynamics (MD) has been widely used to study radiation damage in metals with atomistic resolution over the years~\cite{nordlund_historical_2019}. A significant portion of that work has focused on the description of primary radiation damage~\cite{nordlund_primary_2018}. However, molecular dynamics studies on the effects of cascades overlapping with pre-existing defects, indicate that overlap has a significant impact on cascades behavior and therefore damage production~\cite{gao_computer_1996,granberg_cascade_2019,fellman_radiation_2019,byggmastar_collision_2019}. MD can simulate the accumulation of damage either by explicitly simulating the collision cascades themselves or by means of various accelerated accumulation methods such as Frenkel pair accumulation (FPA)~\cite{limoge1988computer,chartier2005molecular,chartier2019rearrangement,maxwell2020atomistic} that dates back to at least the year 1988~\cite{limoge1988computer}. Similar methods such as the creation relaxation algorithm (CRA)~\cite{derlet_microscopic_2020} have also been widely used in recent years to study defect evolution at the athermal limit. In these methods defects are directly introduced in to the system and allowed to relax. There also exists a host of different somewhat more sophisticated methods that aim to model the damage accumulation by means of simulating the damage formation due to the thermal spike, such as the molten spheres method~\cite{reali2025atomistic}, reduced-order atomistic cascade (ROAC)~\cite{chen2020reduced}, stochastic core–shell (SCS) approximation~\cite{torres2022stochastic} and iterative kinetic approach (IKA)~\cite{issaka2026cascade}. In all these methods the computationally expensive ballistic phase of the collision cascade is essentially skipped. However, these methods usually have to be parametrized based on cascade simulations. Furthermore, care must be taken to ensure that the produced microstructures are actually representative, as it has been noticed the acceleration methods can produce wildly different results compared to explicit cascade simulations~\cite{GRANBERG2023111902}.

In the case of massively overlapping cascade simulations, multiple consecutive collision cascades are explicitly simulated in the same system in order to accumulate damage and achieve higher doses that are more comparable to experiments. Early studies using overlapping cascades investigated the amorphization of semiconductors~\cite{caturla1995recrystallization,weber1997md,NORD2002294,Nord01,gao2002atomic}. Another topic of interest in early cumulative cascade studies was the study of radiation-induced mixing in interfaces~\cite{sule2004cooperative,heinisch2004effects}. Later, overlapping cascades have also been used to investigate defect accumulation in metals and alloys~\cite{yang2008multiple,vortler_effect_2011,granberg_mechanism_2016}. In recent years the majority of the research in metals has focused on nuclear fission and fusion metals such as tungsten~\cite{granberg_molecular_2021,GRANBERG2023111902,boleininger2023microstructure,lindblad2025effect}, nickel~\cite{granberg_mechanism_2016,ullah2016damage,levo_radiation_2017,levo2021temperature,fellman2026effects}, zirconium~\cite{tian2021heavy,torres2022stochastic,li2025molecular} and iron~\cite{byggmastar_effects_2018,granberg_defect_2020,byggmastar_dynamical_2020}. In addition, overlapping cascades have been used to investigate different alloy systems~\cite{vortler_effect_2011,ullah2016damage,levo_radiation_2017,levo2021temperature,li2023irradiation,ma2024effect,wei2024revealing,tk38-t5rr}, with particular interest in finding trends relating to the compositions of alloying elements. For primary radiation damage comprehensive comparative studies have been carried out, looking at the effects of elements and choice of interatomic potentials~\cite{nordlund_defect_1998,becquart_modelling_2021,de2021modelling}. In the case of overlapping cascade simulations, some studies have looked at the effects of the choice of potentials on the predicted defect evolution in specific elements and alloys~\cite{byggmastar_effects_2018,levo2021temperature,granberg_molecular_2021,fellman2026effects}. The general conclusion from these studies is that the choice of interatomic potential can have a significant impact on the predicted defect structures. Despite this substantial body of research done, to the authors' best knowledge there are still many simple elemental metals for which overlapping cascade simulations have not been carried out, even though there are potentially suitable interatomic potentials for them.

Overlapping cascade simulations have been successful in the last decade, especially when combined with complementary experimental studies, to give new insight into not understood phenomena. These simulations have for instance helped understand why certain alloys are fairing better then others when subjected to irradiation, and the key difference was the different dislocation natures~\cite{granberg_mechanism_2016}. Similar simulations also assisted the understanding of a new recovery stage in HCP titanium, due to the atomistic resolution of the defects~\cite{hirst2022revealing}. Overlapping cascades have also been used to explain the experimental RBS/c signal of irradiated materials~\cite{MARKELJ2024119499} and been used to assist with the understanding of PAS signals~\cite{hu2025new}. Not only can the defect structures obtained with this method be directly or indirectly (via some simulation of an experimental technique) be used to explain the phenomena, but the simulations cells themselves can be used as starting points. These highly damaged cells obtained with this technique have been previously used for nano-indentation simulations~\cite{mulewska2023self}, understanding the retention of light elements~\cite{PhysRevMaterials.5.095403} and mechanical testing simulations~\cite{Wang_Molecular_2026}.

Even though the method has been widely used in the last decade, most studies have been on single materials or focused on the alloying effects. No systematic study of a wide variety of materials has been conducted, and possible similarities and differences between elemental materials have not been determined. The aim of this study is the provide a thorough and systematic study of the most common elemental metals, done in an identical manner for basic understanding. We studied, in order of atomic number: Be, Al, Ti, V, Cr, Fe, Co, Ni, Cu, Zr, Nb, Mo, Rh, Pd, Ag, Hf, Ta, W, Pt, Au and Pb. These elements were chosen based on a set of criteria: First, they are common elemental metals with simple lattice structures (FCC, BCC and HCP) and melting temperatures above 300 $\degree$C. Second, the elements were chosen so that reasonable interatomic potentials are available for them that includes some description of short-range repulsive interactions. The main aims of this study are to provide a database of defect evolutions in common metals, a framework for performing overlapping cascade simulations, and to provide insight into the similarities, differences, and correlations between elements in terms of the accumulated defect concentration and microstructure.

\section{Methods}
\label{sec:methods}

\subsection{Interatomic potentials}\label{sec:met:tabgap}

The choice of interatomic potential is essential to provide reliable simulation results. This is especially true for simulations of radiation damage, were far-from-equilibrium phenomena are relevant. A proper description of short-range repulsion is crucial if we intend to use the interatomic potential for radiation damage simulations. One of the most commonly used repulsive pair potentials is the Ziegler-Biersack-Littmark-type (ZBL~\cite{ziegler_stopping_1985}) repulsive potential:
\begin{equation}
E_{\mathrm{rep.}} = \sum_{i<j}^{N} \frac{1}{4 \pi \epsilon_0} \frac{Z_i Z_j e^2}{r_{ij}} \phi (r_{ij}/a) f_{\mathrm{cut}}(r_{ij}),
\end{equation}
where
\begin{equation}
    a = \frac{0.46848}{Z_i^{0.23} + Z_j^{0.23}}.
\end{equation}

The function  $\phi (r_{ij}/a)$ is the screening function, for which a commonly used ''universal'' fit is given as:
\begin{equation}
\begin{split}
    \phi(x) = ~ & 0.1818 \mathrm{e}^{-3.2x} + 0.5099\mathrm{e}^{-0.9423x} \\
    & + 0.2802\mathrm{e}^{-0.4029x} + 0.02817\mathrm{e}^{-0.2016x}.
\end{split}
\end{equation}
However, this function can also be fitted to improve the accuracy of the repulsive pair potential. Recently, the repulsive pair interactions have been calculated using modern quantum chemistry methods~\cite{nordlund2025repulsive}. Furthermore, the authors of that work have provided fits for the screening functions for all element pairs up to $Z = 92$. 

\begin{table*}
 \caption{All interatomic potentials used in the simulations of massively overlapping cascades. }
 \label{tab:potentials_all}
 \begin{threeparttable}
  \begin{tabular*}{\linewidth}{@{\extracolsep{\fill}} llrr }
   \toprule
    \toprule
   &Element & Potential(s) &  Comments\\
   \bottomrule
    FCC & & & \\
    & Al &  Al-EAM~\cite{ercolessi_interatomic_1994,nordlund_defect_1998}, Al-Zhou~\cite{zhou_misfit-energy-increasing_2004}, Al-tabGAP~\cite{fellman2025fast}  & Al-EAM ''hardened'' by K. Nordlund~\cite{nordlund_defect_1998} \\
    & Ni &  Ni-EAM~\cite{stoller_impact_2016}, Ni-Zhou~\cite{zhou_misfit-energy-increasing_2004}, Ni-tabGAP~\cite{fellman2026effects}  & Ni-tabGAP potential is the semicore version~\cite{fellman2026effects} \\
    & Cu &  Cu-EAM~\cite{foiles1985application}, Cu-Zhou~\cite{zhou_misfit-energy-increasing_2004}, Cu-tabGAP~\cite{fellman2025fast}  & Cu-EAM ''hardened'' by K. Nordlund~\cite{nordlund_defect_1998}\\
    & Rh &  Rh-EAM~\cite{sheng_highly_2011} & Rh-EAM ''hardened'' by J. Byggmästar (this work)\\
    & Pd &  Pd-EAM~\cite{foiles1985application,nordlund1998role}, Pd-Zhou~\cite{zhou_misfit-energy-increasing_2004} & Pd-EAM  ''hardened'' by K. Nordlund~\cite{nordlund1998role}\\
    & Ag &  Ag-EAM~\cite{foiles_embedded-atom-method_1986,colla2001ion}, Ag-Zhou~\cite{zhou_misfit-energy-increasing_2004} & Ag-EAM ''hardened'' by K. Nordlund~\cite{colla2001ion}\\
    & Pt &  Pt-EAM~\cite{foiles1985application,nordlund1998role}, Pt-Zhou~\cite{zhou_misfit-energy-increasing_2004} & Pt-EAM ''hardened'' by K. Nordlund~\cite{nordlund1998role}\\
    & Au &  Au-EAM~\cite{kelchner1994construction,henriksson2005fragmentation}, Au-Zhou~\cite{zhou_misfit-energy-increasing_2004} &  Au-EAM ''hardened'' By K.O.E. Henrikson~\cite{henriksson2005fragmentation}\\
    & Pb &  Pb-Zhou~\cite{zhou_misfit-energy-increasing_2004} & \\
    \midrule
    BCC & &  & \\
    & V &  V-EAM~\cite{han2003interatomic}, V-Zhou~\cite{zhao2021defect} , V-tabGAP~\cite{byggmastar2026nine}  &  V added to Zhou, not in original potential\\
    & Cr & Cr-Zhou~\cite{lin2008computational}, Cr-tabGAP~\cite{tk38-t5rr}    &  Cr added to Zhou, Cr-tabGAP from AlCrCuFeNi-tabGAP~\cite{tk38-t5rr} \\
    & Fe &  Fe-EAM~\cite{malerba_comparison_2010,byggmastar_effects_2018}, Fe-Zhou~\cite{zhou_misfit-energy-increasing_2004}, Fe-tabGAP~\cite{byggmastar2022multiscale} & Fe-EAM ''harderned'' by J. Byggmästar~\cite{byggmastar_effects_2018} \\
    & Nb &  Nb-tabGAP~\cite{byggmastar2026nine}   & \\
    & Mo &  Mo-Zhou~\cite{zhou_misfit-energy-increasing_2004}, Mo-tabGAP~\cite{byggmastar2026nine}   & \\
    & Ta &  Ta-Zhou~\cite{zhou_misfit-energy-increasing_2004}, Ta-tabGAP~\cite{byggmastar2026nine}   & \\
    & W &  W-EAM~\cite{mason2023empirical}, W-Zhou~\cite{zhou_misfit-energy-increasing_2004}, W-tabGAP~\cite{byggmastar_simple_2022}  & \\

    \midrule
    HCP & & & \\
    & Be &  Be-Tersoff~\cite{bjorkas_interatomic_2009}  & \\
    & Ti &  Ti-EAM~\cite{ackland1992theoretical}, Ti-Zhou~\cite{zhou_misfit-energy-increasing_2004}, Ti-tabGAP~\cite{byggmastar2026nine}    & Ti-EAM ''hardened'' by G.J. Ackland Ti\_V2.eam.fs NIST~\cite{becker2013considerations}\\
    & Co &  Co-EAM~\cite{pasianot1992embedded,nordlund_inverse_1999}, Co-Zhou~\cite{zhou_misfit-energy-increasing_2004}    & Co-EAM  ''hardened'' by K. Nordlund~\cite{nordlund_inverse_1999}\\
    & Zr &  Zr-EAM~\cite{ackland1995defect}, Zr-Zhou~\cite{zhou_misfit-energy-increasing_2004} Zr-tabGAP~\cite{byggmastar2026nine}  & Zr-EAM ''hardened'' by G.J. Ackland NIST~\cite{becker2013considerations} \\
    & Hf &  Hf-tabGAP~\cite{byggmastar2026nine}    & \\
    \bottomrule
    \bottomrule
  \end{tabular*}
  \begin{tablenotes}
  \end{tablenotes}
 \end{threeparttable}
\end{table*}

Table~\ref{tab:potentials_all} lists all the potentials used for the elements that are considered in this work. The list of potentials is by no means exhaustive and even practical considerations such as availability and familiarity was a consideration in many cases. All potentials used here include a ZBL-like repulsive part with some form of connection or transition to the near-equilibrium energy range. Many of them have been widely used in radiation damage simulations before. However, some of the potentials are quite dated, and in the past the procedures of joining the repulsive interactions (i.e. ``hardening'') were not always very rigorous nor well documented. Therefore, the choices of potentials should not be viewed as endorsements by the authors of the current work. 

The interatomic potentials in this work can be divided into two groups, classical analytical potentials and machine-learned interatomic potentials. The classical analytical potentials all are of the embedded atom method (EAM) type~\cite{daw_embedded-atom_1984,finnis_simple_1984}. The only exception to this is the potential used for beryllium which was a Tersoff-type potential~\cite{tersoff_new_1988}. EAM potentials have been widely used in the simulation of metals as the formalism is based on the fundamental physics of metallic bonding. All the various EAM potentials have been ``hardened'' to include short-range repulsive interactions by various authors or in some cases by the original developers of the potentials. 

The potentials denoted as Zhou potentials in Table~\ref{tab:potentials_all} refers to the Zhou \etal EAM potential, which includes parametrization for a wide variety of metallic elements~\cite{zhou_misfit-energy-increasing_2004}. However, in this work the Zhou potential has been modified in a few key respects: First, the parametrization for Cr~\cite{lin2008computational} and V~\cite{zhao2021defect}, which were developed to be consistent with Zhou EAM functions, have been added to the original potential. Second, the potential has been hardened by joining it with the universal ZBL repulsive pair potential for each element pair. The transition between ZBL and the EAM pair terms was done between $0.4 r_{\mathrm{NN}}-0.7r_{\mathrm{NN}}$, where $r_{\mathrm{NN}}$ is the nearest neighbor distance in each element. The modified Zhou potential was chosen as a point of reference, as each element have been made by the same authors (with the exception of Cr and V) and presumably by similar methodologies. Furthermore, the naive joining of the ''universal'' ZBL might give some indication of the degree to which a proper transition to the repulsive pair interactions is necessary for the simulation of overlapping cascades across multiple elements. 

In addition to the classical interatomic potentials we have simulated the overlapping cascades using a selection of machine-learned potentials. All the ML potentials were tabulated Gaussian approximation potentials (tabGAP)~\cite{byggmastar_simple_2022}. There are two reasons for this choice: First, the authors of this work have been directly involved in the development of all these potentials and they include extra considerations for radiation damage simulations. Second, the tabGAP formalism is computationally efficient compared to most other MLIPs, making it feasible to simulate the system sizes and amount of cascades required by this work with a reasonable computational cost and time. The tabGAP formalism is based on the Gaussian approximation potential (GAP)~\cite{bartok_gaussian_2010}, using low-dimensional descriptors that can be tabulated into grids and efficiently evaluated using cubic-spline interpolations. A more detailed explanation of the tabGAP formalism can be found in Refs.~\citenum{byggmastar_simple_2022,byggmastar2022multiscale} or in the respective publications of the various tabGAP models. All the tabGAPs include ZBL-type repulsive potentials for short-range interactions.

\subsection{Massively overlapping cascade simulations}

Massively overlapping cascades are simulated by performing consecutive collision cascade simulations in the same system, thus accumulating damage in the simulation cell. All individually simulated cascades had a PKA energy of 5 keV. For all elements a system size of around 500 000 atoms was used, corresponding to $50\times50\times50$ FCC unit cells, $63\times63\times63$ BCC unit cells and $70\times40\times44$ orthorhombic unit cells of HCP. This system size was chosen so that even for elements with large spread-out cascades, the self-interaction over the periodic boundary conditions is limited. All MD simulations in this work were run using the \textsc{lammps} simulation package~\cite{thompson_lammps_2022}. All simulation cells were initially relaxed to 300 K in the $NPT$ ensemble before irradiation.

For each potential listed in Table~\ref{tab:potentials_all} three individual runs of 2000 cumulative cascades was performed. The process of simulating one individual cascade is as follows: First, the simulation cell is randomly shifted and the atoms that happen to end up outside of the simulation cell are wrapped back into the simulation cell in accordance with the periodic boundary conditions. The random shift is done to ensure a uniform distribution of damage within the simulation cell. After the random shift a random direction is sampled from a spherically uniform distribution. Then a primary knock-on atom (PKA) is chosen 30 Å distance from the center of the cell in the random direction. The PKA is given a velocity corresponding to 5 keV cascade energy back towards the center of the simulation cell. During the cascade the borders of the simulation cell (thickness of border 5 Å) was coupled to a thermostat to simulate the dissipation of heat to the surrounding material. The center of the simulation cell is kept in the $NVE$ ensemble not to artificially quench the cascade. 

During the cascade an adaptive time step is applied with a maximum allowed movement per time step of 0.1 Å, maximum change of kinetic energy of 300 eV and maximum time step of 0.002 ps for all elements, except for the lightest elements Al and Be, which had maximum time steps of 0.001 ps and 0.0005 ps respectively. Due to the high kinetic energies during cascades, the adaptive time step is used to ensure energy conservation during the MD time integration. Electronic stopping was applied to the atoms by the means of a friction force, pre-computed for each element separately from SRIM calculations~\cite{ziegler_srim_2010}. The electronic stopping was applied to atoms with a kinetic energy of above 5 eV. Each individual cascade event was simulated for 50 ps, after which a relaxation in $NPT$ ensemble was applied for 5000 time steps. The relaxation is performed to ensure that the system relaxes back to the initial temperature of 300 K and zero pressure instead of gradually increasing during the cumulative cascades. After the cascade and relaxation stages are performed, the inverse of the random shift is applied to the cell and the state of the system is stored for analysis, after which the next cascade event is initiated. The script used to launch the \textsc{lammps} overlapping cascade simulations can found from the following repository [LINK TO REPOSITORY TO BE ADDED UPON ACCEPTANCE].

\subsection{Analysis of structures}\label{sec:methods:analysis}

The structures were analysed using the Wigner-Seitz (WS) method~\cite{nordlund_defect_1998} to identify the vacancies and interstitials created during the cascades. We could not use the OVITO~\cite{stukowski2012automated} implementation of WS analysis as it requires a persistent reference system in order to give reliable results. The OVITO implementation starts to quickly break down after a few cascade events, giving unreliable results. Therefore, we employed a WS implementation~\cite{nordlund_defect_1998} used in several previous studies that builds the reference from the current frame, thus eliminating the need for a separate reference system. Unfortunately, this implementation of WS analysis was also not fully robust and could fail to get a reasonable estimation of defects when the simulation cell included considerable amounts of defects, especially large extended defects such as dislocation loops. When the WS analysis fails it predicts orders of magnitude more defects than what is actually present in the system. To mitigate this, different symmetry-preserving coordinate flips are applied to the final structure and WS analysis is performed on each of the flips separately. The flip which predicted the lowest defect amount is used as the best estimate of the defect amount and the inverse flip is applied to the analyzed structure and stored. If however all the flips failed then the analysis results are still stored but the results are ignored in the estimation of defect amounts. 

The failure of WS analysis was most prevalent in systems which produced considerable amounts of defects and in the HCP structures. This is one of the reasons why three individual runs of cascades were performed for each potential and material, as the average result of three different runs is less sensitive to the occasional failure of the analysis of an individual configuration. A further correction can be (and is in these results) applied to the estimation of number of vacancies in the form of a cluster correction. Initially, cluster analysis is performed on both the vacancies and interstitials. The cutoff of the cluster analysis was chosen as the midpoint between the second and third nearest neighbor distance. This is done to identify vacancies and interstitials that have been misidentified by WS analysis, which appear as interstitial-vacancy pairs that are artificially close to each other and, if real, would spontaneously recombine. The falsely identified defects are then removed from the final defect count. The cluster fraction was defined by performing cluster analysis on both the vacancies and interstitials present in the system and counting the number of defects that associated with a cluster. This means that the fraction of monovacancies or single interstitials (including dumbbells) can be found by taking $1-\mathrm{CF}_{\mathrm{def.}}$.

In addition to the WS analysis, DXA~\cite{0965-0393-18-8-085001} analysis was performed, implemented in OVITO~\cite{stukowski2012automated}, in order to find the types and lengths of the dislocations. The vacancy/interstitial nature of the dislocations were determined based on the WS defects associated with the dislocation. For each individual cascade simulation the above mentioned analysis methods are performed and the results are stored. The full atomistic structures are stored every 100 collision cascades and provide convenient checkpoints at various damage doses. These checkpoints can be used to restart the simulations from a certain dose or they can be used as initial conditions for other types of molecular dynamics simulations. The full atomistic structures can be found from [LINK TO REPOSITORY TO BE ADDED UPON ACCEPTANCE].

\section{Results} 

The results in this section are presented as follows. We group each element based on the crystal structure and split these into subsections. The only exception is beryllium that is separated into its own subsection as its results were a clear outlier from the rest. In the defect concentration results (Figs.~\ref{fig:FCC:dc},~\ref{fig:BCC:dc} and~\ref{fig:HCP:dc}) the average of the three individual runs are shown. Furthermore, in these plots a rolling average of 20 cascades is applied for visual clarity. The results without rolling averages are found in the supplementary materials for each element separately. 

\subsection{FCC metals}

\begin{figure*}[t]
    \centering
    \includegraphics[width=0.32\linewidth]{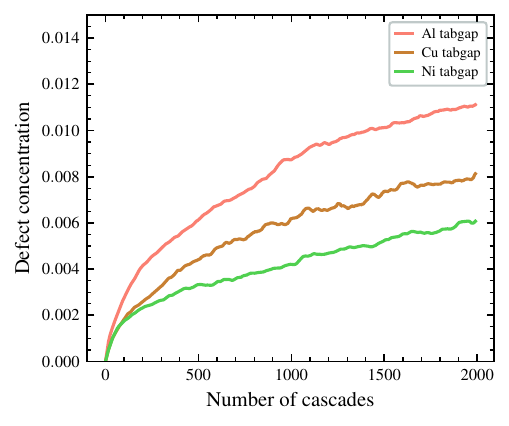}
    \includegraphics[width=0.32\linewidth]{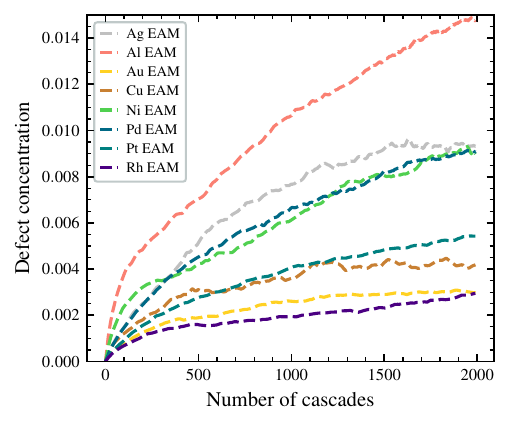}
    \includegraphics[width=0.32\linewidth]{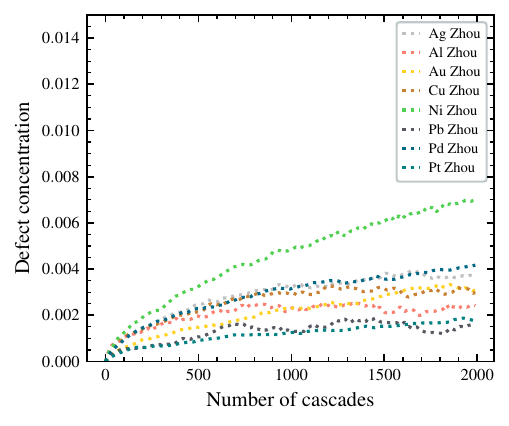}
    \caption{Defect concentration (interstitials and vacancies, $N_{\mathrm{def.}}/N_{\mathrm{tot.}}$) as a function of number of cascades. Left plot includes all FCC elements using tabGAP potentials, middle are the diverse different EAM potentials (see Table~\ref{tab:potentials_all}) and right plot includes overlapping cascades using the modified Zhou et al. potential. In the figures a rolling average is applied with a window of 20 cascades. The scale of the y-axis is kept the same in all subplots.}
    \label{fig:FCC:dc}
\end{figure*}

Fig.~\ref{fig:FCC:dc} shows the defect concentrations as a function of number of cascades for all the FCC elements considered. The different potentials (tabGAPs, EAMs and modified Zhou) are split into separate figures, but the scale of the y-axis is kept the same for all figures for direct comparison. The modified Zhou potential generally predicts lower defect concentrations compared both to the EAM potentials and tabGAPs. In both tabGAP and EAM, Al has the highest overall defect concentration while in the modified Zhou potential Al has a significantly lower defect concentration. Comparing the tabGAPs with EAM and Zhou, the ordering between Al, Cu and Ni is also different in all potentials. This lack of consistency is rather concerning for the interpretation of the overlapping cascade results in FCC metals. Furthermore, we note that Pb and Au are the heaviest of the elements, and have overall very low defect concentrations. However, a clear ordering based on element mass cannot be found from the figures. Some notable exceptions to highlight this point are Al-Zhou, Rh-EAM and Ag-EAM results that clearly do not follow ordering based on mass. Another similar observation can be made based on the element groups (i.e. columns in the periodic table), showing that there is no consistent ordering within or between groups. It is also clear that at 2000 cascades we have not yet reached a steady-state defect concentration in most of the FCC potentials. Only the potentials, which predict the lowest defect concentrations in both EAM and the Zhou potentials, show a degree of saturation of the defect concentrations. 

\begin{figure*}
    \centering
\begin{subfigure}{0.24\textwidth}
    \includegraphics[width=1.1\textwidth]{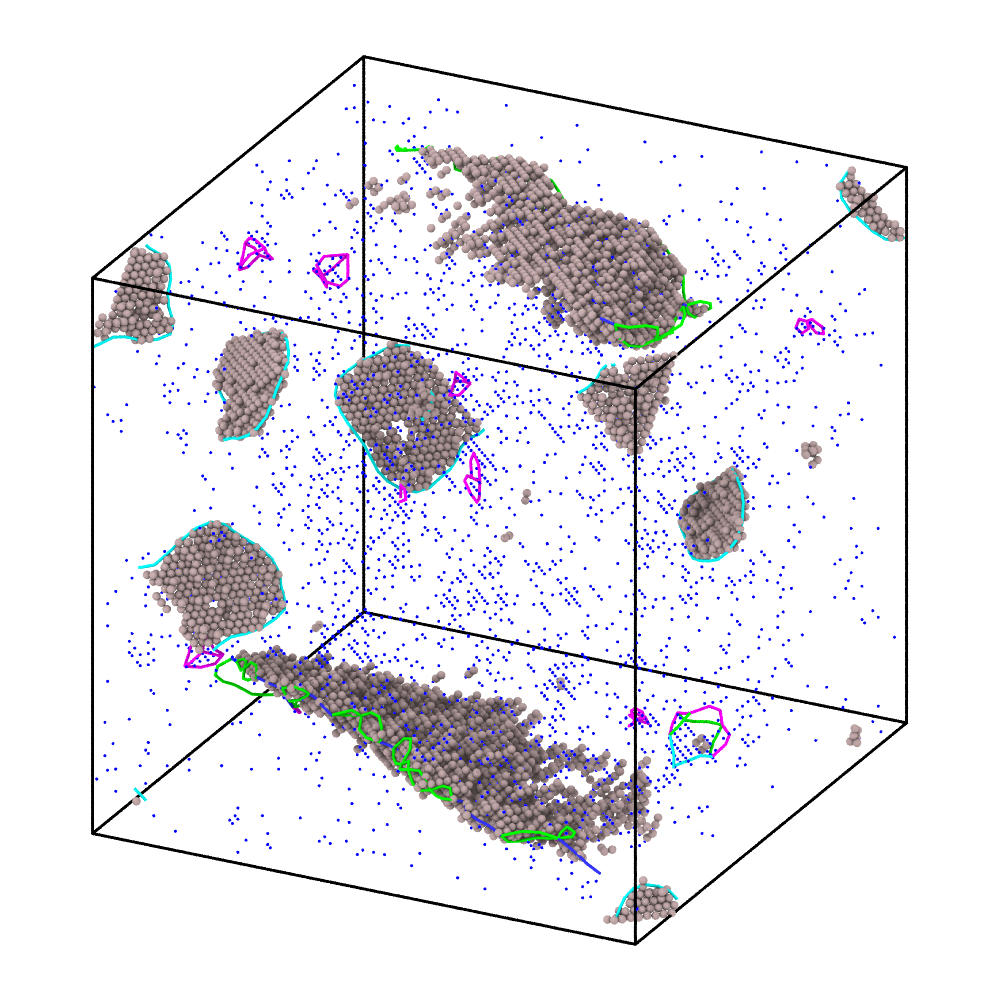}
      \caption{Al tabGAP}
\end{subfigure}
\begin{subfigure}{0.24\textwidth}
    \includegraphics[width=1.1\textwidth]{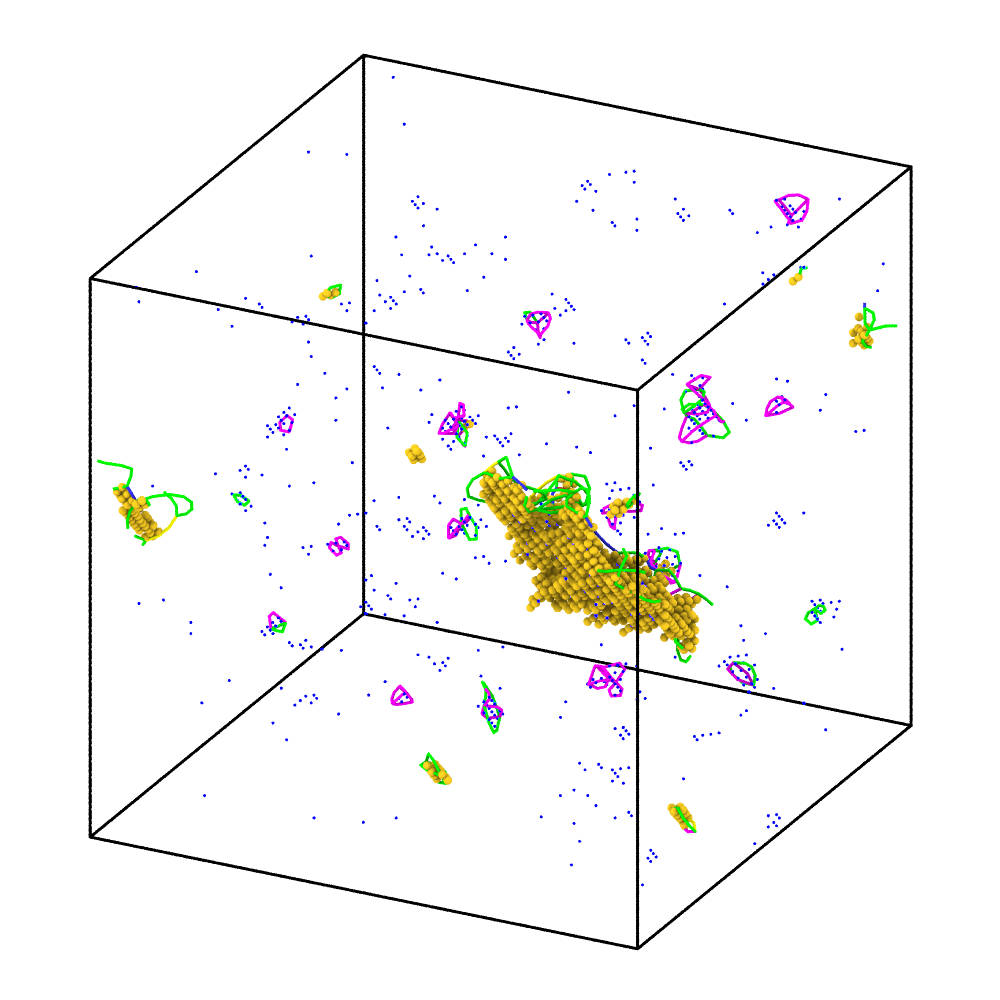}
    \caption{Au EAM}
\end{subfigure}
\begin{subfigure}{0.24\textwidth}
    \includegraphics[width=1.1\textwidth]{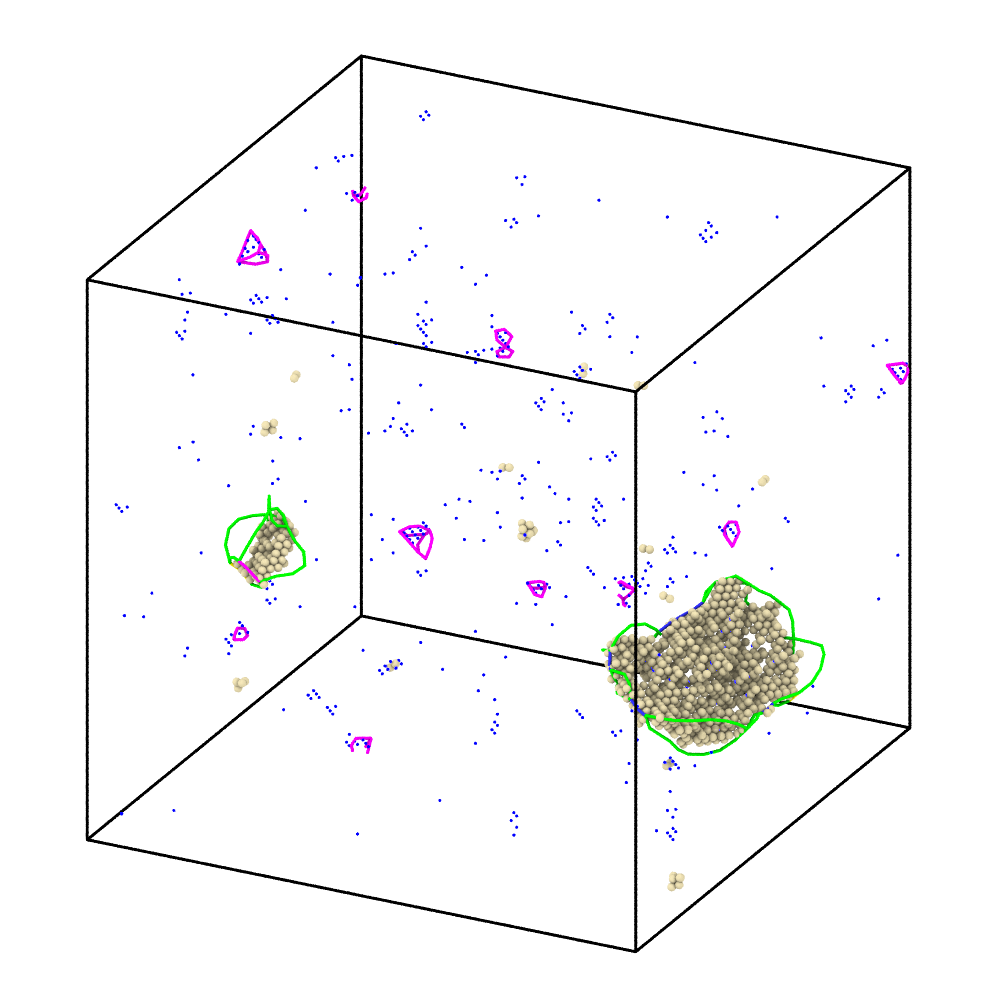}
    \caption{Pt Mod. Zhou EAM}
\end{subfigure}
\begin{subfigure}{0.24\textwidth}
    \includegraphics[width=1.1\textwidth]{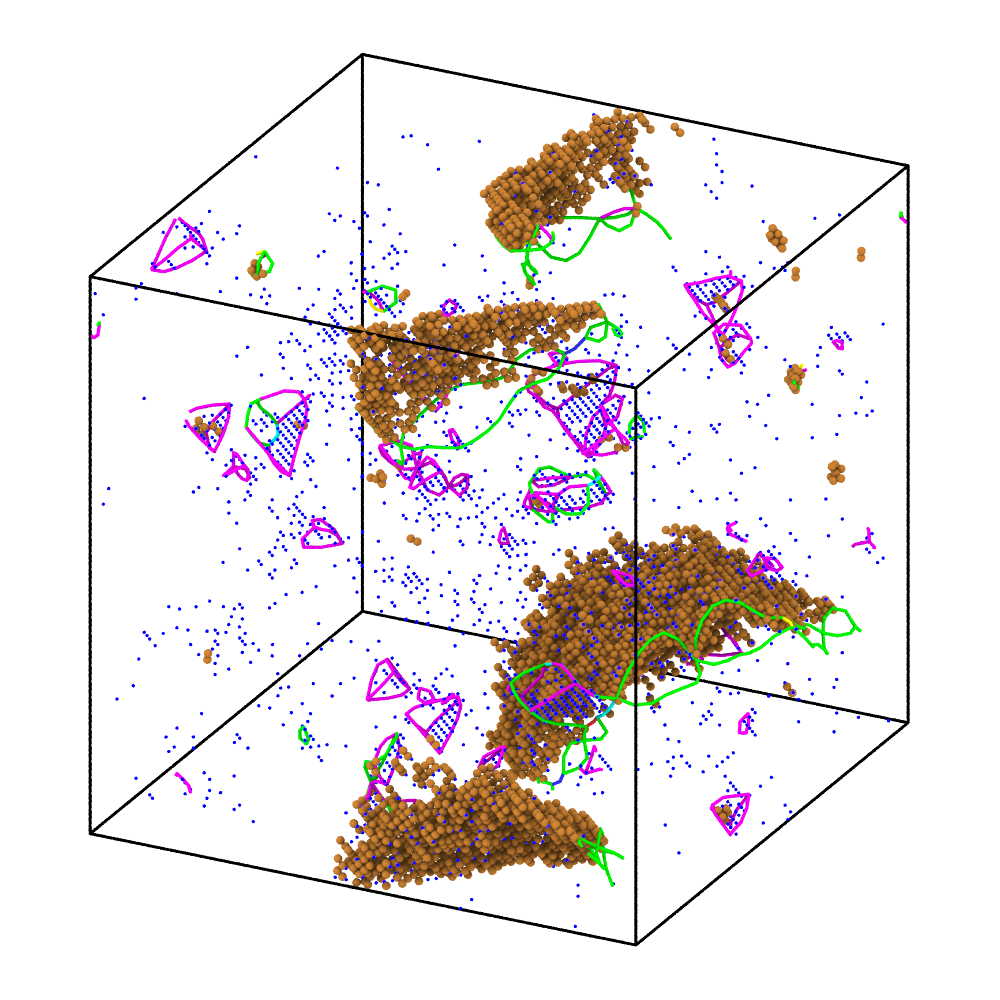}
    \caption{Cu tabGAP}
\end{subfigure}
\begin{subfigure}{0.24\textwidth}
    \includegraphics[width=1.1\textwidth]{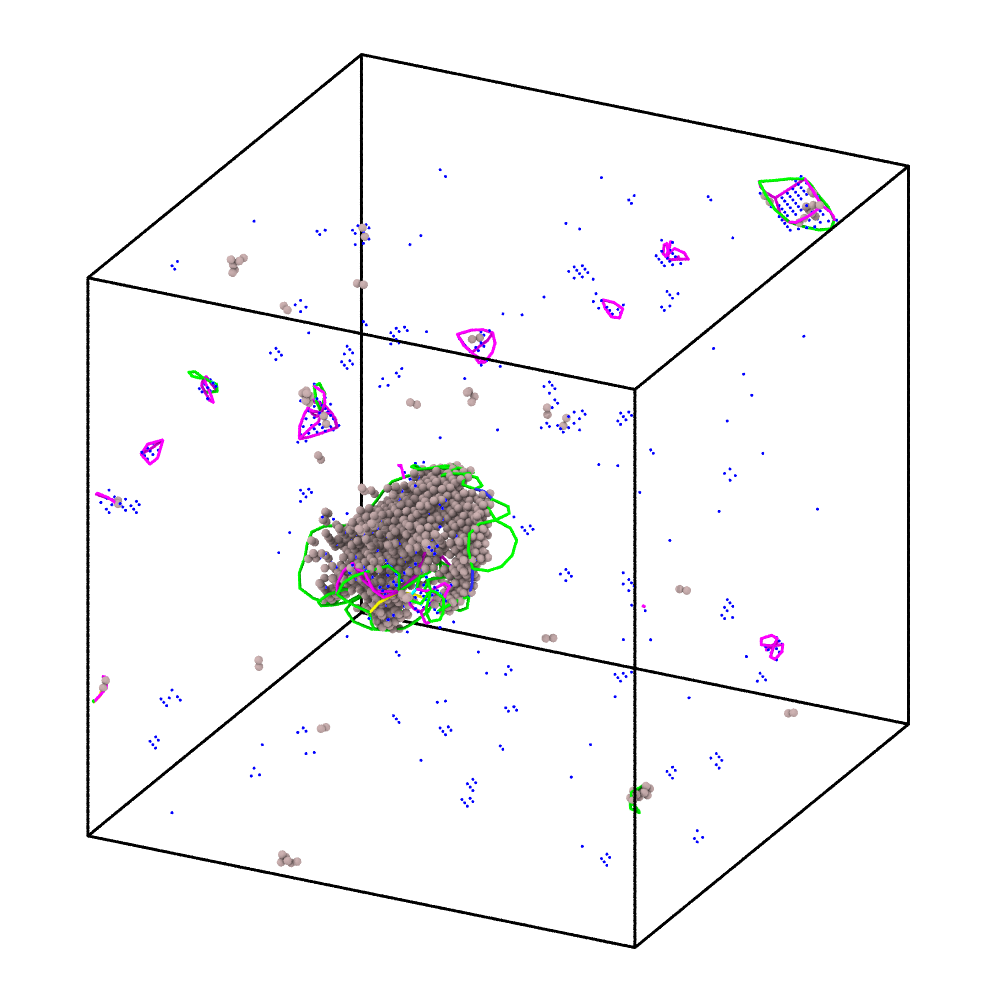}
    \caption{Al Mod. Zhou EAM}
\end{subfigure}
\begin{subfigure}{0.24\textwidth}
    \includegraphics[width=1.1\textwidth]{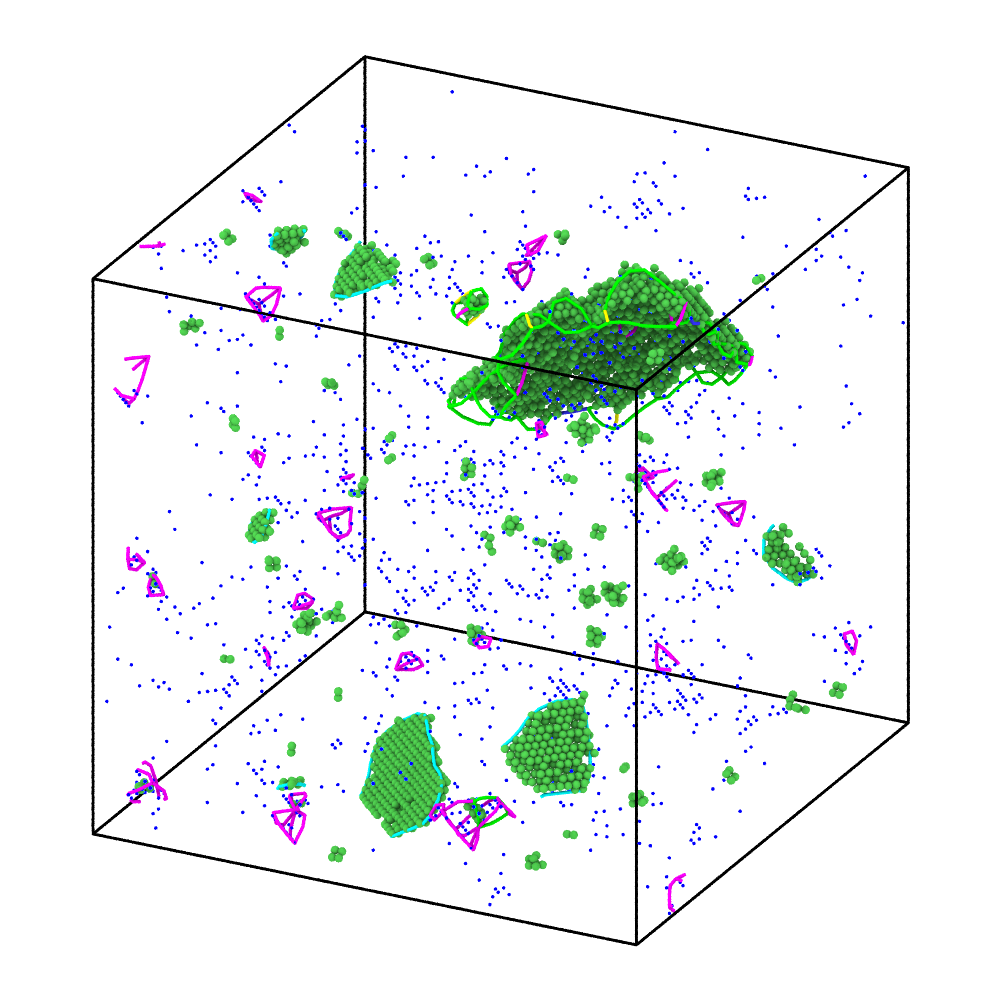}
    \caption{Ni tabGAP}
\end{subfigure}
\begin{subfigure}{0.24\textwidth}
    \includegraphics[width=1.1\textwidth]{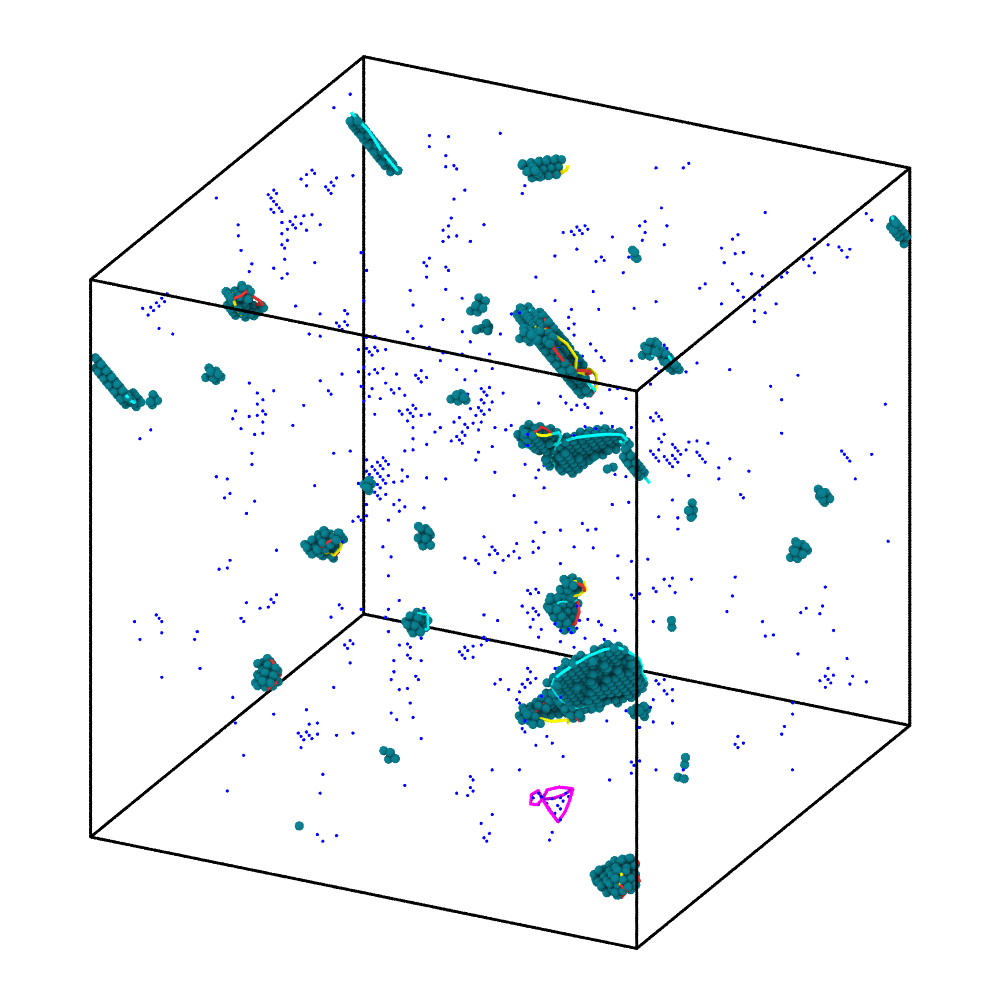}
    \caption{Rh EAM}\label{fig:FCC:conf:Rh}
\end{subfigure}
\begin{subfigure}{0.24\textwidth}
    \includegraphics[width=1.1\textwidth]{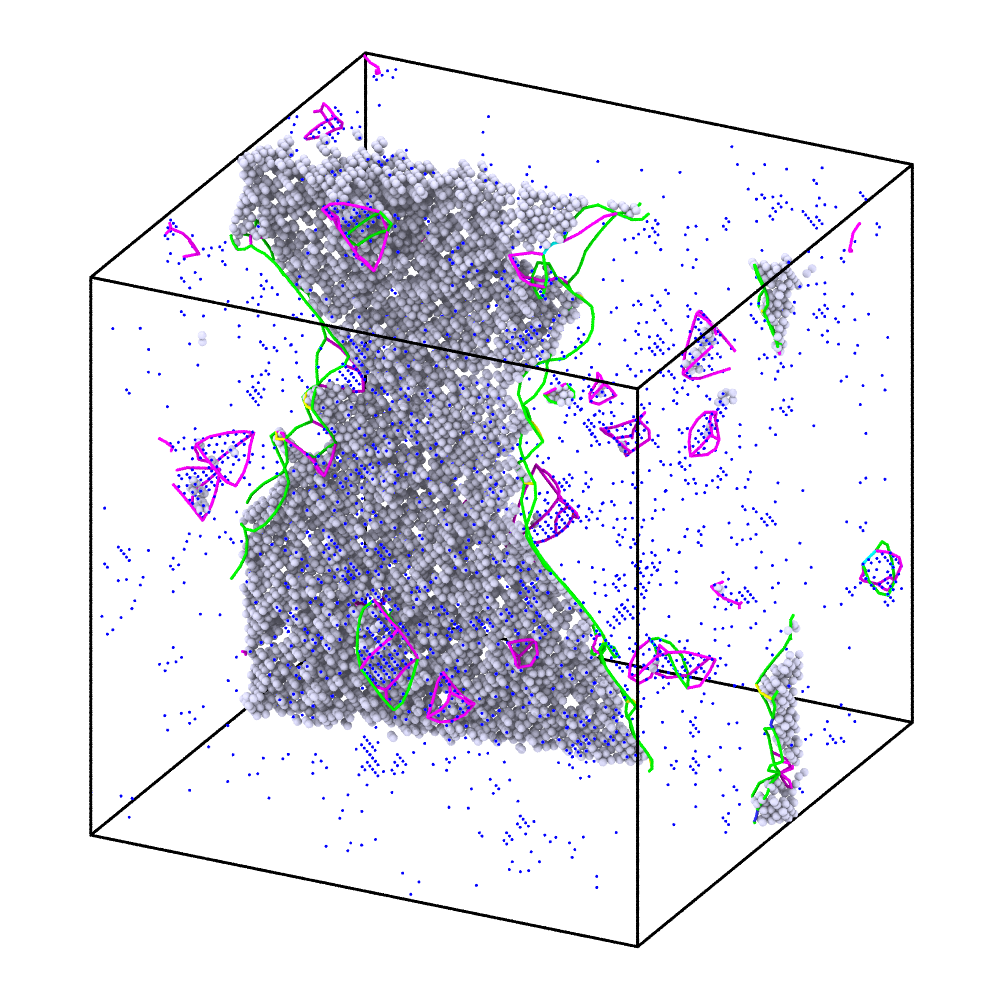}
    \caption{Ag EAM}
\end{subfigure}
    \caption{Final configuration after 2000 overlapping collision cascades. Small blue spheres represent vacancies, the other differently colored spheres represent interstitial atoms, green lines represent Shockley dislocations, blue lines represent perfect disclocations, and cyan lines represent Frank dislocations, yellow lines represent Hirth dislocations, pink lines represent stair-rod dislocations and red lines represent dislocations identified as ''other''.}\label{fig:ff:FCC}
\end{figure*}

Fig.~\ref{fig:ff:FCC} shows a selection of final defect structures visualized using OVITO~\cite{stukowski2012automated}. These particular structures were chosen as a representative overview of the different defect structures that can form in the FCC systems. Visualizations of each element and potential combinations can be found in the supplementary materials. We can see that the most typical dislocations formed are the Shockley partial chains and Frank-type dislocation loops. Another typical structure is stacking fault tetrahedra (SFT), that are present in all the elements in varying numbers. The Rh-EAM (Fig.~\ref{fig:FCC:conf:Rh}) is a clear outlier with regards to SFTs, where they are formed to a much lesser degree than in all the other potentials and elements. Naturally, in the systems with higher defect concentrations, large dislocation structures and numerous SFTs and other vacancy clusters are formed. 

\begin{table*}
 \caption{Cluster fraction and dislocation structures of final configurations of all FCC elements and potentials. CF$_{\mathrm{def.}}$ represents the fraction of defects of a given type that are in clusters, stacking fault tetrahedra (SFT), Perfect dislocations ($1/2$\hkl<110>), Shockley dislocations ($1/6$\hkl<112>), Stair-rod dislocations ($1/6$ \hkl<110>),  Hirth dislocations ($1/3$ \hkl<100>), Frank dislocations (1/3 \hkl<111>) and dislocations identified as ''other'' (other).}
 \label{tab:defects:fcc}
 \begin{threeparttable}
 \small
  \begin{tabular*}{\linewidth}{@{\extracolsep{\fill}} ll|c|c|c|c|c}
   \toprule
    \toprule
  \multicolumn{2}{c}{Element/Potential} & \multicolumn{3}{c}{Vacancy type defects}  &  \multicolumn{2}{c}{Interstitial type defects}\\
    \cmidrule(lr){3-5}
    \cmidrule(lr){6-7}
   \midrule
   & & CF$_{\mathrm{vac.}}$ & SFT &  Vacancy type dislocations & CF$_{\mathrm{int.}}$ & Interstitial type dislocations \\
   \midrule
    
    & Al-tabGAP & 0.449 & X &  & 0.987 & $1/6$\hkl<112> chains,  1/3 \hkl<111> loops  \\
     \midrule
    & Al-EAM & 0.644 & X &  & 0.978& $1/6$\hkl<112> chain, 1/3 \hkl<111> loops  \\
     \midrule
    & Al-Zhou & 0.832 & X &  & 0.960 & $1/6$\hkl<112> chains  \\
     \midrule
    & Ni-tabGAP & 0.510 & X &  & 0.986 &  $1/6$\hkl<112> chain, 1/3 \hkl<111> loops \\
    \midrule
    & Ni-EAM & 0.567 & X &  & 0.976 & $1/6$\hkl<112> chain, 1/3 \hkl<111> loop  \\
    \midrule
    & Ni-Zhou & 0.821 & X &  &0.987 & 1/3 \hkl<111> loops \\
    \midrule
    & Cu-tabGAP & 0.629 & X & & 0.977 & $1/6$\hkl<112> chain  \\
    \midrule
    & Cu-EAM &  0.677 & X&  & 0.983 & $1/6$\hkl<112> chain  \\
    \midrule
    & Cu-Zhou & 0.811 & X & & 0.983 & $1/6$\hkl<112> chain   \\
    \midrule
    & Rh-EAM & 0.489 & X& & 0.994 &  $1/3$ \hkl<100>--``other'' chain, 1/3 \hkl<111> loops  \\
    \midrule
    & Pd-EAM & 0.674 & X& &  0.985 & $1/6$\hkl<112> chain, 1/3 \hkl<111> loops  \\
    \midrule
    & Pd-Zhou & 0.685 & X&  & 0.985 &  $1/6$\hkl<112> chain \\ 
    \midrule
    & Ag-EAM & 0.647 & X&  & 0.985 &  $1/6$\hkl<112> chain,  1/3 \hkl<111> loop \\
    \midrule
    & Ag-Zhou &  0.699 & X&  & 0.980 & $1/6$\hkl<112> chain  \\
    \midrule
    & Pt-EAM & 0.538 & X&  & 0.990 &  $1/6$\hkl<112> chain  \\
    \midrule
    & Pt-Zhou & 0.672 & X&  & 0.975 & $1/6$\hkl<112> chain  \\
    \midrule
    & Au-EAM & 0.675 & X&  & 0.994 & $1/6$\hkl<112> chain  \\
    \midrule
    & Au-Zhou & 0.598 & X&  & 0.987 & $1/6$\hkl<112> chain, 1/3 \hkl<111> loop  \\
    \midrule
    & Pb-Zhou & 0.684 & X&  & 0.949 & $1/6$\hkl<112> chain  \\
    \bottomrule
        \bottomrule
  \end{tabular*}
  \begin{tablenotes}
  \end{tablenotes}
 \end{threeparttable}
\end{table*}

In Table~\ref{tab:defects:fcc} we show the cluster fraction of both vacancy- and interstitial-type defects as well as a description of the dislocations present in the final defect structures. The table only represents what is present in the final defect structures. The Burgers vectors listed gives the dominant types of dislocations present in the final structures based on a visual analysis of each run separately, therefore there is certain degree of subjectivity in the table that the reader has to keep in mind. The use of plural form in reference to dislocations loops or Shockley partial chains in the table indicates that there were typically several of these dislocation structures present in the final structures. Furthermore, in all the individual runs performed in the FCC elements SFTs were present (in various amounts), while they can be identified by the stair-rod dislocations they have been separated in to their own column, where the X represents that they are present. The purpose of the table is to give a brief description of the typical defect structures predicted by the potentials, for a more complete picture the final structures and the dislocation densities of each dislocation type as a function of the number of cascades are given in the supplementary materials for each element separately. Furthermore, since the table only describes the defect structure of the final state, it does not tell that if Frank loops (for example) are not listed there that they could not form or that they had not formed at an earlier stage and subsequently transformed into something else. From the table we can see that all dislocation loops are of interstitial type and no vacancy type loops could be found in the final configurations. All potentials produce SFTs in FCC elements. Additionally, we also identify that the Zhou potentials have a higher tendency for clustering of vacancies than the other potentials (except in Au). The Rh-EAM is also an interesting exception as it produces dislocation structures which DXA identifies as Hirth and ``other'' types of dislocation segments, which also can be seen in the final structure shown in Fig.~\ref{fig:FCC:conf:Rh}. 

Previous work on overlapping cascades in FCC materials has mainly focused on Ni and Ni-based alloys~\cite{levo2021temperature,ullah2016damage,fellman2026effects}. Comparing the Ni results from the current work with those found in the literature, we find that the defect structures are qualitatively similar. As a side note, the system sizes were smaller in the previous work for Ni, therefore in order to perform a direct comparison of the defect concentrations in Fig.~\ref{fig:FCC:dc} we would need to convert the results to a function of dose. This can be achieved via the use of a damage model such as the NRT-dpa equation~\cite{nordlund_primary_2018}, but this requires us to have an estimation of the threshold displacement energy of each potential separately, which is outside the scope of this work. We can make further qualitative comparisons to past results in Al~\cite{issaka2026cascade}, Cu~\cite{wu2018radiation, chen2020reduced,reali2025atomistic,10.1063/5.0314764}, Pt~\cite{barr2022irradiation} and Au~\cite{fu2012surface,vizoso2021size}, where overlapping cascade simulations and/or damage accumulation studies are available. Although, some of these are simulated on grain boundaries, nano-foams or nano-wires, we can still get some indication on types of defect structures to be expected in MD simulations of FCC metals. The general conclusion of this comparison is that there is a qualitative consistency in the types of defects that are seen in overlapping cascade simulations in the literature. However, quantitative comparisons are challenging due to different simulated structures, PKA energies, temperatures, system sizes and so forth.

\subsection{BCC metals}

\begin{figure*}[th]
    \centering
    \includegraphics[width=0.32\linewidth]{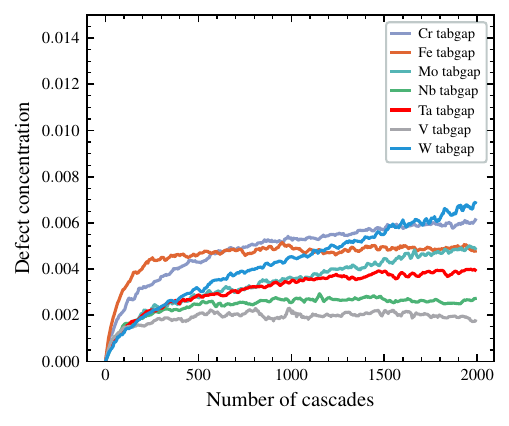}
    \includegraphics[width=0.32\linewidth]{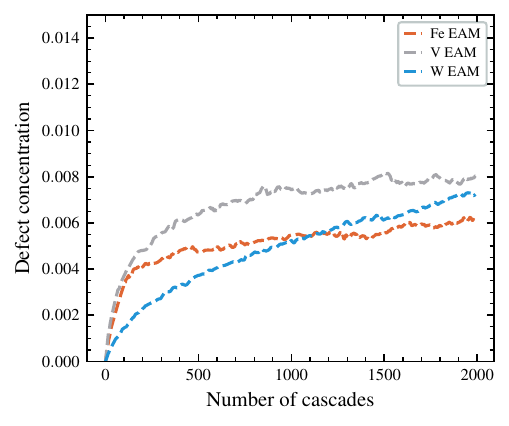}
    \includegraphics[width=0.32\linewidth]{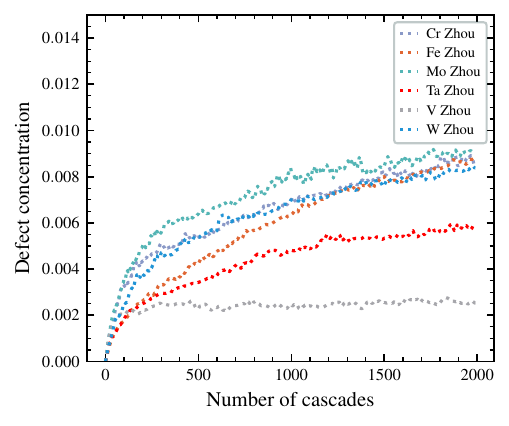}
    \caption{Defect  concentration (interstitials and vacancies) as a function of number of cascades. Left plot includes all BCC elements using tabGAP potentials, middle are the diverse different EAM potentials (see Table~\ref{tab:potentials_all}) and right plot includes overlapping cascades using the modified Zhou et al. potential. In the figures a rolling average is applied with a window of 20 cascades. The scale of the y-axis is kept the same in all subplots.}
    \label{fig:BCC:dc}
\end{figure*}

\begin{figure*}[th]
    \centering
\begin{subfigure}{0.24\textwidth}
    \includegraphics[width=1.1\textwidth]{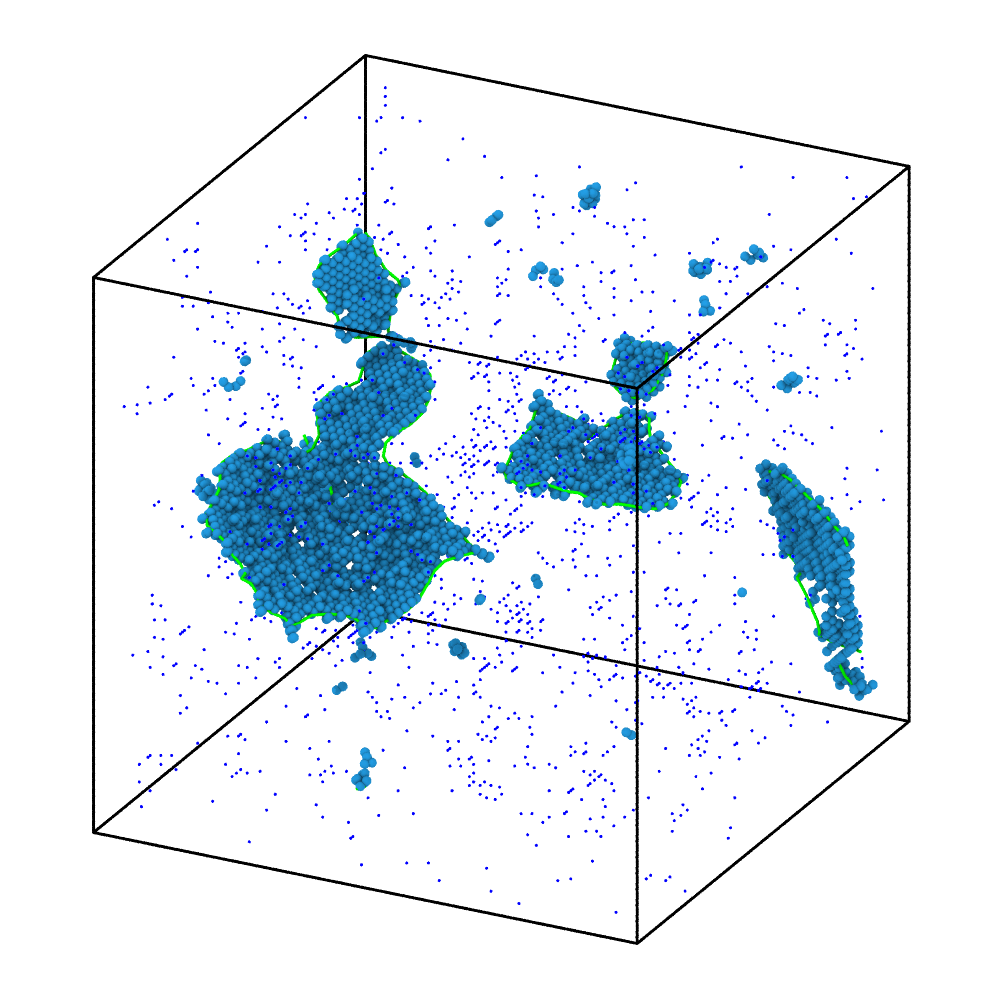}
      \caption{W tabGAP}
\end{subfigure}
\begin{subfigure}{0.24\textwidth}
    \includegraphics[width=1.1\textwidth]{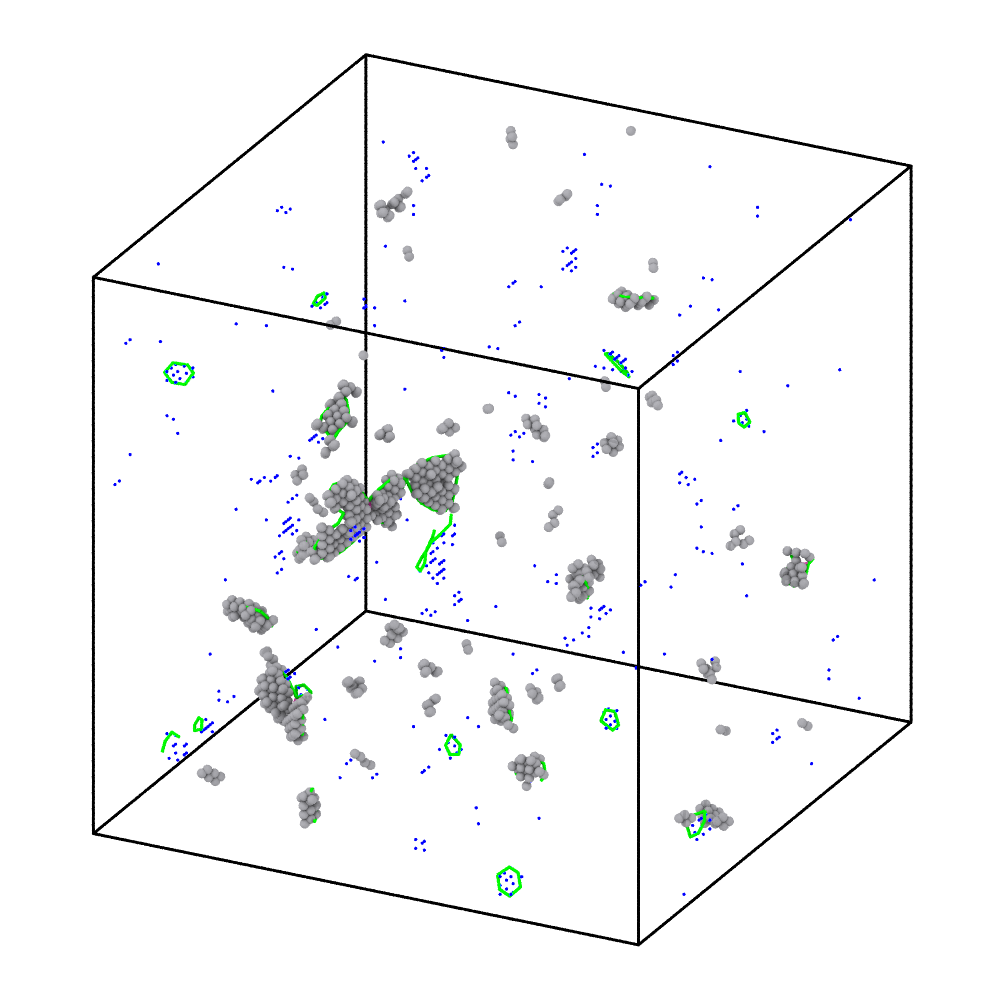}
    \caption{V tabGAP}
\end{subfigure}
\begin{subfigure}{0.24\textwidth}
    \includegraphics[width=1.1\textwidth]{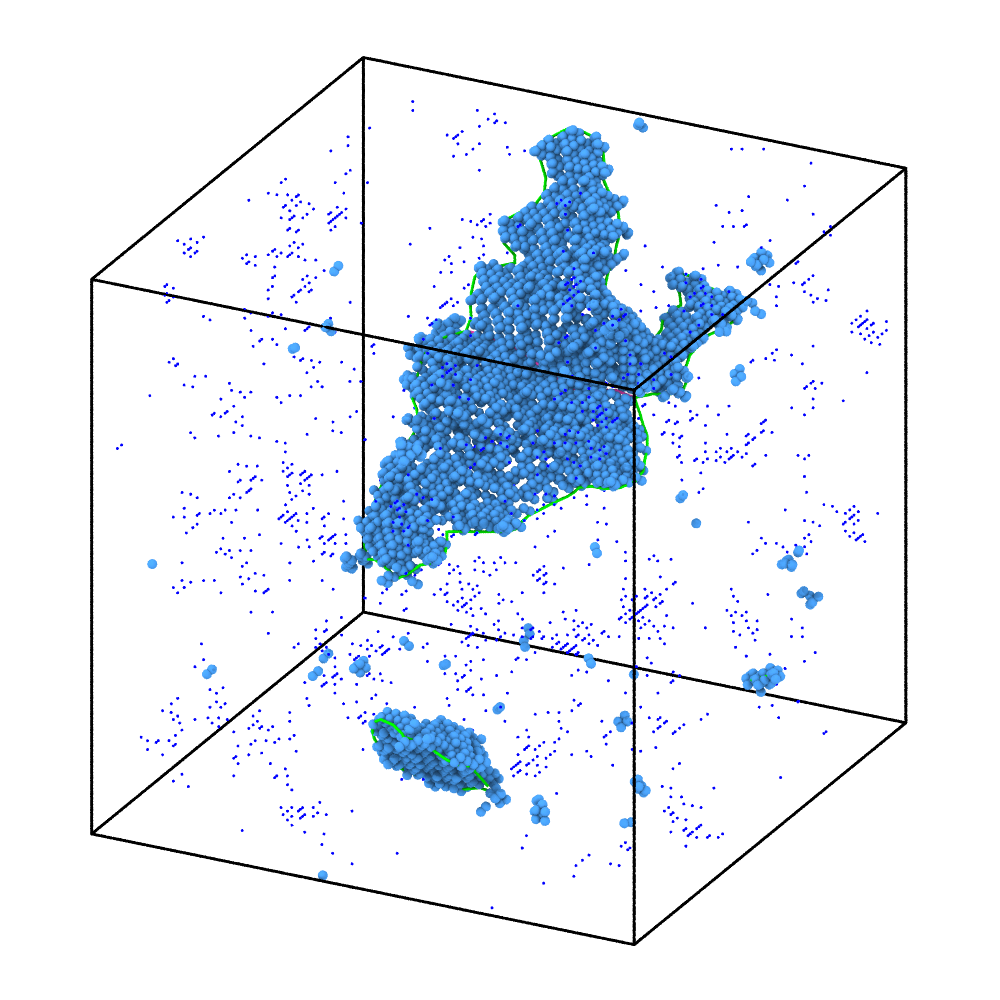}
    \caption{Ta Mod. Zhou EAM}
\end{subfigure}
\begin{subfigure}{0.24\textwidth}
    \includegraphics[width=1.1\textwidth]{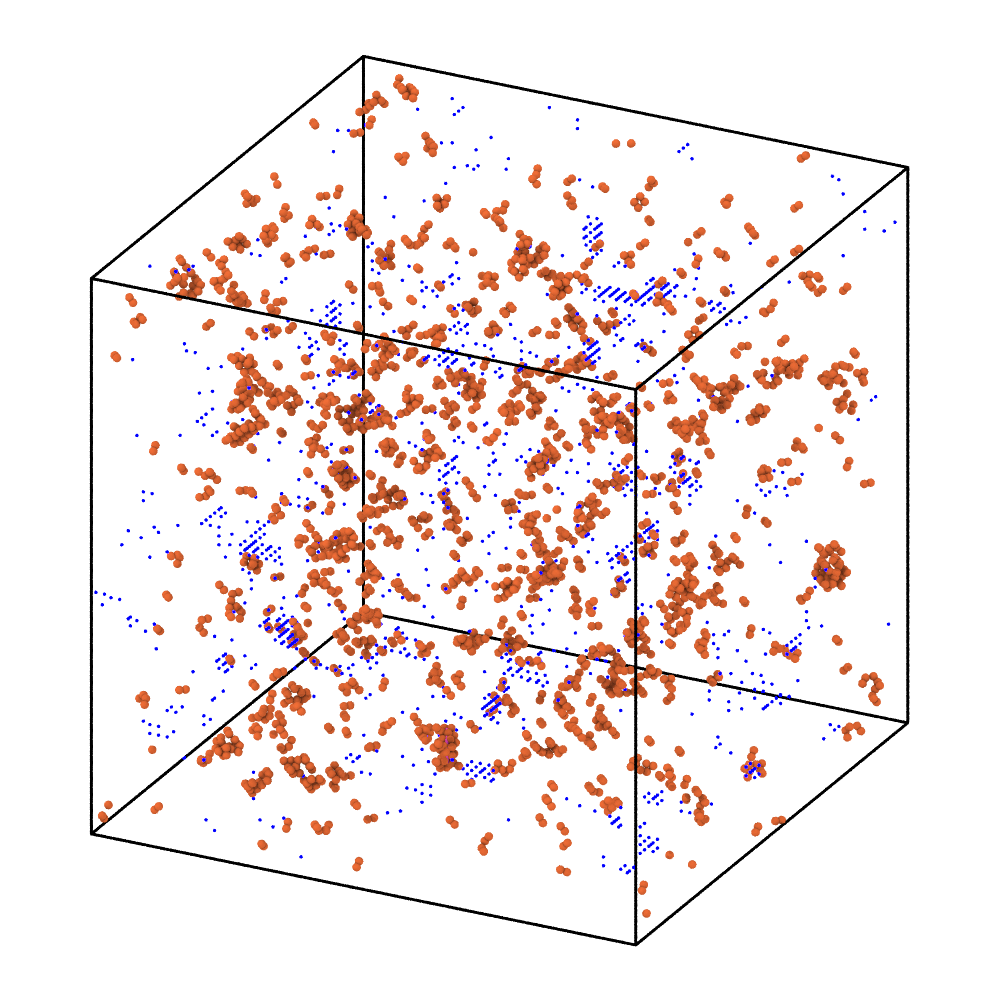}
    \caption{Fe tabGAP}
\end{subfigure}
\begin{subfigure}{0.24\textwidth}
    \includegraphics[width=1.1\textwidth]{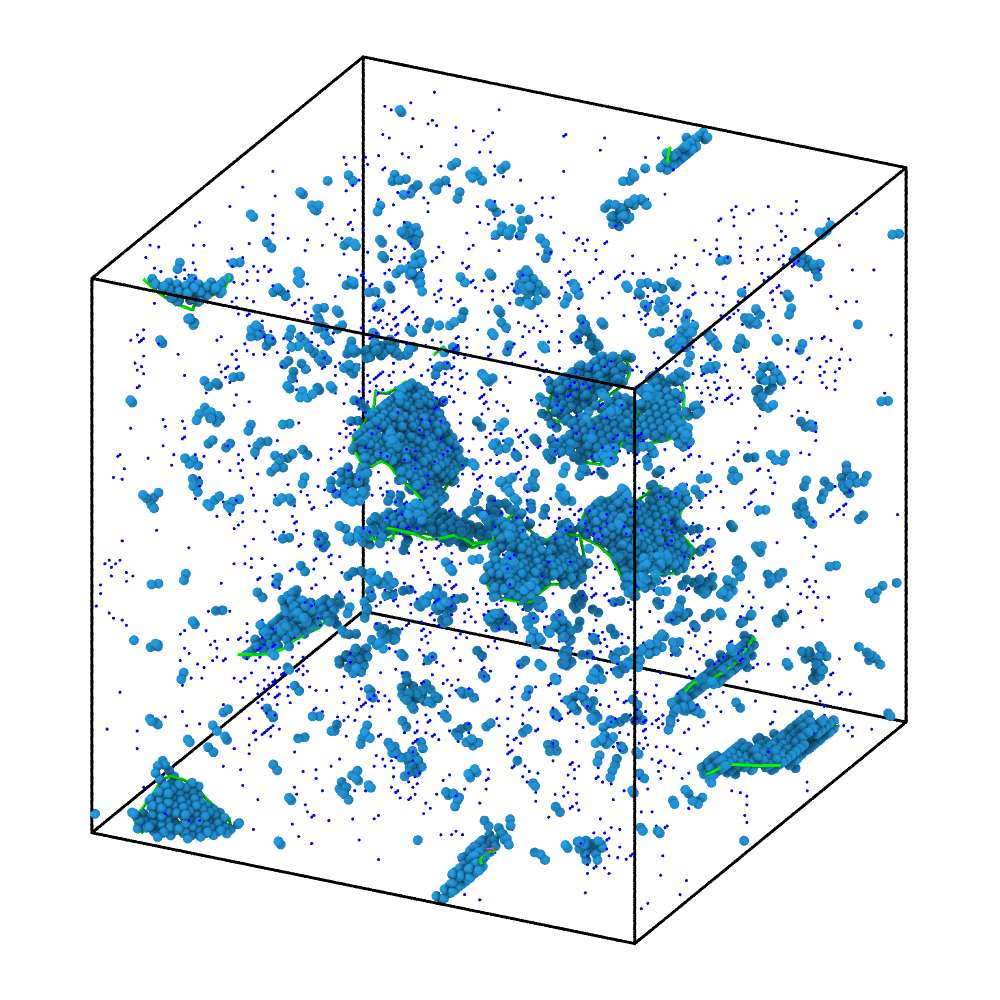}
    \caption{W Mod. Zhou EAM}
\end{subfigure}
\begin{subfigure}{0.24\textwidth}
    \includegraphics[width=1.1\textwidth]{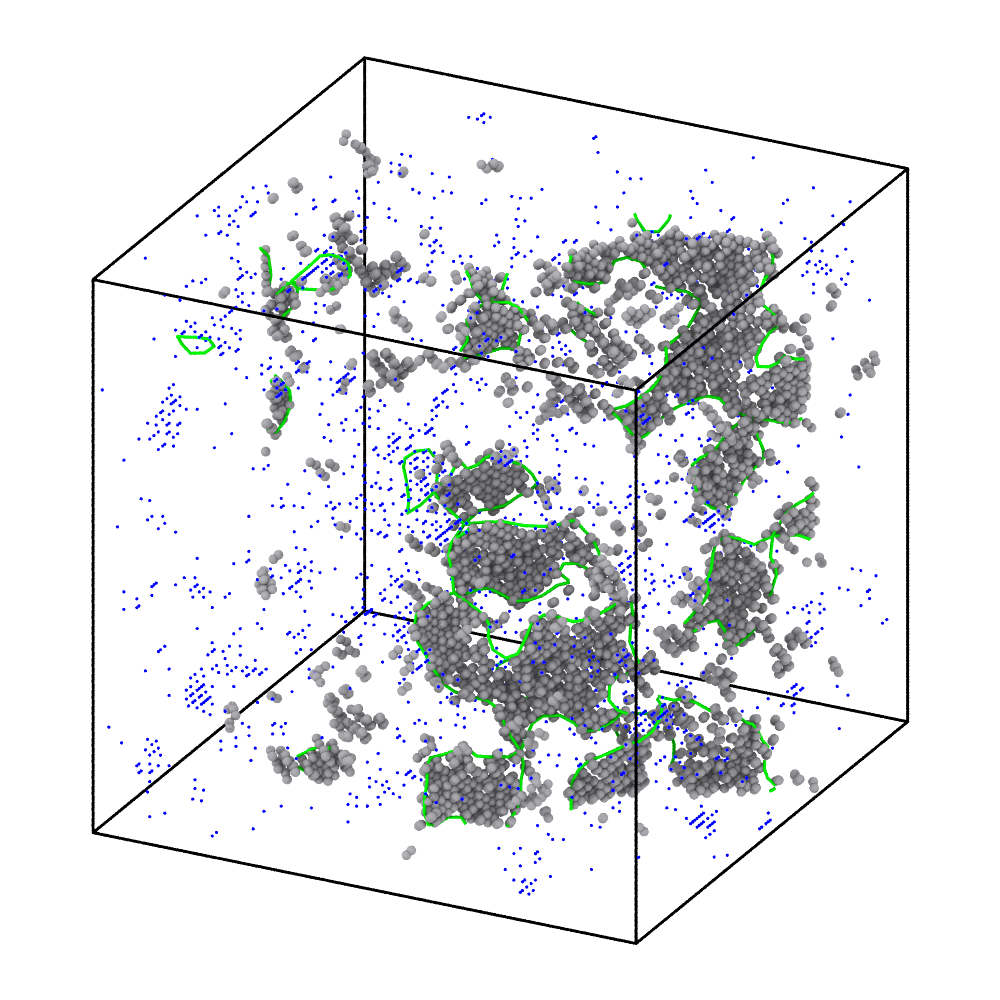}
    \caption{V EAM}
\end{subfigure}
\begin{subfigure}{0.24\textwidth}
    \includegraphics[width=1.1\textwidth]{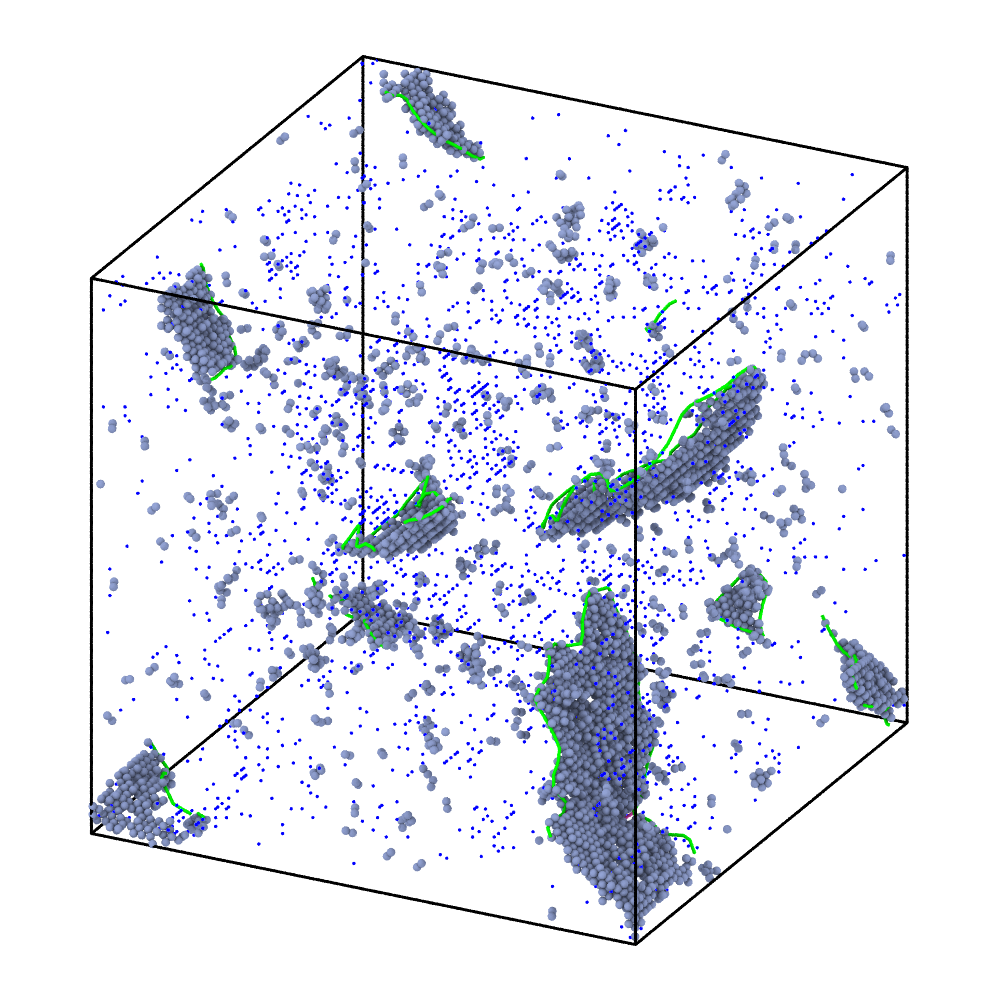}
    \caption{Cr Mod. Zhou EAM }
\end{subfigure}
\begin{subfigure}{0.24\textwidth}
    \includegraphics[width=1.1\textwidth]{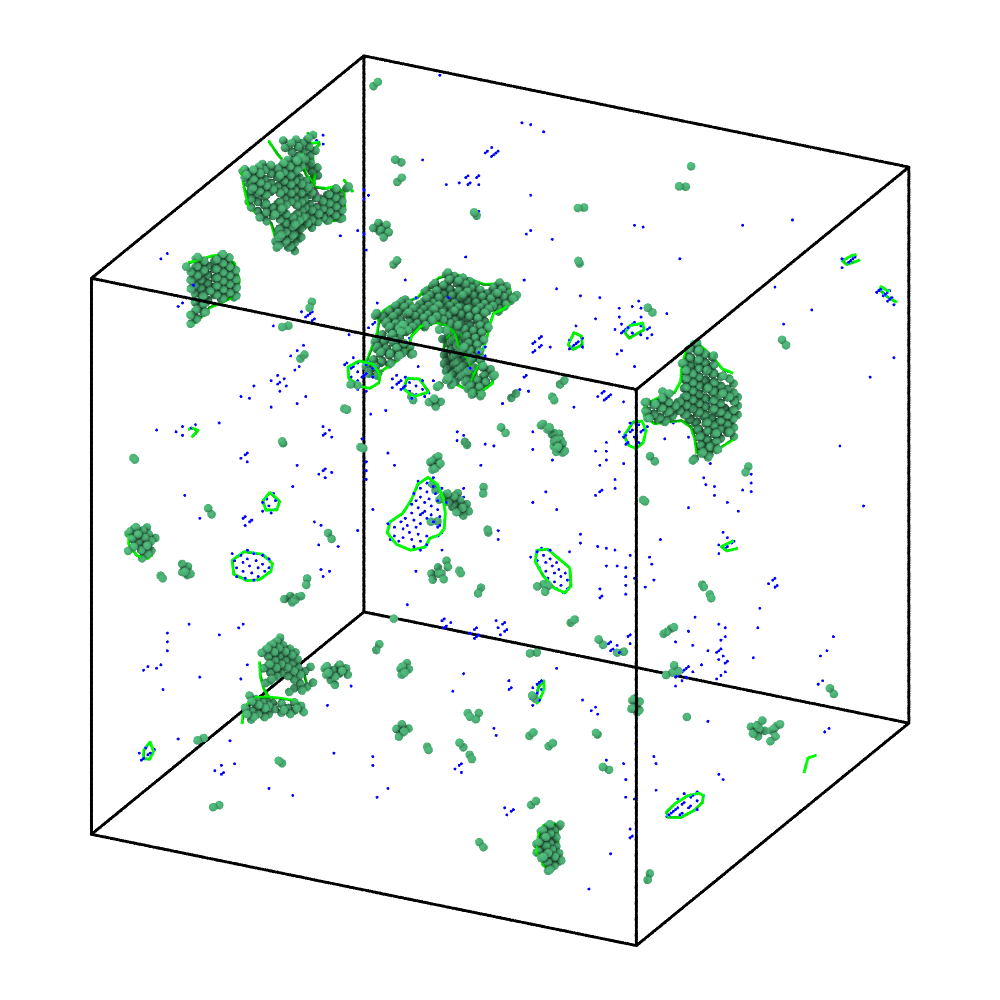}
    \caption{Nb tabGAP}
\end{subfigure}
    \caption{Final configuration after 2000 overlapping collision cascades. Small blue spheres represent vacancies, the other differently colored spheres represent interstitial atoms, green lines represent 1/2 \hkl<111> dislocations, and pink lines \hkl<100> dislocations.}\label{fig:BCC:fc}
\end{figure*}

\begin{table*}
 \caption{Description of the typical final defect structures of each potential in BCC elements. Following is the description of abbreviations used in the table: $N_\mathrm{ico.}^{\mathrm{avg.}}$ represents the average number (rounded up to nearest integer) of icosahedral atoms as identified by polyhedral template matching (PTM) as implemented in OVITO, CF$_{\mathrm{def.}}$ represents the fraction of defects of a given type that are in clusters, Dislocations: $1/2$\hkl<111>, \hkl<100>, \hkl<110>, and dislocations identified as ''other'' (other).}
 \label{tab:defects:bcc}
 \begin{threeparttable}
  \begin{tabular*}{\linewidth}{@{\extracolsep{\fill}} ll|c|c|c|c|c }
   \toprule
    \toprule
  \multicolumn{3}{c}{Element/Potential}  & \multicolumn{2}{c}{Vacancy type defects}  &  \multicolumn{2}{c}{Interstitial type defects}\\
    \cmidrule(lr){4-5}
    \cmidrule(lr){6-7}
   \midrule
   & & $N_\mathrm{ico.}^{\mathrm{avg.}}$ & CF$_{\mathrm{vac.}}$ & Vacancy type dislocations & CF$_{\mathrm{int.}}$ & Interstitial type dislocations \\
   \midrule
    & V-tabGAP & 0 & 0.708 & $1/2$\hkl<111> loops & 0.872 &  $1/2$\hkl<111> loops  \\
     \midrule
    & V-EAM & 32 & 0.545  & $1/2$\hkl<111> loops  & 0.903 &  $1/2$\hkl<111> loops  \\
     \midrule
    & V-Zhou & 1 & 0.682 & $1/2$\hkl<111> loops  & 0.807 &  $1/2$\hkl<111> loops  \\
    \midrule
    & Cr-tabGAP & 1 & 0.490 &  & 0.884 &  $1/2$\hkl<111> loops  \\
    \midrule
    & Cr-Zhou & 27 & 0.412 &  & 0.886 &  $1/2$\hkl<111> loops  \\
    \midrule
    & Fe-tabGAP & 238 & 0.665 & & 0.789 & $1/2$\hkl<111> loops~\footnote{only a couple of small dislocations in two of the runs.}  \\
    \midrule
    & Fe-EAM & 274 & 0.607 &  & 0.869 &  $1/2$\hkl<111> loops  \\
    \midrule
    & Fe-Zhou & 4 & 0.538 & & 0.968 &   $1/2$\hkl<111> loop  \\
    \midrule
    & Nb-tabGAP & 0 & 0.778 & $1/2$\hkl<111> loops & 0.908 &  $1/2$\hkl<111> loops  \\
    \midrule
    & Mo-tabGAP & 0 & 0.599 & & 0.931 &  $1/2$\hkl<111> loop  \\
    \midrule
    & Mo-Zhou & 143 & 0.594 &  & 0.886 &  $1/2$\hkl<111> loops  \\ 
    \midrule
    & Ta-tabGAP & 2 & 0.712 &  & 0.934 &  $1/2$\hkl<111> loops  \\
    \midrule
    & Ta-Zhou & 7 & 0.662 & & 0.952 &  $1/2$\hkl<111> loops  \\
    \midrule
    & W-tabGAP & 1 & 0.277 &  & 0.995 &  $1/2$\hkl<111> loops  \\
    \midrule
    & W-EAM & 11 & 0.358 &  & 0.982 &  $1/2$\hkl<111> loops  \\
    \midrule
    & W-Zhou & 107 & 0.530 &  & 0.883 &  $1/2$\hkl<111> loops  \\
    \bottomrule
        \bottomrule
  \end{tabular*}
  \begin{tablenotes}
  \end{tablenotes}
 \end{threeparttable}
\end{table*}

Fig.~\ref{fig:BCC:dc} shows the defect concentration for all the BCC elements. Note that the y-axis scale has been kept the same as for the FCC elements in Fig.~\ref{fig:FCC:dc}. Interesting here is that while in FCC the modified Zhou potentials showed consistently lower defect concentrations, in BCC they show similar or even somewhat higher defect concentrations than the other potentials. Another observation is that in tabGAP and modified Zhou, V shows the lowest defect concentration, while in the EAM case it is the highest. Comparing element-wise, the tabGAP potentials show consistently lower defect concentrations compared to both EAM and the modified Zhou potentials. Furthermore, the defect concentrations for W are quite similar for all the potentials. We also note that the Cr-tabGAP potential is not specifically made for simulation of elemental Cr and the amount of elemental Cr in its training data is quite limited, so its accuracy in simulating elemental Cr is uncertain.

Like for FCC, Fig.~\ref{fig:BCC:fc} shows a selection of final defect structures for the BCC metals. It is clear that irrespective of potential or element the 1/2 \hkl<111> dislocation loop is the dominant form of dislocation formed in the BCC materials. The Fe-tabGAP is an outlier as it resisted the formation of dislocations and tends to prefer the formation of C15 Laves-phase clusters. Furthermore, the figures can explain the high defect concentration of V-EAM by the formation of large interstitial type dislocations. Table~\ref{tab:defects:bcc} lists the typical defects structures of the BCC materials like in the case of FCC materials in Table~\ref{tab:defects:fcc}. The table highlights that in all potentials interstitial-type dislocations of various sizes are formed. Another interesting observation is that vacancy-type dislocations are only observed in V and Nb. 

Occasional $1/2$\hkl<111> loops are formed in the Fe-tabGAP, which can be identified during the simulations, but the lack of larger more stable dislocation loops is characteristic of the potential. This is likely due to the fact that C15 clusters are particularly stable in iron and the high stability of small C15 clusters and nonparallel precursor clusters are well reproduced by the Fe-tabGAP~\cite{byggmastar2022multiscale}. The C15 clusters can be identified by the presence of atoms with icosahedral coordination. Therefore, we can use the number of icosahedral atoms as a proxy for the prevalence of C15 clusters. Looking at these numbers we found that Fe-tabGAP and Fe-EAM potentials have the largest amount of icosahedral atoms. The Fe-Zhou potential has very few icosahedral atoms, because it does not correctly stabilise C15 clusters, whereas it predicts considerable amounts for Mo-Zhou and W-Zhou. In addition to these, the V-EAM, W-EAM and Cr-Zhou show a few icosahedral atoms. The other potentials show only single-digit numbers or none at all. However, one must keep in mind that in the cases with only a few icosahedral atoms they can be due to temporary fluctuations or displacements of atoms that are associated with dislocations and not C15 clusters. 

In BCC materials the vast majority of work on overlapping cascades has focused on W~\cite{granberg_molecular_2021,GRANBERG2023111902,PhysRevMaterials.5.095403,boleininger2023microstructure,PhysRevMaterials.5.095403,BRUNCRONA2026114820,lindblad2025effect,feichtmayer2024fast} and Fe~\cite{yang2008multiple,granberg_cascade_2017,byggmastar_effects_2018,granberg_defect_2020,lai2020synergistic,li2023comprehensive,mulewska2023self,BRUNCRONA2026114820}. The results here are largely consistent with the previous literature, mainly because we use the same well-established EAM potentials for W and Fe and W-tabGAP (while Fe-tabGAP has not been used in overlapping cascades before). The formation of large and continuously growing 1/2 \hkl<111> dislocations in single-crystal W is consistent well known from previous work. Similarly, the frequent formation of C15 clusters is consistent with previous work where potentials that correctly stabilise C15 clusters are used.

In addition to W and Fe, overlapping cascade simulations and/or damage accumulation studies have previously been performed in at least V~\cite{gayathri2022post}, Cr~\cite{yan2025study} and Nb~\cite{chen2019irradiation,chen2020reduced,vizoso2021size}. Unfortunately, the V simulations in the literature are so small in scale and do not report any dislocation information to make a meaningful comparison. However, in another study of primary cascades in V the formation of $1/2$\hkl<111> vacancy loops is reported~\cite{QIU2023101394}. Additionally, in that work the formation energies of dislocation loops are reported to be very close to that of voids in the Han \etal potential (same as V-EAM in this work). In the case of Cr, the formation of interstitial $1/2$\hkl<111> loops are reported after continued irradiation~\cite{yan2025study}. In Nb, both $1/2$\hkl<111> loops and \hkl<100> dislocations are reported~\cite{chen2019irradiation}, but the majority of loops are of $1/2$\hkl<111>-type, and \hkl<100> dislocations appear at higher doses in the form of dislocation segments connecting portions of a dislocation network mainly consisting of $1/2$\hkl<111> dislocation segments. We could not determine from the previous results if any of the $1/2$\hkl<111> loops in Nb were of vacancy nature, as is the case in the results presented in this work. In general however, the defect structures generated in previous simulations of overlapping cascades in BCC metals are for the most part consistent with the results presented in this work.

\subsection{HCP metals}

\begin{figure*}[t]
    \centering
    \includegraphics[width=0.32\linewidth]{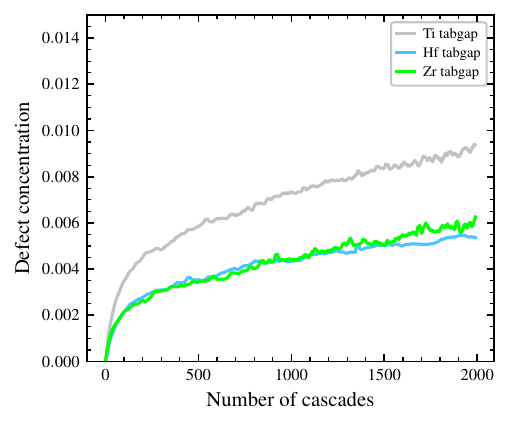}
    \includegraphics[width=0.32\linewidth]{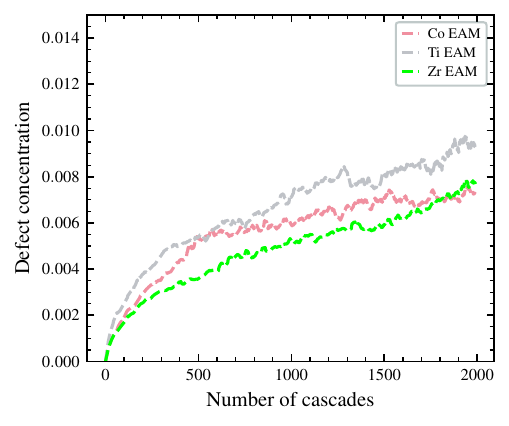}
    \includegraphics[width=0.32\linewidth]{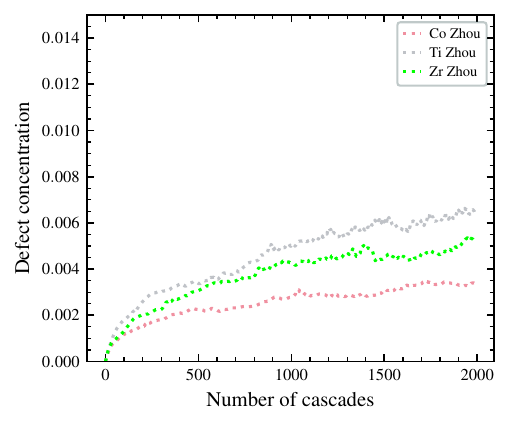}
    \caption{Defect concentration (interstitials and vacancies) as a function of number of cascades. Left plot includes all HCP elements using tabGAP potentials, middle are the diverse different EAM potentials (see Table~\ref{tab:potentials_all}) and right plot includes overlapping cascades using the modified Zhou et al. potential. In the figures a rolling average is applied with a window of 20 cascades. The scale of the y-axis is kept the same in all subplots.}
    \label{fig:HCP:dc}
\end{figure*}

\begin{figure*}
    \centering
\begin{subfigure}{0.24\textwidth}
    \includegraphics[width=1.1\textwidth]{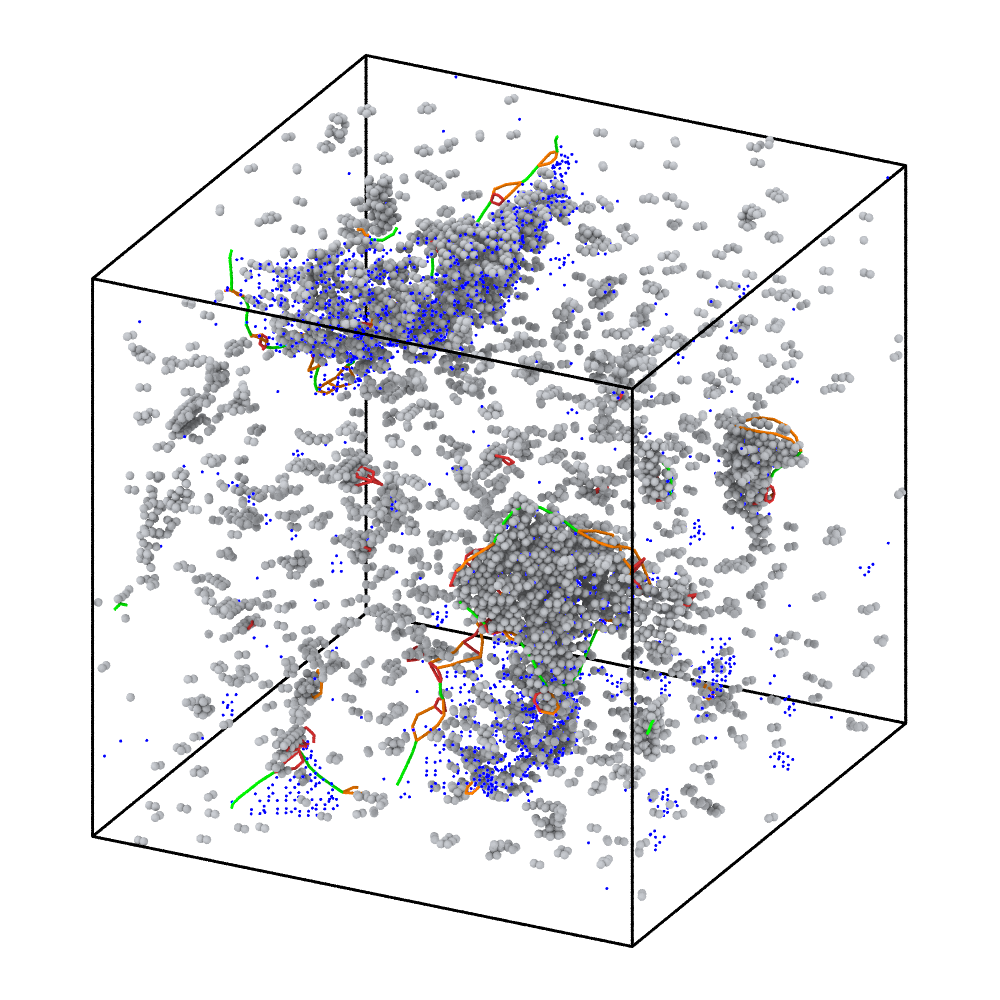}
      \caption{Ti tabGAP}
\end{subfigure}
\begin{subfigure}{0.24\textwidth}
    \includegraphics[width=1.1\textwidth]{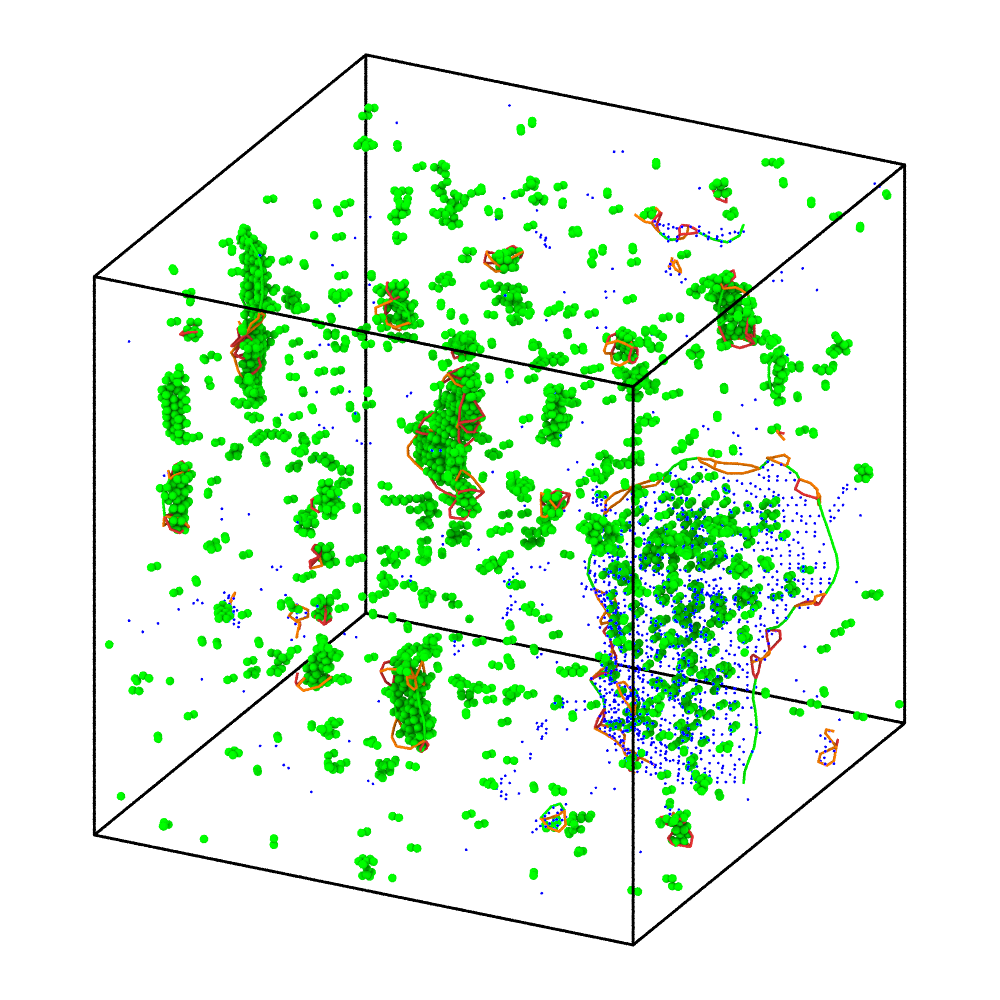}
    \caption{Zr tabGAP}
\end{subfigure}
\begin{subfigure}{0.24\textwidth}
    \includegraphics[width=1.1\textwidth]{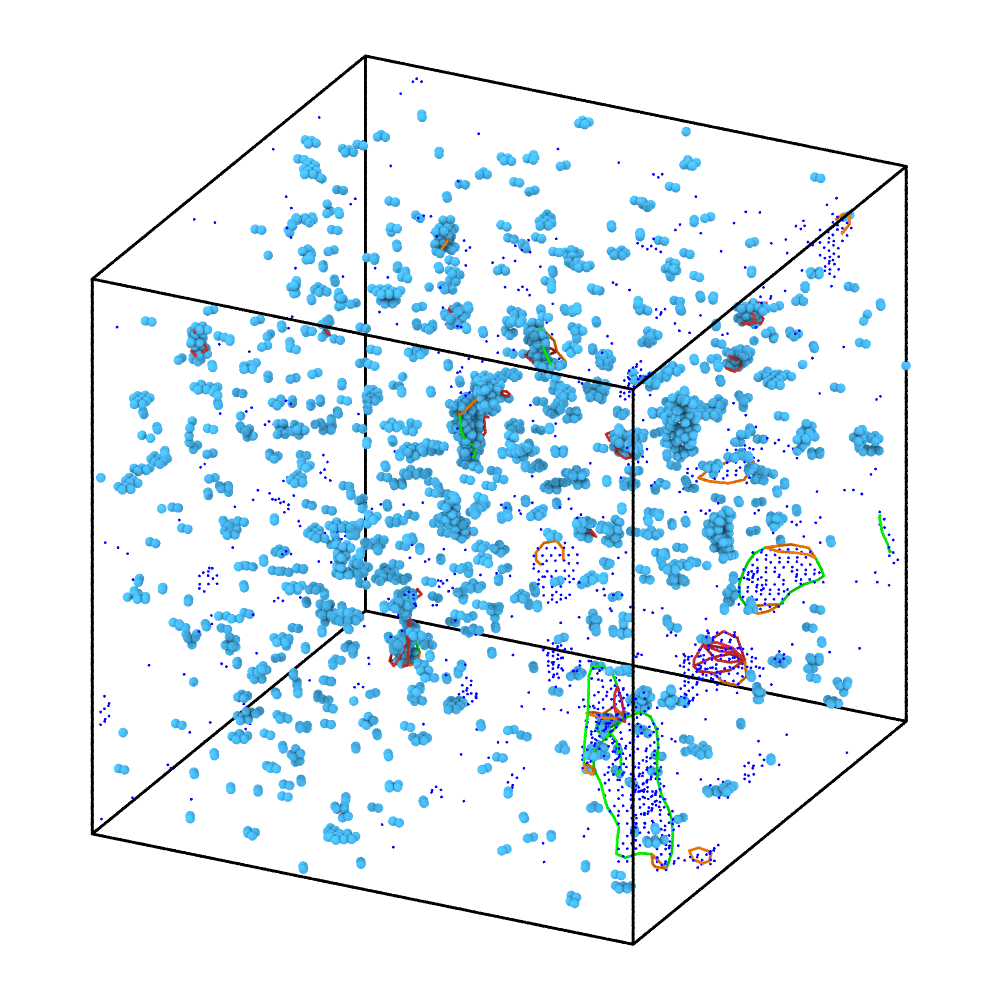}
    \caption{Hf tabGAP}
\end{subfigure}
\begin{subfigure}{0.24\textwidth}
    \includegraphics[width=1.1\textwidth]{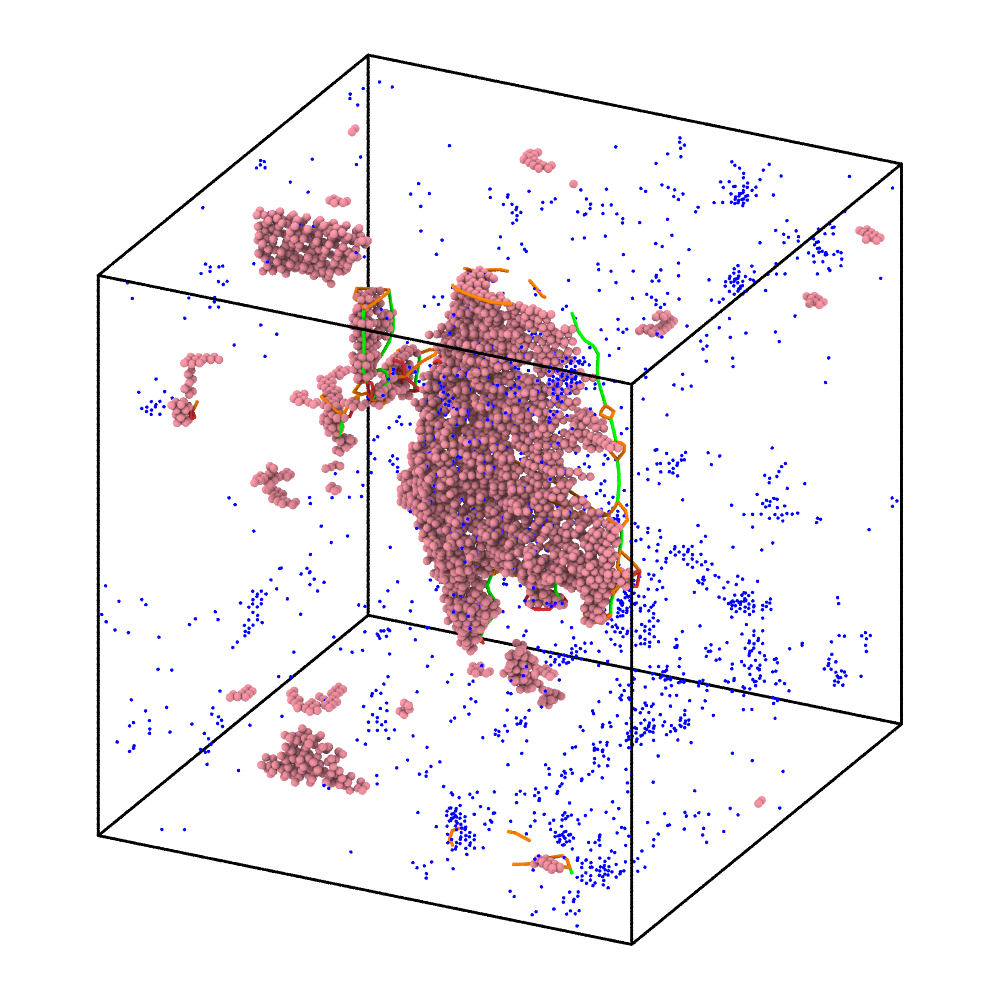}
    \caption{Co EAM}
\end{subfigure}
\begin{subfigure}{0.24\textwidth}
    \includegraphics[width=1.1\textwidth]{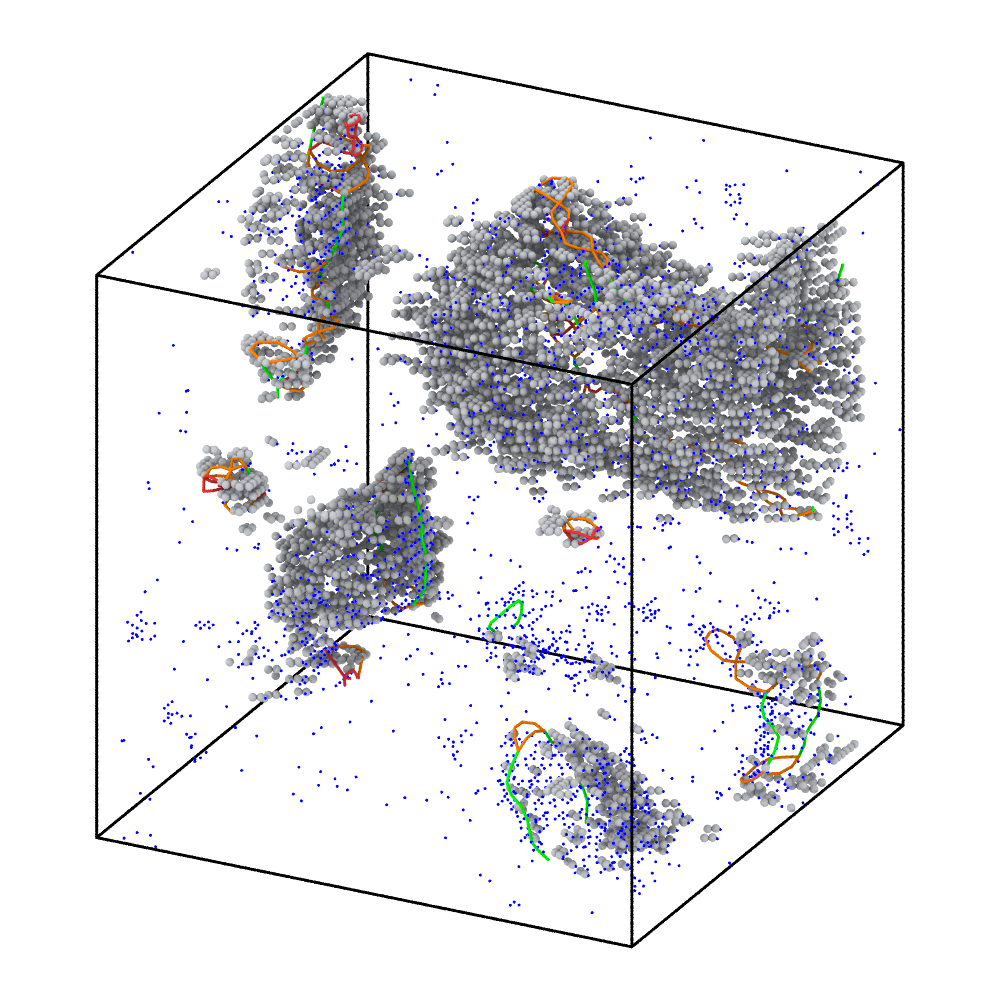}
    \caption{Ti Mod. Zhou EAM}
\end{subfigure}
\begin{subfigure}{0.24\textwidth}
    \includegraphics[width=1.1\textwidth]{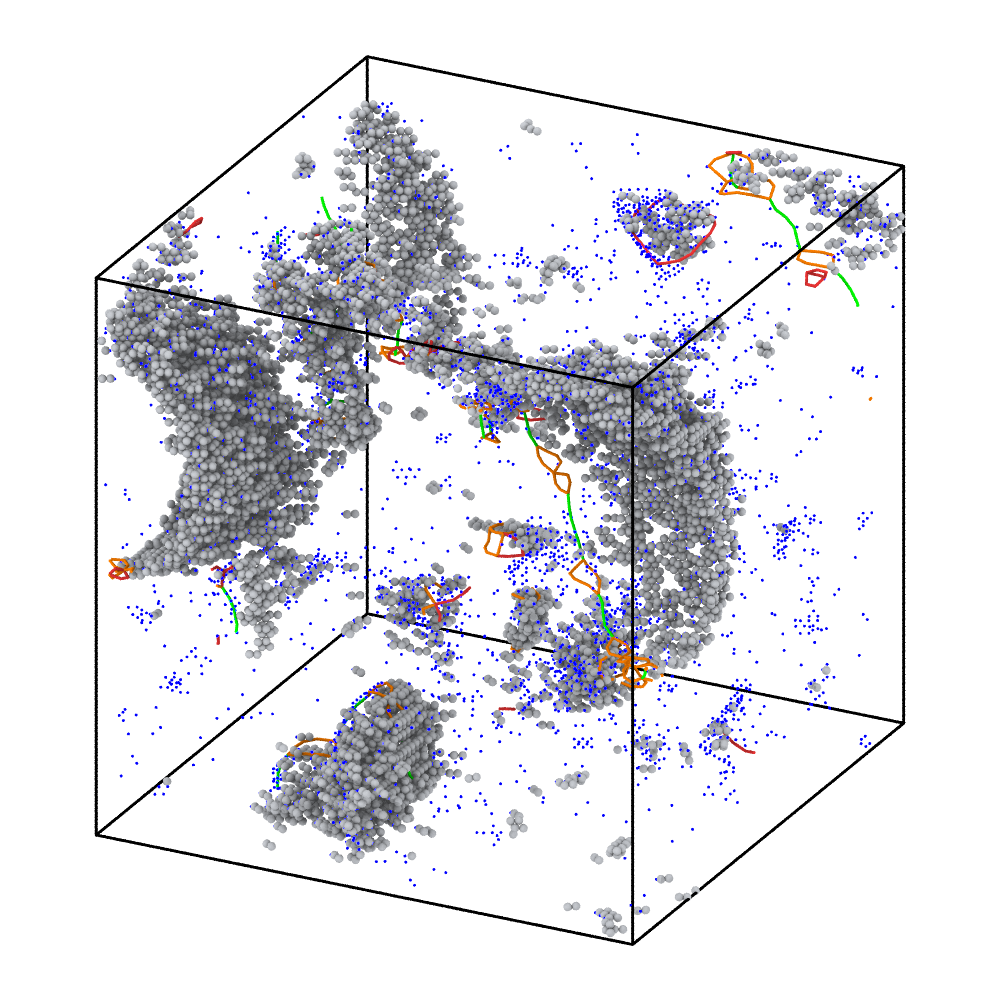}
    \caption{Ti EAM}\label{fig:HCP:fc:f}
\end{subfigure}
\begin{subfigure}{0.24\textwidth}
    \includegraphics[width=1.1\textwidth]{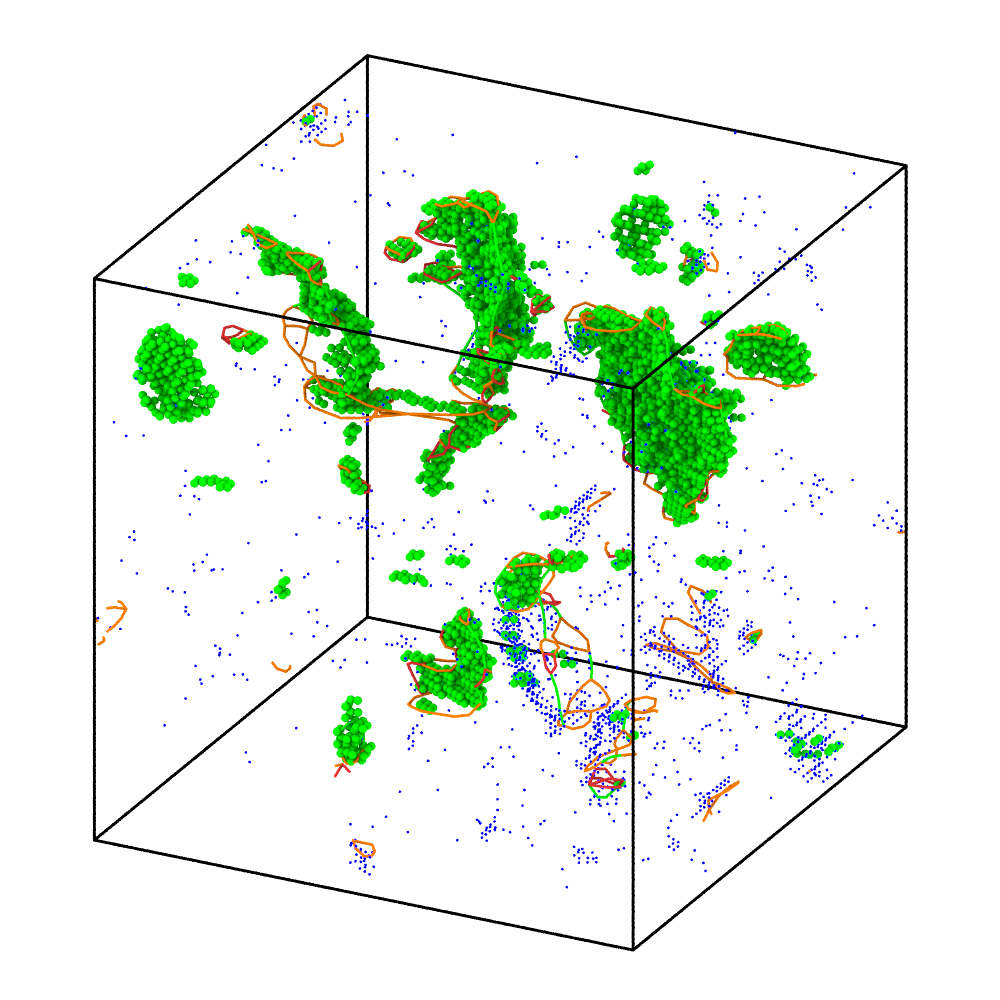}
    \caption{Zr Mod. Zhou EAM }
\end{subfigure}
\begin{subfigure}{0.24\textwidth}
    \includegraphics[width=1.1\textwidth]{figures/frames/frame-Co-zhou-run-2-2000.png}
    \caption{Co Mod. Zhou EAM}
\end{subfigure}
    \caption{Final configuration after 2000 overlapping collision cascades. Small blue spheres represent vacancies, the other differently colored spheres represent interstitial atoms, green lines are $1/3$ \hkl<1-210> dislocations, blue lines are \hkl<0001> dislocations, magenta lines are \hkl<1-100> dislocations, orange lines are $1/3$\hkl<1-100> dislocations, yellow lines are $1/3$ \hkl<1-213> dislocations and red lines are dislocations with type ''other'' as identified by DXA analysis.} \label{fig:HCP:fc}
\end{figure*}

\begin{table*}
 \caption{Description of the typical final defect structures of each potential in HCP elements. Following is the description of abbreviations used in the table: CF$_{\mathrm{def.}}$ represents the fraction of defects of a given type that are in clusters, Dislocations: $a$-type and $c$-type.}
 \label{tab:defects:hcp}
 \begin{threeparttable}
  \begin{tabular*}{\linewidth}{@{\extracolsep{\fill}} ll|c|c|c|c}
   \toprule
    \toprule
  \multicolumn{2}{c}{Element/Potential} & \multicolumn{2}{c}{Vacancy type defects}  &  \multicolumn{2}{c}{Interstitial type defects}\\
    \cmidrule(lr){3-4}
    \cmidrule(lr){5-6}
   \midrule
   & & CF$_{\mathrm{vac.}}$  & Vacancy type dislocations & CF$_{\mathrm{int.}}$ & Interstitial type dislocations \\
   \midrule
    & Ti-tabGAP & 0.856 & $a$-type & 0.801 &  $a$-type  \\ 
     \midrule
    & Ti-EAM & 0.651 &  $a$-type, $c$-type & 0.885 &  $a$-type \\
     \midrule
    & Ti-Zhou & 0.635 &  $a$-type  & 0.886 &  $a$-type  \\
     \midrule
    & Co-EAM & 0.598 &  & 0.983 &   $a$-type  \\
    \midrule
    & Co-Zhou & 0.681 & $a$-type   & 0.987 &   $a$-type\\
    \midrule
    & Zr-tabGAP & 0.841 & $a$-type & 0.801 &  $a$-type  \\
    \midrule
    & Zr-EAM & 0.676 & $a$-type, $c$-type  & 0.920 & $a$-type  \\
    \midrule
    & Zr-Zhou & 0.698 & $a$-type  & 0.967 &  $a$-type  \\
    \midrule
    & Hf-tabGAP & 0.761 & $a$-type    & 0.642 & $a$-type  \\ 
    \bottomrule
        \bottomrule
  \end{tabular*}
  \begin{tablenotes}
  \end{tablenotes}
 \end{threeparttable}
\end{table*}

Fig.~\ref{fig:HCP:dc} shows the defect concentration in the HCP metals. In this case the different potentials are in more agreement with each other than in either FCC or BCC, however there are a lot fewer HCP elements simulated, therefore the agreement can very well be coincidental. Like in the BCC metals, the modified Zhou potentials predict somewhat lower defect concentrations compared to the other potentials. Another observation is the fact that Ti has the highest defect concentration in all the different potentials. Furthermore, none of the potentials have reached a steady-state defect concentration after 2000 cascades. 

HCP materials form generally $a$-type dislocations which consist of Burgers vectors $1/3$\hkl<1-100> and $1/3$\hkl<1-210>. We also find rather odd Burgers vectors (e.g. $1/18$\hkl<-2-463> and $1/18$\hkl<-4043>) that the DXA analysis groups in the type ''other'', these odd Burgers vectors have been reported in literature~\cite{chen2019irradiation,warwick2021microstructural,tian2021heavy}. Chen \etal even states that the Burgers vectors could be classed as $c$-component type dislocations~\cite{chen2019irradiation}. Fig.~\ref{fig:HCP:fc} shows a selection of final configurations of the HCP materials. Looking at the defect structures we can see the source of the challenges of identifying point defects in HCP materials, which is clearest in the case of Ti, where we see that the defects associated with the dislocations are seen to be bulging (looking like a cloud) thus resulting in a lot of falsely identified point defects. This is the reason why the cluster correction is applied as it can be used to identify these false defects and remove them from the count of the total number of defects. Note that in all the visualizations all the defects identified by the WS analysis are shown. 

The dislocations that we observed in our HCP simulations were mostly interstitial $1/3$\hkl<1-100> and type ''other'' chains linked with $1/3$\hkl<1-210> dislocations. Furthermore, we can also observe vacancy-type loops. We also observe small ill-defined clusters (both vacancy and interstitial-type) that consist of several dislocation segments of type ''other'' in the different HCP materials. In one of the runs in Ti-EAM we observe a well-defined vacancy $c$-type loop with Burgers vector 1/2\hkl<000-1> and a dislocation line length of 117 Å. This loop can be seen in Fig.~\ref{fig:HCP:fc:f}, but a more clear illustration can be found in the Supplementary materials after the structure has been conjugate gradient relaxed. After the relaxation a part of the loop dissociated into segments. Furthermore, in Ti-EAM we also see large dislocations that span the entire simulation cell and consist of 1/3\hkl<1-100> partials. In Zr-EAM we see in one of the runs a $c$-type cluster consisting of 1/2\hkl<000-1> and 1/6\hkl<0-22-3> dislocation segments. After conjugate gradient relaxation the cluster dissociates, with only two short 1/6\hkl<0-22-3> segments with the rest becoming Burgers vectors similar to $1/18$\hkl<-2-463> and 1/9\hkl<-12-10> etc. Another striking observation is the fact that the Hf-tabGAP potential seems to resist the formation of large interstitial type loops and seems to form larger vacancy loops. Conversely, the Co-EAM formed exclusively interstitial-type dislocations, while the vacancies (although somewhat clustered) resisted the formation of dislocations. However, we again remind that we are analysing mainly the final defect structures.

Table~\ref{tab:defects:hcp} lists the typical defect structures that are formed in the HCP metals. Here we only distinguish between $a$-type and $c$-type dislocations as the dislocations seen in Fig.~\ref{fig:HCP:fc} are often quite complicated in nature and a consistent classification is otherwise challenging. Additionally, a potential is classified as having $c$-type loops if a distinct loop is formed, we do not include the small clusters that form complex balls of ''other'' segments to could according to Chen \etal be classed as having $c$-component-type. From the table we can see that all the potentials formed $a$-type interstitial loops. Additionally, we only observe $c$-type vacancy loops in Ti-EAM and Zr-EAM that we could identify unequivocally. Although in both cases only singular instances were identified and were not a common feature of the microstructure. We also report the cluster fractions as was done for the FCC and BCC metals, and we can notice that the tabGAP potentials show the strongest clustering of vacancies and conversely the lowest clustering of interstitials, this being for Ti and Zr, where such comparisons can be made due to limited amount of potentials. The Co-EAM which was the only one that did not form any vacancy-type dislocations and also had the lowest vacancy cluster fraction.

Like in the case of FCC and BCC metals, the vast majority of previous simulation literature has focused on a select few elements. For HCP metals the vast majority of work in overlapping cascade and damage accumulation studies has been done in Zr~\cite{chen2019irradiation,warwick2021microstructural,tian2021heavy,tian2021radiation,wang2022interaction,barzdajn2024molecular,li2025molecular,yan2025study}, but simulations exist also for Ti~\cite{hirst2022revealing}. Again, we can make a general statement that the dislocation structures that we observe are similar in nature to those previously reported in literature.

\subsection{Beryllium}

\begin{figure}[ht!]
    \centering
    \includegraphics[width=\linewidth]{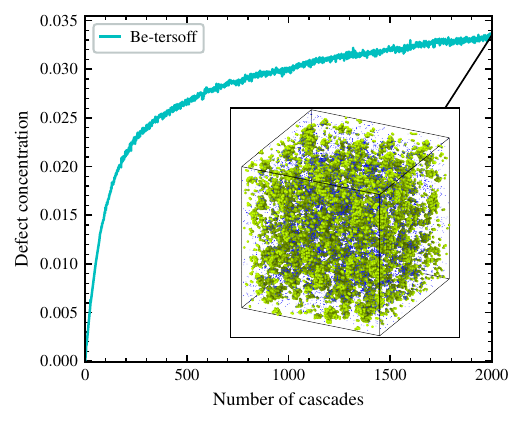}
    \caption{Defect concentration (interstitials and vacancies) as a function of number of cascades in beryllium. The inset image shows the final configuration after 2000 overlapping collision cascades. Blue spheres represent vacancies, the larger green
spheres represent interstitial atoms.}\label{fig:Be:dc}
\end{figure}

We have separated beryllium from the other HCP results as it was clearly distinct from all the other elements simulated. It is also by far the lightest element that was simulated. Fig.~\ref{fig:Be:dc} shows the defect concentration in Be. Here the striking observation is that the defect concentration is significantly higher than in all the other simulated materials. The defect concentrations is about double of the Al-EAM results which shows the second-highest defect concentration. Unsurprisingly also the volume change is significantly higher than the other elements. Also shown in Fig.~\ref{fig:Be:dc} is the final defect structure of one of the runs of the Be overlapping cascades. There we can see that entire simulation cell is distributed with defects, but no dislocations are seen. The defect structure looks like a partial amorphization. Neutron irradiation experiments observe the formation of numerous interstitial dislocations, bubbles and the formation of complex dislocation networks~\cite{CHAKIN2002657,wang2025review}. Furthermore, both $a$-type and $c$-type loops are expected in ion irradiated beryllium~\cite{liu2019microstructure}. All of these experimentally observed features are missing from our simulation results. Additionally, looking at previous studied molecular dynamics collision cascades in beryllium~\cite{borodin2016damage}, there is no indication of the strong disordering that we observe. Hence, our conclusion is that either there is something in the simulation methodology that causes issues in the case of beryllium or, more likely, there are problems with the interatomic potential used.

\section{Discussion}

\subsection{Comparison to experiments}

There exists a vast body of literature on the experimental characterization of radiation damage in elemental metals under various conditions. While we do not attempt here a quantitative comparison with any specific experiment or experimental condition, we will discuss in general terms how the microstructures we observe compare with the overall experimental literature.

Starting with the FCC metals, FCC materials are known to form both faulted loops (also called Frank loops) as well as perfect loops~\cite{zinkle20121} during irradiation. In our MD simulations we do not really see pure perfect loops, but they are dissociated into the Shockley partial chains, which are energetically more favorable~\cite{wu2010ab}. The dislocation analysis tool DXA in OVITO has an option to generate perfect dislocations, in which case these Shockley partial chains appear as perfect dislocations in the identification. Another typical feature of FCC metals is the formation of stacking fault tetrahedra~\cite{Eyre_1973,schaublin2005irradiation}. Experimental literature on the observations of SFTs exists in all the simulated FCC metal except Rh (Al~\cite{PhysRevB.69.094108}, \{Ni, Cu, Ag, Au\}~\cite{Kojima01031989}, Pt~\cite{10.1063/1.5016822}, Pd~\cite{schaublin2005irradiation} and Pb~\cite{Urban01121974}). In general, experimental literature of characterization of radiation damage in elemental Rh is scarce. Therefore, we are not sure whether our observation of SFTs in Rh is realistic or not. It is sometimes argued that a high stacking fault energy would make the formation of SFTs less favourable~\cite{PhysRevB.69.094108}. While this is not a complete explanation (take Ni for example where we see lots of SFTs), it might partially explain why Al and Pt eluded experimental observations of SFTs in the past~\cite{Eyre_1973,schaublin2005irradiation,https://doi.org/10.1002/pssb.19670210212}. Given that Al, Pt and Rh have high stacking fault energies~\cite{10.1063/1.4870462}, the fact that they have relatively few SFTs in our simulation results is sensible. Furthermore, in the literature there is mention that the SFTs are observed in significant amounts in Ni, Cu, Ag and Au~\cite{ZINKLE1995123, schaublin2005irradiation,kiritani1997story}, and our results seem to also reflect this to some degree, keeping in mind that there can be significant difference in the amount of overall defects produced by the different potentials. At medium temperatures (between defect recovery stage III and stage V) it is expected that vacancies in FCC metals are associated with SFTs~\cite{zinkle20121} and as the PKA energy in the current study is low enough that vacancy loops are not directly formed due to cascade collapse it is not surprising that the dislocations we observe in the FCC metals are of interstitial types. 

In BCC materials the typical dislocations are 1/2\hkl<111> and \hkl<100>. Experimentally, we expect the majority of the dislocation to be of 1/2\hkl<111> type~\cite{phythian1988irradiation,yi_situ_2013,hu2022recent}, which is also what we observe. In the case of C15, their small size makes them challenging to observe experimentally, but based on DFT calculations at least in iron the C15 clusters should be energetically favorable~\cite{marinica_irradiation-induced_2012}. Iron is also a unique BCC metal due to magnetism. Low-temperature ferromagnetic Fe favours nonparallel interstitial clusters (like C15 and its precursors~\cite{malerba_comparison_2010,marinica_irradiation-induced_2012}) at small cluster sizes and 1/2\hkl<111> loops at larger sizes. At higher temperatures, \hkl<100> loops become energetically more stable than 1/2\hkl<111> loops~\cite{dudarev_effect_2008} and are consequently frequently observed in experiments~\cite{masters_dislocation_1965,yao_temperature_2010}. However, none of the potentials used here can explicitly reproduce the magnetic and temperature-dependent effects in Fe, hence we mainly observe C15 clusters and 1/2\hkl<111> loops. The latter consistent with room and low-temperature experiments~\cite{yao_temperature_2010} while C15 clusters are yet to be experimentally reported.

HCP materials can form both $a$-type and $c$-type dislocations. One characteristic feature of HCP is the different dislocation evolution on the basal and prism planes~\cite{zinkle20121}. Experimentally, this can result in anisotropic growth in the different directions. It has been reported that in pure HCP metals the majority of the observed dislocation loops are of $a$-type~\cite{choi2013radiation,griffiths1993evolution} and at lower temperatures interstitial loops are dominant~\cite{choi2013radiation}. $c$-type loops are also observed, but typically at higher fluences and generally a high $a$-type loop density is needed before they are observed~\cite{griffiths1993evolution}. Contrasting against this, in our simulations we have not formed saturated networks of $a$-type dislocations and given that experimental observations of $c$-type loops were challenging in the past, the fact that we observe them infrequently in our simulations seems also sensible. 

\subsection{Underlying assumptions and possible improvements}

There are some underlying approximations and assumptions that need to be discussed in order to put the presented results in the proper context. First, there are some uncertainty about the reliability of the older potentials used in this work. While these potentials have been used in previous works in studying radiation damage, the fact remains that the coupling with the repulsive interactions were not always made very rigorously. This can be seen from dimer curves and quasi-static drag curves that can be found in the supplementary materials. Previously it has been reported in the case of iron that slight modifications to the repulsive interactions did not result in significant changes in the final defect structure in overlapping cascades~\cite{byggmastar_effects_2018}. However, in a more recent work by the authors of this work, modifications of the repulsive interactions were carried out in nickel, where significant changes in the defect concentrations were observed~\cite{fellman2026effects}. Therefore, it is likely for some of the potentials a more rigorous coupling to repulsive interactions would change the results significantly. Another key finding in Ref.~\citenum{fellman2026effects} was the fact that even potentials that have similar repulsive interactions can give wildly different defect structures. What is evident from the results in this work is that one cannot just naively connect a repulsive pair potential to any arbitrary potential and expect consistent results in overlapping cascade simulations. Care must be taken to validate the potentials suitability for the use in radiation damage simulations. 

Another key implicit approximation in overlapping cascade simulations is neglecting the long-term defect evolution. In the literature this is often followed with the statement that the defect migration in the material is sufficiently slow at the simulated temperature so that long-term defect evolution can be assumed to be limited. This is true for elements such as tungsten, where the vacancy migration barrier is high, but is not true in general for materials with low vacancy and self-interstitial migration barriers. Since in the current work the long-term defect evolution is neglected, this obviously calls into question how realistic the defect structures presented in this work are. While this question is completely warranted, the current work does not aim to resolve the issue. In recent years, new approaches which aim to introduce the long-term defect evolution into MD simulations have started to appear~\cite{wu2026unveiling}.

There are also certain methodological choices made in the current work that needs mentioning. First, all the cascades were simulated with the same PKA energy of 5 keV. This energy has been used frequently in past works, mainly because 5 keV cascades do not require large simulation cells for the cascades to be fully contained within them. In the literature it is also reported that instead of running mono-energetic cascades, sampling cascade energies from a distribution changes the observed defect structure, even if the average PKA energy remains the same~\cite{BRUNCRONA2026114820}. Another practical choice is the simulation temperature of 300 K. Levo \etal studied the temperature dependency of overlapping cascades in both Ni and Ni-based alloys and concluded that the defect concentrations are temperature-dependent, with higher temperatures resulting in lower defect concentrations~\cite{levo2021temperature}. This means that when comparing the current results with those from the past (and also in future comparisons) one must keep in mind both the choices of PKA energy and temperature. For example, the formation of vacancy-type loops might be increased by the increase of the cascade energy. When the cascade core is sufficiently large and contains sufficient amount of vacancies, it becomes increasingly probable that the cascade core can collapse into a stable vacancy loop~\cite{kirk_collapse_1987,byggmastar2026four}. The temperatures and energies where such changes in the defect structures occur are of course element specific. 

The electronic stopping used in the work was implemented as a friction force that was computed from SRIM calculations. In all cases, the friction is applied to atoms with kinetic energies above 5 eV. However, this cutoff should in principle be optimized for each element separately. In the literature, 1 eV threshold is recommended for Fe and 10 eV for W~\cite{bjorkas_assessment_2009,sand_high-energy_2013}, and comparisons with ion beam mixing experiment suggest 10 eV cutoff for Ni, Pd and Pt~\cite{SAND201599}. However, there are many elements in this work for which we do not know the optimal value for the cutoff, therefore we opted to use the same cutoff in all cases for the sake of consistency. Furthermore, there are also more sophisticated approaches available for the implementation of electronic stopping into cascades simulations, such as the two-temperature model~\cite{zarkadoula2014electronic} or the EPH model~\cite{tamm_role_2019}. In these cases, the parametrizations exist for some elements but not for all the elements considered in this work. 

Another aspect of possible improvements could be in the defect analysis methods themselves. In section~\ref{sec:methods:analysis} we already touched upon the main challenges with the defect analysis. WS analysis is a well established method to identify point defects. However, WS implementations have some challenges when it comes to analyzing highly damaged structures. First, in the case where we need an persistent external reference, like in OVITO, we quickly run into the issue where the reference start to diverge from the structure we are analyzing, making the results unreliable. There are implementations of WS~\cite{nordlund_defect_1998} that do not rely on an external reference (like the one used in this work) and as such can build the reference from the structure it is analyzing. We found that these work very well for FCC and BCC systems, but struggle in HCP systems. This is likely due to the anisotropic growth in HCP materials which make building of the reference lattice unreliable. Recently, more advanced methods have started to appear in the literature for the identification of point defects~\cite{DELFRE2026114653,shin2026improved,PhysRevMaterials.5.095403}. The checkpoints (for which the full atomistic structures were saved) provided by this work constitute an excellent test set to validate the robustness and computational performance of future defect analysis implementations across a wide range of damage states and materials.  

\subsection{Overall trends}

\begin{figure*}[ht!]
    \centering
    \includegraphics[width=\linewidth]{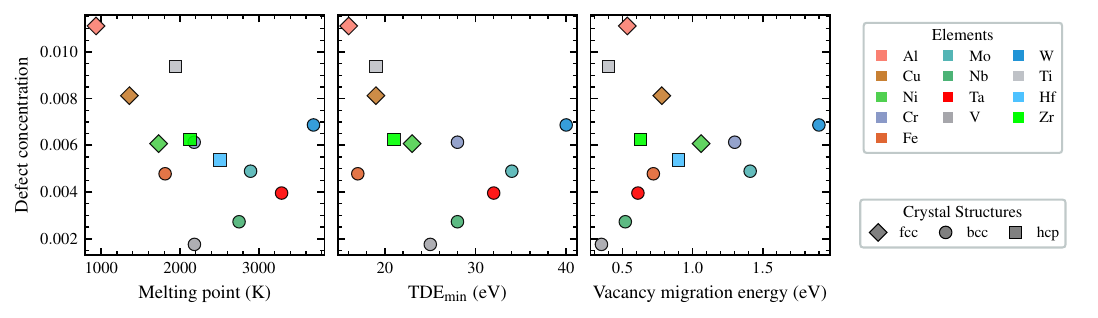}
    \caption{Final defect concentrations of tabGAP potentials (mean value of last 20 cascades) after 2000 5 keV cascades correlated with; experimental melting point, minimum threshold displacement energy as reported in experiments~\cite{Jung01011978} and vacancy migration energies as predicted by the different tabGAP potentials.}\label{fig:comp}
    \label{fig:corr}
\end{figure*}

Experimentally, significant effort has been made to find correlations between damage production and various physical properties, but this has turned out to be challenging~\cite{JENKINS199316,KLABUNDE1982183}. Nonetheless, we have also attempted this with the dataset generated in this work. We compared the defect concentrations (mean value of the last 20 data points) with properties such as: Atomic mass, melting point, density, minimum threshold displacement energy, stacking fault energy (in the FCC materials), temperatures of defect recovery stage I and III temperatures and vacancy migration energies. These comparisons can be found in the supplementary materials, but the main conclusion is that in general clear trends are difficult to find, especially if we look at all the different potentials. However, if we only focus on the tabGAP potentials, which should be most consistent with each other, we find a curious correlation with vacancy migration energies which is shown in Fig.~\ref{fig:corr}. In the figure the final defect concentration compared with the vacancy migration energies predicted by each of the potentials is shown. The vacancy migration energies are taken from the respective publications of the potentials, with the exception of Cr that was calculated in this work. What is striking in the figure is that there are actually two opposite trends; In BCC the defect concentration increases as the vacancy migration energy increases, but in FCC and HCP it decreases. The same trend can also be seen if we plot the concentrations with respect to the defect recovery stage III temperatures predicted by experiments (found in the supplementary materials), which makes sense as it indicates the onset of vacancy defect mobility. However, looking at the EAM and modified Zhou potentials compared with recovery stage III we do not see such a clear trend. Another correlation is seen if we compare the defect production with respect to minimum threshold displacement energy as reported in experiments~\cite{Jung01011978}. Furthermore, similar trends are also seen when compared to the experimental melting points of each material.

The defect accumulation in cumulative cascade damage is essentially a competition between generation of new defects and recombination of the defects present. The fact that there are correlations with the particular properties considered is thus reasonable. For example the threshold displacement energy can be correlated with the amount of defects produced by individual cascades via damage models such as the commonly used arc-dpa model~\cite{nordlund_improving_2018}. Furthermore, recently Boleininger \etal have proposed a model to describe the saturated vacancy concentration based on damage models and the size of the molten region during the heat spike~\cite{boleininger2023microstructure}. Thus also the correlation with melting point is sensible.  

Now the question arises that why do we see opposing trends for FCC and HCP compared to BCC. The obvious statement here is that FCC and HCP are close-packed crystal structures while BCC has slightly smaller packing density. However, currently we do not have a mechanistic understanding of where these trends arise from. Furthermore, there are a few caveats that increase the uncertainty of the trends. First, some of the elements have reached steady-state saturation while others have not. Second, for FCC and HCP we have a limited amount of elements simulated using the machine-learned models. Nonetheless, these curious trends warrant further investigations.

\section{Conclusions}

We have in this work studied massively overlapping cascades using various interatomic potentials in Be, Al, Ti, V, Cr, Fe, Co, Ni, Cu, Zr, Nb, Mo, Rh, Pd, Ag, Hf, Ta, W, Pt, Au and Pb. These elements represent a varied selection of elemental metals with FCC, BCC and HCP crystal structures. The main conclusion is that the results of the overlapping cascade simulations are highly dependent on the choice of interatomic potential. Some general trends was seen for specific crystal structures and some correlations between properties could be observed.

In all FCC materials the final defect structures include stacking fault tetrahedra. Furthermore, no vacancy-type dislocations were identified in the final structures of the FCC materials at the 300 K temperature and 5 keV cascade energy used in all the simulations. While heavier FCC elements tended to have less defects, the trend was not consistent across all the FCC elements and potentials. In all BCC elements the 1/2\hkl<111> interstitial dislocations were the dominant dislocation structures observed (although in Fe interstitial-type C15 Laves clusters are more prevalent than dislocations). In V and Nb also vacancy type 1/2\hkl<111> dislocations were observed in the final defect structures. HCP elements tend to form complex dislocations of mainly $a$-type that can be of either vacancy or interstitial type. Overlapping cascades in Be were also attempted, but resulted in defect structures that did not include any dislocations and were very disordered. It was concluded that the conflicting results were likely due to limitations of the Be interatomic potential.

We also observed interesting trends with regards to defect concentrations when comparing against melting points, minimum threshold displacement energies and vacancy migration energies, in the tabGAP potentials. In this case opposing trends were seen between the different lattice structures (BCC vs. FCC and HCP).

\section*{Acknowledgments}

The authors wish to thank the Finnish Computing Competence Infrastructure (FCCI) and CSC -- IT Center for Science for supporting this project with computational and data storage resources. FG acknowledges that this work has been partially carried out within the framework of the EUROfusion Consortium, funded by the European Union via the Euratom Research and Training Programme (Grant Agreement No 101052200 — EUROfusion). Views and opinions expressed are however those of the author(s) only and do not necessarily reflect those of the European Union or the European Commission. Neither the European Union nor the European Commission can be held responsible for them. AF acknowledges funding from the Jane and Aatos Erkko foundation (Grant No 250022). JB acknowledges funding from the Research council of Finland through the OCRAMLIP project, grant number 354234.

\bibliography{mybib}

\end{document}


\title{Supplemental material for: Comprehensive study of massively overlapping cascades in common elemental metals}

\author{A. Fellman}
\thanks{Corresponding author}
\email{aslak.fellman@helsinki.fi}
\affiliation{Department of Physics, P.O. Box 43, FI-00014 University of Helsinki, Finland}
\author{J. Byggmästar}
\affiliation{Department of Physics, P.O. Box 43, FI-00014 University of Helsinki, Finland}
\author{F. Granberg}
\affiliation{Department of Physics, P.O. Box 43, FI-00014 University of Helsinki, Finland}


\maketitle
\section*{Table of contents}
\makeatletter
\@starttoc{toc}
\makeatother
\clearpage

\section{Dimer curves}

\begin{figure}[ht!]
    \centering
    \includegraphics[width=0.24\linewidth]{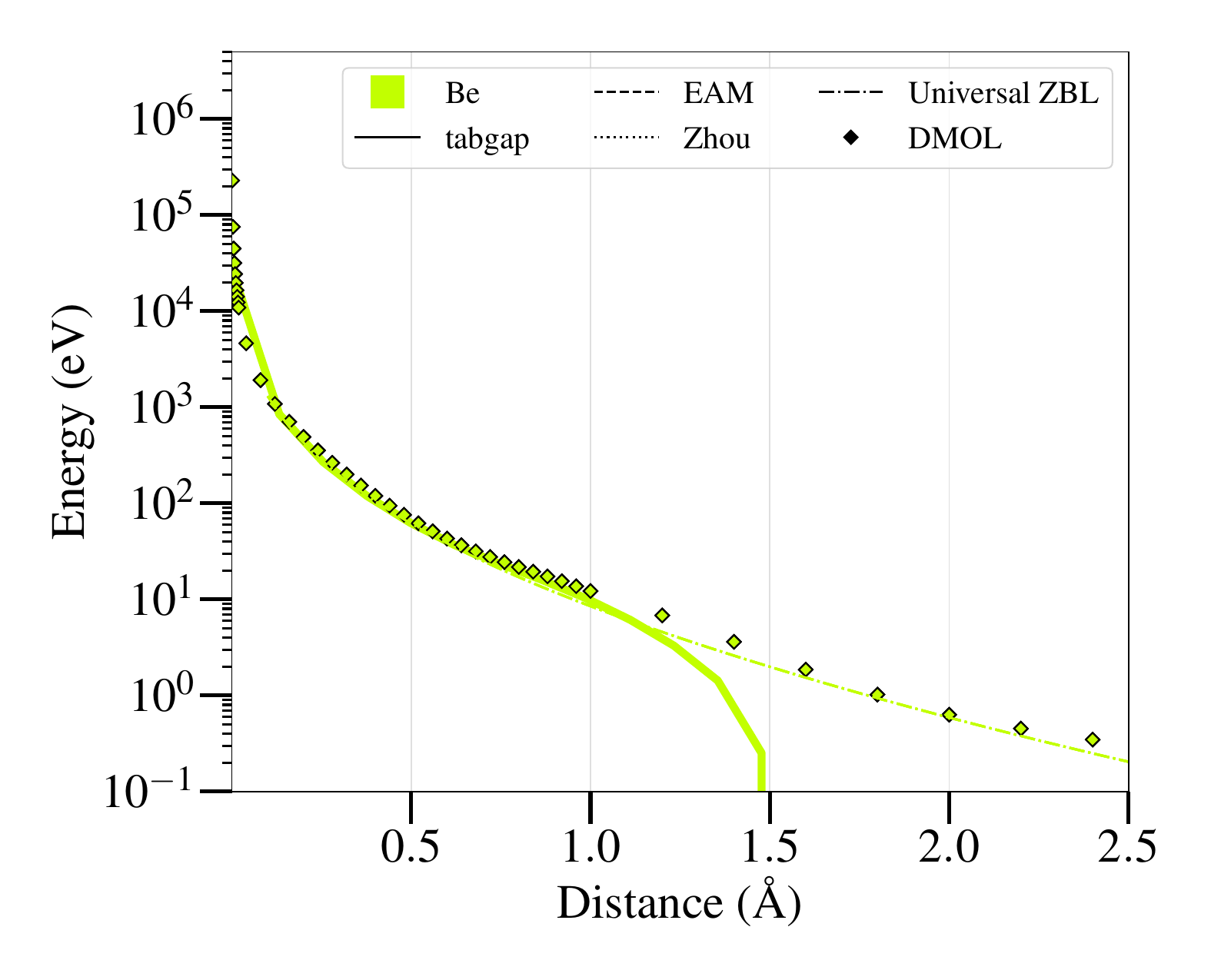}
    \includegraphics[width=0.24\linewidth]{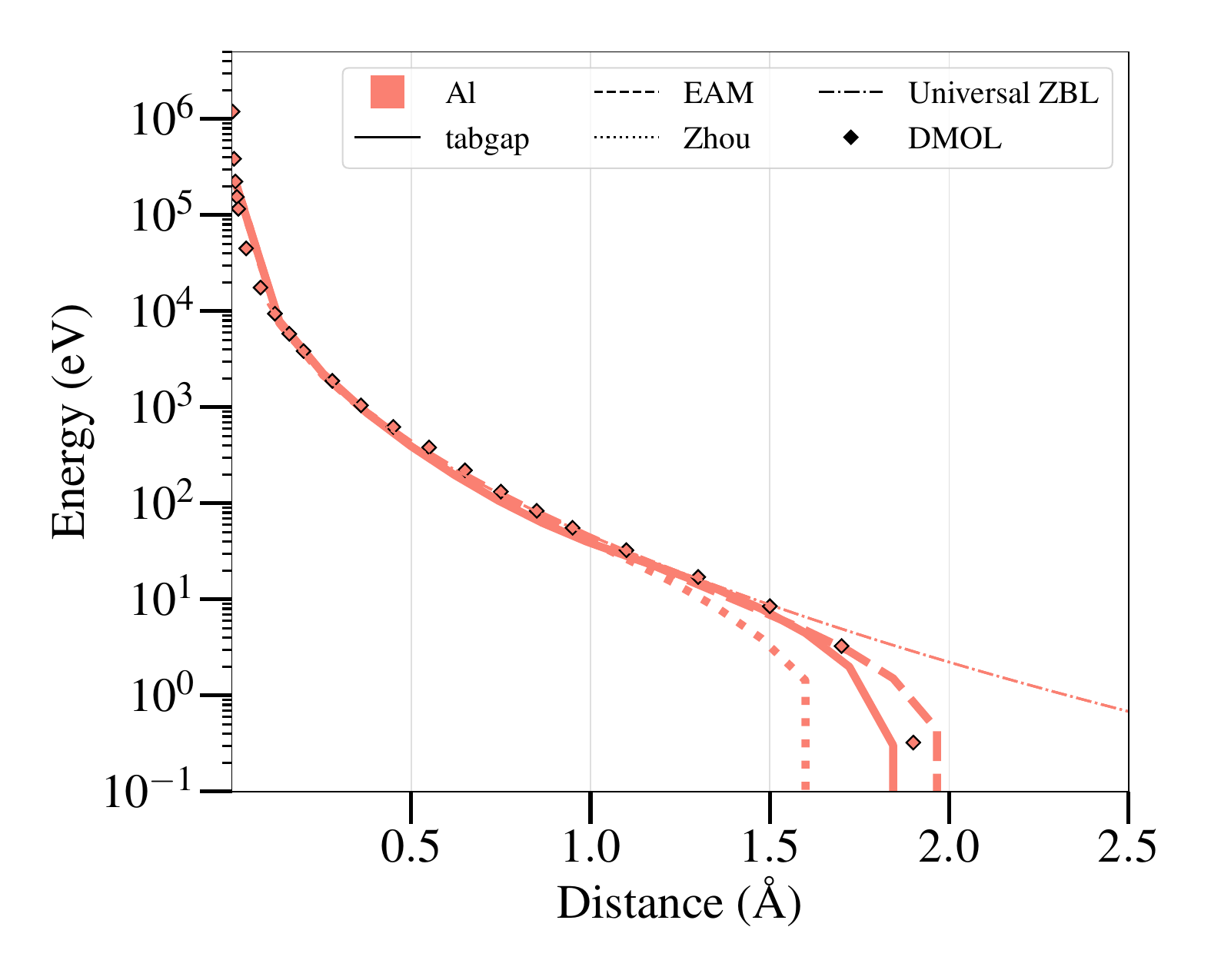}
    \includegraphics[width=0.24\linewidth]{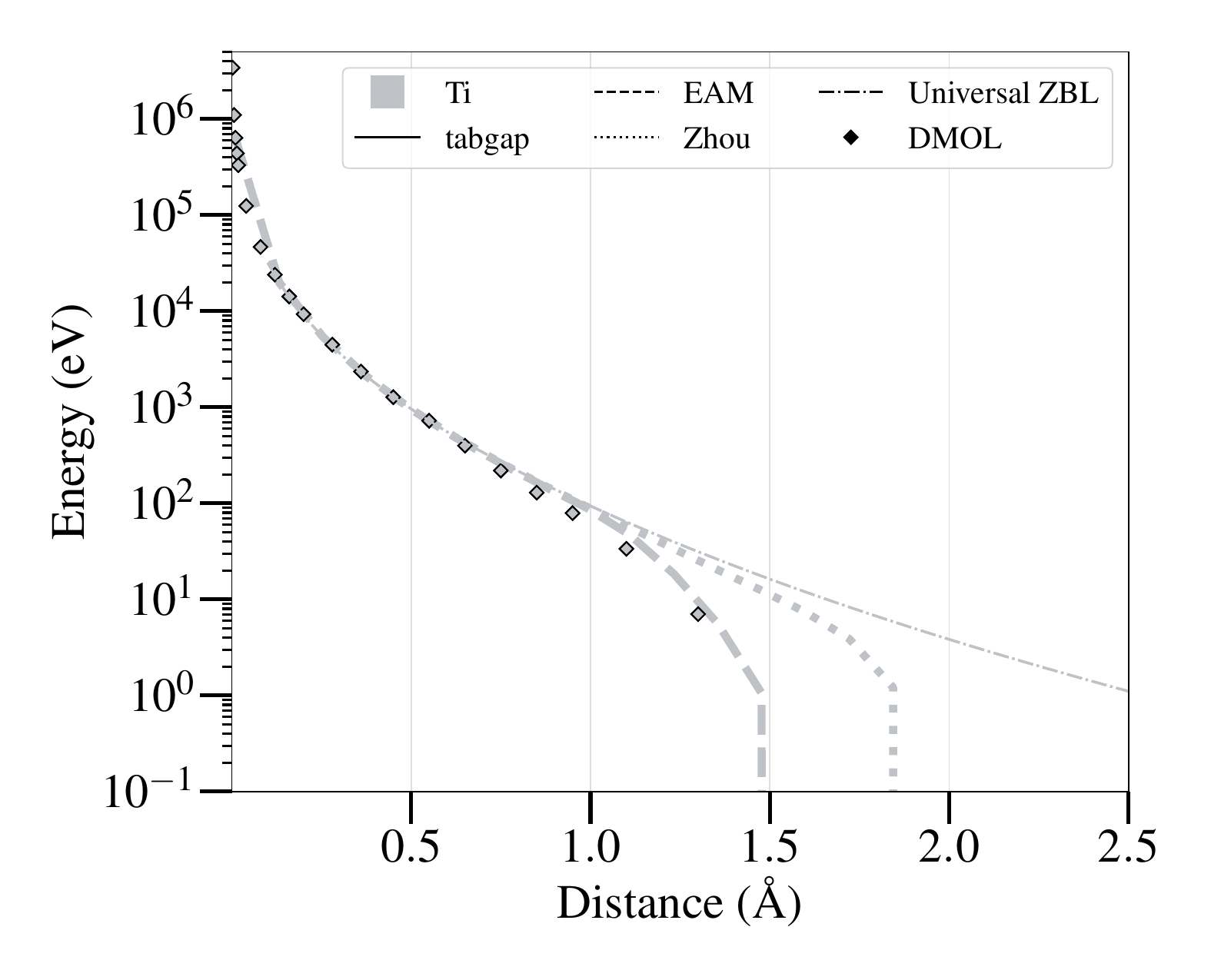}
    \includegraphics[width=0.24\linewidth]{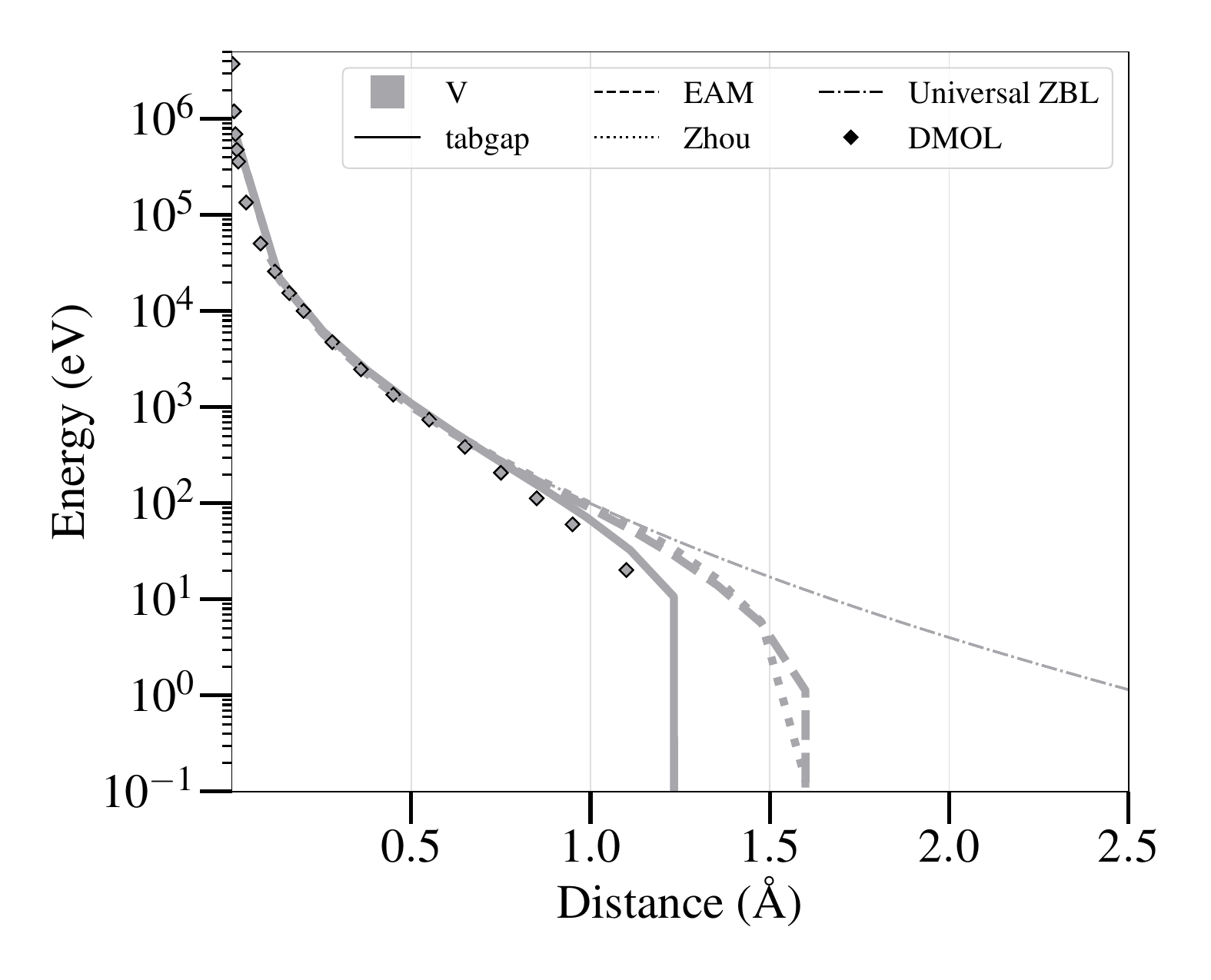}
    \includegraphics[width=0.24\linewidth]{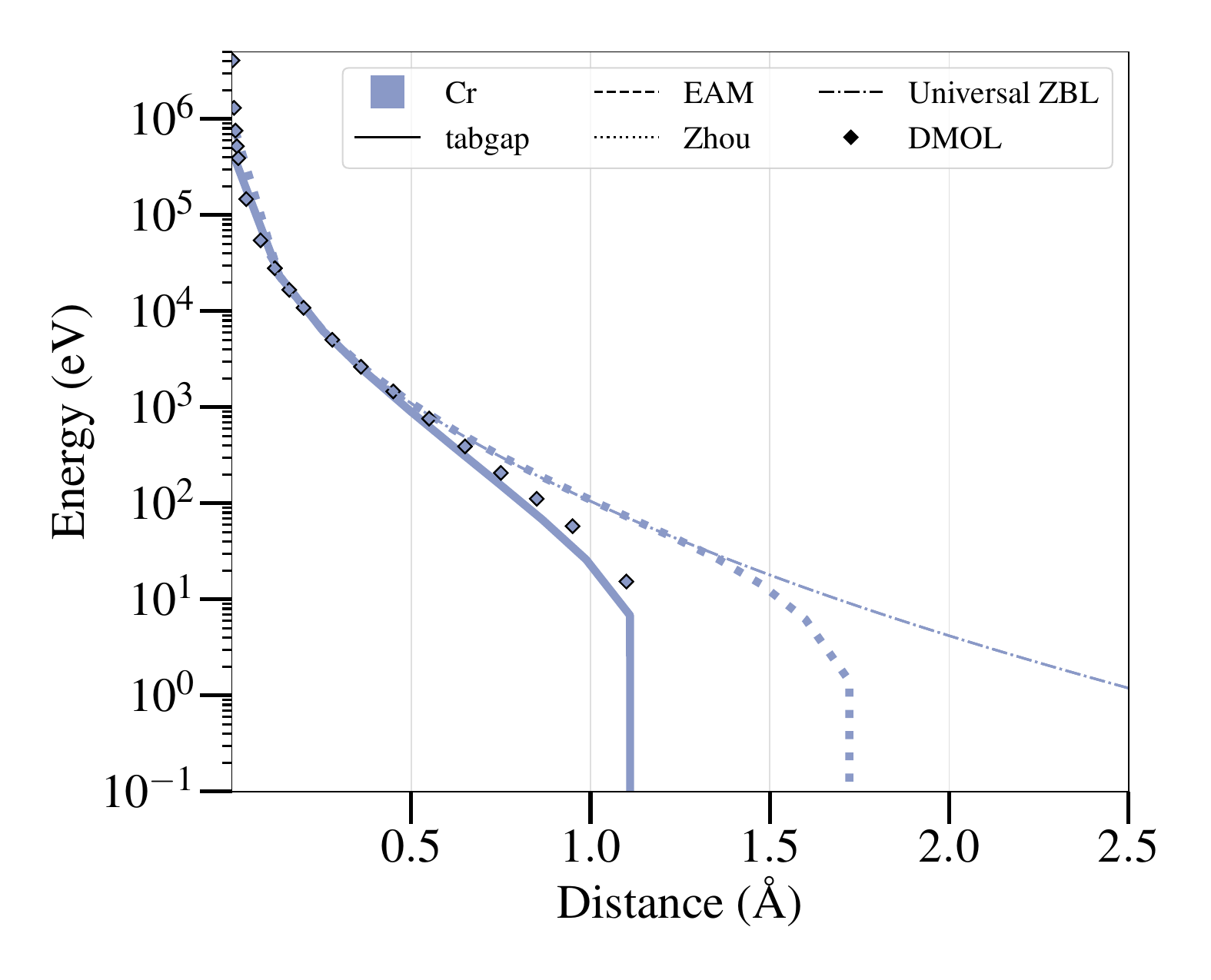}
    \includegraphics[width=0.24\linewidth]{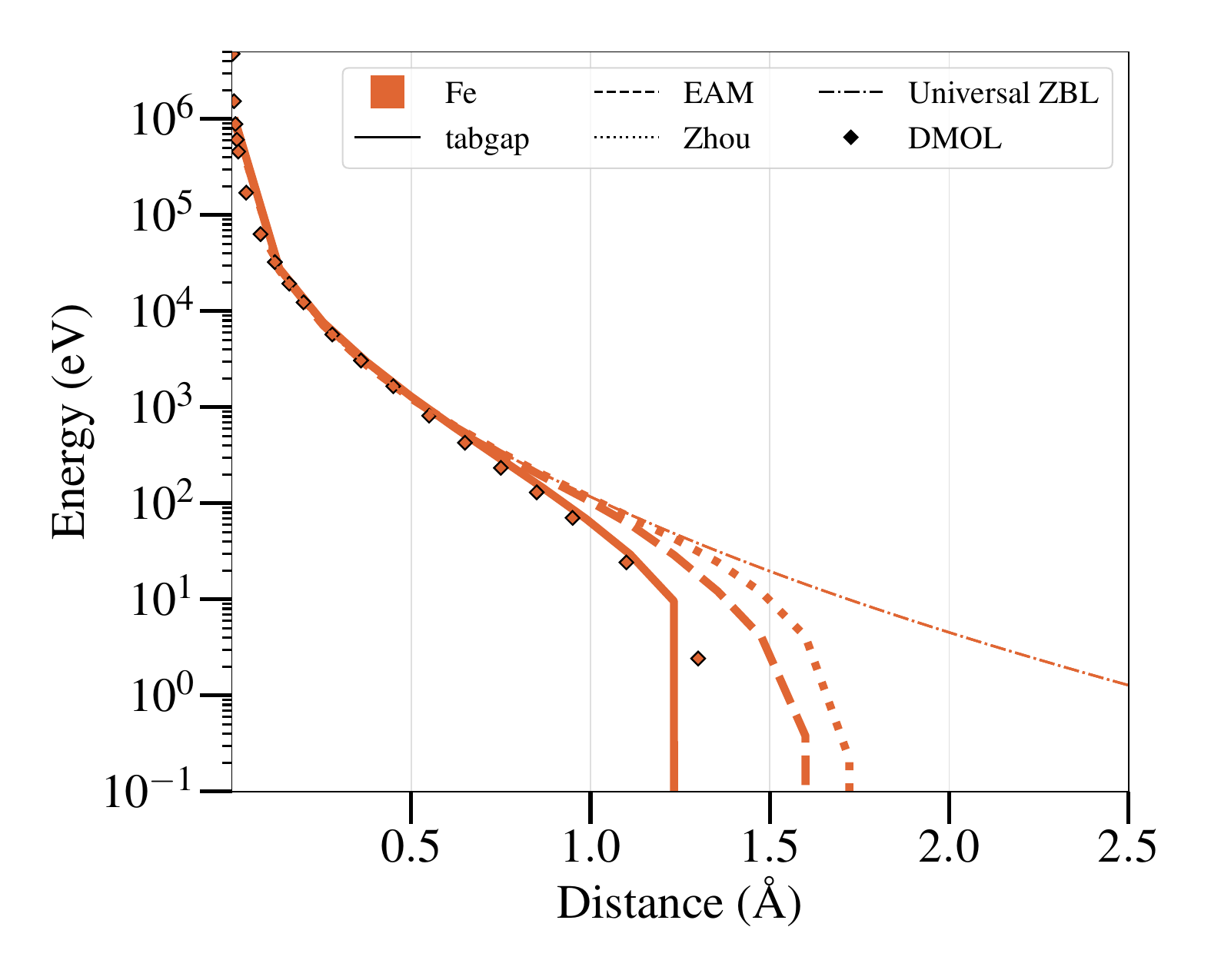}
    \includegraphics[width=0.24\linewidth]{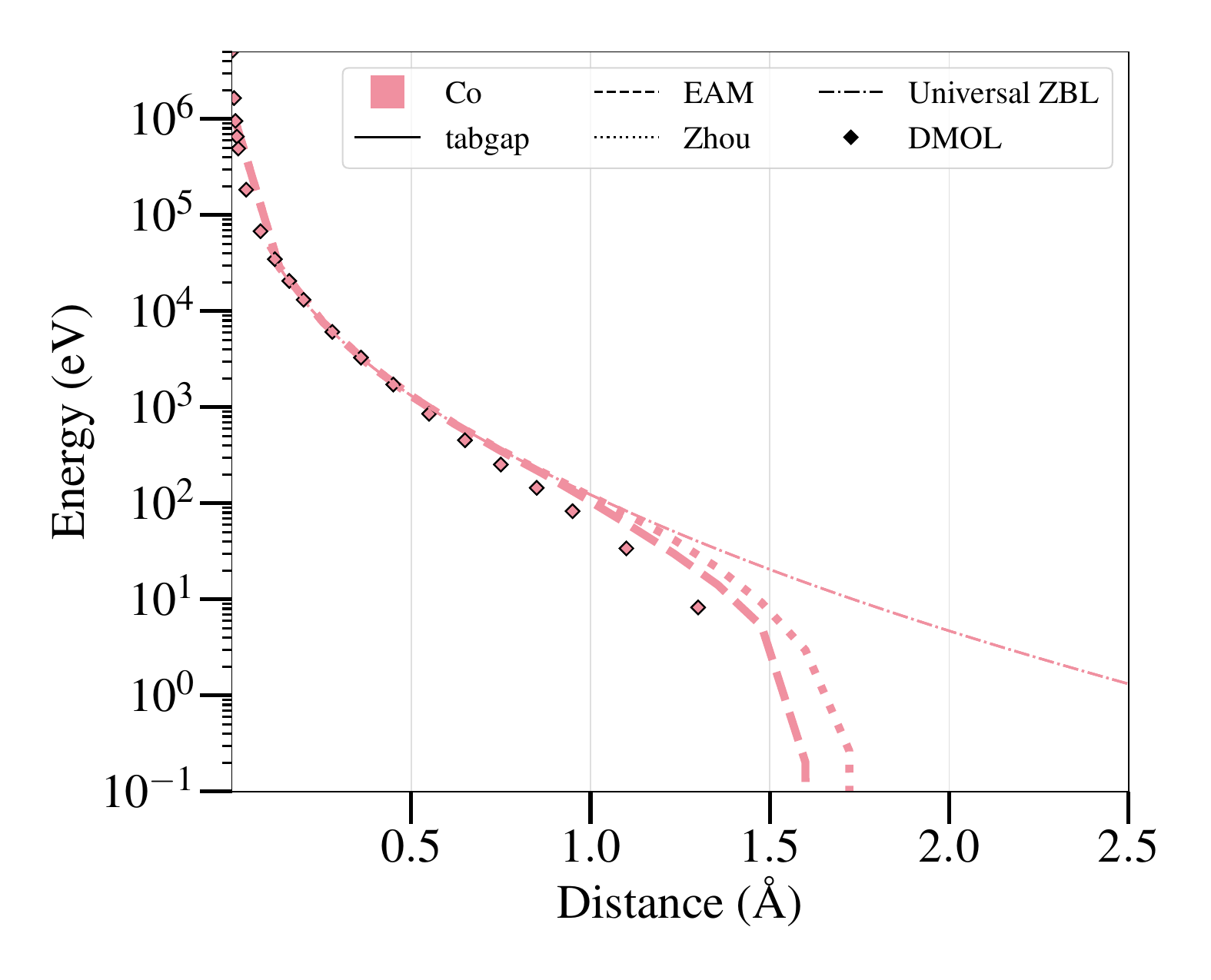}
    \includegraphics[width=0.24\linewidth]{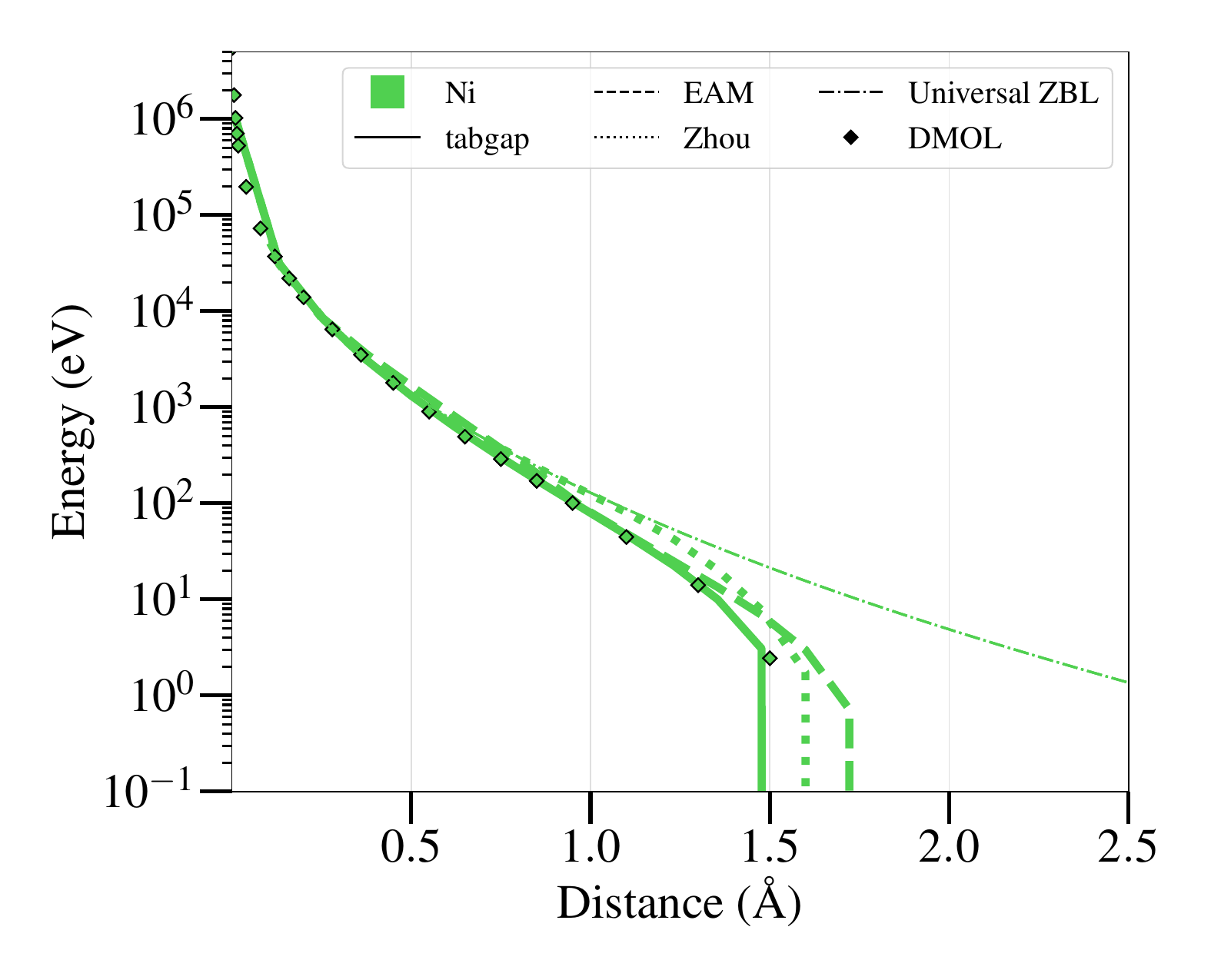}
    \includegraphics[width=0.24\linewidth]{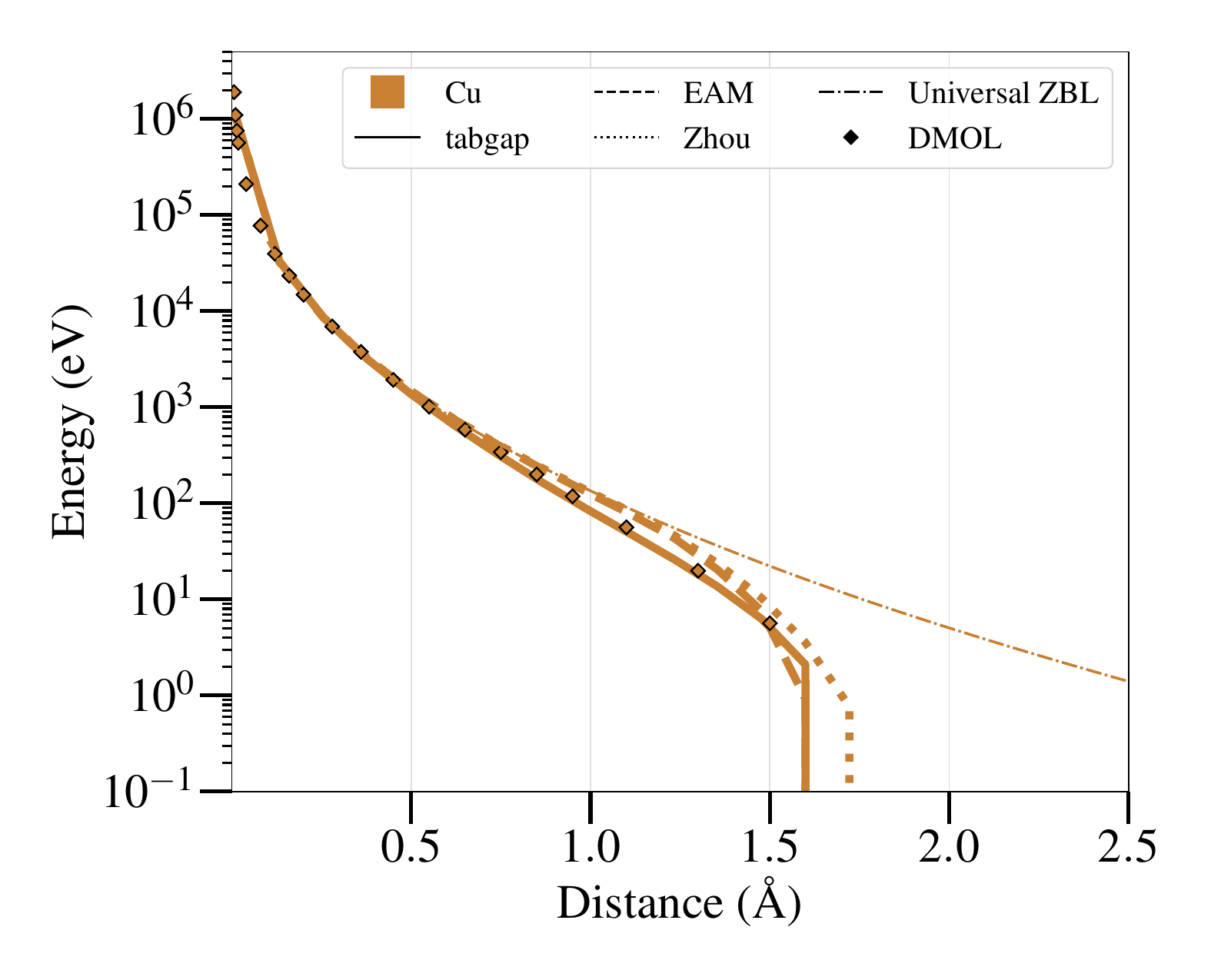}
    \includegraphics[width=0.24\linewidth]{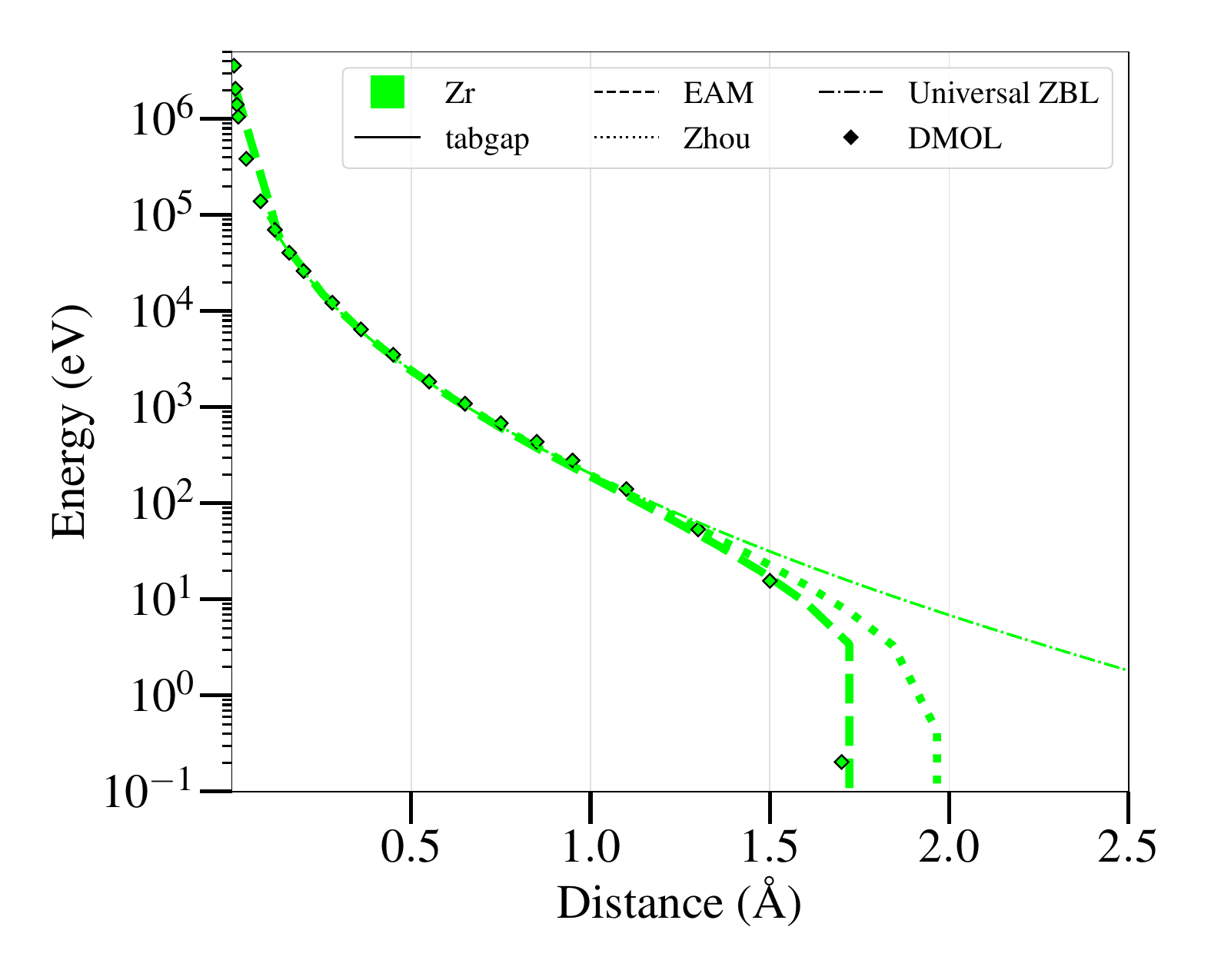}
    \includegraphics[width=0.24\linewidth]{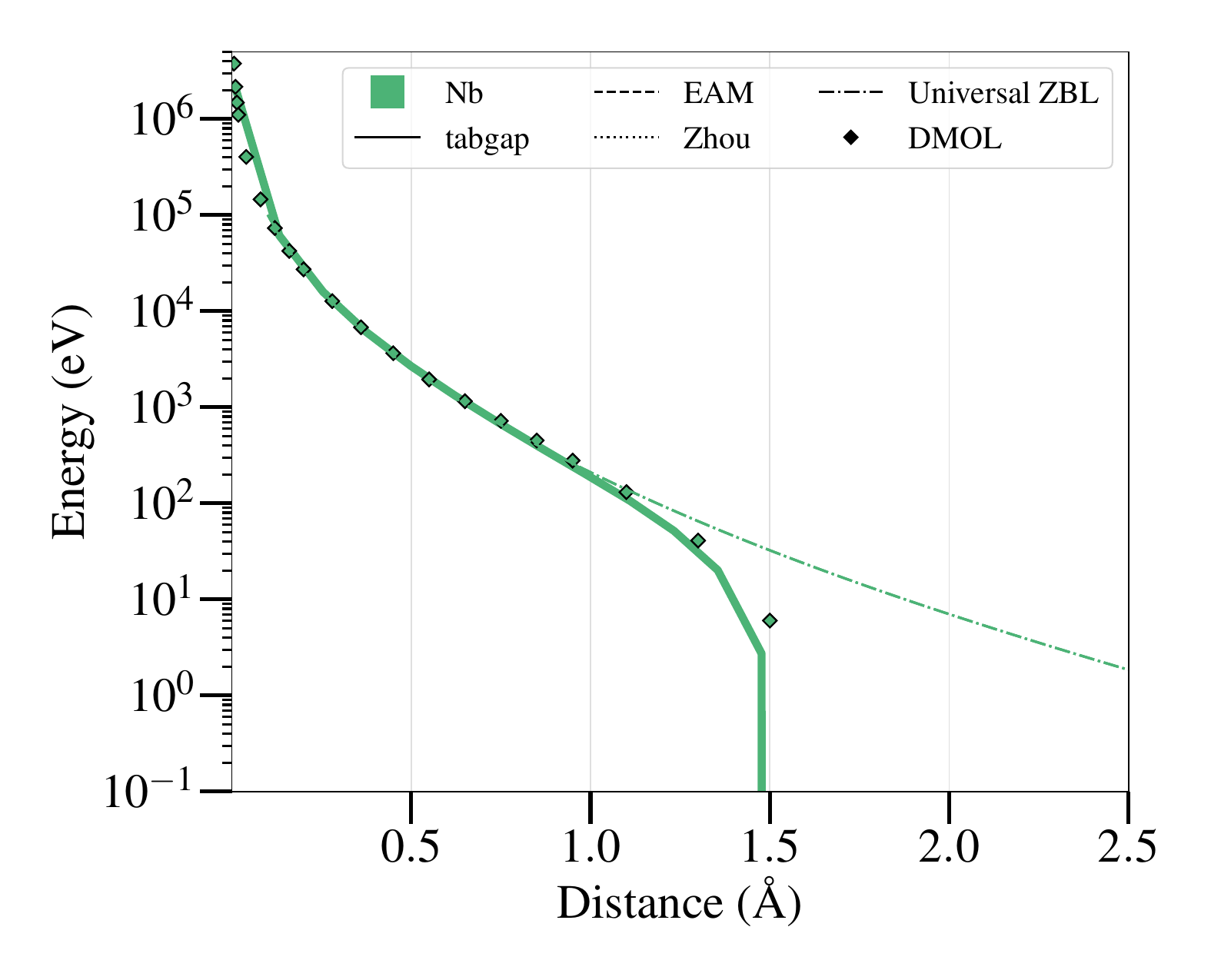}
    \includegraphics[width=0.24\linewidth]{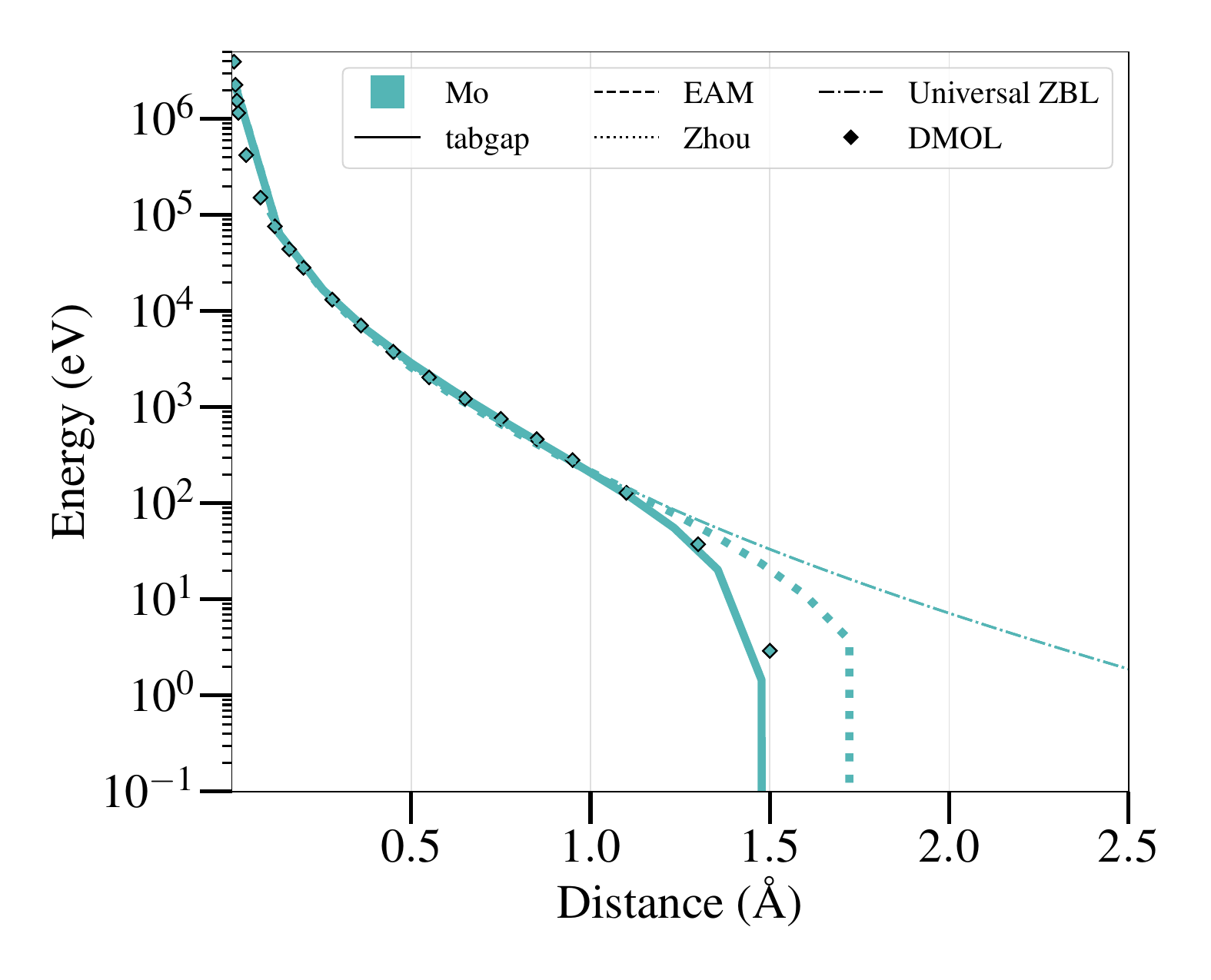}
    \includegraphics[width=0.24\linewidth]{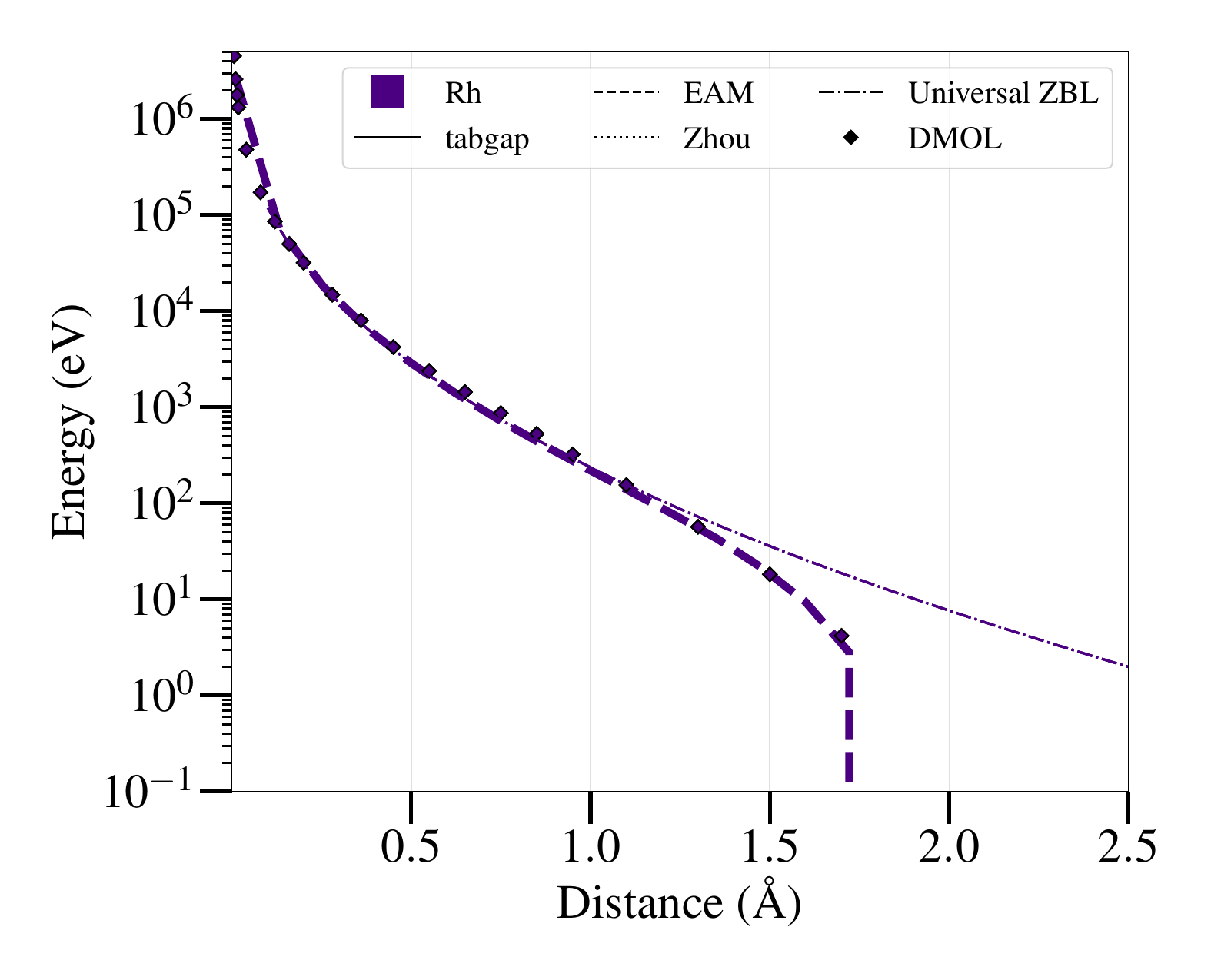}
    \includegraphics[width=0.24\linewidth]{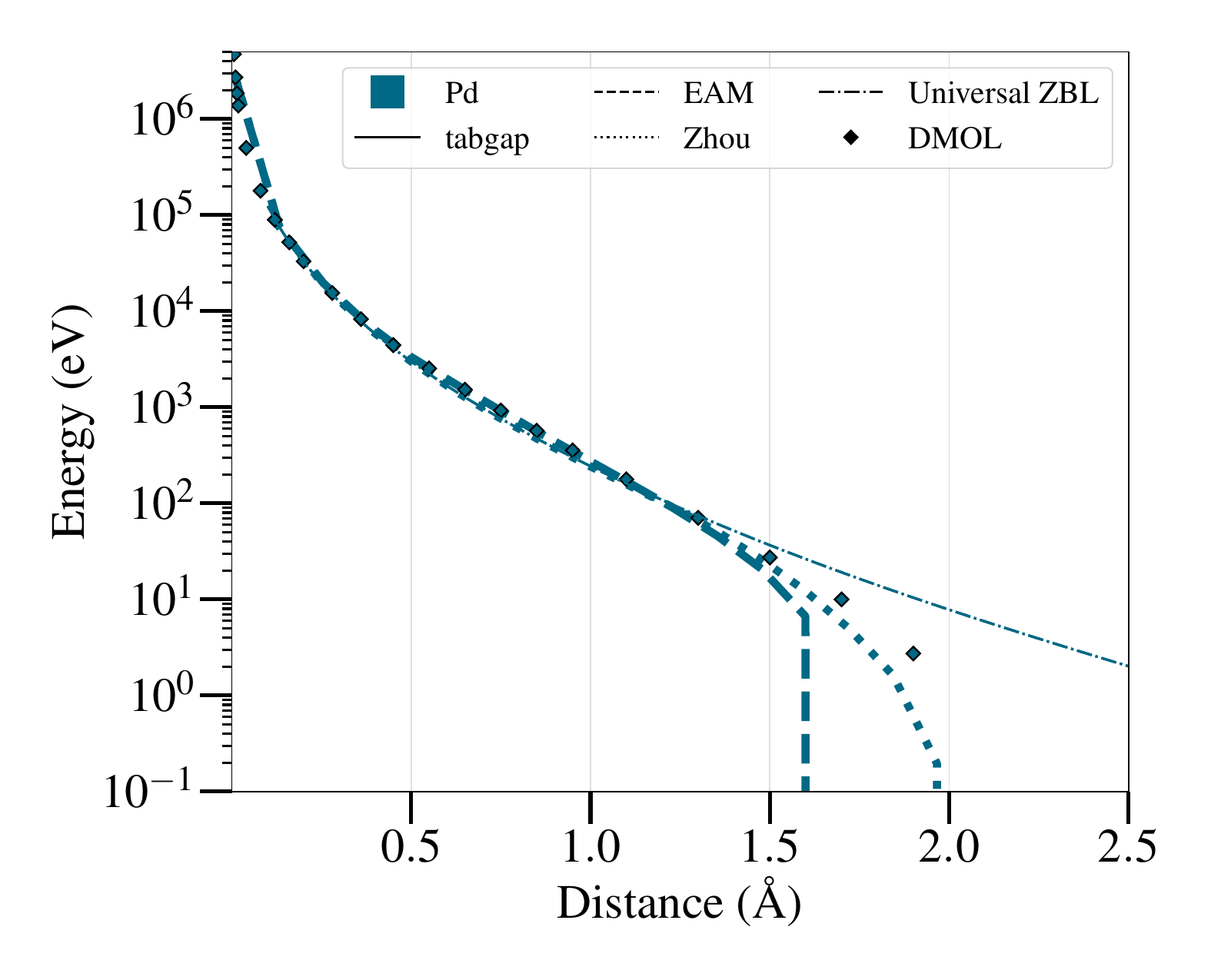}
    \includegraphics[width=0.24\linewidth]{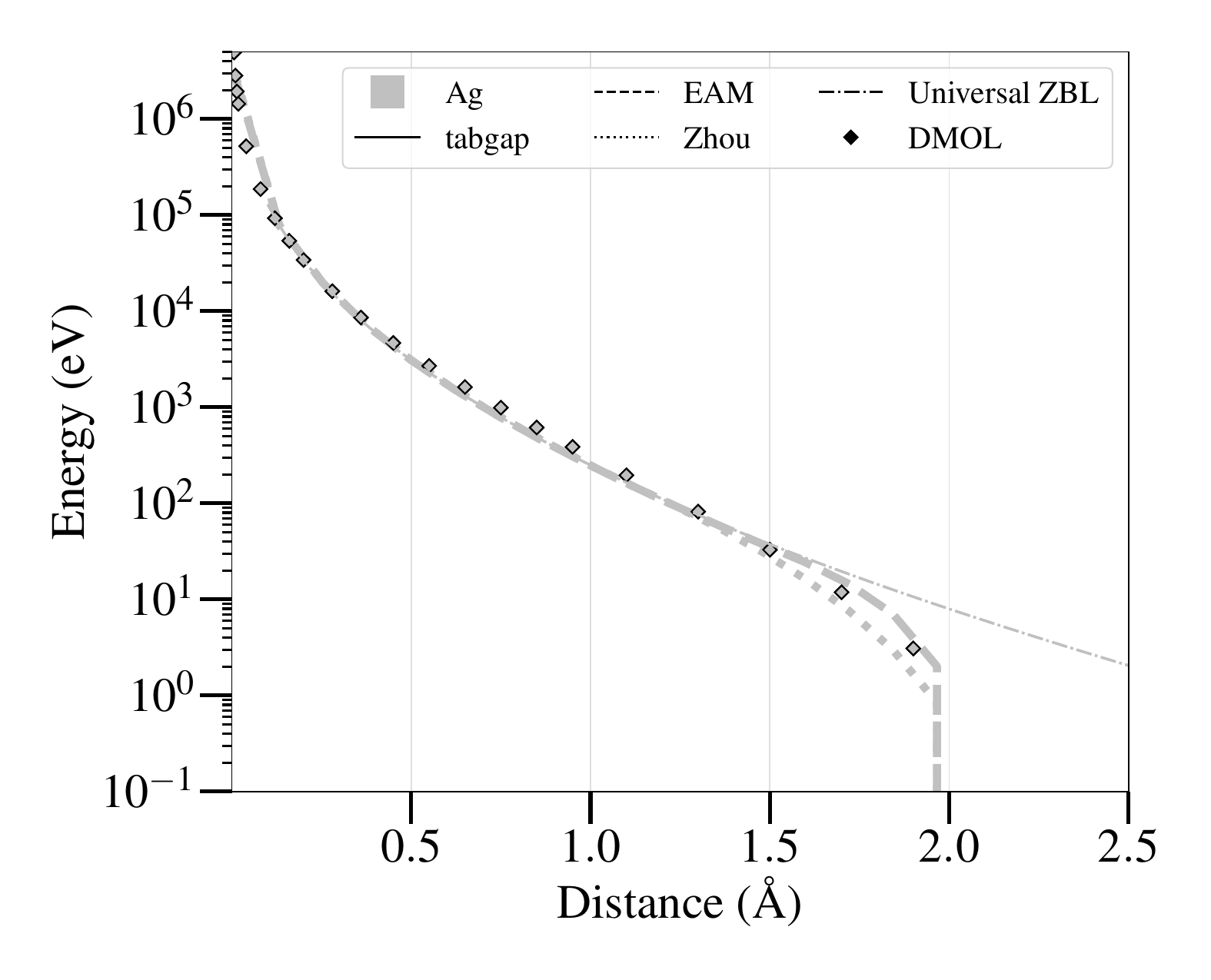}
    \includegraphics[width=0.24\linewidth]{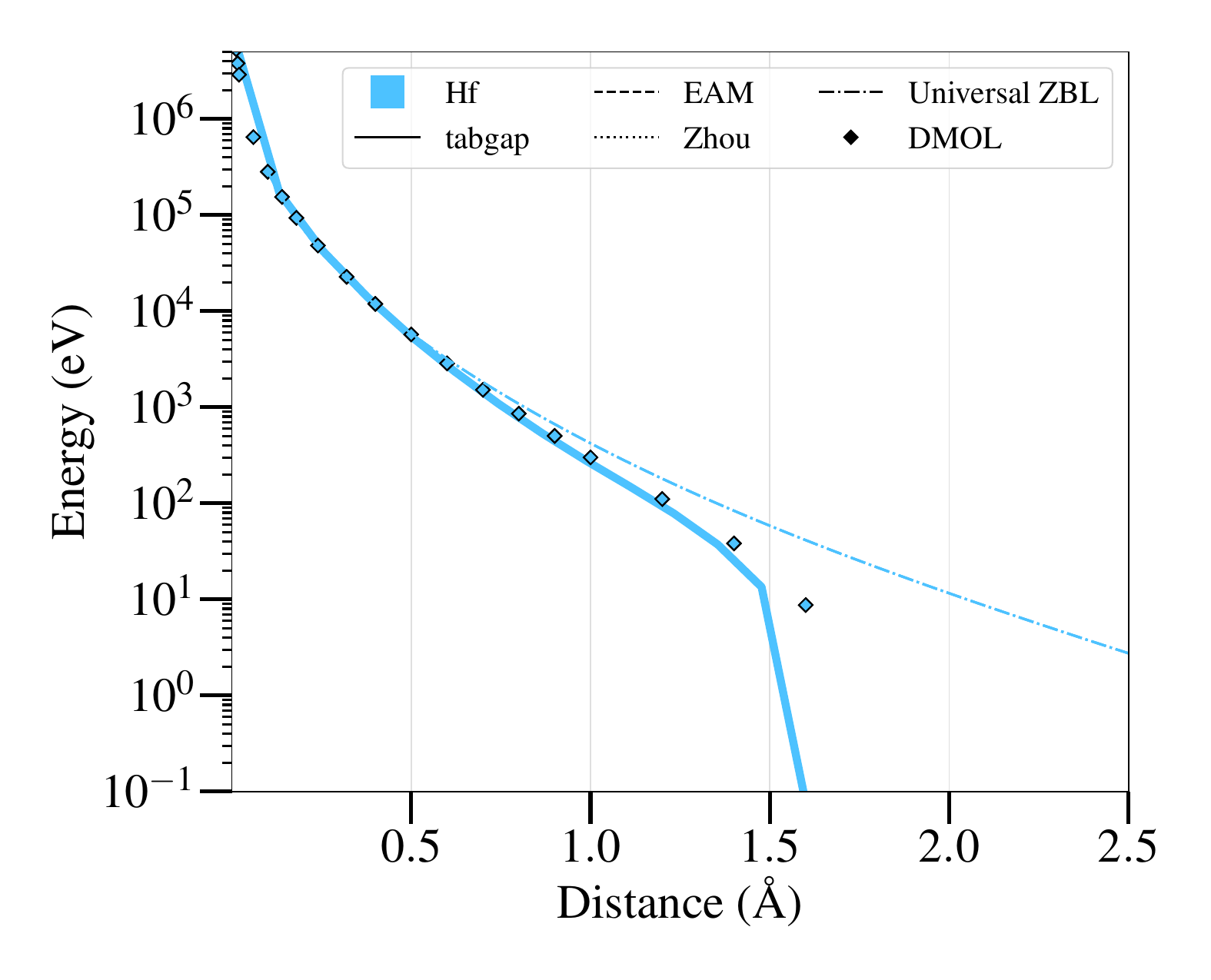}
    \includegraphics[width=0.24\linewidth]{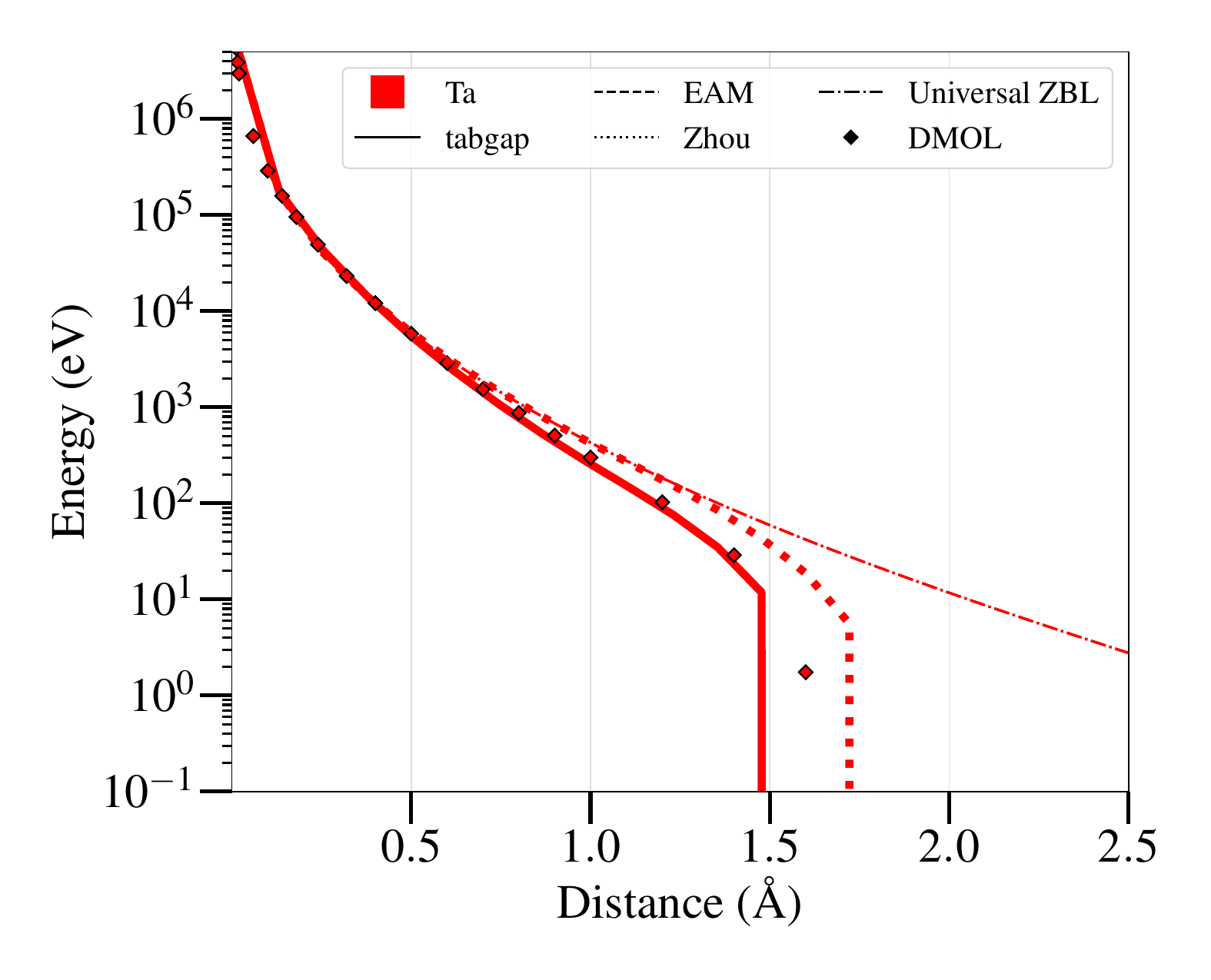}
    \includegraphics[width=0.24\linewidth]{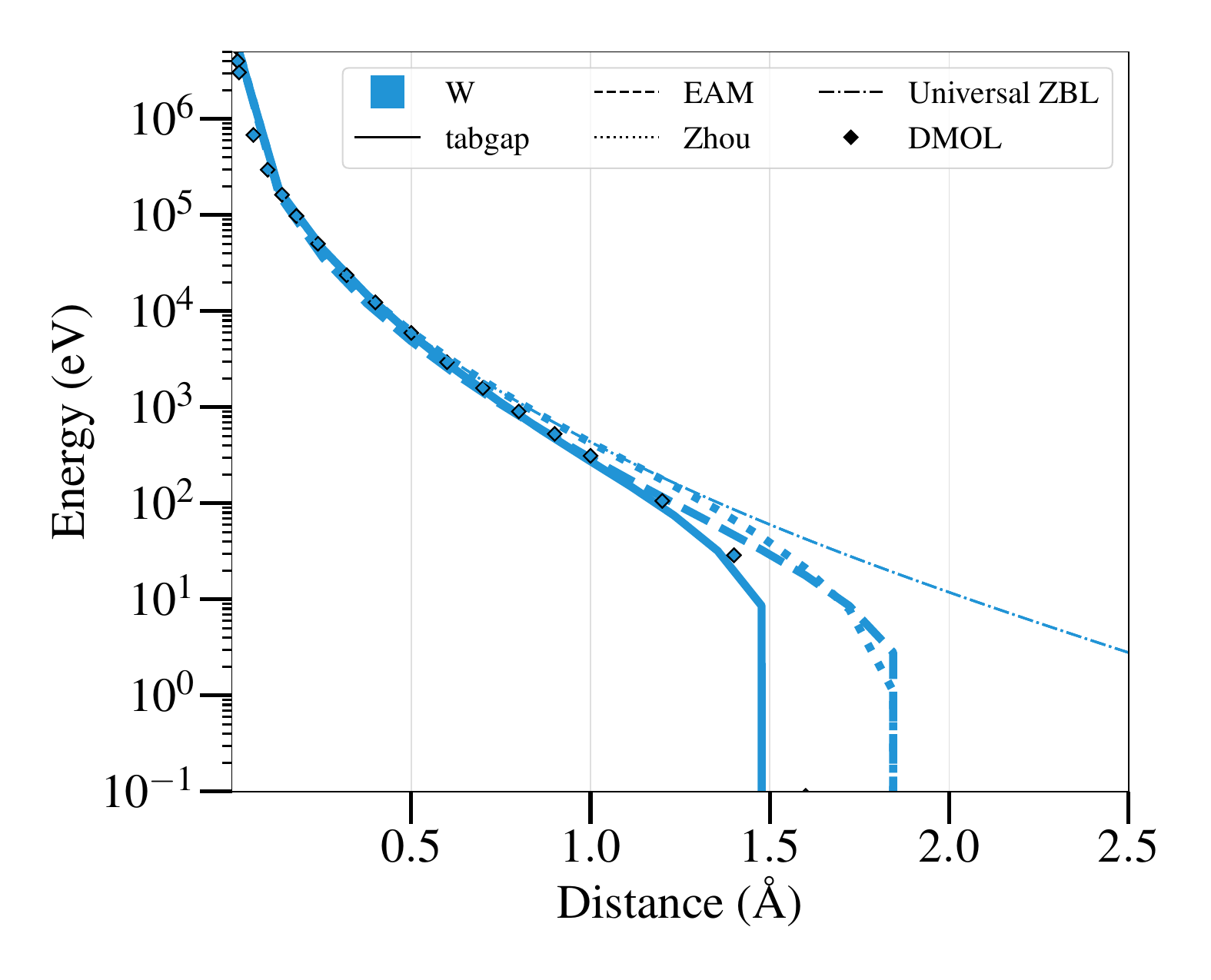}
    \includegraphics[width=0.24\linewidth]{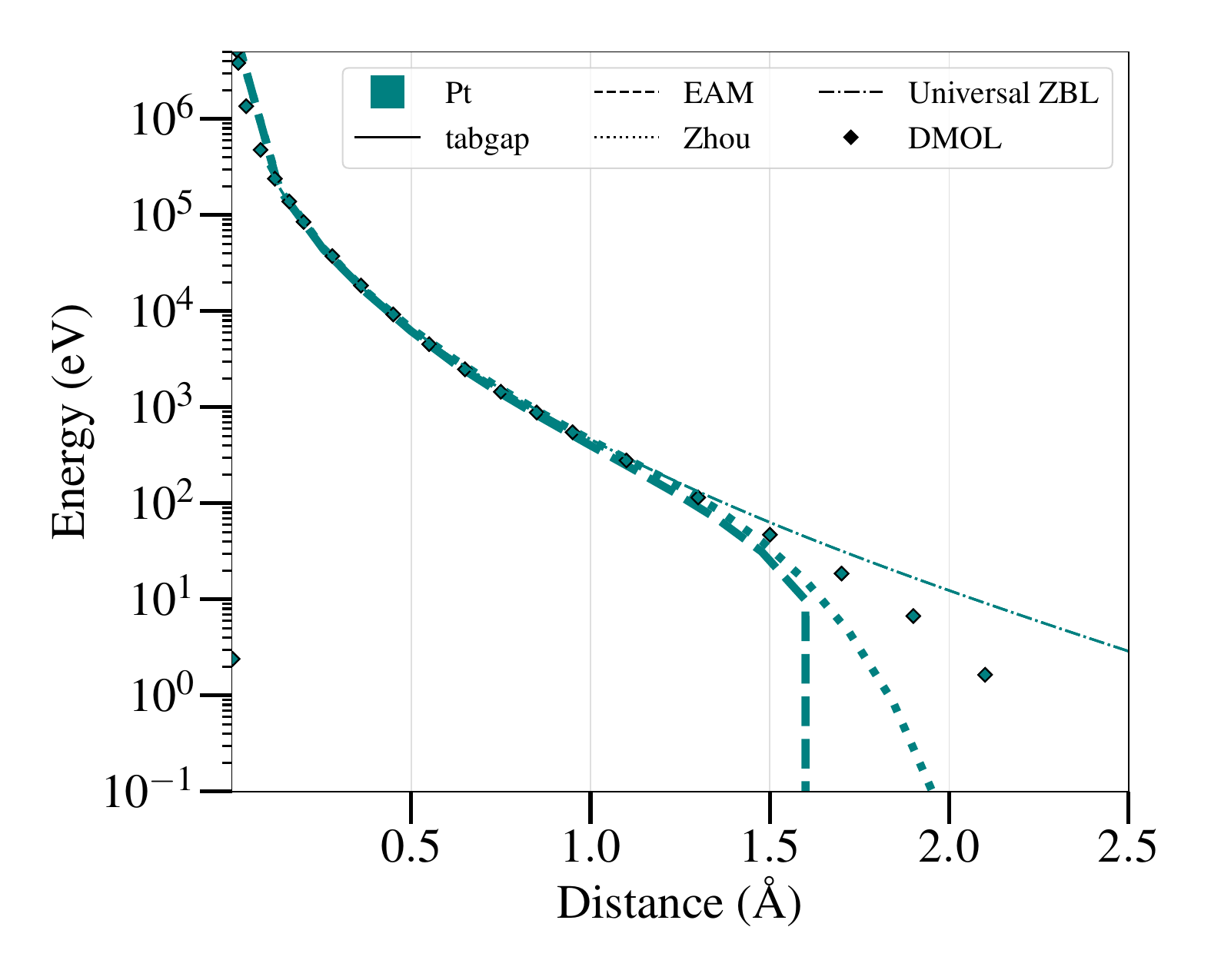}
    \includegraphics[width=0.24\linewidth]{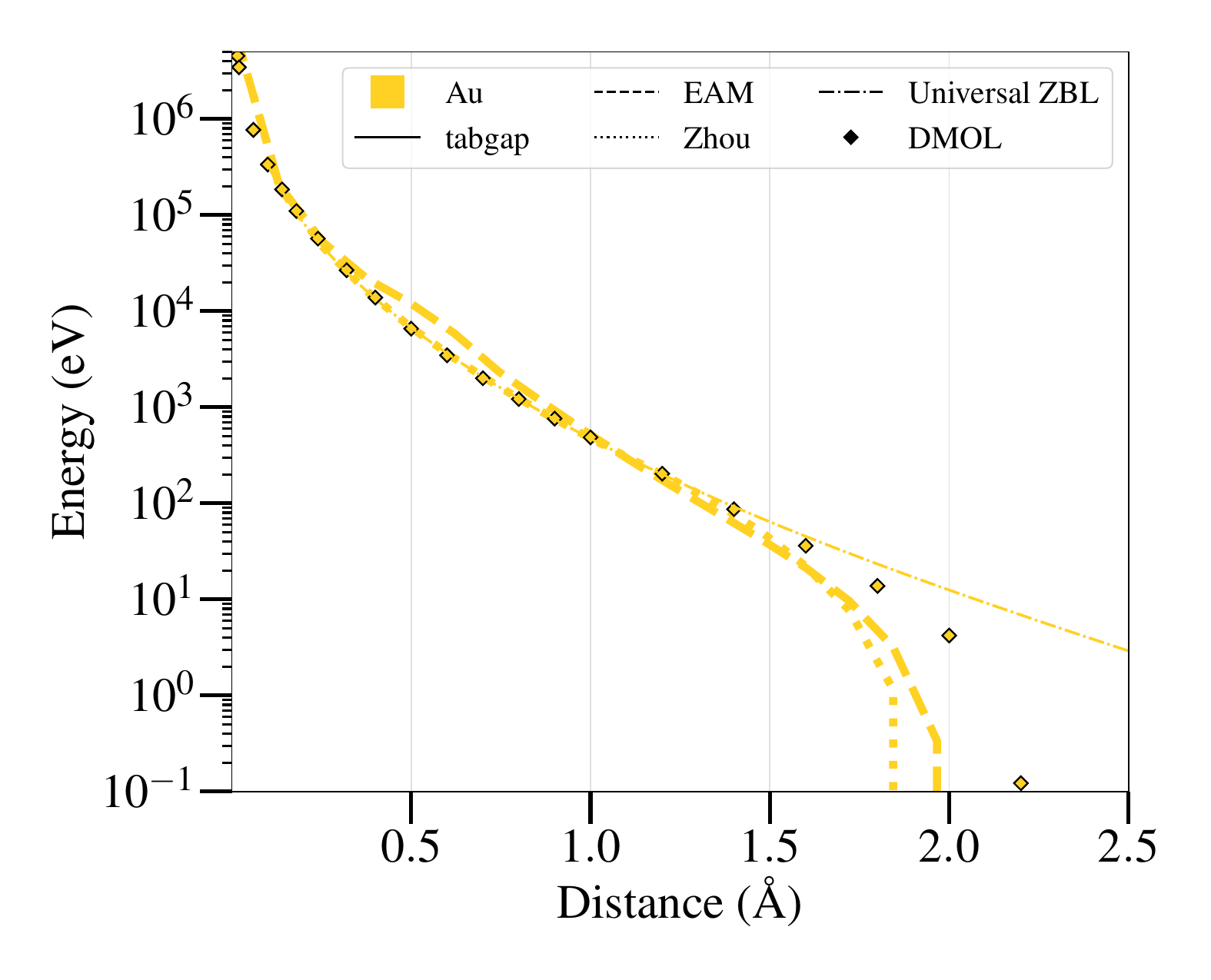}
    \includegraphics[width=0.24\linewidth]{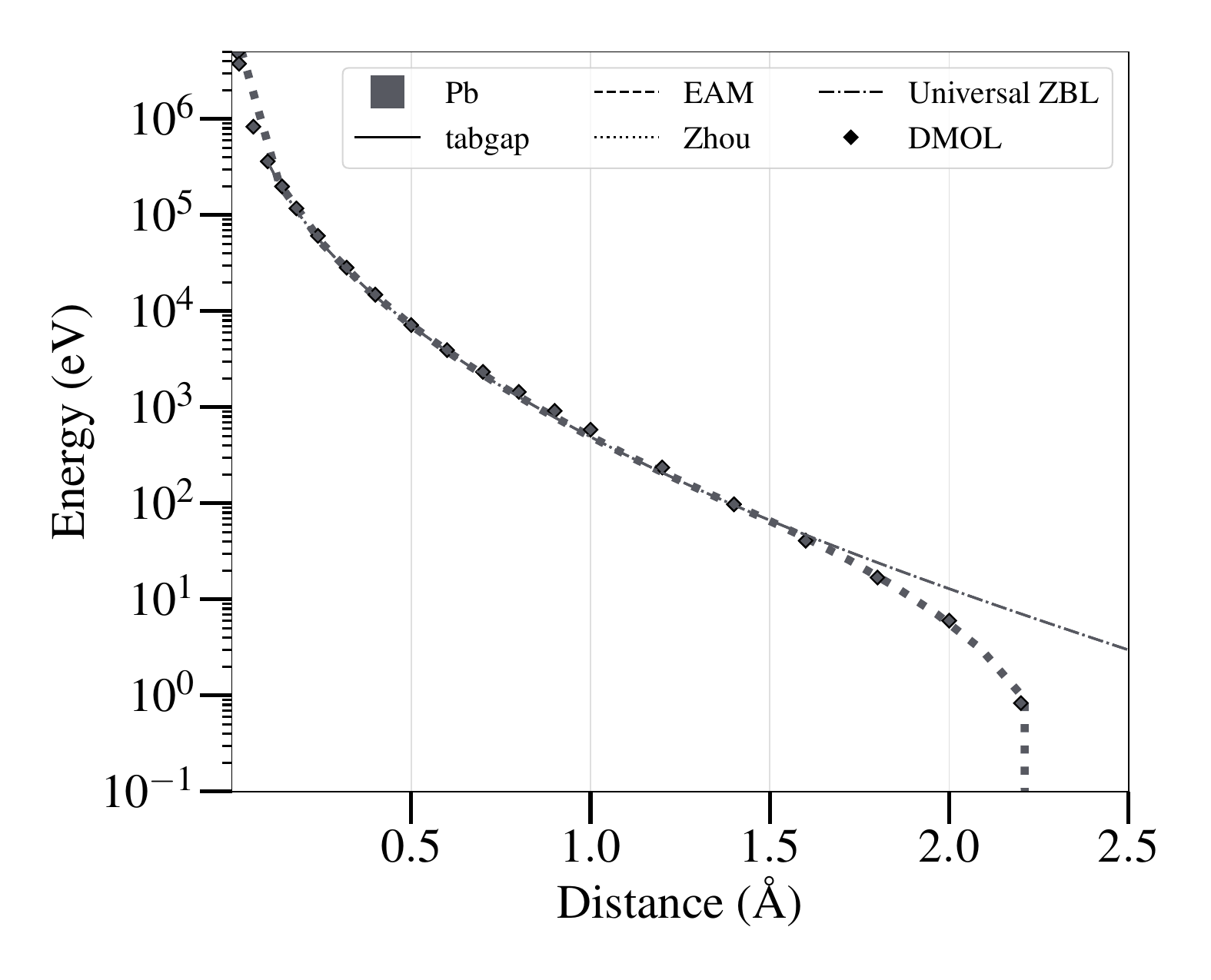}
    \caption{Dimer curves for each element. In the plots also the "universal" ZBL repulsive potentials and all-electron dimer data from Ref.~\cite{nordlund2025repulsive} as well as the different potentials themselves are given.}
\end{figure}

\clearpage

\section{Quasi static drag}

\begin{figure}[ht!]
    \centering
    \includegraphics[width=0.3\linewidth]{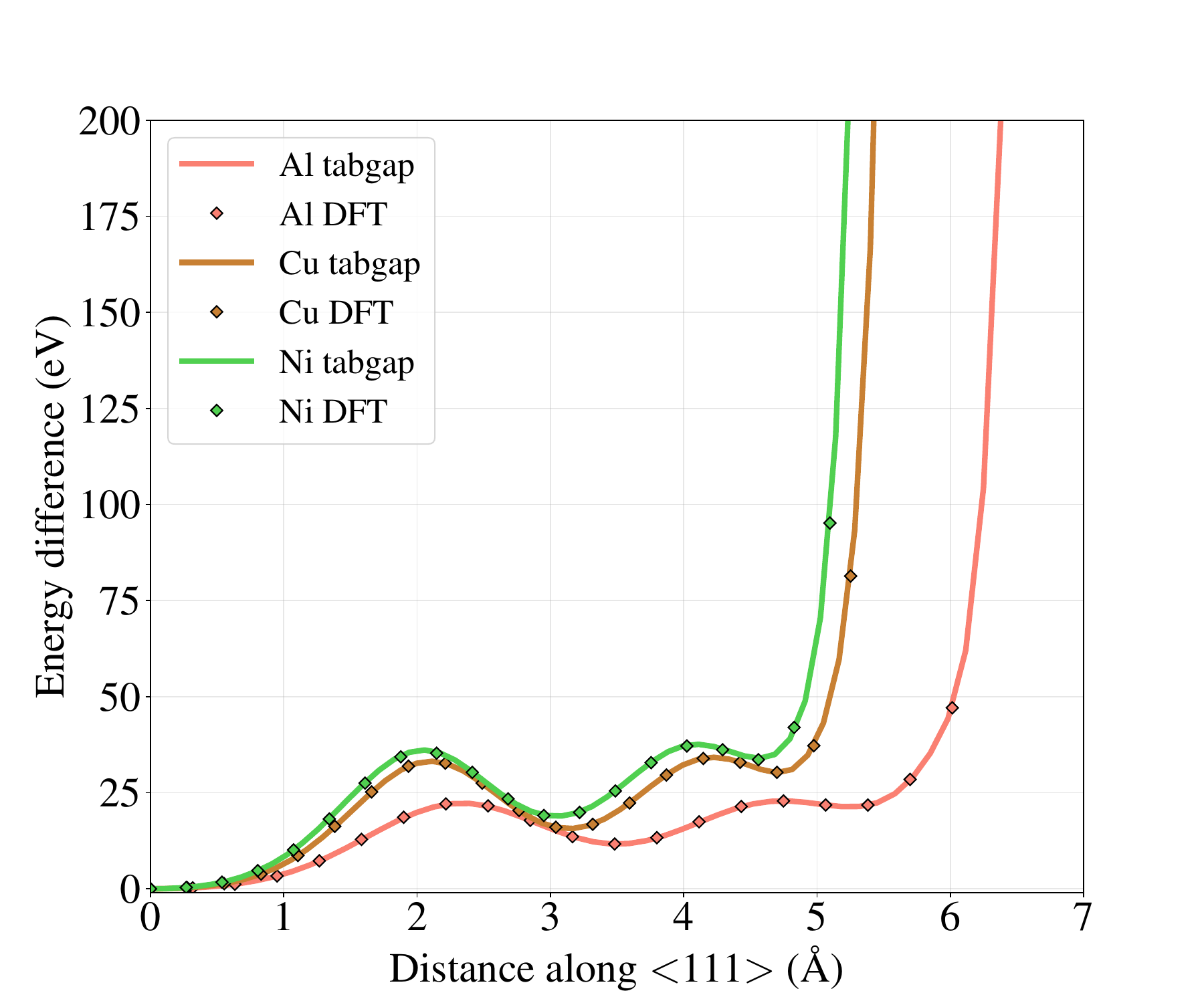}
    \includegraphics[width=0.3\linewidth]{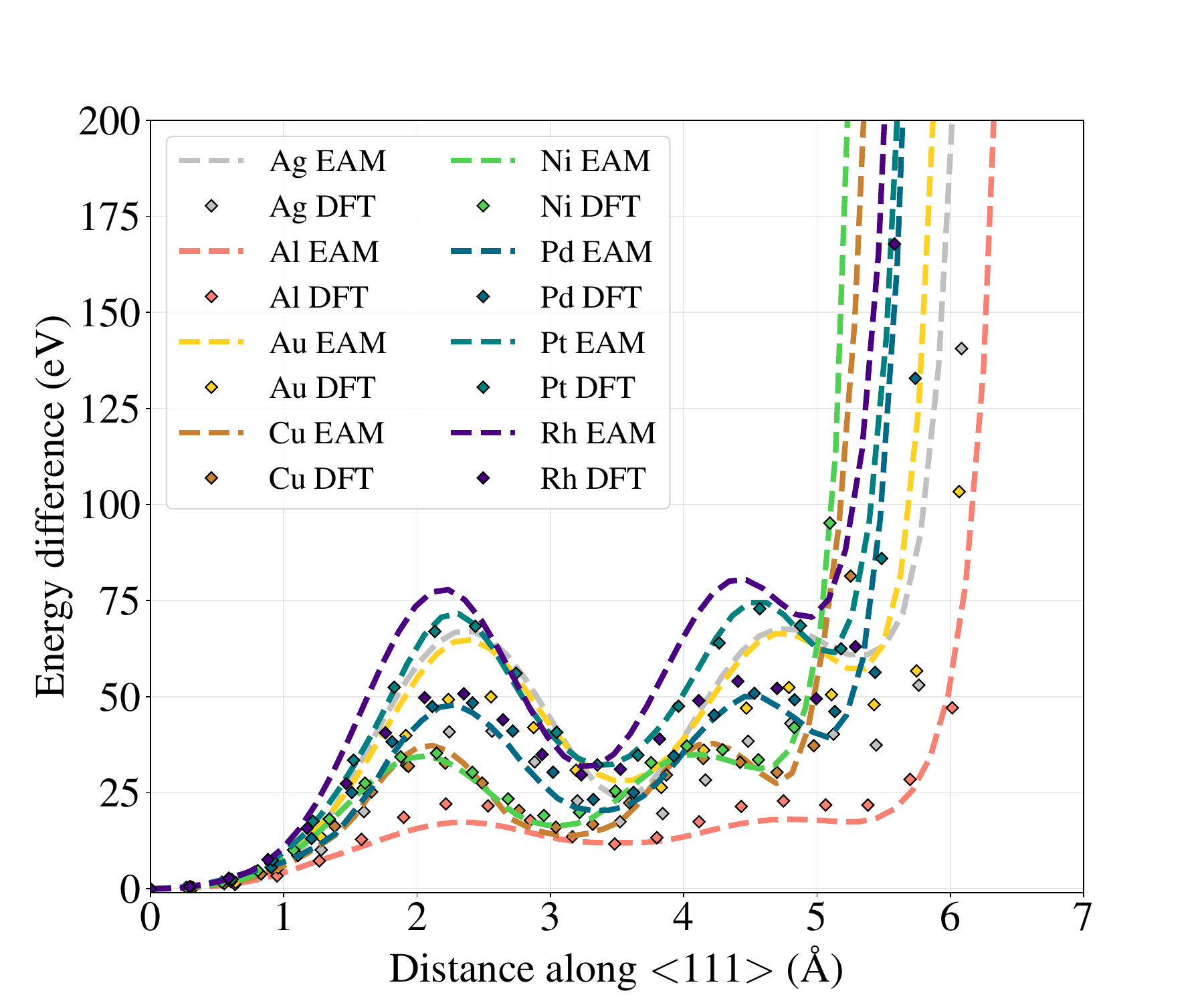}
    \includegraphics[width=0.3\linewidth]{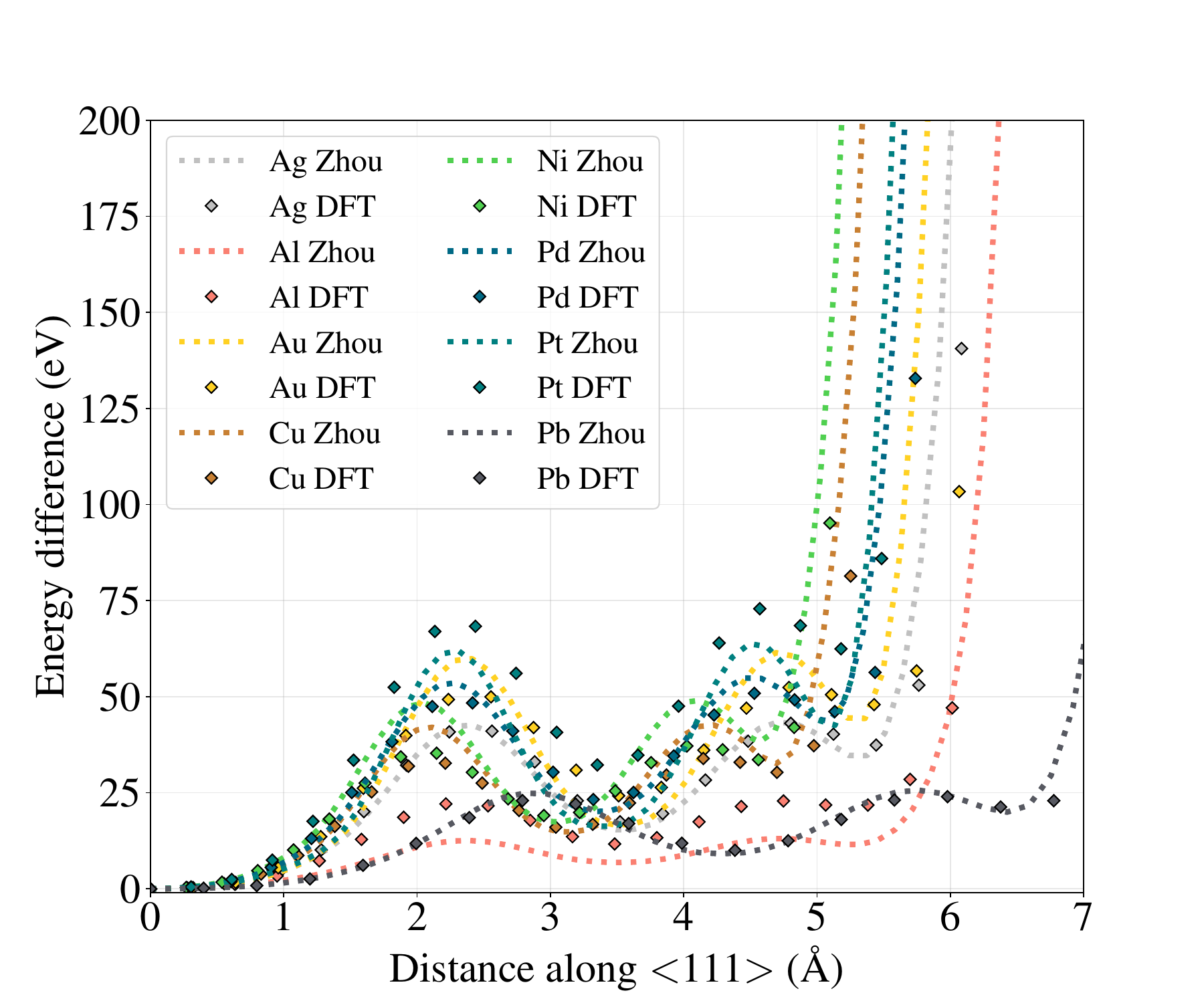}
    \includegraphics[width=0.3\linewidth]{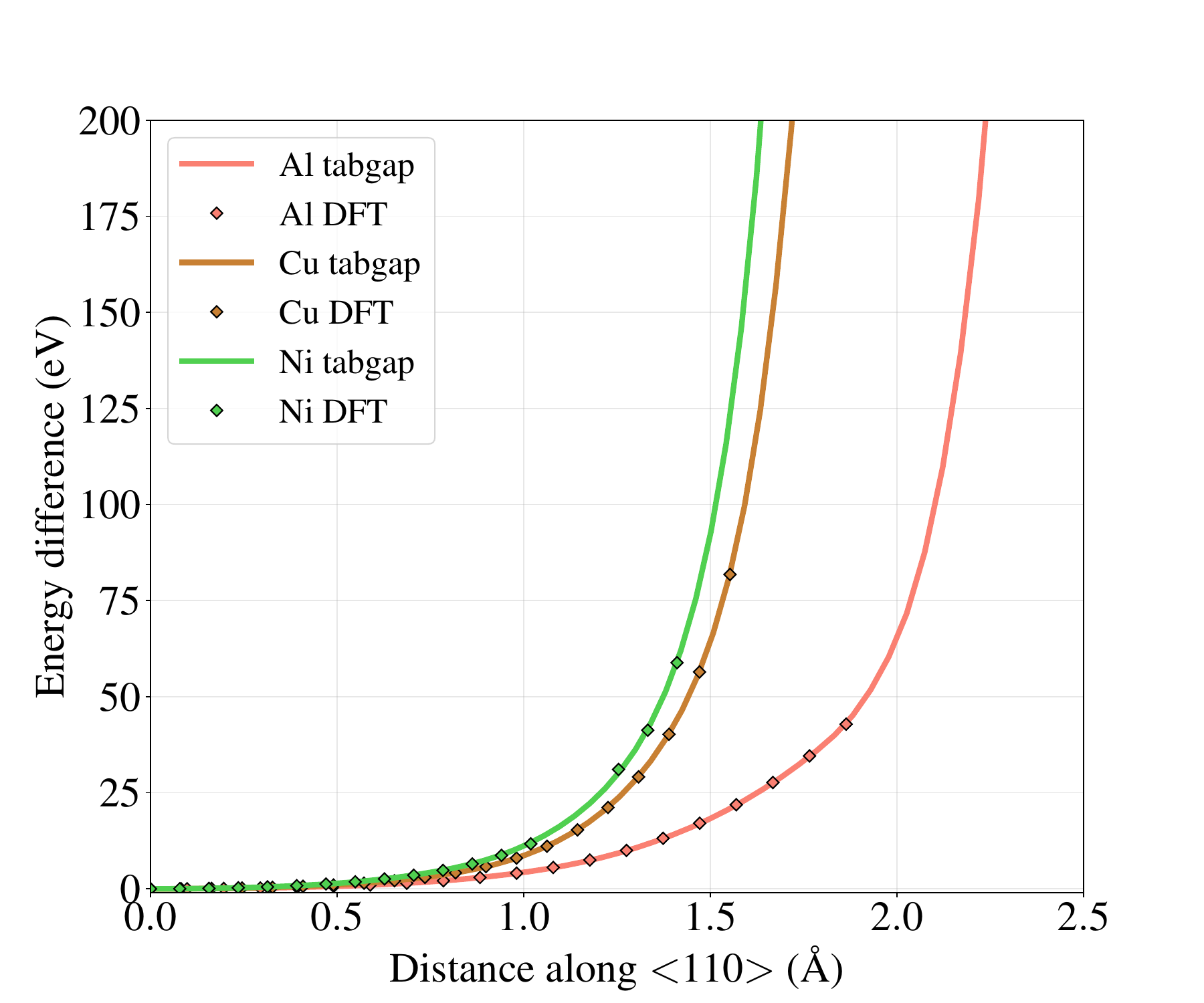}
    \includegraphics[width=0.3\linewidth]{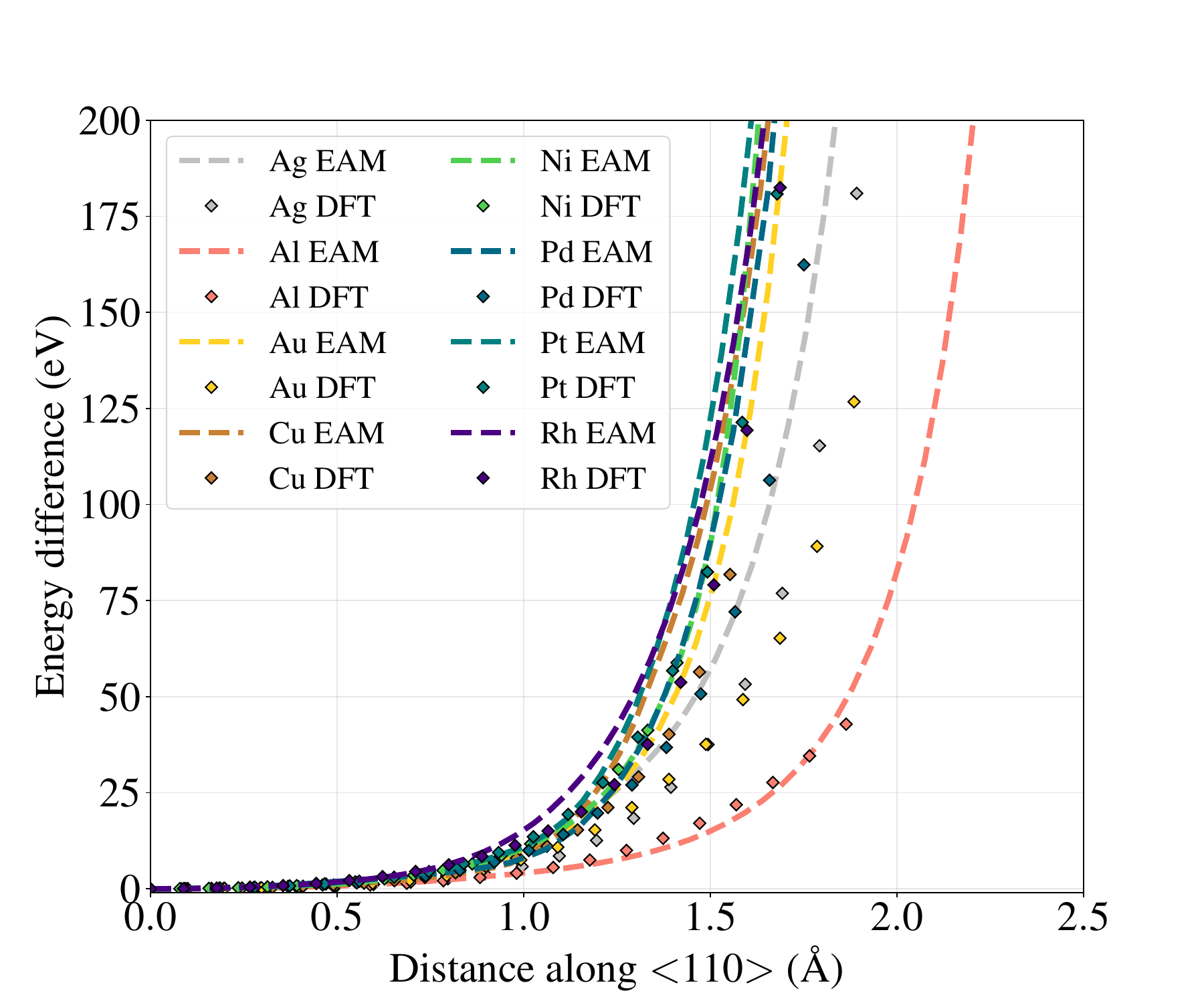}
    \includegraphics[width=0.3\linewidth]{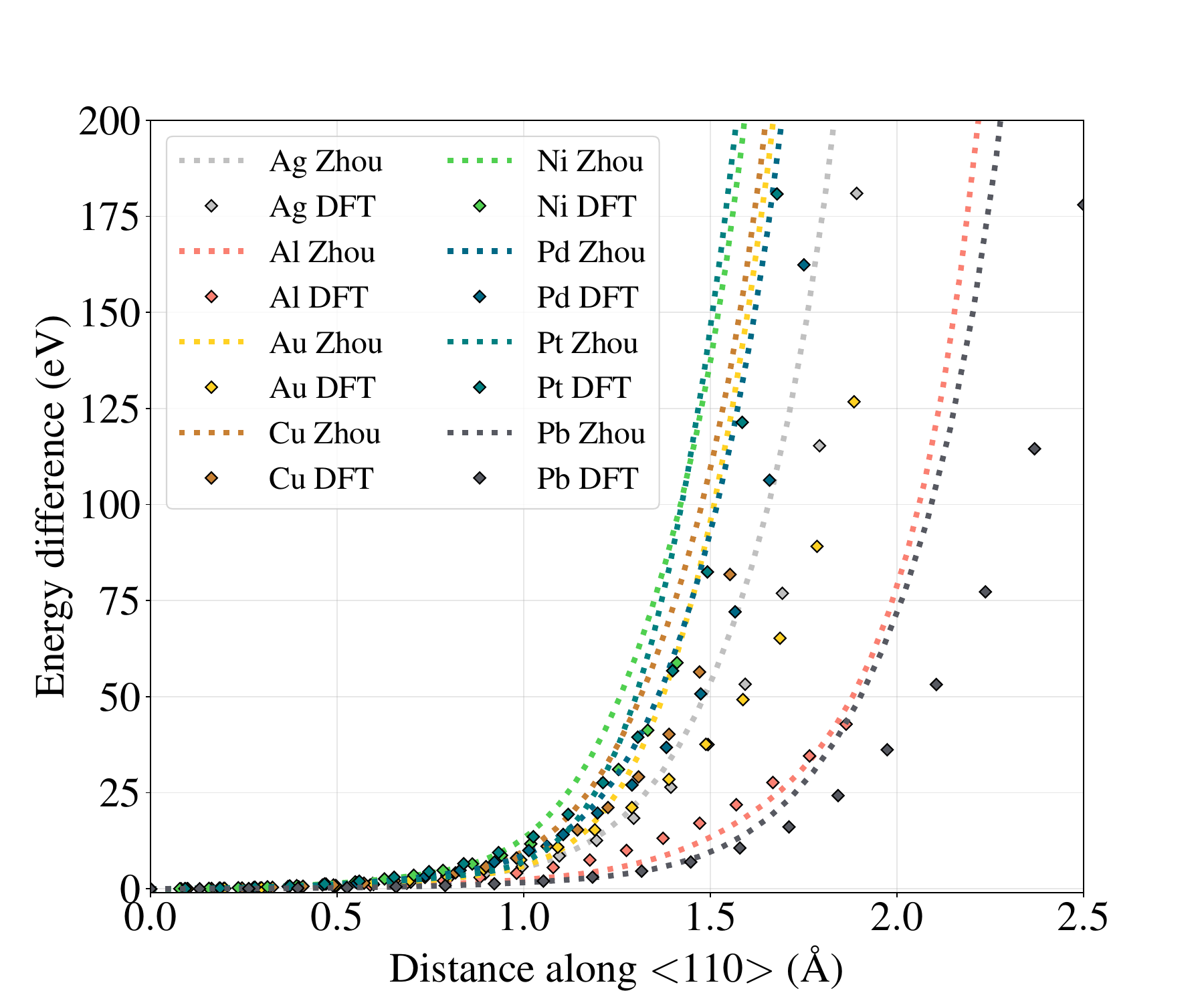}
    \includegraphics[width=0.3\linewidth]{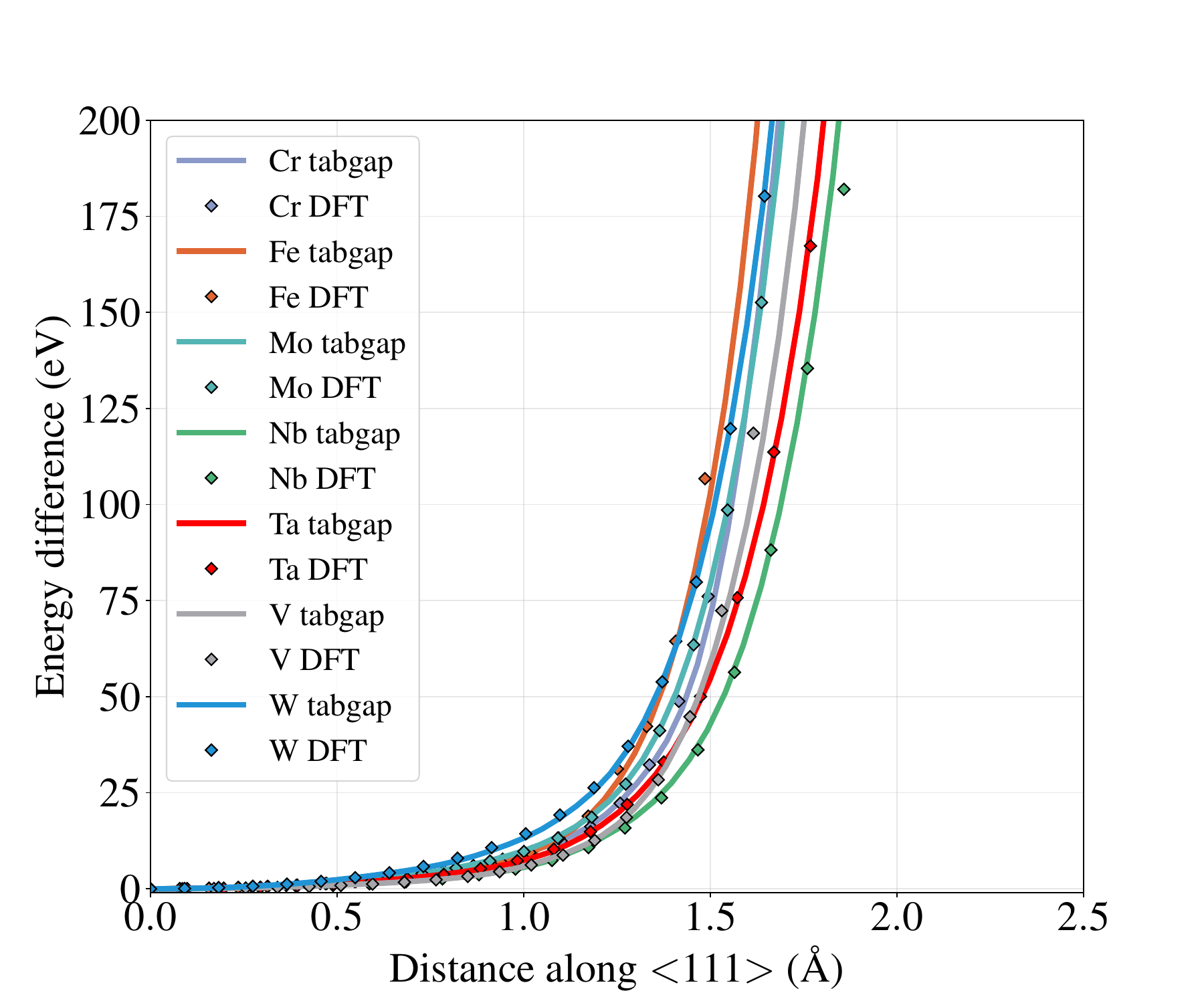}
    \includegraphics[width=0.3\linewidth]{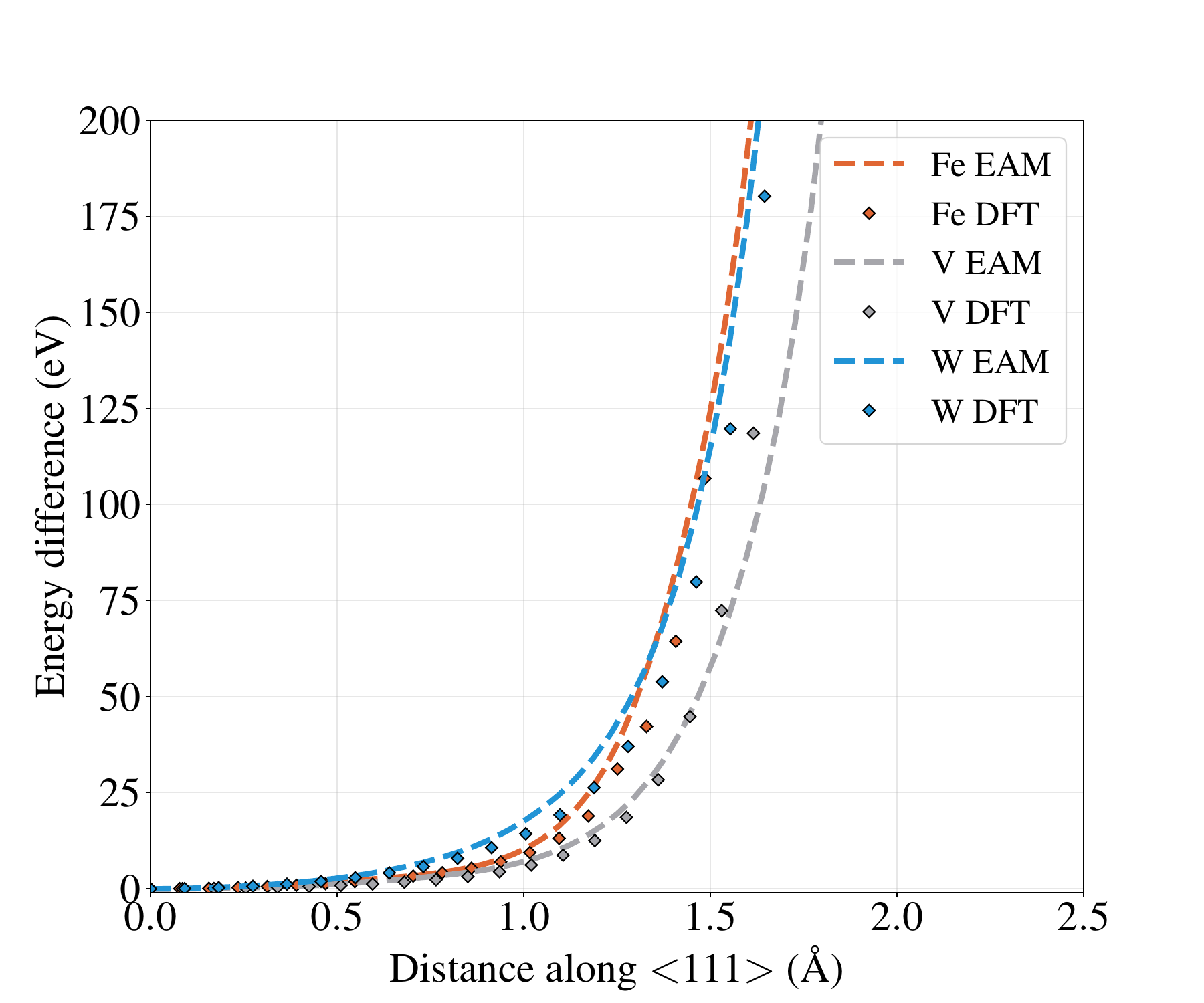}
    \includegraphics[width=0.3\linewidth]{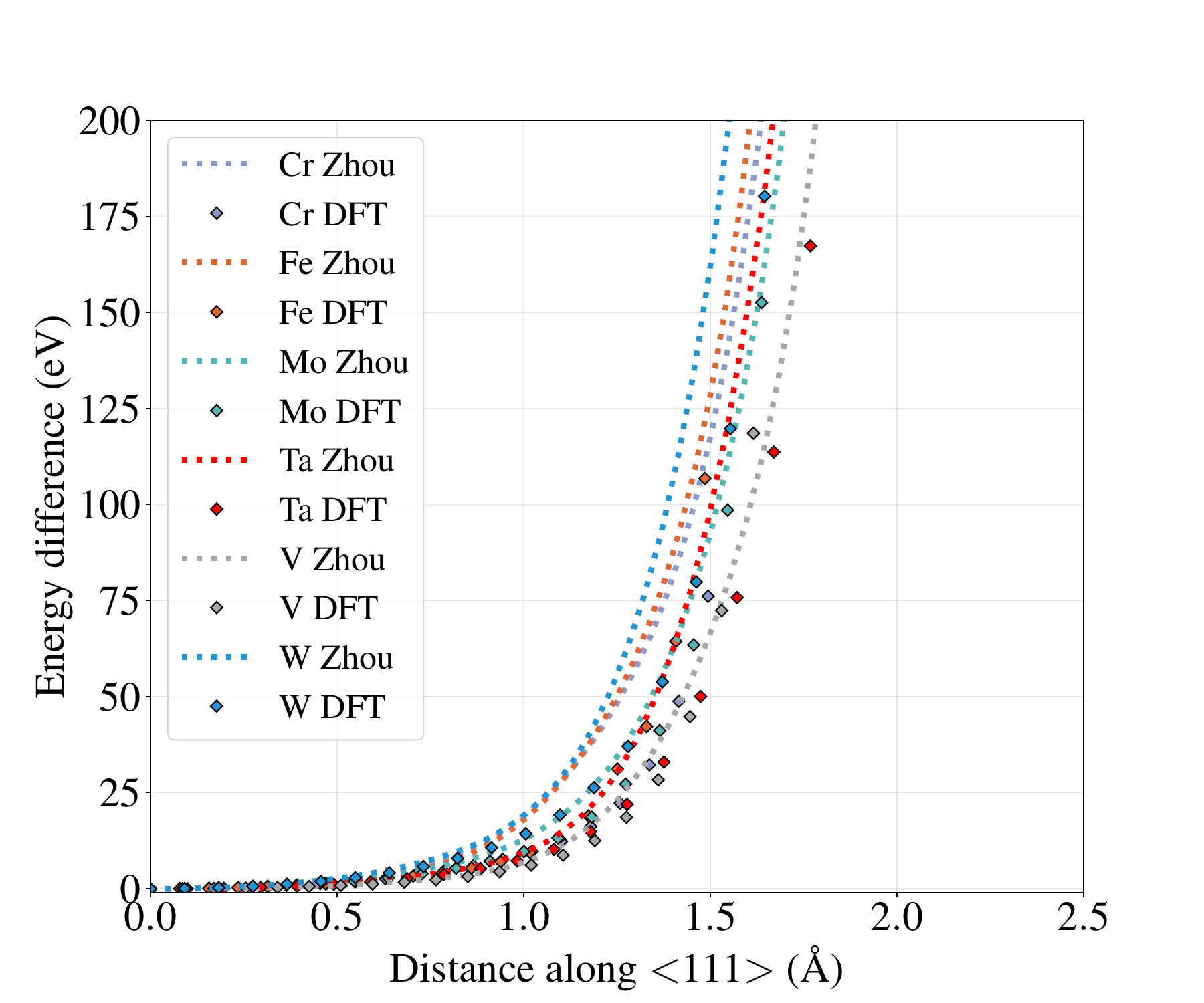}
    \includegraphics[width=0.3\linewidth]{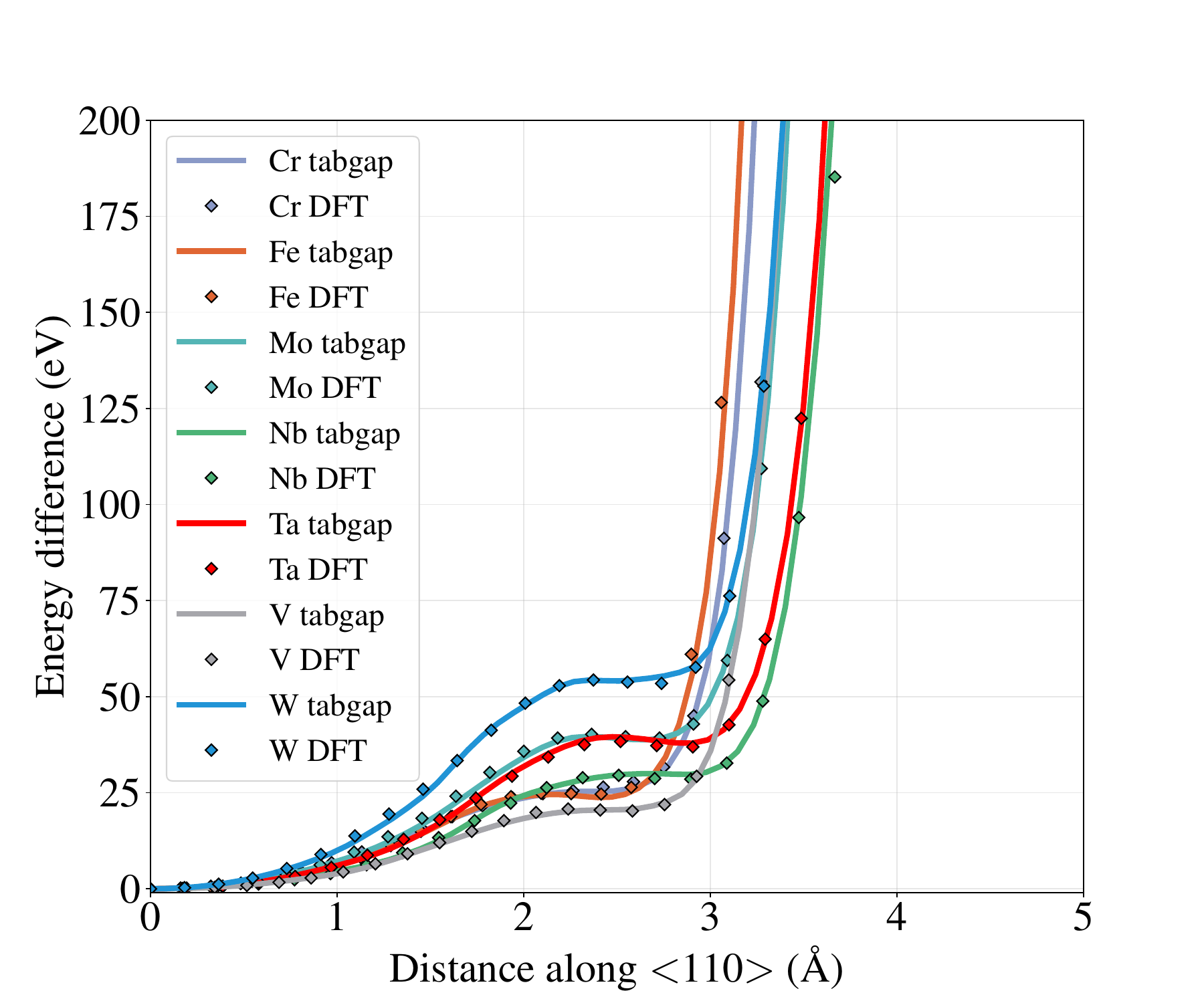}
    \includegraphics[width=0.3\linewidth]{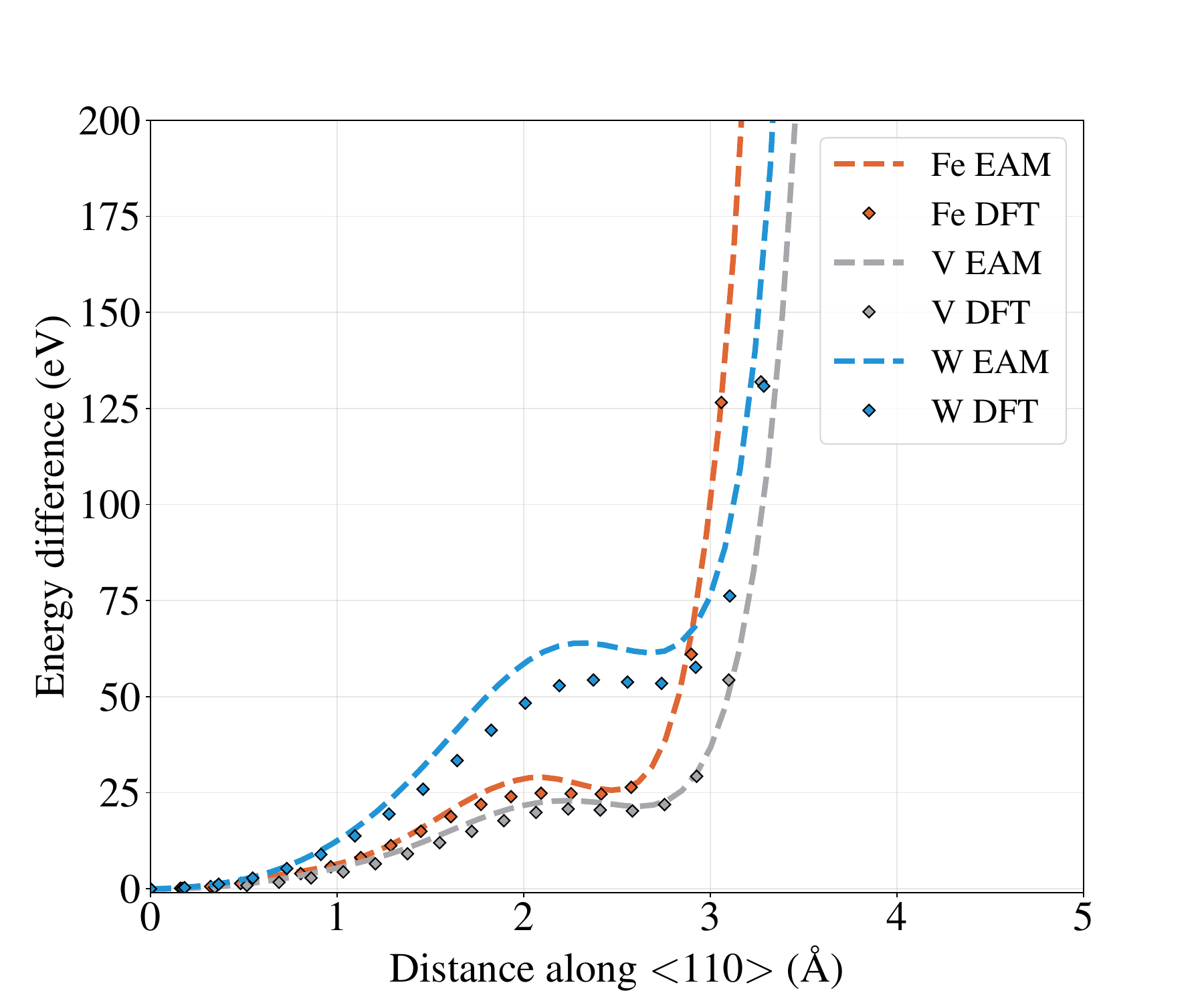}
    \includegraphics[width=0.3\linewidth]{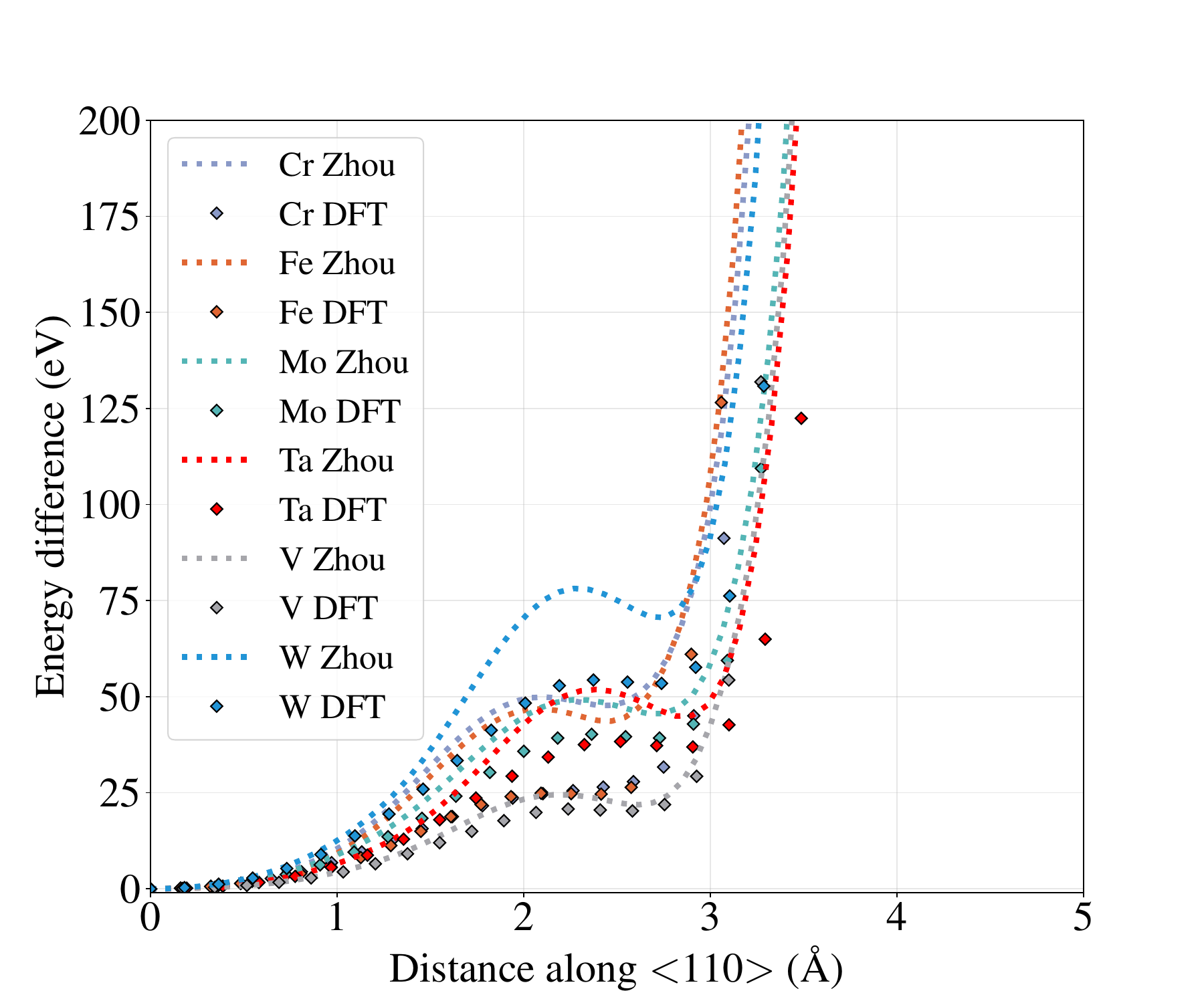}
    \includegraphics[width=0.3\linewidth]{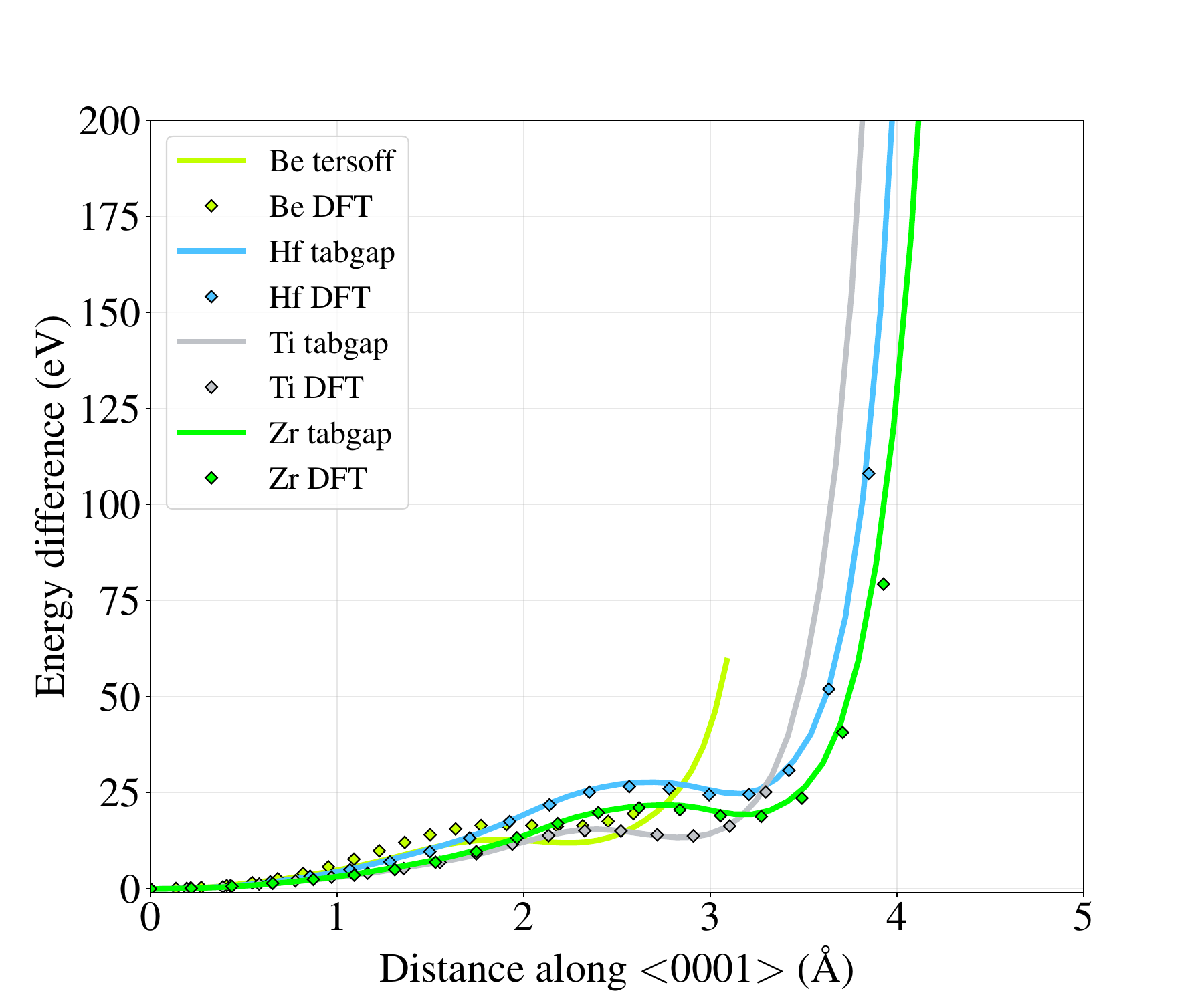}
    \includegraphics[width=0.3\linewidth]{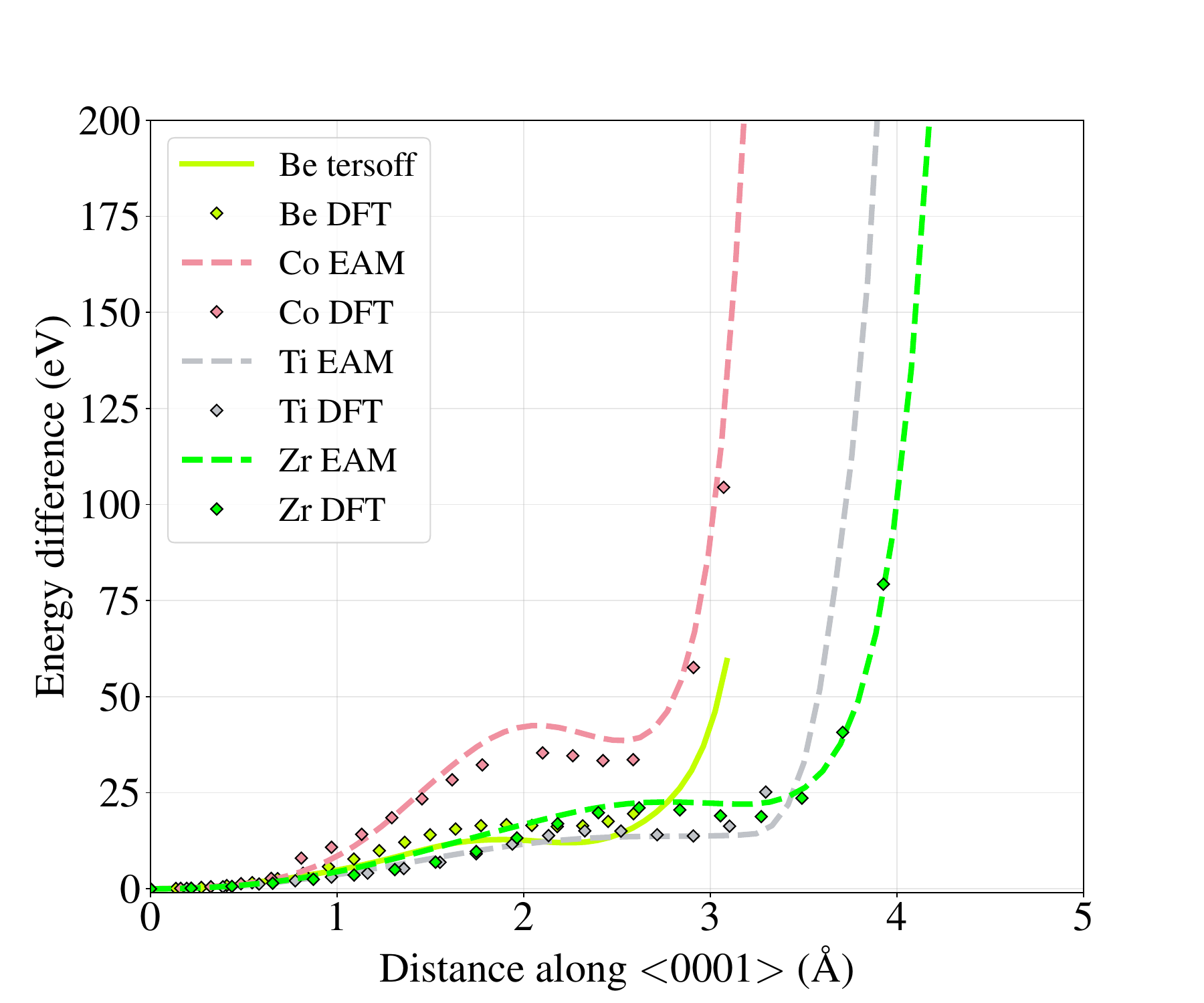}
    \includegraphics[width=0.3\linewidth]{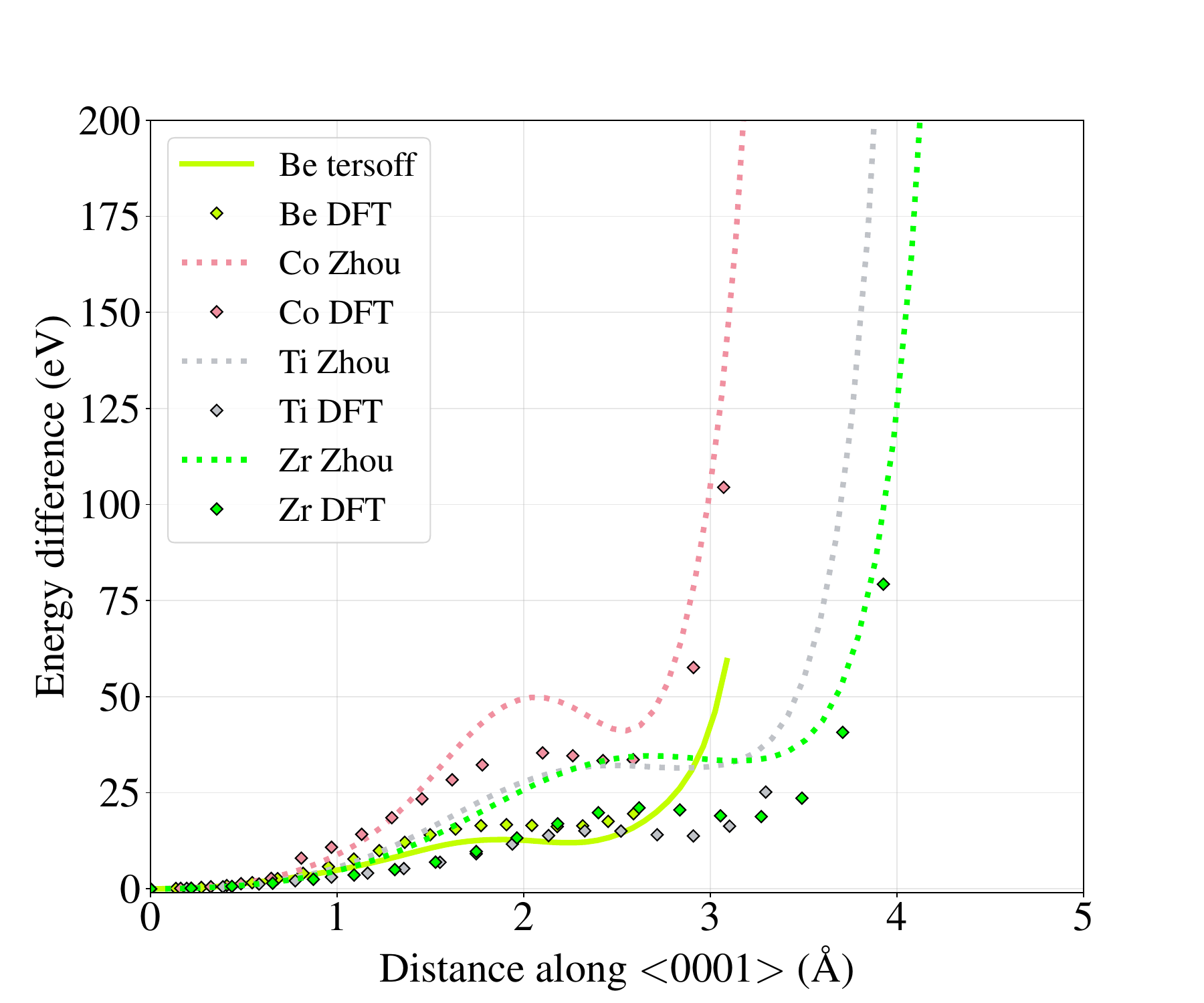}
    \caption{Quasi static drag (QSD) curves (moving an atom in a high symmetry direction while keeping all other atoms stationary) in the different potentials and in different directions. The DFT reference points were calculated using the \textsc{VASP} DFT code~\cite{kresse_ab_1993,kresse_ab_1994,kresse_efficiency_1996,kresse_efficient_1996}. The calculations used the PBE GGA exchange-correlation functional~\cite{perdew_generalized_1996} and projector augmented wave (PAW) pseudopotentials (\texttt{Be\_sv}, \texttt{Al}, \texttt{Ti\_sv}, \texttt{V\_sv}, \texttt{Cr\_pv}, \texttt{Fe\_sv}, \texttt{Co\_sv}, \texttt{Ni\_pv}, \texttt{Cu\_pv}, \texttt{Zr\_sv}, \texttt{Nb\_sv}, \texttt{Mo\_sv}, \texttt{Rh\_pv}, \texttt{Pd\_pv}, \texttt{Ag\_pv}, \texttt{Hf\_sv}, \texttt{Ta\_pv}, \texttt{W\_sv}, \texttt{Pt\_pv}, \texttt{Au}, \texttt{Pb\_d}). The plane-wave expansion energy cutoff was 650 eV for all elements. $k$-points were defined using $\Gamma$-centered Monkhorst–Pack grids~\cite{monkhorst_special_1976} with maximum $k$-spacing of 0.15 Å$^{-1}$. Additionally, first order Methfessel-Paxton smearing of 0.1 eV was applied~\cite{methfessel_high-precision_1989}. Structure optimization was performed initially before QSD. }
\end{figure}

\clearpage

\section{Crude estimation of primary cascade damage}

\begin{table}[ht!]
    \centering
    \begin{tabular}{c|c|c}
    \toprule
    \toprule
    Element & potential & $N_\mathrm{vac.}$ \\
    \midrule
Ag & EAM & 8.667 $\pm$ 2.494 \\
Ag & Zhou & 2.667 $\pm$ 1.247 \\
Al & EAM & 24.000 $\pm$ 1.414 \\
Al & Zhou & 16.333 $\pm$ 5.249 \\
Al & tabgap & 15.667 $\pm$ 4.028 \\
Au & EAM & 4.333 $\pm$ 4.028 \\
Au & Zhou & 2.333 $\pm$ 1.247 \\
Cu & EAM & 7.333 $\pm$ 3.771 \\
Cu & Zhou & 4.333 $\pm$ 1.247 \\
Cu & tabgap & 11.333 $\pm$ 1.700 \\
Ni & EAM & 7.333 $\pm$ 1.700 \\
Ni & Zhou & 7.000 $\pm$ 2.449 \\
Ni & tabgap & 9.000 $\pm$ 1.414 \\
Pb & Zhou & 7.000 $\pm$ 1.414 \\
Pd & EAM & 3.667 $\pm$ 0.471 \\
Pd & Zhou & 4.000 $\pm$ 0.816 \\
Pt & EAM & 5.667 $\pm$ 1.247 \\
Pt & Zhou & 3.333 $\pm$ 1.247 \\
Rh & EAM & 5.667 $\pm$ 3.300 \\
Cr & Zhou & 13.667 $\pm$ 2.625 \\
Cr & tabgap & 14.000 $\pm$ 3.266 \\
Fe & EAM & 17.667 $\pm$ 0.471 \\
Fe & Zhou & 7.667 $\pm$ 3.771 \\
Fe & tabgap & 18.333 $\pm$ 0.943 \\
Mo & Zhou & 9.000 $\pm$ 1.414 \\
Mo & tabgap & 6.333 $\pm$ 1.700 \\
Nb & tabgap & 7.667 $\pm$ 1.247 \\
Ta & Zhou & 11.333 $\pm$ 2.357 \\
Ta & tabgap & 7.333 $\pm$ 0.943 \\
V & EAM & 21.333 $\pm$ 0.471 \\
V & Zhou & 17.000 $\pm$ 2.944 \\
V & tabgap & 8.667 $\pm$ 2.357 \\
W & EAM & 5.667 $\pm$ 0.943 \\
W & Zhou & 8.333 $\pm$ 2.625 \\
W & tabgap & 6.667 $\pm$ 0.943 \\
Co & EAM & 13.000 $\pm$ 3.742 \\
Co & Zhou & 9.000 $\pm$ 2.160 \\
Ti & EAM & 16.667 $\pm$ 1.247 \\
Ti & Zhou & 9.000 $\pm$ 1.633 \\
Ti & tabgap & 21.667 $\pm$ 2.867 \\
Hf & tabgap & 9.667 $\pm$ 2.055 \\
Zr & EAM & 11.333 $\pm$ 0.471 \\
Zr & Zhou & 13.000 $\pm$ 0.816 \\
Zr & tabgap & 15.333 $\pm$ 2.055 \\
Be & tersoff & 61.333 $\pm$ 7.363 \\
    \bottomrule
    \bottomrule
    \end{tabular}
    \caption{Number of vacancies produced in a single 5 keV cascade in pristine material. Here the value is calculated as the average of the absolutely first cascades from the three different runs thus ensuring no cascade overlap. This gives us a crude first degree approximation of the number of vacancies produced by primary damage in 5 keV cascades in each potential. Obviously the statistics here is very limited and as such these values should not be viewed as more than indicative. For a proper estimation one should perform several more primary cascades. These values given here mainly due to curiosity.}
    \label{tab:prim}
\end{table}

\clearpage

\section{Icosahedral atoms in BCC materials}

\begin{table}[ht!]
    \centering
    \begin{tabular}{c|c|c|c|c|c}
    \toprule
    \toprule
    Element & potential & $N_\mathrm{ico.}^{\mathrm{run-1}}$ & $N_\mathrm{ico.}^{\mathrm{run-2}}$ & $N_\mathrm{ico.}^{\mathrm{run-3}}$ & $N_\mathrm{ico.}^{\mathrm{avg.}}$ \\
    \midrule
Cr & Zhou & 18 & 43 & 20 & 27 \\
Cr & tabgap & 0 & 3 & 0 & 1 \\
Fe & EAM & 255 & 339 & 228 & 274 \\
Fe & Zhou & 0 & 0 & 12 & 4 \\
Fe & tabgap & 199 & 237 & 278 & 238 \\
Mo & Zhou & 178 & 143 & 107 & 143 \\
Mo & tabgap & 0 & 0 & 0 & 0 \\
Nb & tabgap & 0 & 0 & 0 & 0  \\
Ta & Zhou & 3 & 2 & 15 & 7 \\
Ta & tabgap & 3 & 1 & 2 & 2 \\
V & EAM & 32 & 39 & 25 & 32 \\
V & Zhou & 1 & 0 & 0 & 1 \\
V & tabgap & 0 & 0 & 0 & 0 \\
W & EAM & 21 & 0 & 11 & 11 \\
W & Zhou & 109 & 80 & 132 & 107 \\
W & tabgap & 0 & 0 & 3 & 1 \\
    \bottomrule
    \bottomrule
    \end{tabular}
    \caption{Number of icosahedral atoms as identified by polyhedral template matching (PTM) as implemented in OVITO. The last column represents the average number of icosahedral atom rounded up to the nearest integer (ceil).}
    \label{tab:ico}
\end{table}

\clearpage

\section{$c$-type dislocations in Ti  and Zr}

\begin{figure}[ht!]
    \centering
\begin{subfigure}{0.29\textwidth}
    \includegraphics[width=1.1\textwidth]{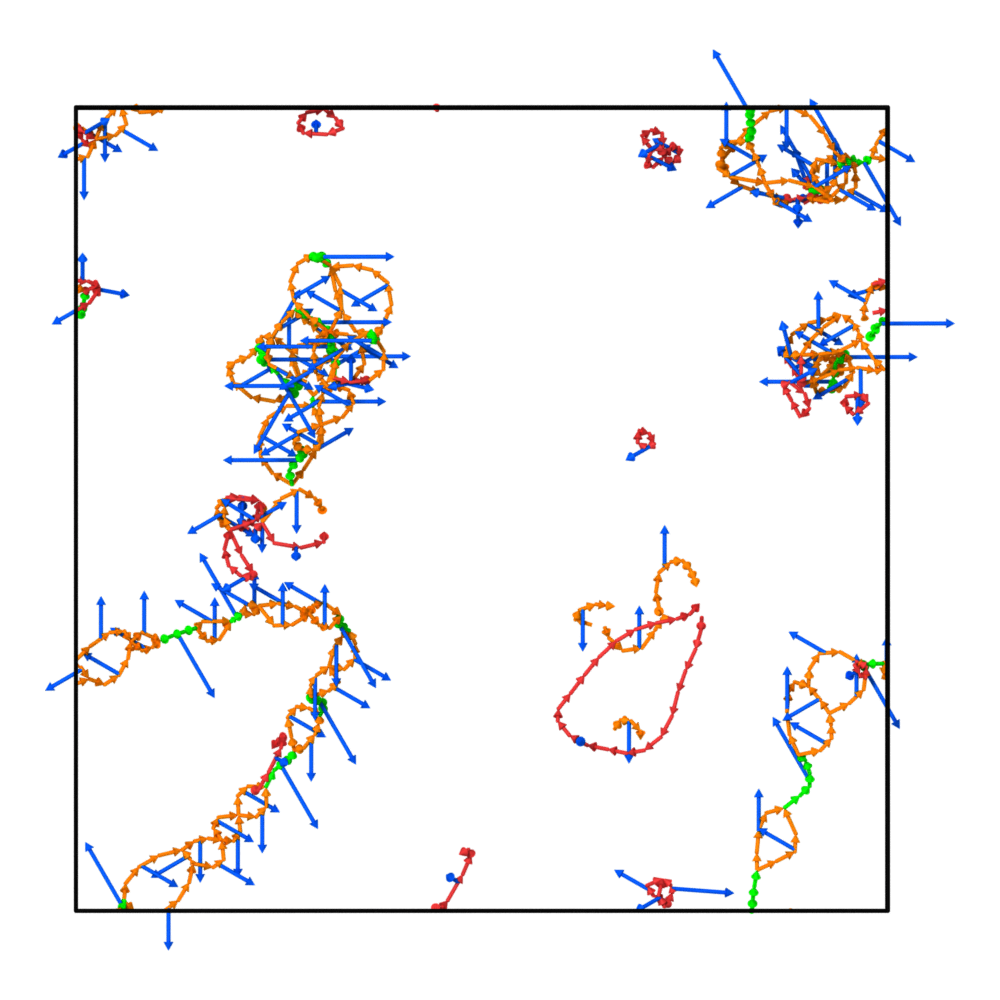}
      \caption{Final frame run \#1 no conjugate gradient minimization}
\end{subfigure}
\begin{subfigure}{0.29\textwidth}
    \includegraphics[width=1.1\textwidth]{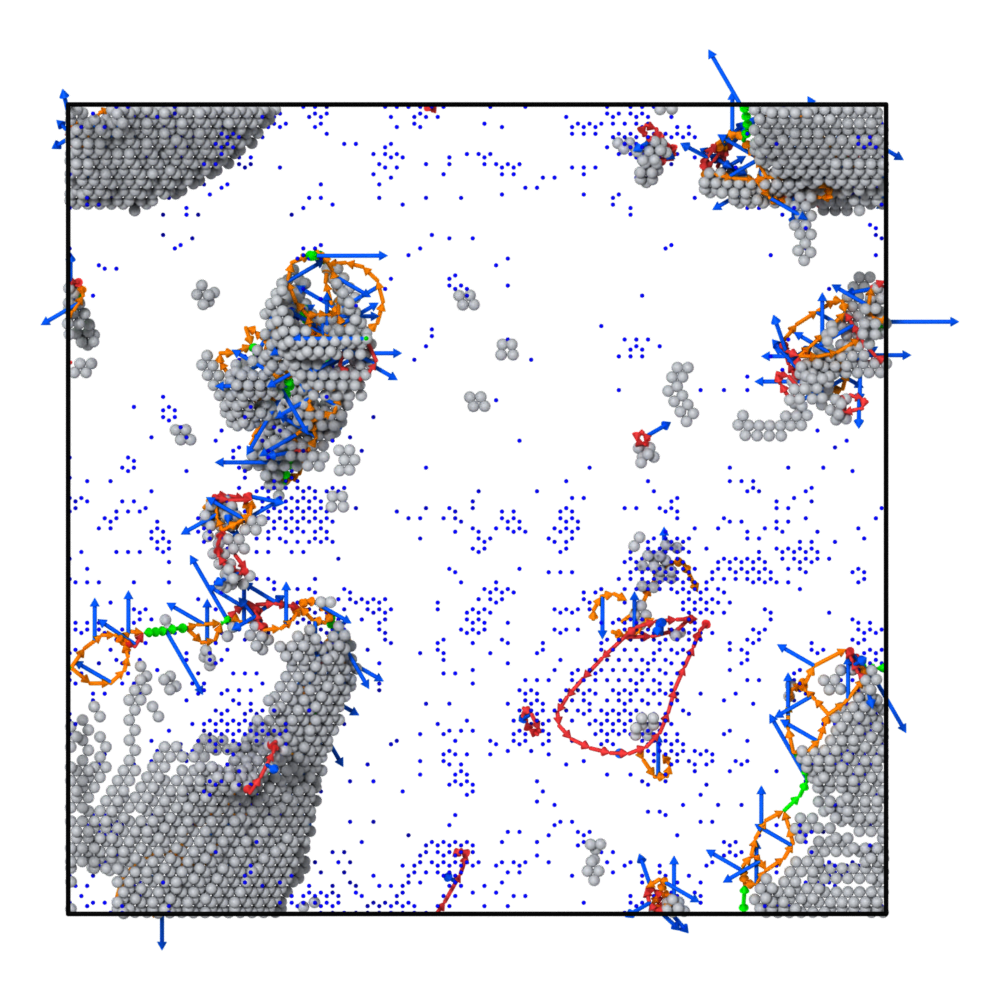}
    \caption{Final frame run \#1 after conjugate gradient minimization}
\end{subfigure}
    \caption{Final configuration after 2000 overlapping collision cascades. Small blue spheres represent vacancies, gray spheres represent interstitial atoms, green lines are $1/3$ \hkl<1-210> dislocations, blue lines are \hkl<0001> dislocations, magenta lines are \hkl<1-100> dislocations, orange lines are $1/3$\hkl<1-100> dislocations, yellow lines are $1/3$ \hkl<1-213> dislocations and red lines are dislocations with type ''other'' as identified by DXA analysis. Blue arrows illustrate the Burgers vectors. (a) only DXA without minimization (b) DXA and point defects with minimization}
\end{figure}

\begin{figure}[ht!]
    \centering
\begin{subfigure}{0.25\textwidth}
    \includegraphics[width=1.1\textwidth]{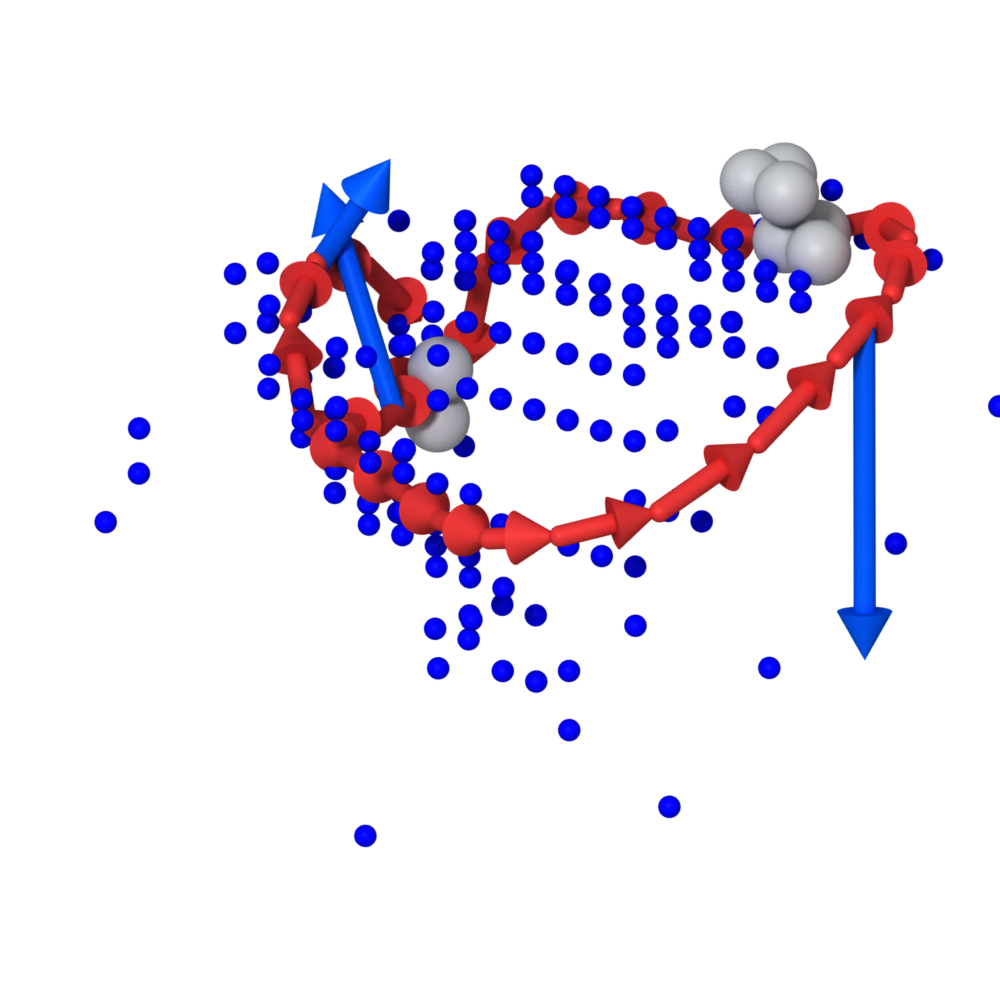}
      \caption{Ortho}
\end{subfigure}
\begin{subfigure}{0.25\textwidth}
    \includegraphics[width=1.1\textwidth]{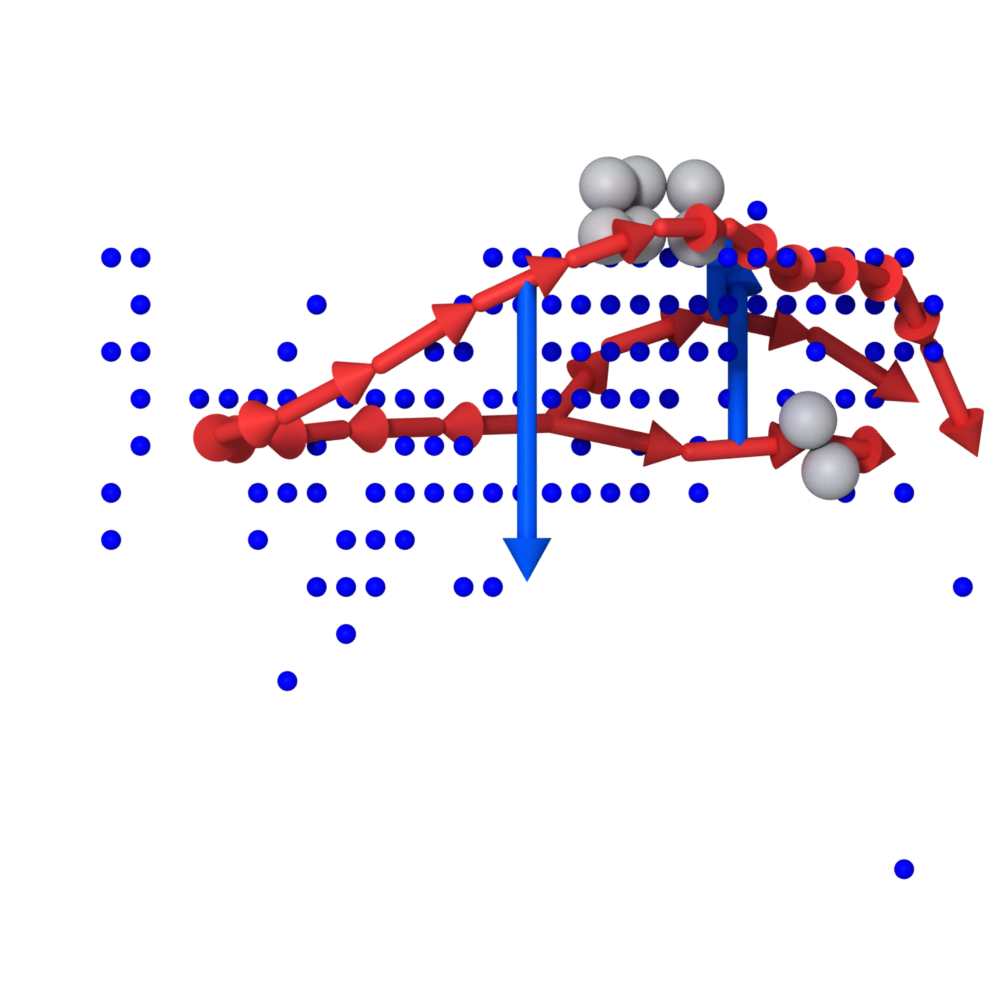}
      \caption{Front}
\end{subfigure}
\begin{subfigure}{0.25\textwidth}
    \includegraphics[width=1.1\textwidth]{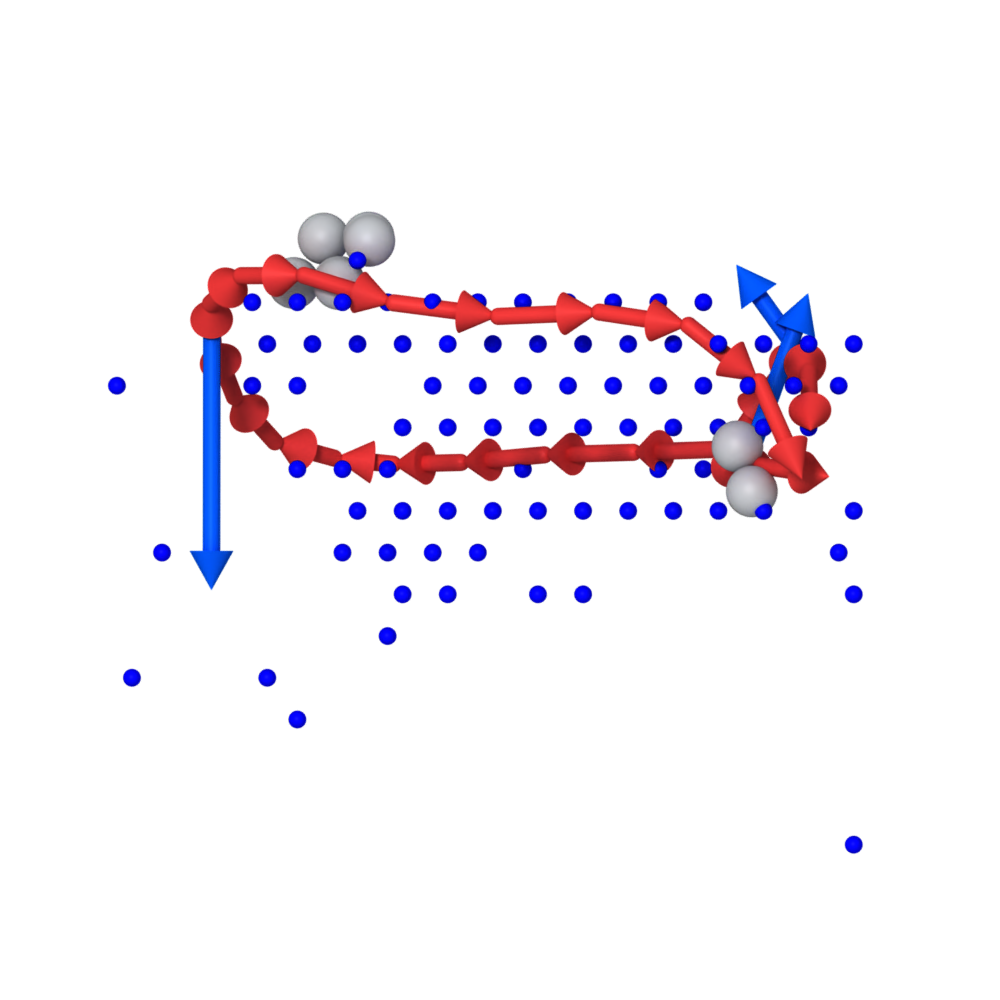}
      \caption{Right}
\end{subfigure}
    \caption{$c$-type dislocation in Ti from different viewing angles after conjugate gradient minimization. Small blue spheres represent vacancies, gray spheres represent interstitial atoms. Red lines are dislocations with type ''other'' as identified by DXA analysis. Blue arrows illustrate the Burgers vectors.}
\end{figure}

\begin{figure}[ht!]
    \centering
\begin{subfigure}{0.25\textwidth}
    \includegraphics[width=1.1\textwidth]{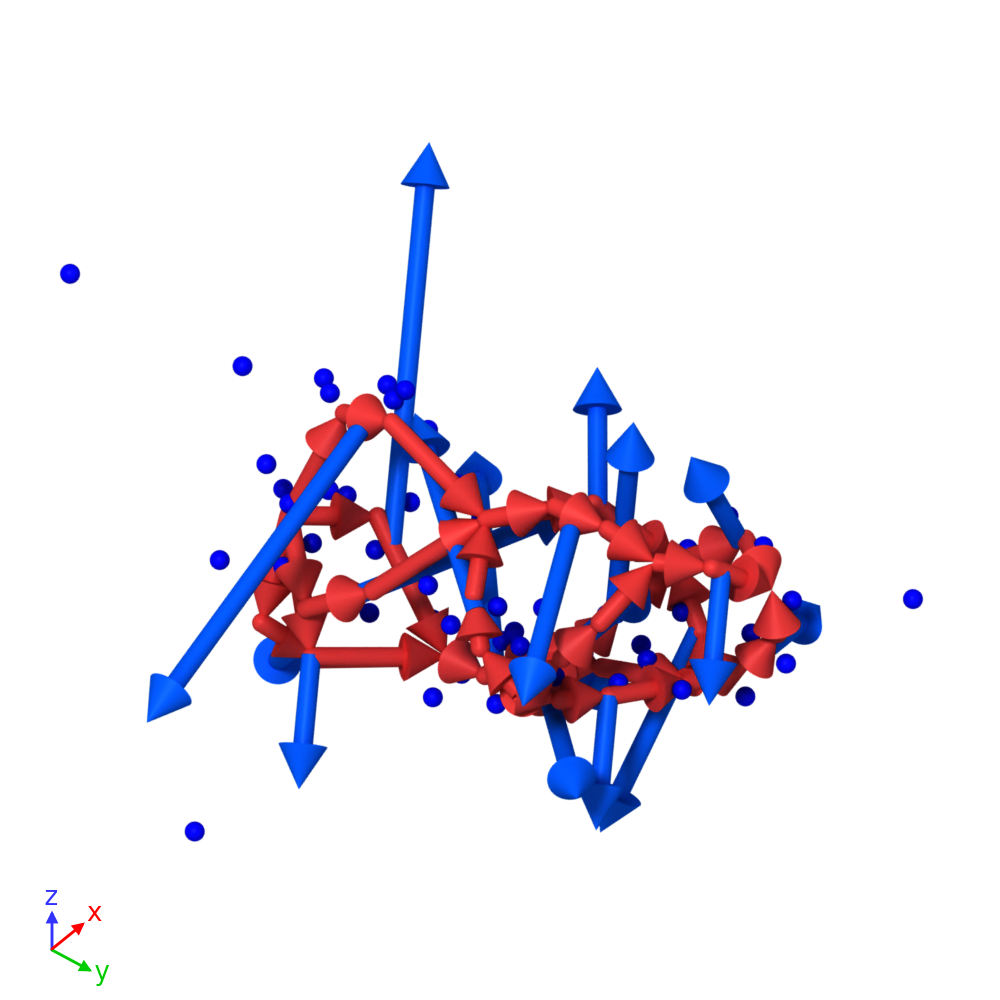}
      \caption{Ortho}
\end{subfigure}
\begin{subfigure}{0.25\textwidth}
    \includegraphics[width=1.1\textwidth]{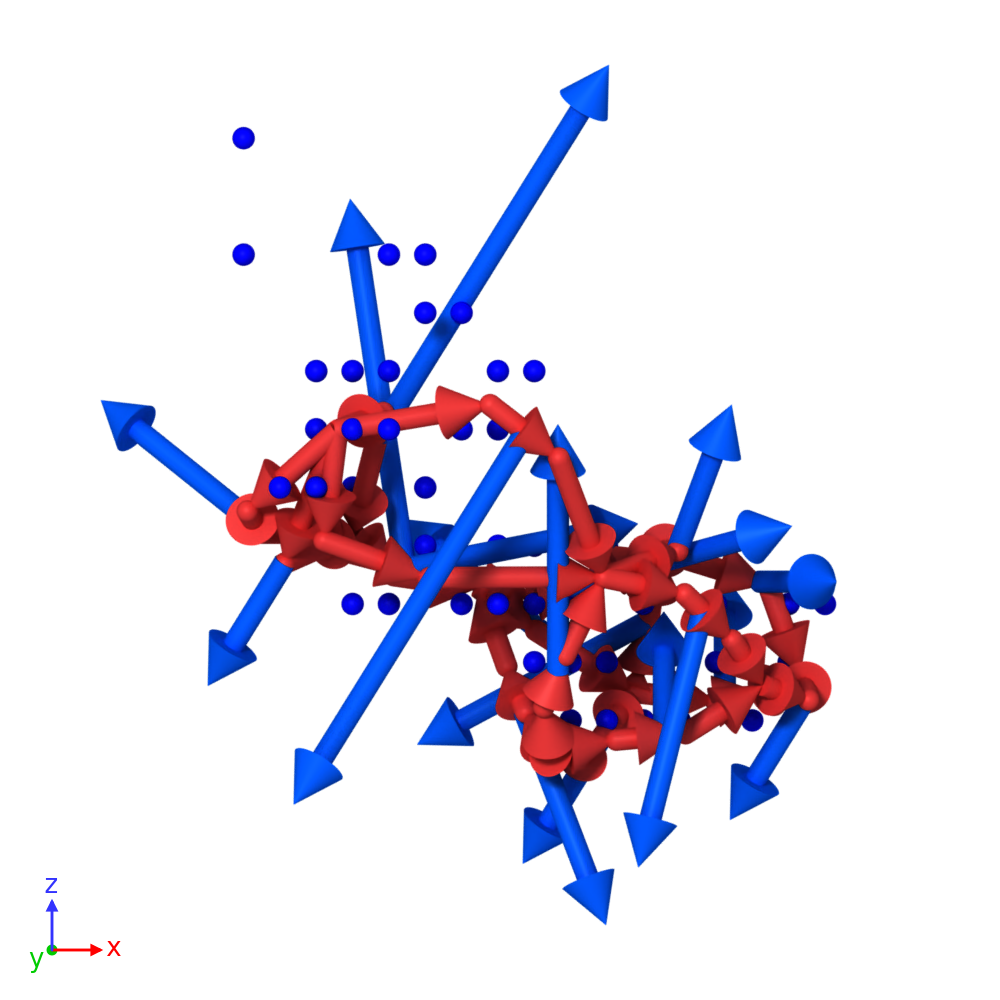}
      \caption{Front}
\end{subfigure}
\begin{subfigure}{0.25\textwidth}
    \includegraphics[width=1.1\textwidth]{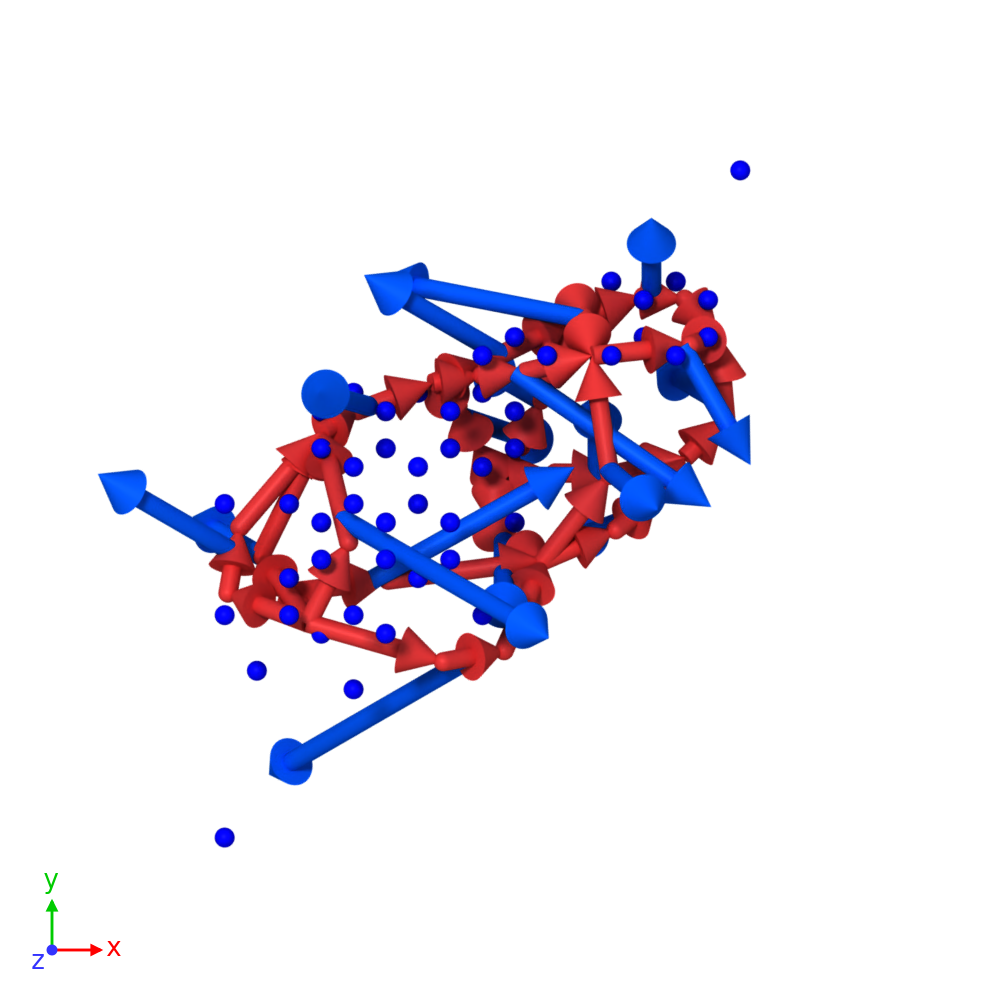}
      \caption{Top}
\end{subfigure}
    \caption{$c$-type dislocation in Zr from different viewing angles after conjugate gradient minimization. Small blue spheres represent vacancies, green spheres represent interstitial atoms. Red lines are dislocations with type ''other'' as identified by DXA analysis. Blue arrows illustrate the Burgers vectors.}
\end{figure}

\clearpage
\section{Comparison with material properties}

In this section we have compared the final defect concentrations (mean value of the last 20 data points) with various material properties such as: Atomic mass, melting point, density, minimum threshold displacement energy, stacking fault energy, temperatures of defect recovery stage I and III. The mass, melting point and density are taken from Ref.~\cite{Kim2021}. Experimental minimum threshold displacement energies are from Ref.~\cite{Jung01011978}. The stacking fault energies are from Ref.~\cite{10.1063/1.4870462}.  Recovery stage temperatures are from Ref.~\cite{zinkle20121}.

\begin{figure}[ht!]
    \centering
    \includegraphics[width=\linewidth]{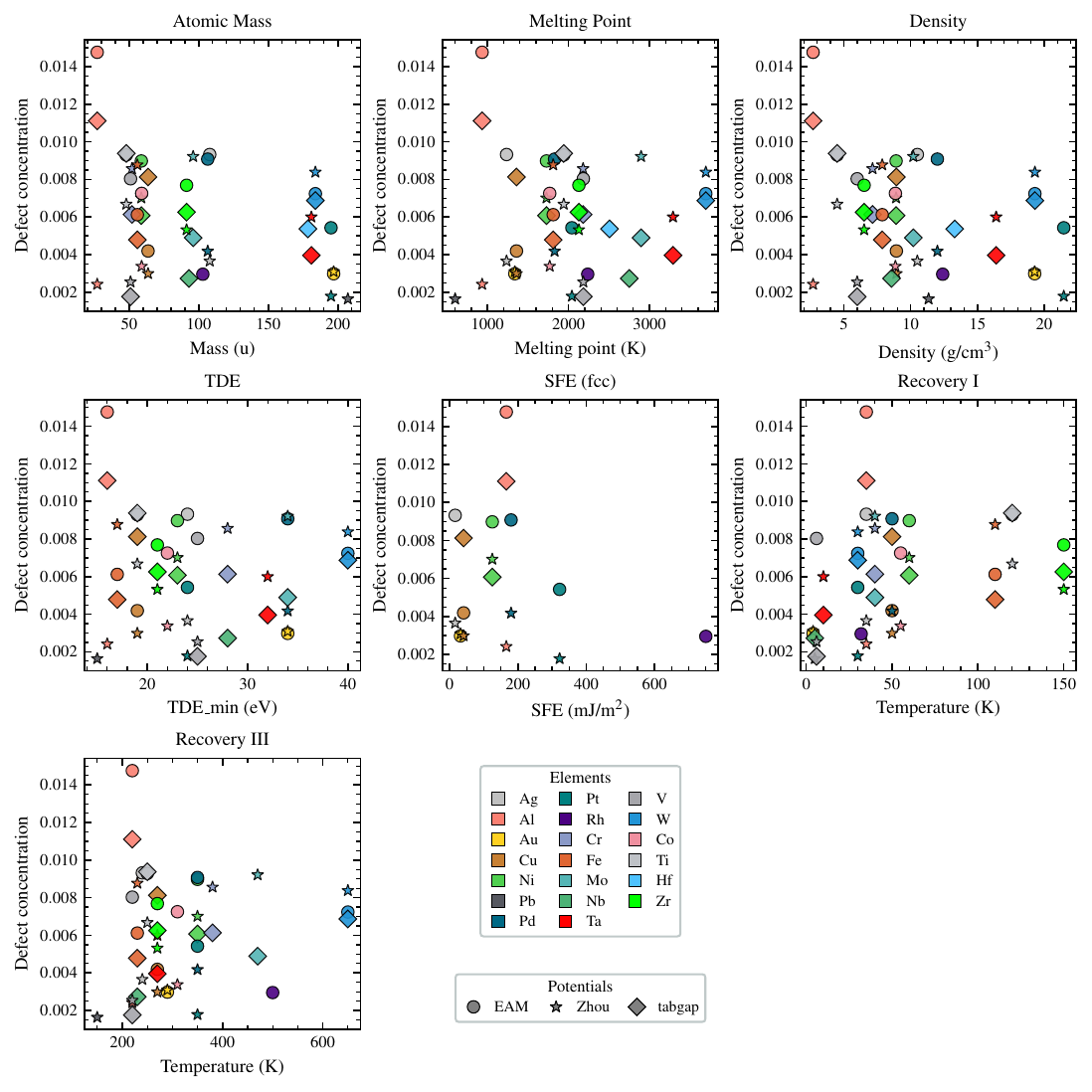}
    \caption{Comparison between final defect concentration (mean value of last 20 datapoints) with various material properties in all the potentials. The different marker styles indicate different potential chategories.}
    \label{fig:props}
\end{figure}

\clearpage
\section{Comparison with material properties: tabGAP}

In this section we have compared the final defect concentrations in the tabGAP potentials (mean value of the last 20 data points) with various material properties such as: Atomic mass, melting point, density, minimum threshold displacement energy, stacking fault energy, temperatures of defect recovery stage I and III. The mass, melting point and density are taken from Ref.~\cite{Kim2021}. Experimental minimum threshold displacement energies are from Ref.~\cite{Jung01011978}. The stacking fault energies are from Ref.~\cite{10.1063/1.4870462} .  Recovery stage temperatures are from Ref.~\cite{zinkle20121}. Vacancy migration energies are from the the respective publications of the potentials, with the exception of Cr that was calculated in this work. 

\begin{figure}[ht!]
    \centering
    \includegraphics[width=\linewidth]{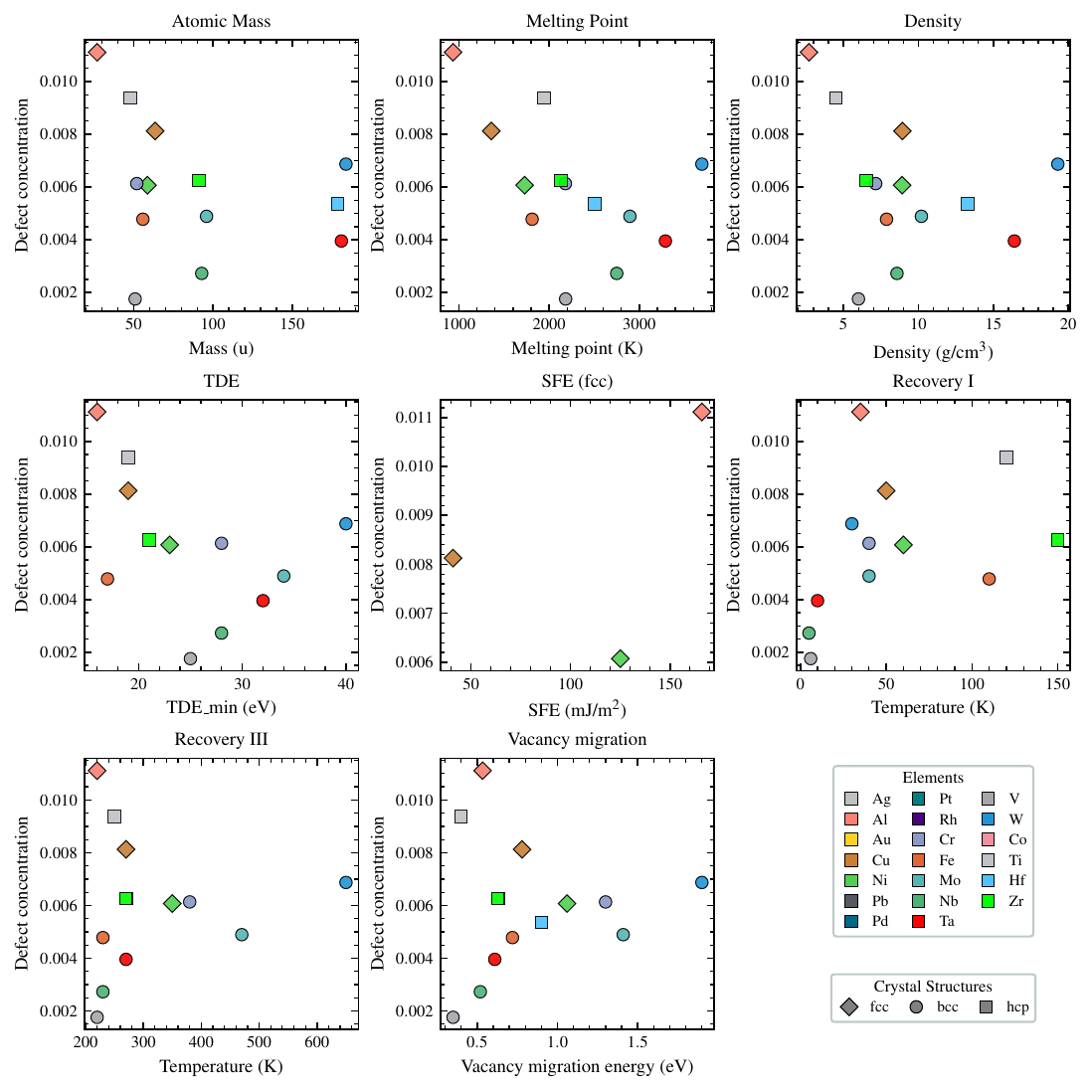}
    \caption{Comparison between final defect concentration (mean value of last 20 datapoints) with various material properties in the tabgap potentials. Different marker styles indicate different crystal structures.}
    \label{fig:props}
\end{figure}

\clearpage

\section{Overlapping cascades: Beryllium $\text{Be}~~Z=4$}
\begin{table}[ht!]
    \centering
    \begin{tabular}{c|c|c}
        \toprule
        Potential & Ref. & \hspace{5cm} notes \hspace{5cm} \\
        \midrule
        Be-Tersoff & \cite{bjorkas_interatomic_2009} & \\
        \midrule
        \bottomrule
    \end{tabular}
    \caption{Details of potentials}
\end{table}
\vspace{-0.5cm}
\begin{figure}[ht!]
    \centering
        \includegraphics[width=0.35\linewidth]{figures/elemental/Defect_concentration_plot_notitle-Be.pdf}
    \includegraphics[width=0.35\linewidth]{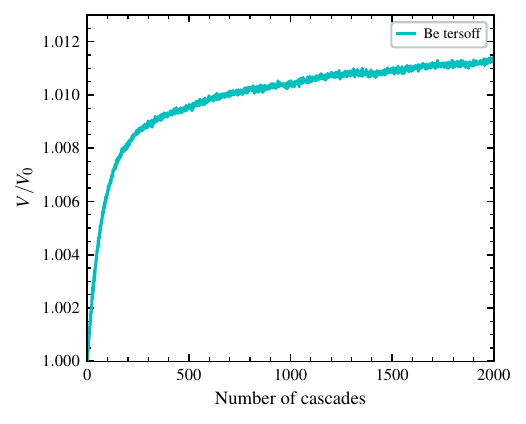}
    \caption{Left is the defect  concentration (interstitials and vacancies) as a function of number of cascades. Right is the volume change as a function of number of cascades.}
\end{figure}

\vspace{-1cm}

\begin{figure}[ht!]
    \centering
\begin{subfigure}{0.29\textwidth}
    \includegraphics[width=1.1\textwidth]{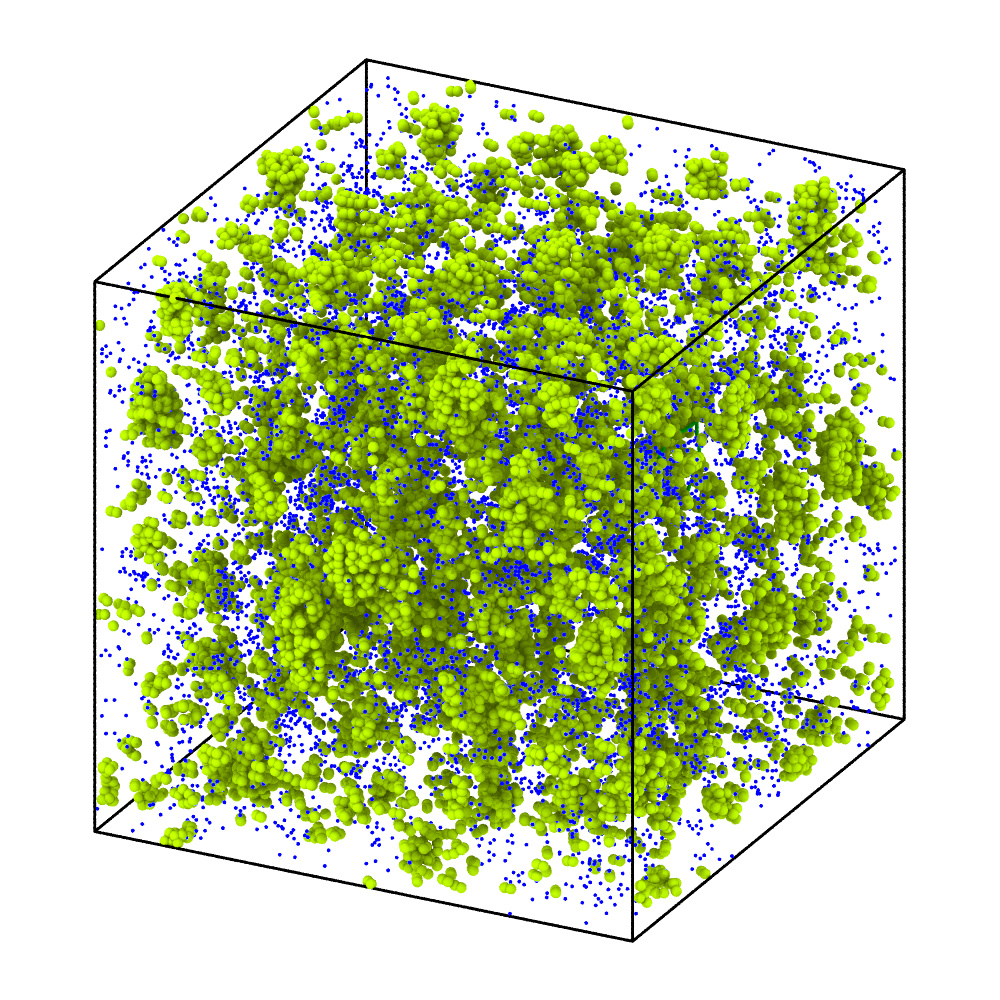}
      \caption{Be tersoff run 1}
\end{subfigure}
\begin{subfigure}{0.29\textwidth}
    \includegraphics[width=1.1\textwidth]{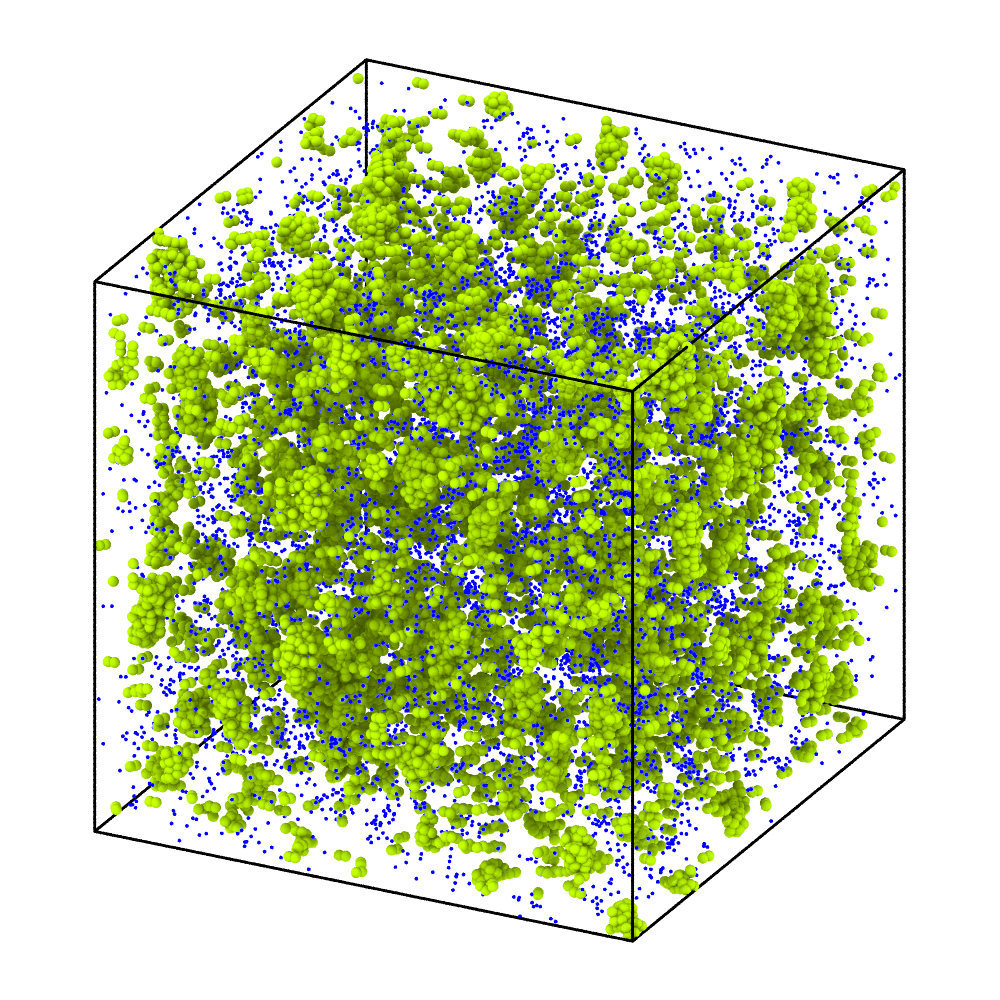}
    \caption{Be tersoff run 2}
\end{subfigure}
\begin{subfigure}{0.29\textwidth}
    \includegraphics[width=1.1\textwidth]{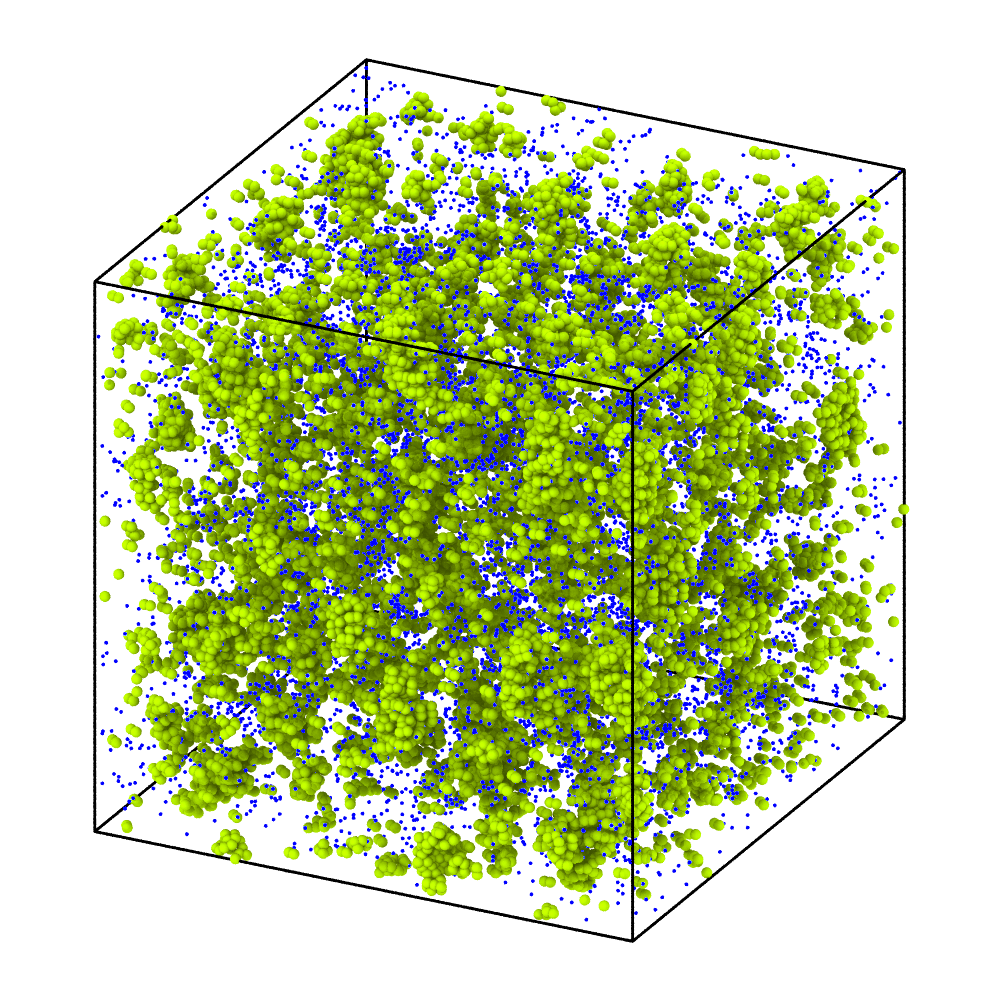}
    \caption{Be tersoff run 3}
\end{subfigure}
    \caption{Final configuration after 2000 overlapping collision cascades. blue spheres represent vacancies, colored spheres represent interstitial atoms, red dislocation is classified as ''other'' by DXA analysis.}
\end{figure}
\clearpage

\section{Overlapping cascades: Aluminum $\text{Al}~~Z=13$}

\begin{table}[ht!]
    \centering
    \begin{tabular}{c|c|c}
        \toprule
        Potential & Ref. & \hspace{5cm} notes \hspace{5cm} \\
        \midrule
        Al-tabGAP & \cite{fellman2025fast} & \\
        \midrule
        Al-EAM & \cite{ercolessi_interatomic_1994,nordlund_defect_1998}  &  Al-EAM ''hardened'' by K. Nordlund \cite{nordlund_defect_1998}\\
        \midrule
        Al-Zhou &\cite{zhou_misfit-energy-increasing_2004} &  \\
        \bottomrule
    \end{tabular}
    \caption{Details of potentials}
\end{table}
\vspace{-0.5cm}
\begin{figure}[ht!]
    \centering
    \includegraphics[width=0.35\linewidth]{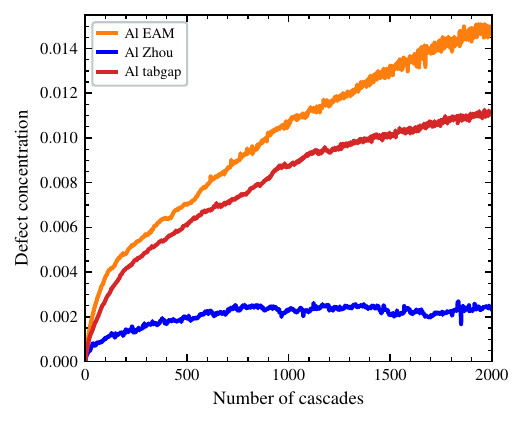}
    \includegraphics[width=0.35\linewidth]{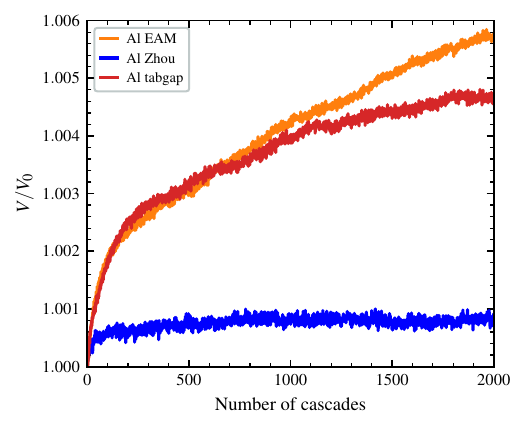}
    \caption{Left is the defect  concentration (interstitials and vacancies) as a function of number of cascades. Right is the volume change as a function of number of cascades.}
\end{figure}

\vspace{-1cm}

\begin{figure}[ht!]
    \centering
\begin{subfigure}{0.29\textwidth}
    \includegraphics[width=1.1\textwidth]{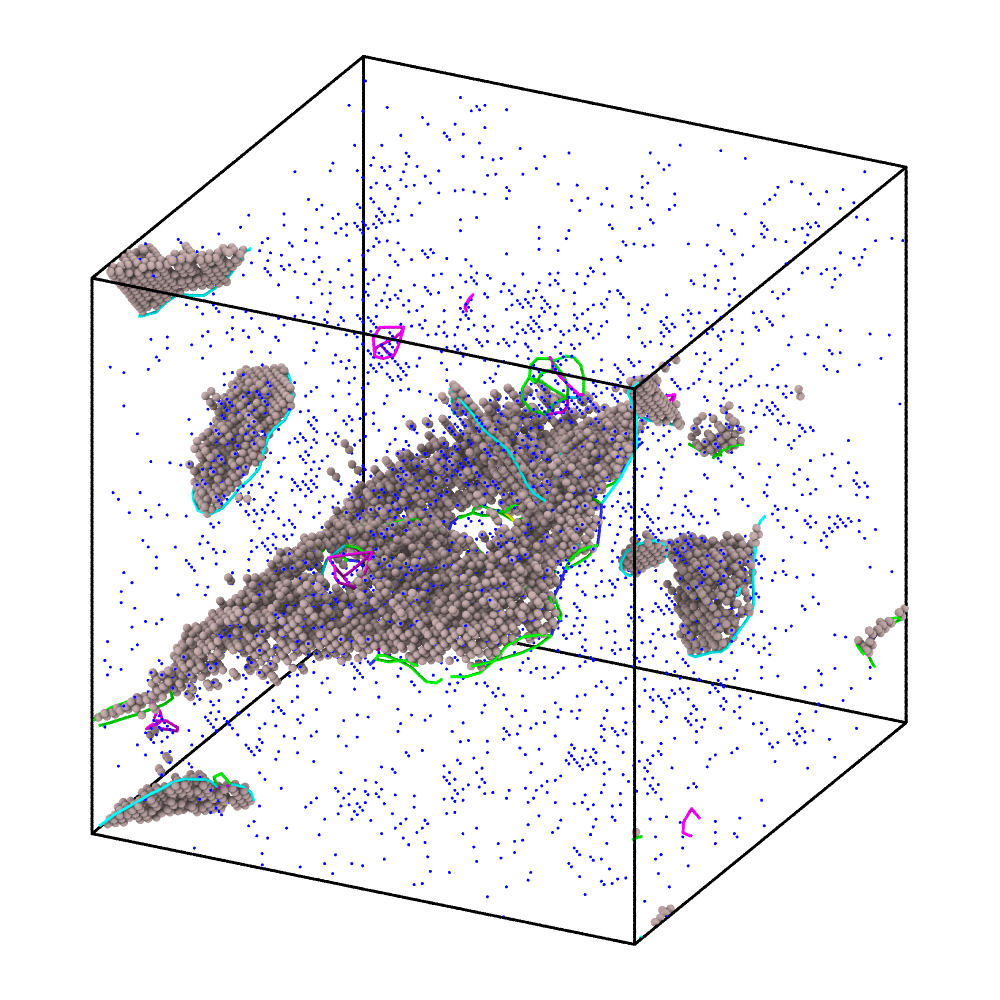}
      \caption{tabGAP}
\end{subfigure}
\begin{subfigure}{0.29\textwidth}
    \includegraphics[width=1.1\textwidth]{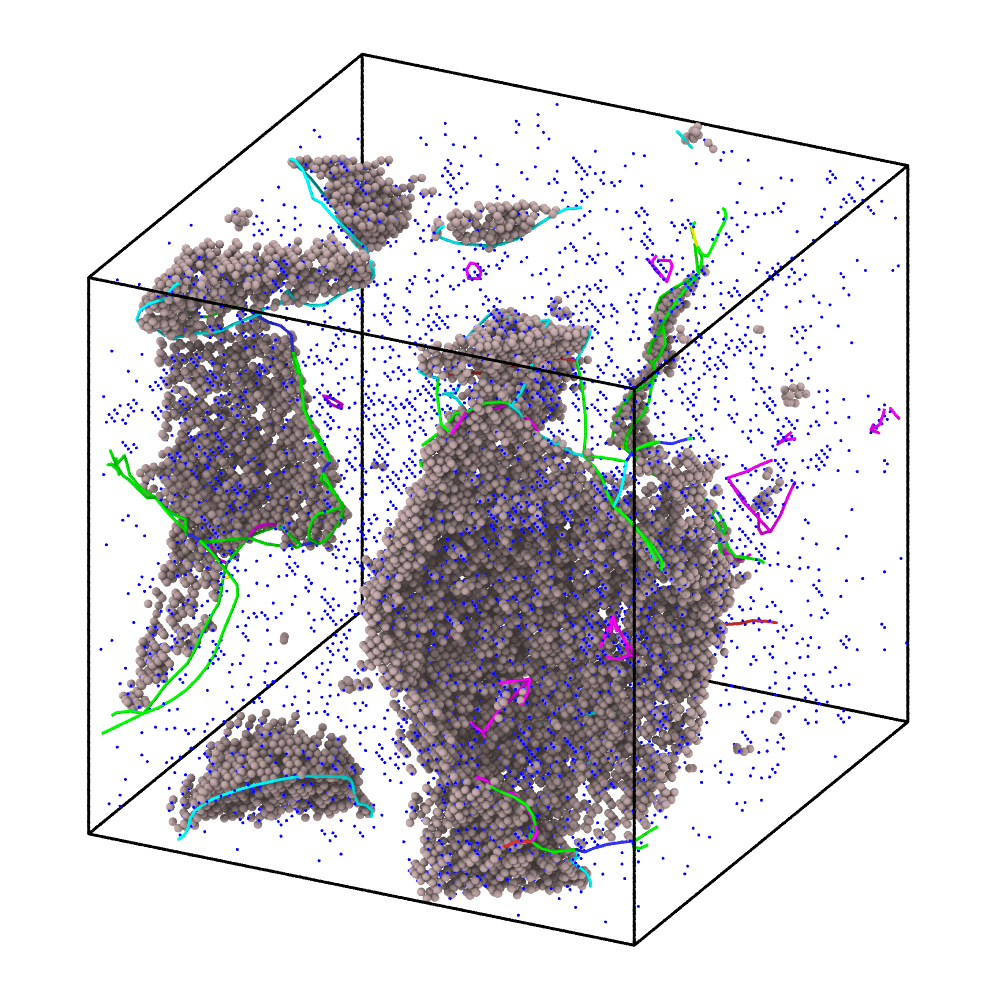}
    \caption{EAM}
\end{subfigure}
\begin{subfigure}{0.29\textwidth}
    \includegraphics[width=1.1\textwidth]{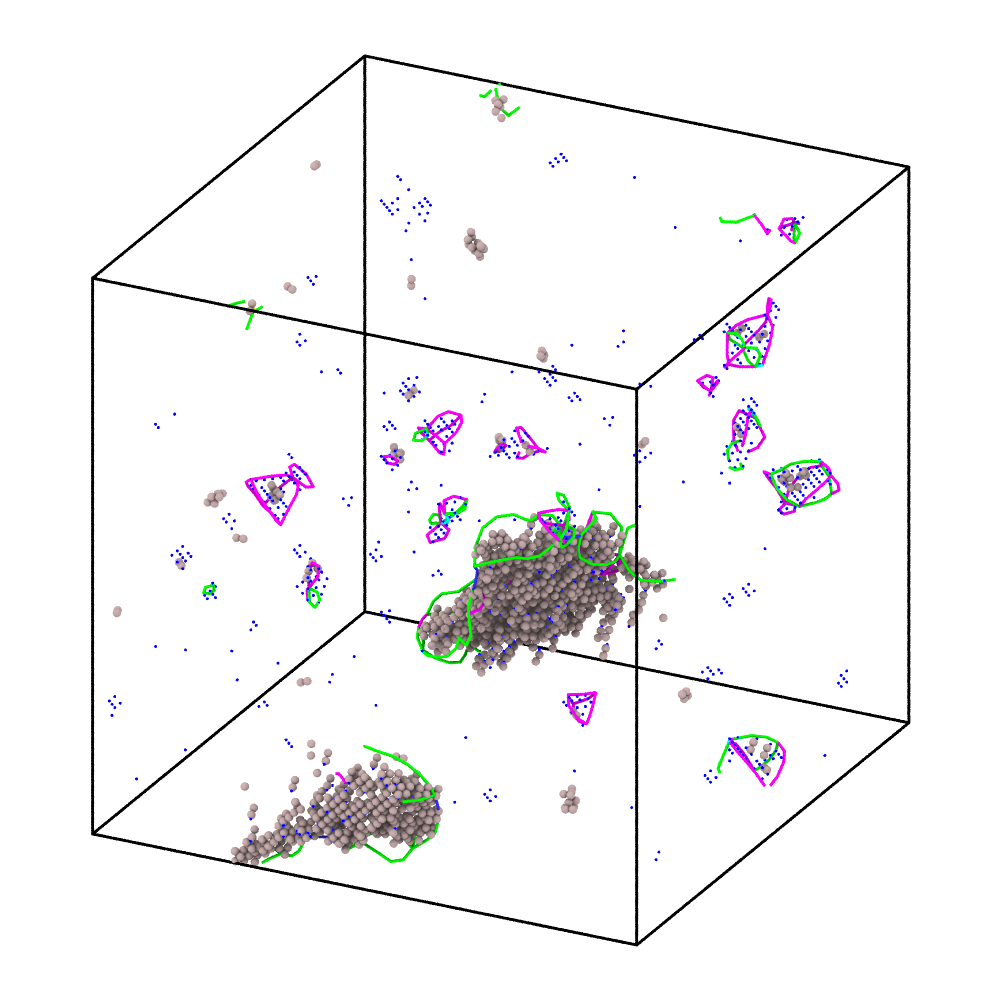}
    \caption{Modified Zhou EAM}
\end{subfigure}
    \caption{Final configuration after 2000 overlapping collision cascades. blue spheres represent vacancies, colored spheres represent interstitial atoms, green lines represent Shockley dislocations, and cyan lines represent Frank dislocations and pink lines represent stair-rod dislocations.}
\end{figure}

\begin{figure}[ht!]
    \centering
    \includegraphics[width=0.2\linewidth]{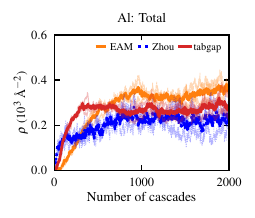}
    \includegraphics[width=0.2\linewidth]{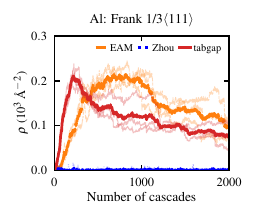}
    \includegraphics[width=0.2\linewidth]{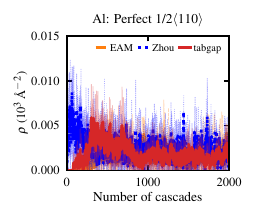}
    \includegraphics[width=0.2\linewidth]{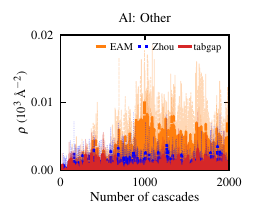}
    \includegraphics[width=0.2\linewidth]{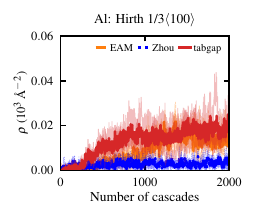}
    \includegraphics[width=0.2\linewidth]{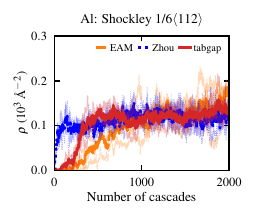}
    \includegraphics[width=0.2\linewidth]{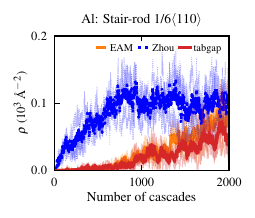}
    \caption{Dislocation density obtained from DXA analysis. Faint curves are individual runs and bold is the average.}
\end{figure}

\clearpage

\section{Overlapping cascades: Titanium $\text{Ti}~~Z=22$}

\begin{table}[ht!]
    \centering
    \begin{tabular}{c|c|c}
        \toprule
        Potential & Ref. & \hspace{5cm} notes \hspace{5cm} \\
        \midrule
        Ti-tabGAP & \cite{byggmastar2026nine} &  \\
        \midrule
        Ti-EAM & \cite{ackland1992theoretical}  &  Ti-EAM ''hardened'' by G.J. Ackland Ti\_V2.eam.fs NIST \\
        \midrule
        Ti-Zhou & \cite{zhou_misfit-energy-increasing_2004} &  \\
        \bottomrule
    \end{tabular}
    \caption{Details of potentials}
\end{table}
\vspace{-0.5cm}
\begin{figure}[ht!]
    \centering
    \includegraphics[width=0.33\linewidth]{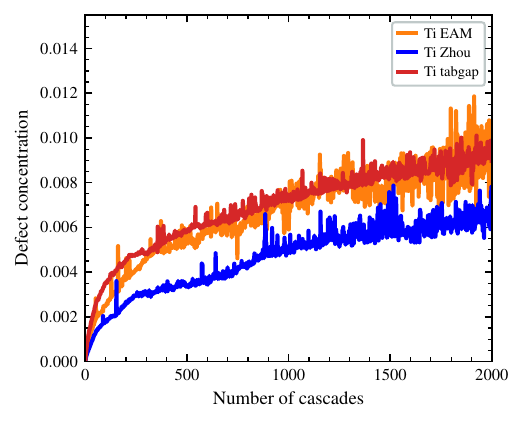}
    \includegraphics[width=0.33\linewidth]{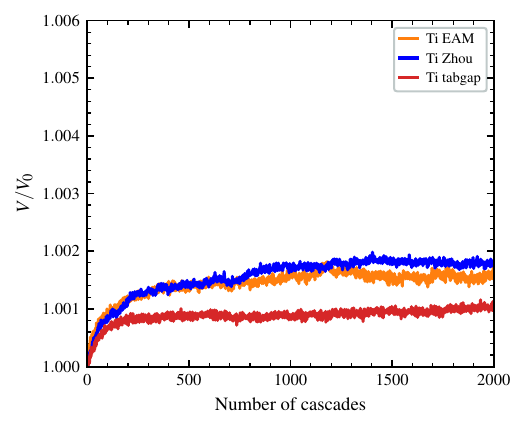}
    \caption{Left is the defect  concentration (interstitials and vacancies) as a function of number of cascades. Right is the volume change as a function of number of cascades.}
\end{figure}

\vspace{-1cm}

\begin{figure}[ht!]
    \centering
\begin{subfigure}{0.29\textwidth}
    \includegraphics[width=1.1\textwidth]{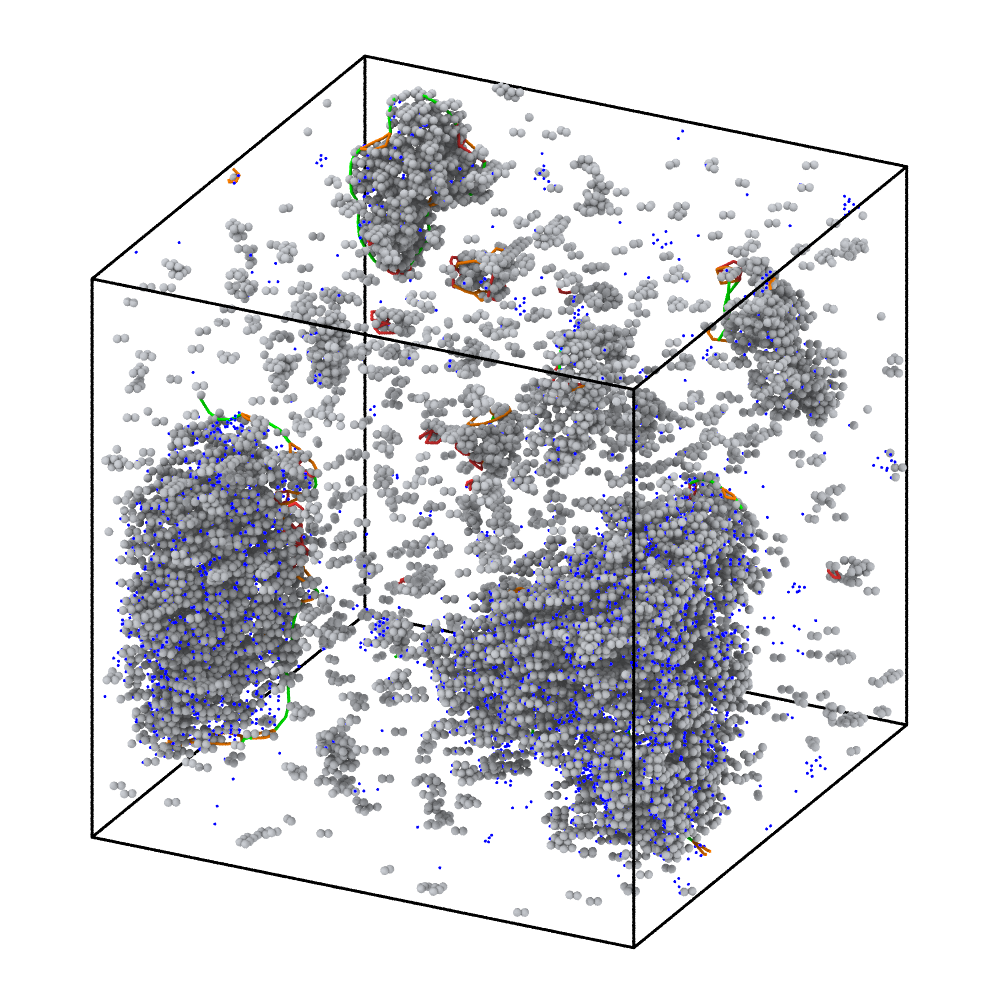}
      \caption{tabGAP}
\end{subfigure}
\begin{subfigure}{0.29\textwidth}
    \includegraphics[width=1.1\textwidth]{figures/frames/frame-Ti-EAM-1-run-1-2000.png}
    \caption{EAM}
\end{subfigure}
\begin{subfigure}{0.29\textwidth}
    \includegraphics[width=1.1\textwidth]{figures/frames/frame-Ti-Zhou-run-2-2000.png}
    \caption{Modified Zhou EAM}
\end{subfigure}
    \caption{Final configuration after 2000 overlapping collision cascades. small blue spheres represent vacancies, colored spheres represent interstitial atoms, green lines are $1/3$ \hkl<1-210> dislocations, blue lines are \hkl<0001> dislocations, magenta lines are \hkl<1-100> dislocations, orange lines are $1/3$\hkl<1-100> dislocations, yellow lines are $1/3$ \hkl<1-213> dislocations and red lines are dislocations with type ''other'' as identified by DXA analysis.}
\end{figure}

\begin{figure}[ht!]
    \centering
    \includegraphics[width=0.2\linewidth]{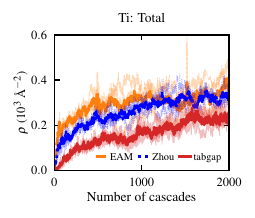}
    \includegraphics[width=0.2\linewidth]{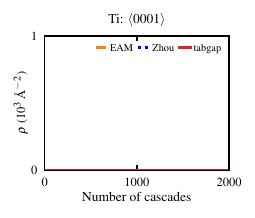}
    \includegraphics[width=0.2\linewidth]{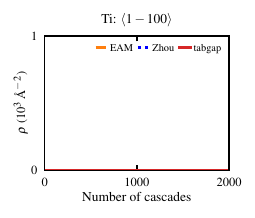}
    \includegraphics[width=0.2\linewidth]{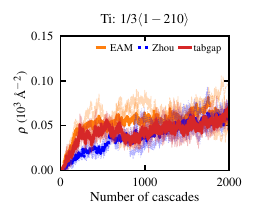}
    \includegraphics[width=0.2\linewidth]{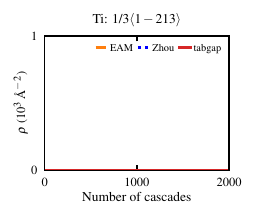}
    \includegraphics[width=0.2\linewidth]{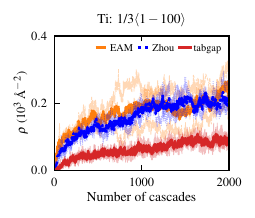}
    \includegraphics[width=0.2\linewidth]{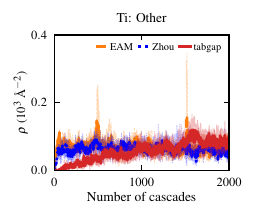}
    \caption{Dislocation density obtained from DXA analysis. Faint curves are individual runs and bold is the average.}
\end{figure}

\clearpage

\section{Overlapping cascades: Vanadium $\text{V}~~Z=23$}

\begin{table}[ht!]
    \centering
    \begin{tabular}{c|c|c}
        \toprule
        Potential & Ref. & \hspace{5cm} notes \hspace{5cm} \\
        \midrule
        V-tabGAP & \cite{byggmastar2026nine} &  \\
        \midrule
        V-EAM & \cite{han2003interatomic} & \\
        \midrule
        V-Zhou & \cite{zhao2021defect} &  V added to Zhou, not in original potential \\
        \bottomrule
    \end{tabular}
    \caption{Details of potentials}
\end{table}
\vspace{-0.5cm}
\begin{figure}[ht!]
    \centering
    \includegraphics[width=0.35\linewidth]{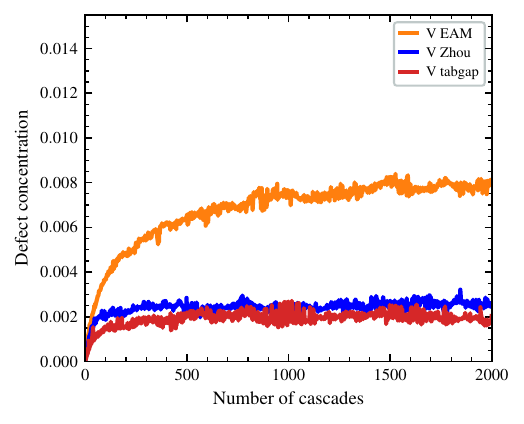}
    \includegraphics[width=0.35\linewidth]{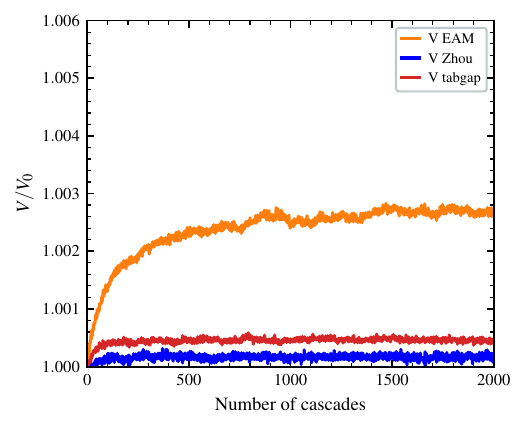}
    \caption{Left is the defect  concentration (interstitials and vacancies) as a function of number of cascades. Right is the volume change as a function of number of cascades.}
\end{figure}

\vspace{-1cm}

\begin{figure}[ht!]
    \centering
\begin{subfigure}{0.29\textwidth}
    \includegraphics[width=1.1\textwidth]{figures/frames/frame-V-tabgap-run-2-2000.png}
      \caption{tabGAP}
\end{subfigure}
\begin{subfigure}{0.29\textwidth}
    \includegraphics[width=1.1\textwidth]{figures/frames/frame-V-EAM-1-run-3-2000.png}
    \caption{EAM}
\end{subfigure}
\begin{subfigure}{0.29\textwidth}
    \includegraphics[width=1.1\textwidth]{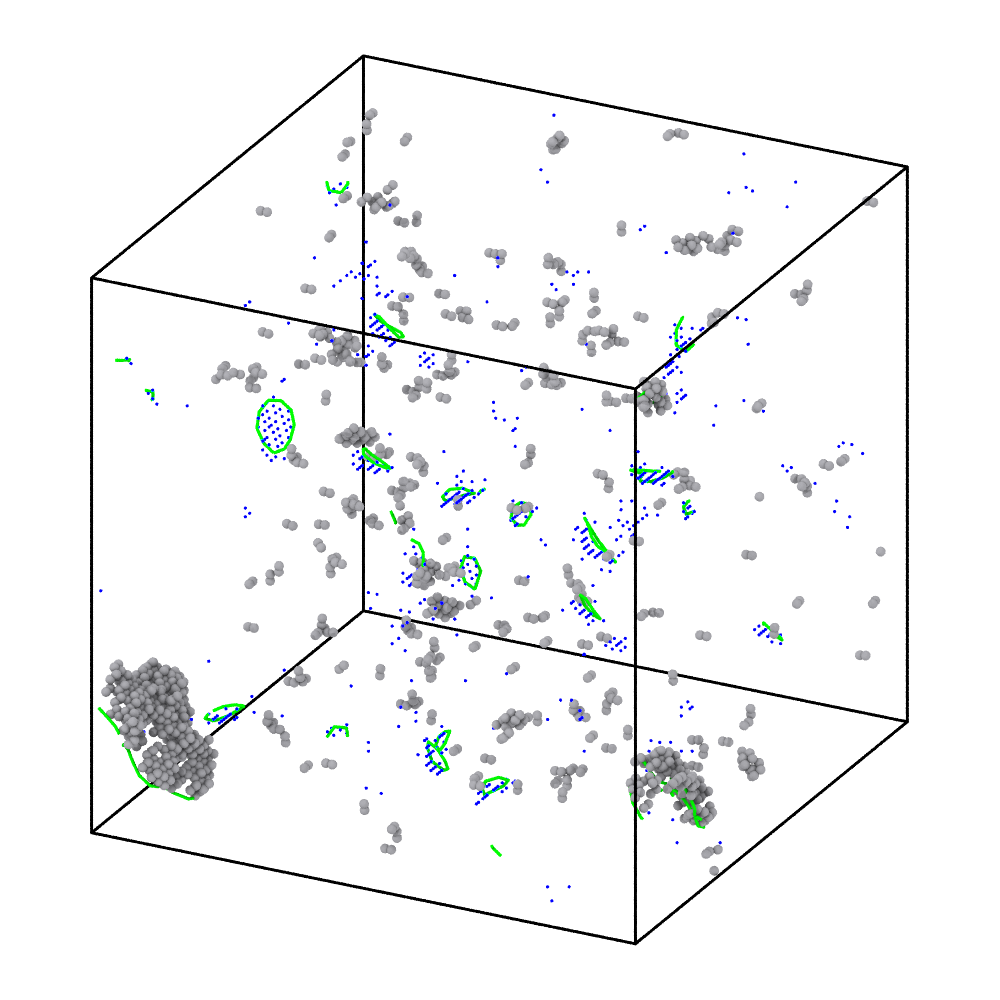}
    \caption{Modified Zhou EAM}
\end{subfigure}
    \caption{Final configuration after 2000 overlapping collision cascades. small blue spheres represent vacancies, colored spheres represent interstitial atoms, green lines represent 1/2 \hkl<111> dislocations, and pink lines \hkl<100> dislocations.}
\end{figure}

\begin{figure}[ht!]
    \centering
    \includegraphics[width=0.2\linewidth]{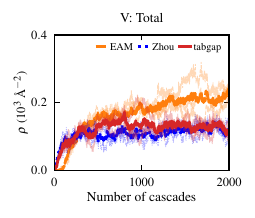}
    \includegraphics[width=0.2\linewidth]{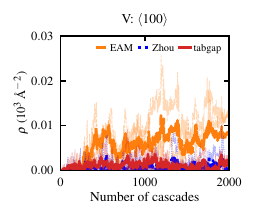}
    \includegraphics[width=0.2\linewidth]{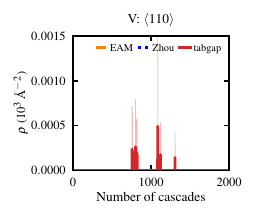}
    \includegraphics[width=0.2\linewidth]{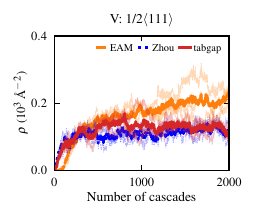}
    \includegraphics[width=0.2\linewidth]{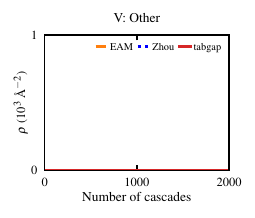}
    \caption{Dislocation density obtained from DXA analysis. Faint curves are individual runs and bold is the average.}
\end{figure}

\clearpage

\section{Overlapping cascades: Chromium $\text{Cr}~~Z=24$}

\begin{table}[ht!]
    \centering
    \begin{tabular}{c|c|c}
        \toprule
        Potential & Ref. & \hspace{5cm} notes \hspace{5cm} \\
        \midrule
        Cr-tabGAP & \cite{fellman2025radiation} & Cr-tabGAP from AlCrCuFeNi-tabGAP. Not specifically made for simulation of Cr.\\
        \midrule
        Cr-Zhou & \cite{lin2008computational} &  Cr added to Zhou, not in original potential \\
        \bottomrule
    \end{tabular}
    \caption{Details of potentials}
\end{table}
\vspace{-0.5cm}
\begin{figure}[ht!]
    \centering
    \includegraphics[width=0.35\linewidth]{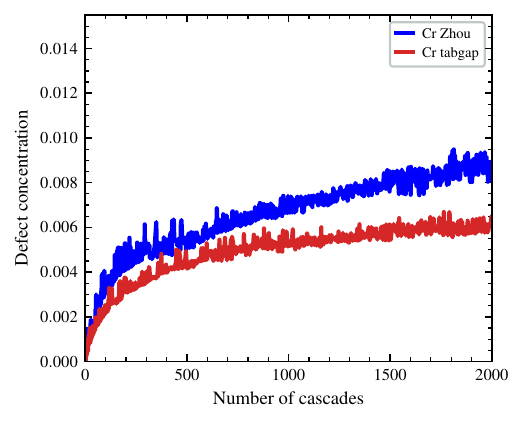}
    \includegraphics[width=0.35\linewidth]{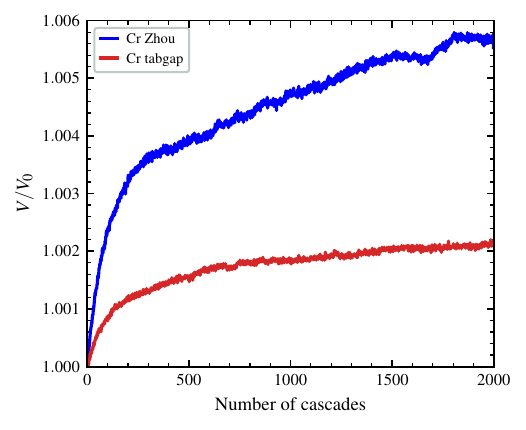}
    \caption{Left is the defect  concentration (interstitials and vacancies) as a function of number of cascades. Right is the volume change as a function of number of cascades.}
\end{figure}

\vspace{-1cm}

\begin{figure}[ht!]
    \centering
\begin{subfigure}{0.29\textwidth}
    \includegraphics[width=1.1\textwidth]{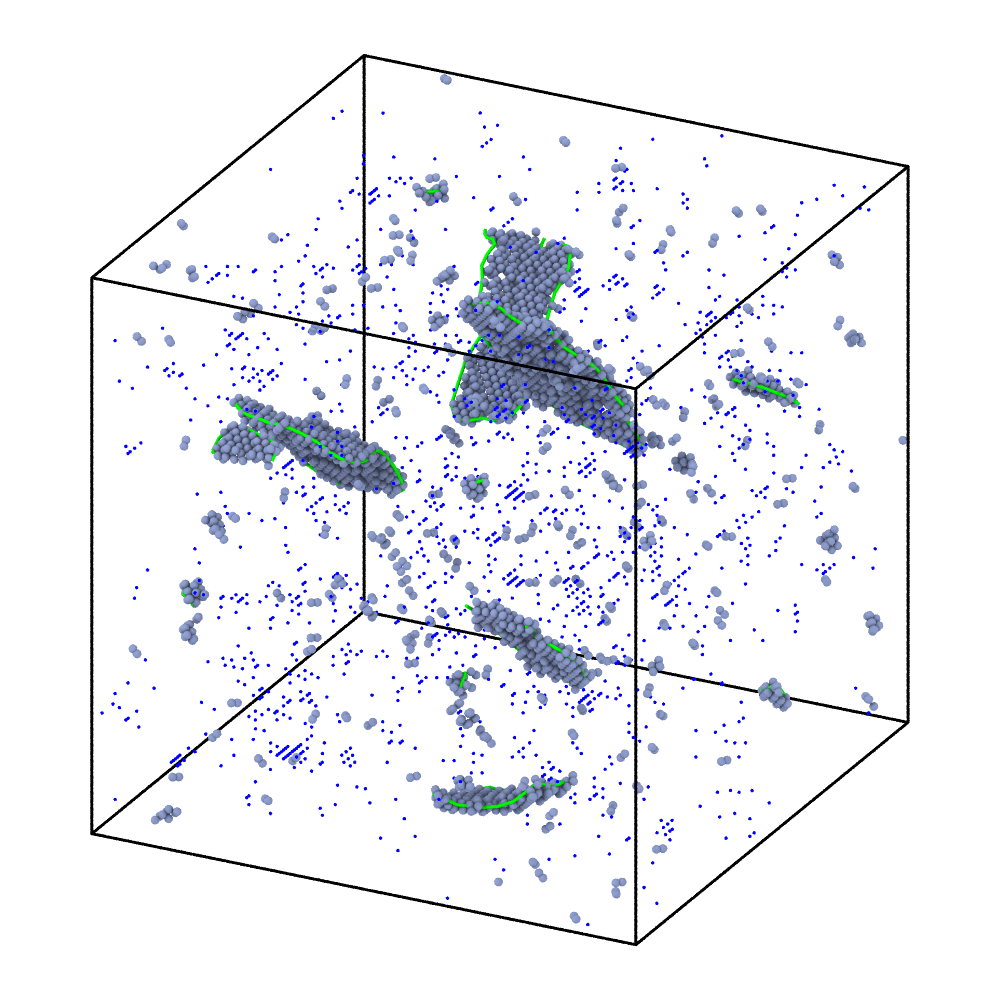}
      \caption{tabGAP}
\end{subfigure}
\begin{subfigure}{0.29\textwidth}
\label{fig:final:Ni}
    \includegraphics[width=1.1\textwidth]{figures/frames/frame-Cr-Zhou-run-2-2000.png}
    \caption{Modified Zhou EAM}
\end{subfigure}
    \caption{Final configuration after 2000 overlapping collision cascades. small blue spheres represent vacancies, colored spheres represent interstitial atoms, green lines represent 1/2 \hkl<111> dislocations, and pink lines \hkl<100> dislocations.}
    \label{fig:final}
\end{figure}

\begin{figure}[ht!]
    \centering
    \includegraphics[width=0.2\linewidth]{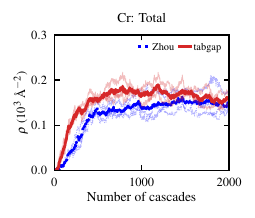}
    \includegraphics[width=0.2\linewidth]{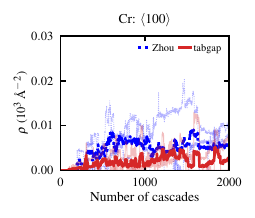}
    \includegraphics[width=0.2\linewidth]{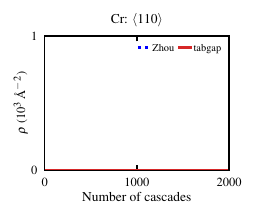}
    \includegraphics[width=0.2\linewidth]{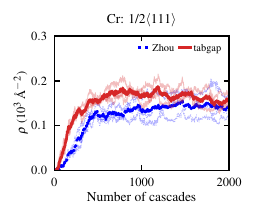}
    \includegraphics[width=0.2\linewidth]{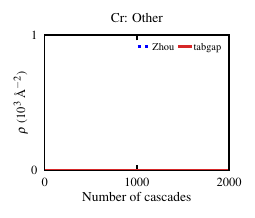}
    \caption{Dislocation density obtained from DXA analysis. Faint curves are individual runs and bold is the average.}
    \label{fig:FCC:dc}
\end{figure}

\clearpage

\section{Overlapping cascades: Iron $\text{Fe}~~Z=26$}

\begin{table}[ht!]
    \centering
    \begin{tabular}{c|c|c}
        \toprule
        Potential & Ref. & \hspace{5cm} notes \hspace{5cm} \\
        \midrule
        Fe-tabGAP & \cite{byggmastar2022multiscale} & \\
        \midrule
        Fe-EAM & \cite{malerba_comparison_2010,byggmastar_effects_2018} & Fe-EAM ''hardened'' by J. Byggmästar~\cite{byggmastar_effects_2018}\\
        \midrule
        Fe-Zhou & \cite{zhou_misfit-energy-increasing_2004} & \\
        \bottomrule
    \end{tabular}
    \caption{Details of potentials}
\end{table}
\vspace{-0.5cm}
\begin{figure}[ht!]
    \centering
    \includegraphics[width=0.35\linewidth]{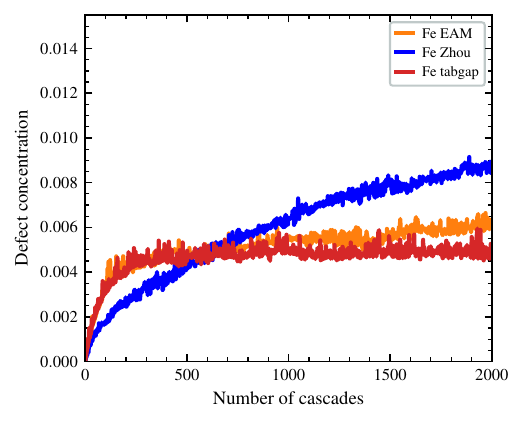}
    \includegraphics[width=0.35\linewidth]{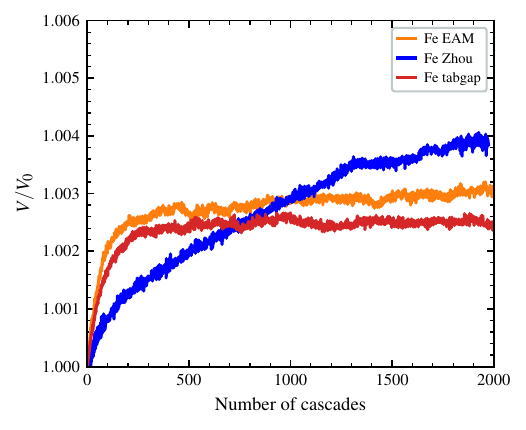}
    \caption{Left is the defect  concentration (interstitials and vacancies) as a function of number of cascades. Right is the volume change as a function of number of cascades.}
\end{figure}

\vspace{-1cm}

\begin{figure}[ht!]
    \centering
\begin{subfigure}{0.29\textwidth}
    \includegraphics[width=1.1\textwidth]{figures/frames/frame-Fe-tabgap-run-3-2000.png}
      \caption{tabGAP}
\end{subfigure}
\begin{subfigure}{0.29\textwidth}
    \includegraphics[width=1.1\textwidth]{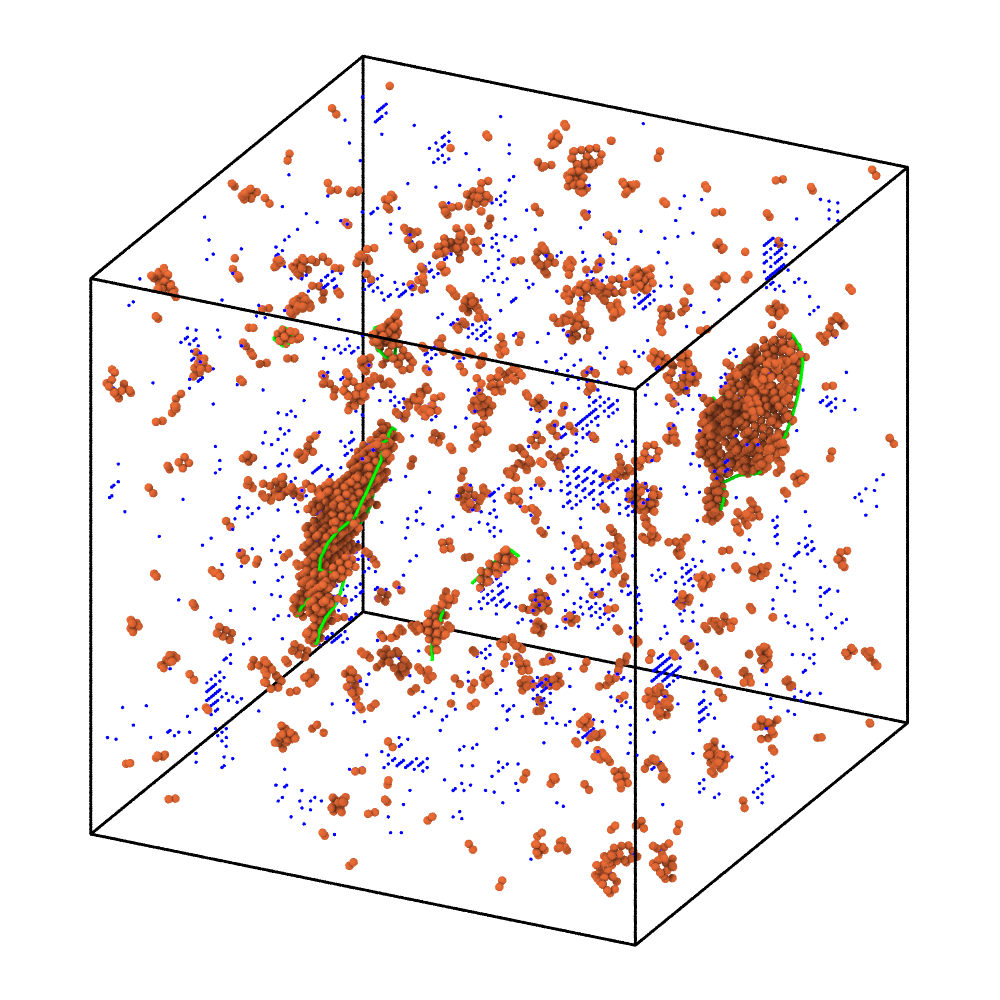}
    \caption{EAM}
\end{subfigure}
\begin{subfigure}{0.29\textwidth}
    \includegraphics[width=1.1\textwidth]{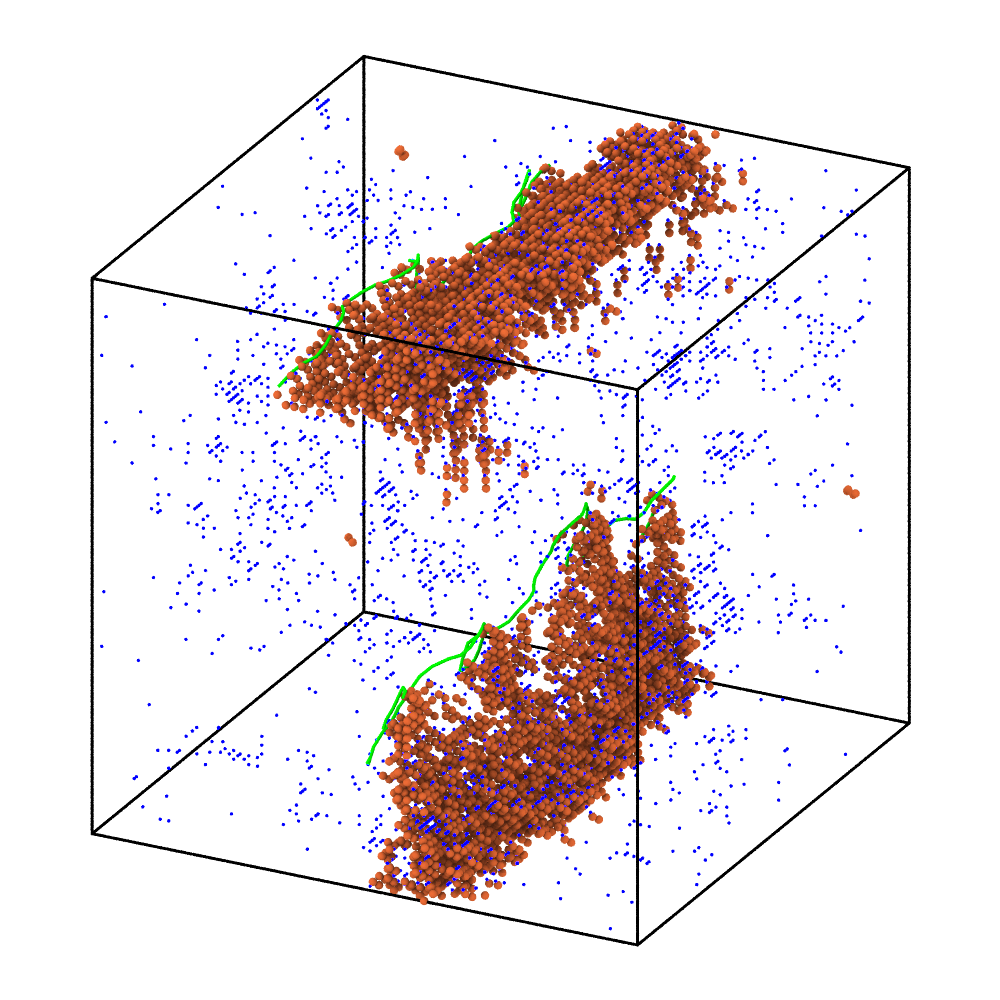}
    \caption{Modified Zhou EAM}
\end{subfigure}
    \caption{Final configuration after 2000 overlapping collision cascades. small blue spheres represent vacancies, colored spheres represent interstitial atoms, green lines represent 1/2 \hkl<111> dislocations, and pink lines \hkl<100> dislocations.}
\end{figure}

\begin{figure}[ht!]
    \centering
    \includegraphics[width=0.2\linewidth]{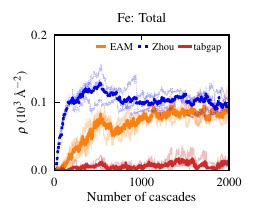}
    \includegraphics[width=0.2\linewidth]{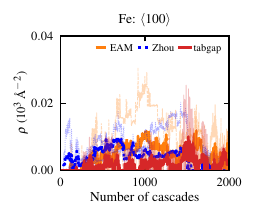}
    \includegraphics[width=0.2\linewidth]{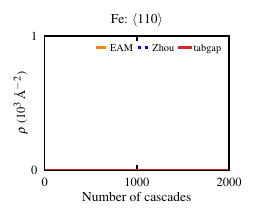}
    \includegraphics[width=0.2\linewidth]{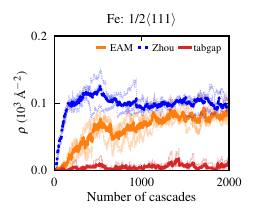}
    \includegraphics[width=0.2\linewidth]{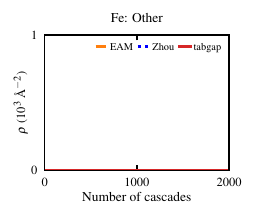}
    \caption{Dislocation density obtained from DXA analysis. Faint curves are individual runs and bold is the average.}
\end{figure}

\clearpage

\section{Overlapping cascades: Cobalt $\text{Co}~~Z=27$}

\begin{table}[ht!]
    \centering
    \begin{tabular}{c|c|c}
        \toprule
        Potential & Ref. & \hspace{5cm} notes \hspace{5cm} \\
        \midrule
        Co-EAM & \cite{pasianot1992embedded,nordlund_inverse_1999}  &  Co-EAM  ''hardened'' by K. Nordlund~\cite{nordlund_inverse_1999}\\
        \midrule
        Co-Zhou & \cite{zhou_misfit-energy-increasing_2004} &  \\
        \bottomrule
    \end{tabular}
    \caption{Details of potentials}
\end{table}
\vspace{-0.5cm}
\begin{figure}[ht!]
    \centering
    \includegraphics[width=0.35\linewidth]{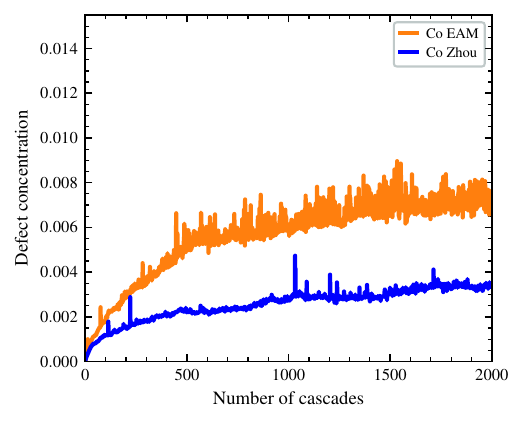}
    \includegraphics[width=0.35\linewidth]{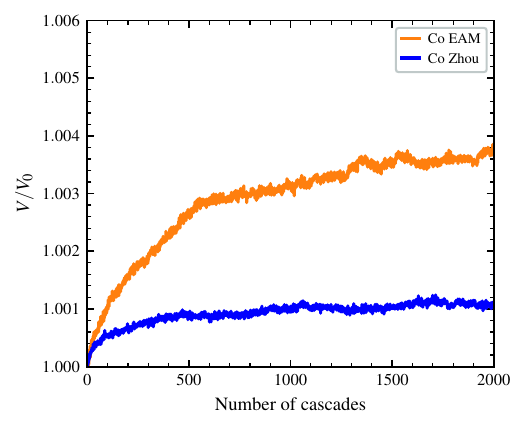}
    \caption{Left is the defect  concentration (interstitials and vacancies) as a function of number of cascades. Right is the volume change as a function of number of cascades.}
\end{figure}

\vspace{-1cm}

\begin{figure}[ht!]
    \centering
\begin{subfigure}{0.29\textwidth}
    \includegraphics[width=1.1\textwidth]{figures/frames/frame-Co-EAM-1-run-1-2000.png}
    \caption{EAM}
\end{subfigure}
\begin{subfigure}{0.29\textwidth}
    \includegraphics[width=1.1\textwidth]{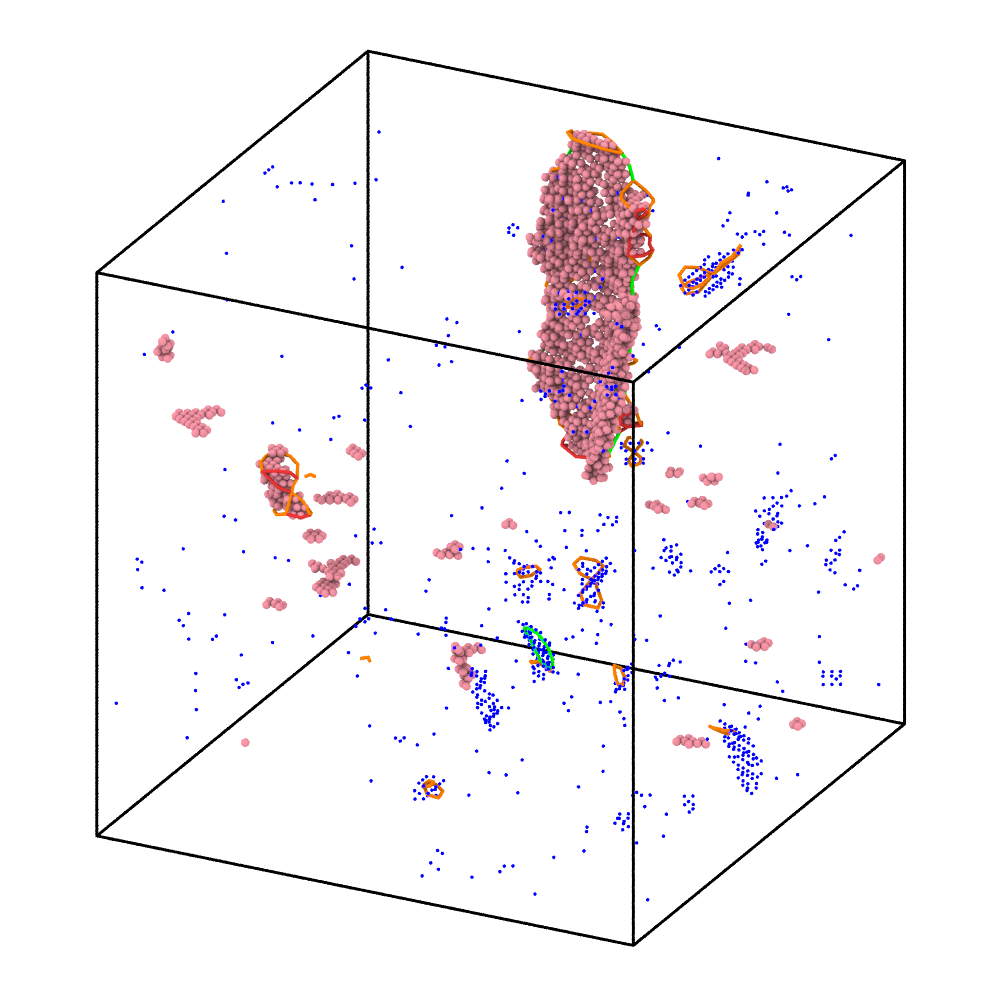}
    \caption{Modified Zhou EAM}
\end{subfigure}
    \caption{Final configuration after 2000 overlapping collision cascades. small blue spheres represent vacancies, colored spheres represent interstitial atoms, green lines are $1/3$ \hkl<1-210> dislocations, blue lines are \hkl<0001> dislocations, magenta lines are \hkl<1-100> dislocations, orange lines are $1/3$\hkl<1-100> dislocations, yellow lines are $1/3$ \hkl<1-213> dislocations and red lines are dislocations with type ''other'' as identified by DXA analysis.}
\end{figure}

\begin{figure}[ht!]
    \centering
    \includegraphics[width=0.2\linewidth]{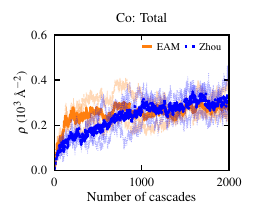}
    \includegraphics[width=0.2\linewidth]{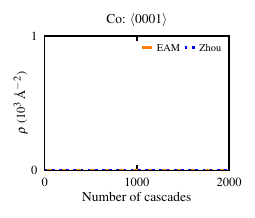}
    \includegraphics[width=0.2\linewidth]{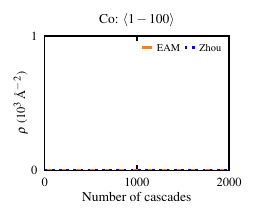}
    \includegraphics[width=0.2\linewidth]{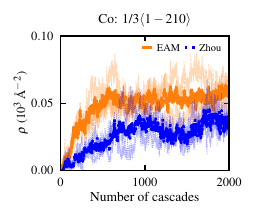}
    \includegraphics[width=0.2\linewidth]{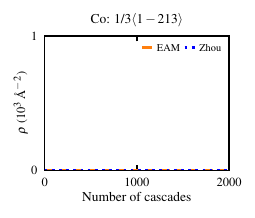}
    \includegraphics[width=0.2\linewidth]{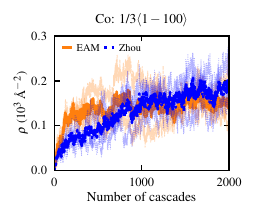}
    \includegraphics[width=0.2\linewidth]{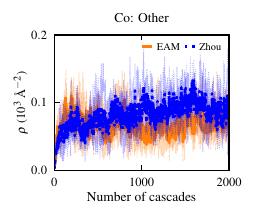}
    \caption{Dislocation density obtained from DXA analysis. Faint curves are individual runs and bold is the average.}
\end{figure}

\clearpage

\section{Overlapping cascades: Nickel $\text{Ni}~~Z=28$}

\begin{table}[ht!]
    \centering
    \begin{tabular}{c|c|c}
        \toprule
        Potential & Ref. & \hspace{5cm} notes \hspace{5cm} \\
        \midrule
        Ni-tabGAP & \cite{fellman2026effects}  & Ni-tabGAP potential is the semicore version in~\cite{fellman2026effects} \\
        \midrule
        Ni-EAM & \cite{stoller_impact_2016} & \\
        \midrule
        Ni-Zhou & \cite{zhou_misfit-energy-increasing_2004} & \\
        \bottomrule
    \end{tabular}
    \caption{Details of potentials}
\end{table}
\vspace{-0.5cm}
\begin{figure}[ht!]
    \centering
    \includegraphics[width=0.35\linewidth]{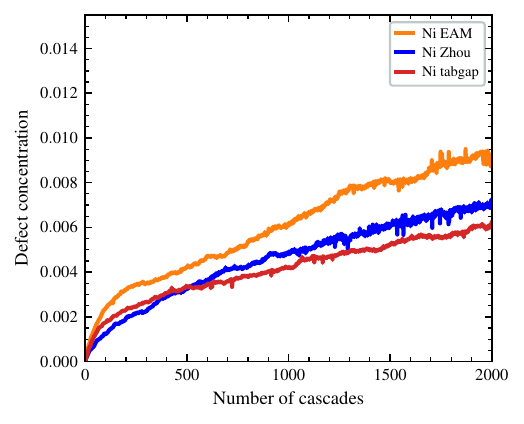}
    \includegraphics[width=0.35\linewidth]{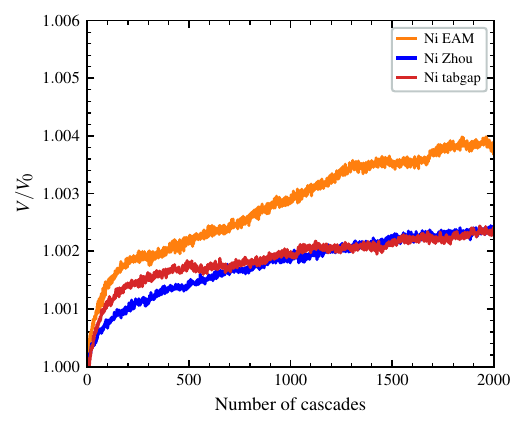}
    \caption{Left is the defect  concentration (interstitials and vacancies) as a function of number of cascades. Right is the volume change as a function of number of cascades.}
\end{figure}

\vspace{-1cm}

\begin{figure}[ht!]
    \centering
\begin{subfigure}{0.29\textwidth}
    \includegraphics[width=1.1\textwidth]{figures/frames/frame-Ni-tabgap-run-3-2000.png}
      \caption{tabGAP}
\end{subfigure}
\begin{subfigure}{0.29\textwidth}
    \includegraphics[width=1.1\textwidth]{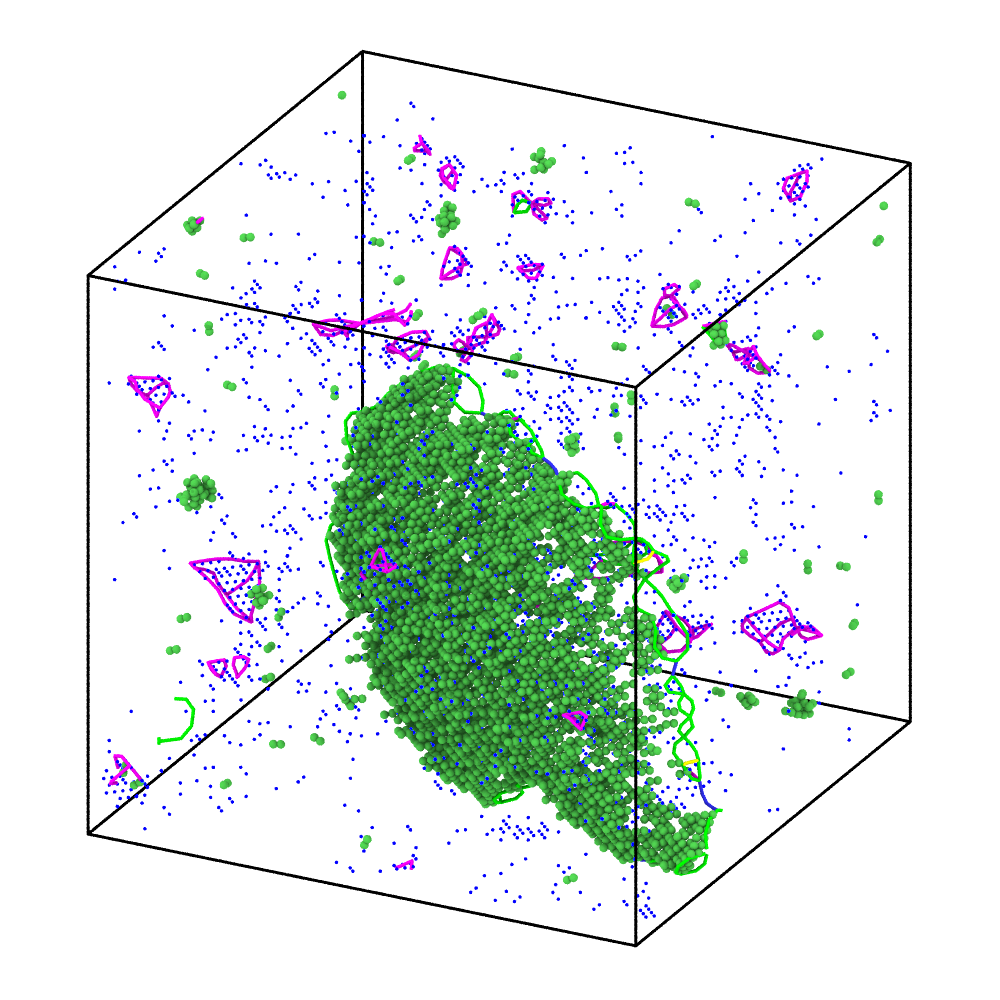}
    \caption{EAM}
\end{subfigure}
\begin{subfigure}{0.29\textwidth}
    \includegraphics[width=1.1\textwidth]{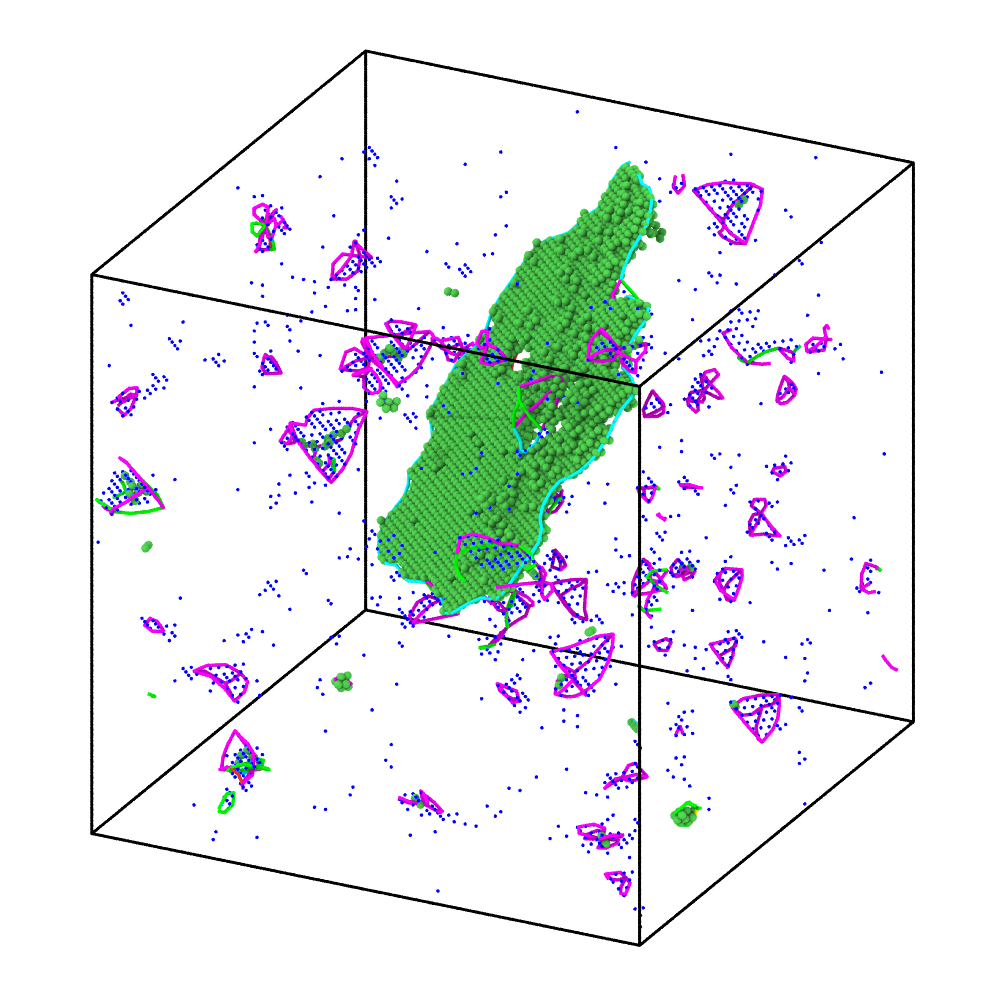}
    \caption{Modified Zhou EAM}
\end{subfigure}
    \caption{Final configuration after 2000 overlapping collision cascades. blue spheres represent vacancies, colored spheres represent interstitial atoms, green lines represent Shockley dislocations, and cyan lines represent Frank dislocations and pink lines represent stair-rod dislocations.}
\end{figure}

\begin{figure}[ht!]
    \centering
    \includegraphics[width=0.2\linewidth]{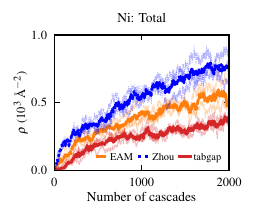}
    \includegraphics[width=0.2\linewidth]{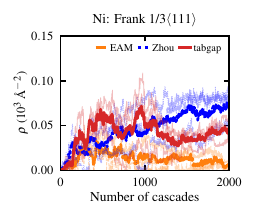}
    \includegraphics[width=0.2\linewidth]{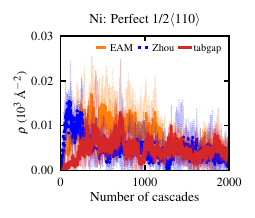}
    \includegraphics[width=0.2\linewidth]{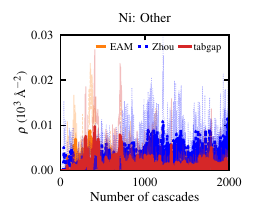}
    \includegraphics[width=0.2\linewidth]{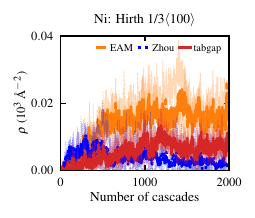}
    \includegraphics[width=0.2\linewidth]{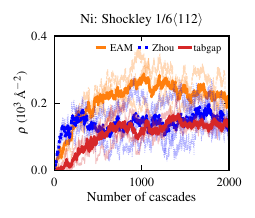}
    \includegraphics[width=0.2\linewidth]{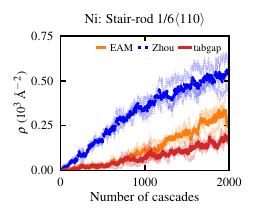}
    \caption{Dislocation density obtained from DXA analysis. Faint curves are individual runs and bold is the average.}
\end{figure}

\clearpage

\section{Overlapping cascades: Copper $\text{Cu}~~Z=29$}

\begin{table}[ht!]
    \centering
    \begin{tabular}{c|c|c}
        \toprule
        Potential & Ref. & \hspace{5cm} notes \hspace{5cm} \\
        \midrule
        Cu-tabGAP & \cite{fellman2025fast} & \\
        \midrule
        Cu-EAM & \cite{foiles1985application} & Cu-EAM ''hardened'' by K. Nordlund~\cite{nordlund_defect_1998}\\
        \midrule
        Cu-Zhou & \cite{zhou_misfit-energy-increasing_2004} & \\
        \bottomrule
    \end{tabular}
    \caption{Details of potentials}
\end{table}
\vspace{-0.5cm}
\begin{figure}[ht!]
    \centering
    \includegraphics[width=0.35\linewidth]{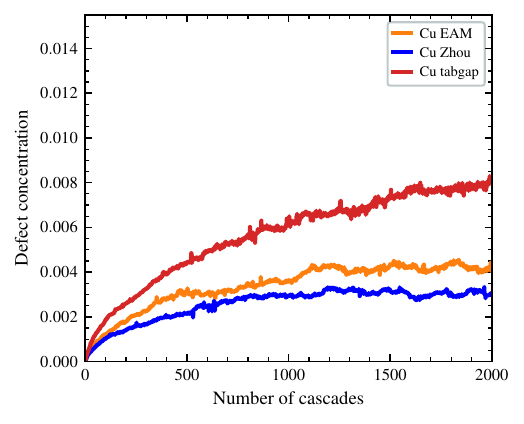}
    \includegraphics[width=0.35\linewidth]{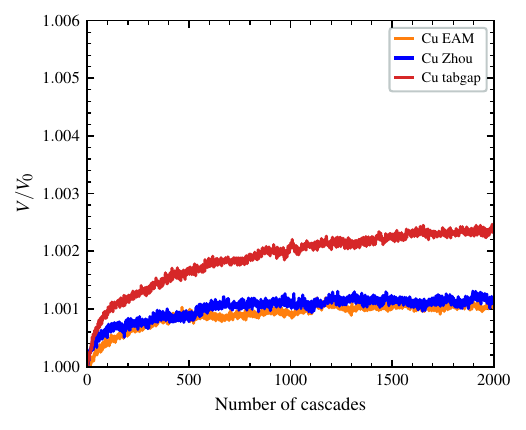}
    \caption{Left is the defect  concentration (interstitials and vacancies) as a function of number of cascades. Right is the volume change as a function of number of cascades.}
\end{figure}

\vspace{-1cm}

\begin{figure}[ht!]
    \centering
\begin{subfigure}{0.29\textwidth}
    \includegraphics[width=1.1\textwidth]{figures/frames/frame-Cu-tabgap-run-2-2000.png}
      \caption{tabGAP}
\end{subfigure}
\begin{subfigure}{0.29\textwidth}
    \includegraphics[width=1.1\textwidth]{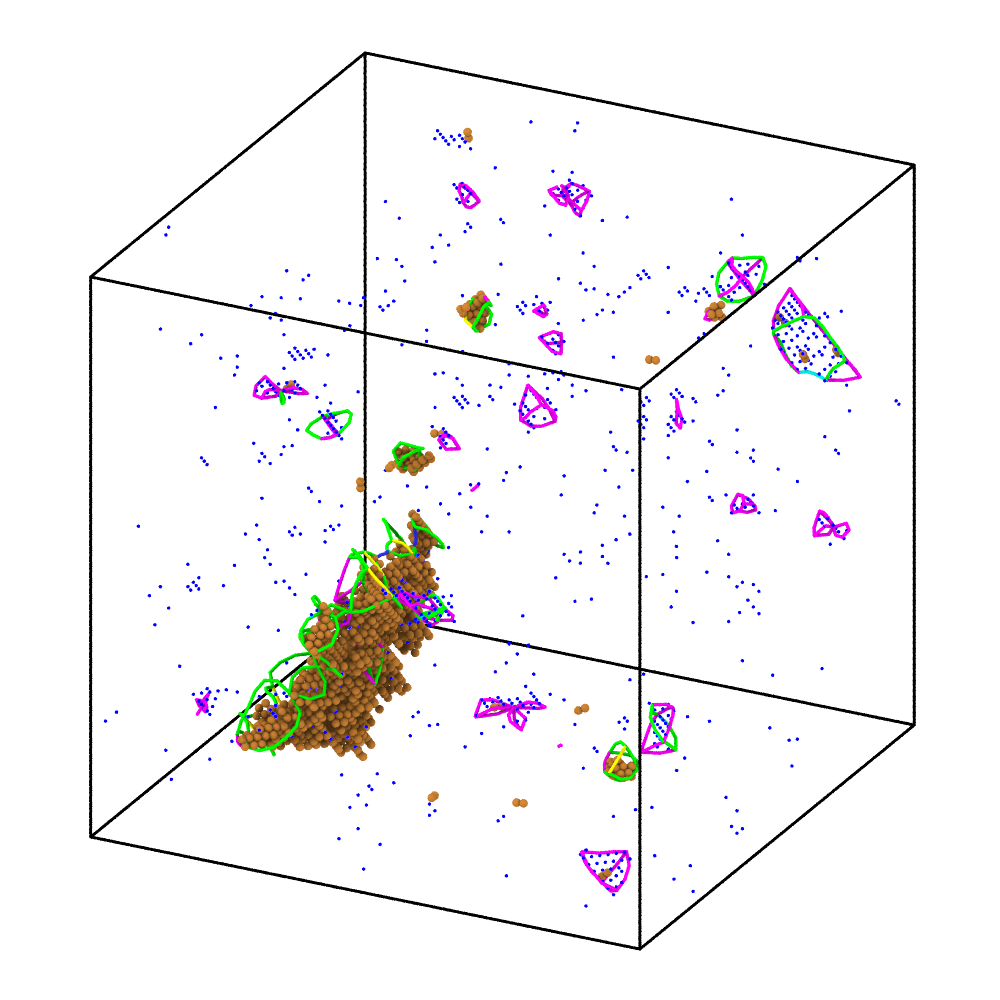}
    \caption{EAM}
\end{subfigure}
\begin{subfigure}{0.29\textwidth}
    \includegraphics[width=1.1\textwidth]{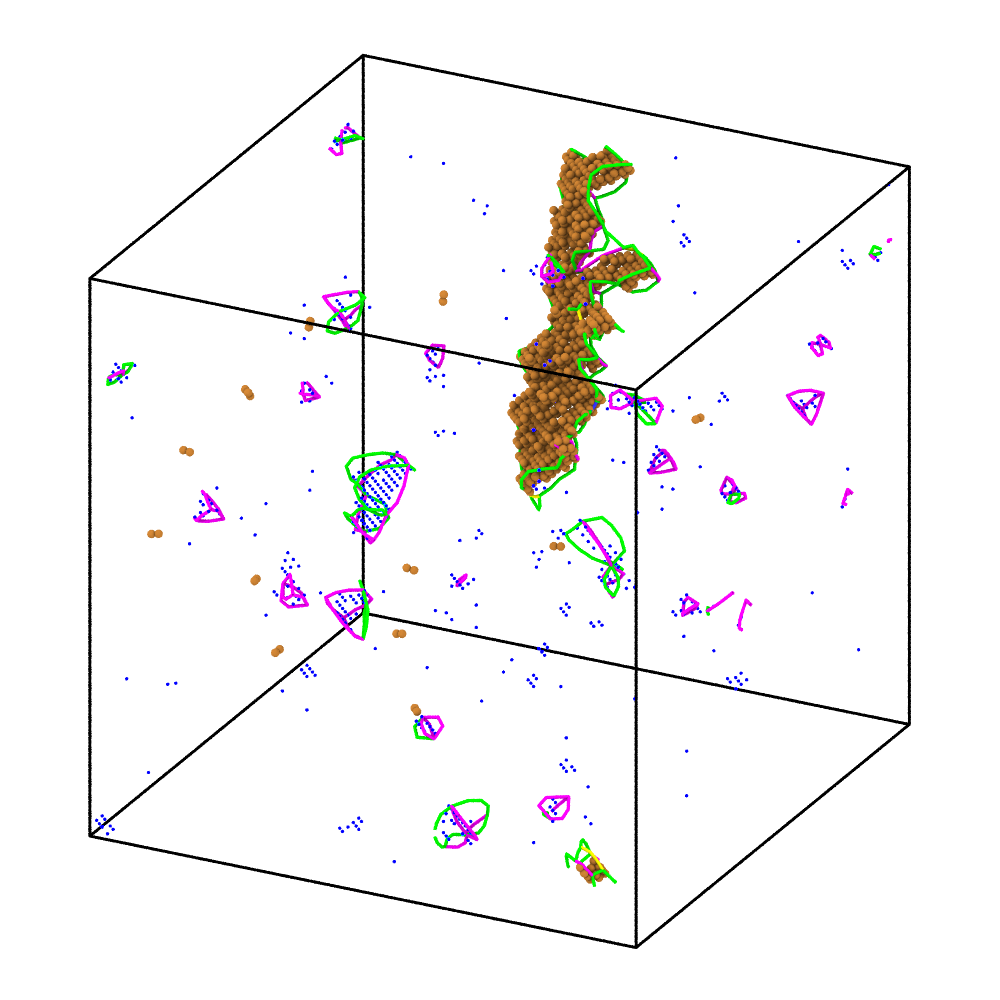}
    \caption{Modified Zhou EAM}
\end{subfigure}
    \caption{Final configuration after 2000 overlapping collision cascades. blue spheres represent vacancies, colored spheres represent interstitial atoms, green lines represent Shockley dislocations, and cyan lines represent Frank dislocations and pink lines represent stair-rod dislocations.}
\end{figure}

\begin{figure}[ht!]
    \centering
    \includegraphics[width=0.2\linewidth]{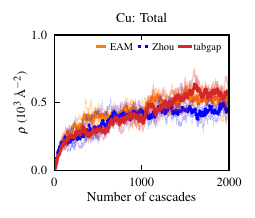}
    \includegraphics[width=0.2\linewidth]{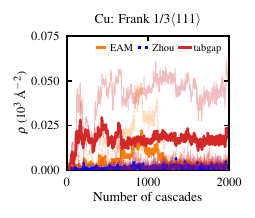}
    \includegraphics[width=0.2\linewidth]{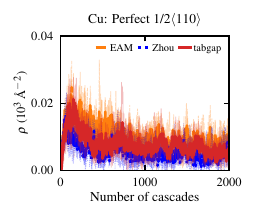}
    \includegraphics[width=0.2\linewidth]{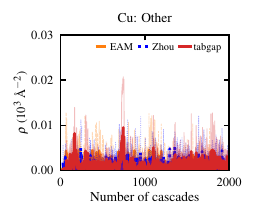}
    \includegraphics[width=0.2\linewidth]{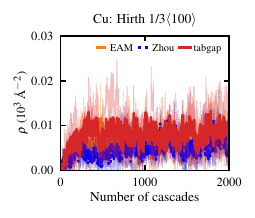}
    \includegraphics[width=0.2\linewidth]{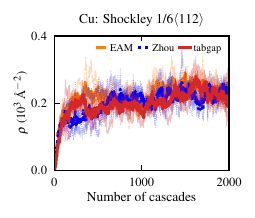}
    \includegraphics[width=0.2\linewidth]{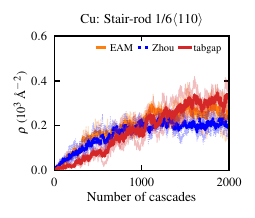}
    \caption{Dislocation density obtained from DXA analysis. Faint curves are individual runs and bold is the average.}
\end{figure}

\clearpage

\section{Overlapping cascades: Zirconium $\text{Zr}~~Z=40$}

\begin{table}[ht!]
    \centering
    \begin{tabular}{c|c|c}
        \toprule
        Potential & Ref. & \hspace{5cm} notes \hspace{5cm} \\
        \midrule
        Zr-tabGAP & \cite{byggmastar2026nine} &   \\
        \midrule
        Zr-EAM & \cite{ackland1995defect} & \\
        \midrule
        Zr-Zhou & \cite{zhou_misfit-energy-increasing_2004} & \\
        \bottomrule
    \end{tabular}
    \caption{Details of potentials}
\end{table}
\vspace{-0.5cm}
\begin{figure}[ht!]
    \centering
    \includegraphics[width=0.33\linewidth]{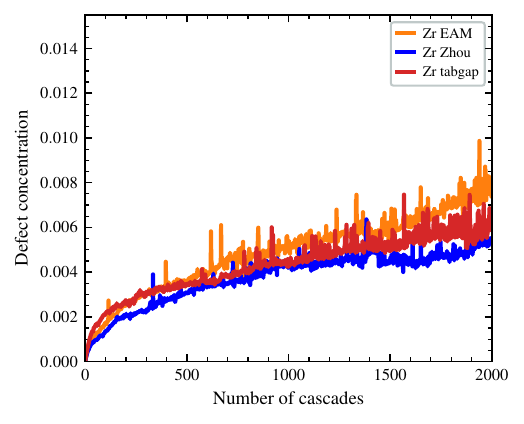}
    \includegraphics[width=0.33\linewidth]{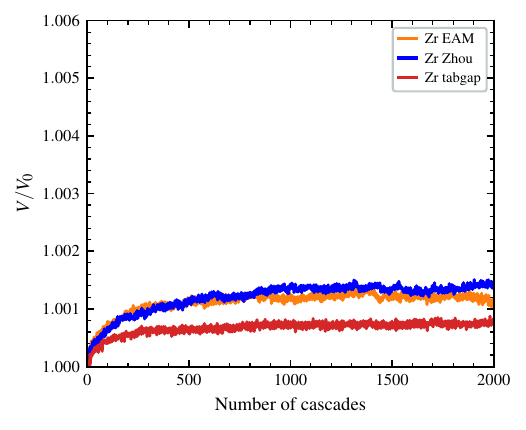}
    \caption{Left is the defect  concentration (interstitials and vacancies) as a function of number of cascades. Right is the volume change as a function of number of cascades.}
\end{figure}

\vspace{-1cm}

\begin{figure}[ht!]
    \centering
\begin{subfigure}{0.29\textwidth}
    \includegraphics[width=1.1\textwidth]{figures/frames/frame-Zr-tabgap-run-1-2000.png}
      \caption{tabGAP}
\end{subfigure}
\begin{subfigure}{0.29\textwidth}
    \includegraphics[width=1.1\textwidth]{figures/frames/frame-Zr-EAM-1-run-3-2000.png}
    \caption{EAM}
\end{subfigure}
\begin{subfigure}{0.29\textwidth}
    \includegraphics[width=1.1\textwidth]{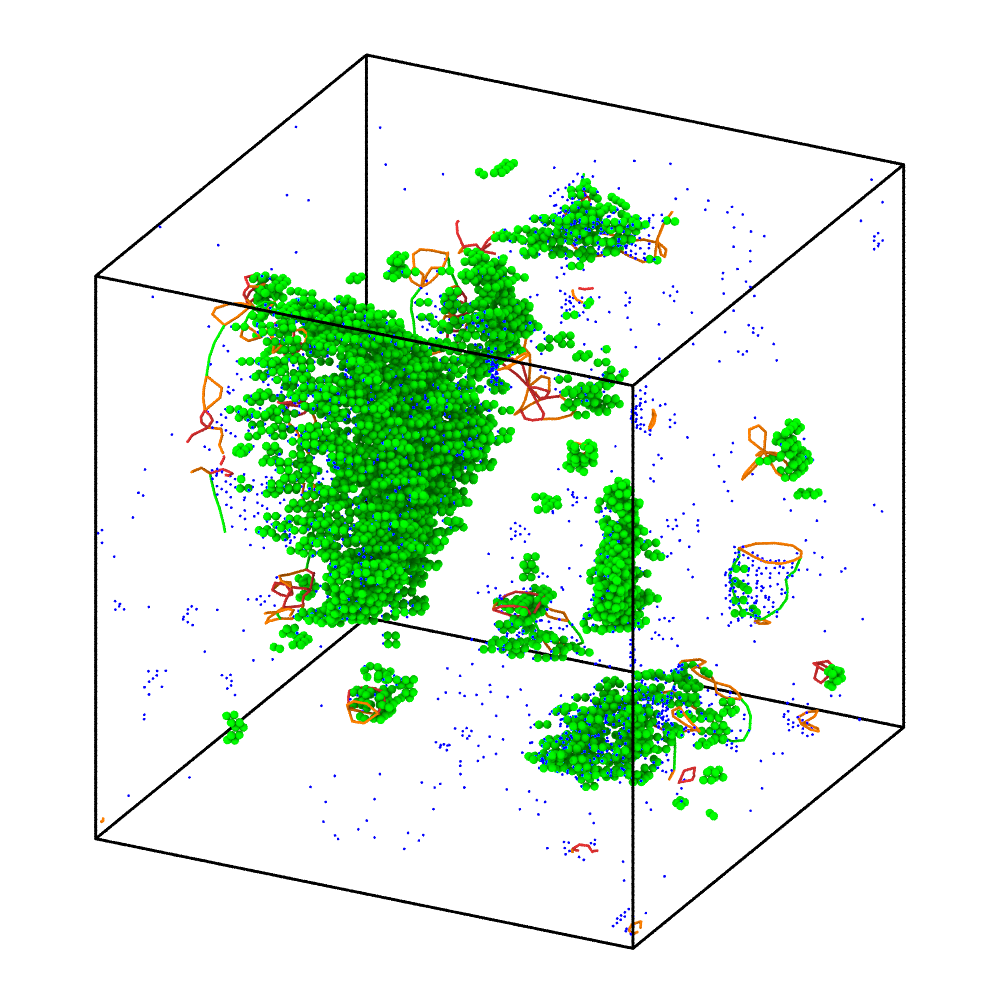}
    \caption{Modified Zhou EAM}
\end{subfigure}
    \caption{Final configuration after 2000 overlapping collision cascades. small blue spheres represent vacancies, colored spheres represent interstitial atoms, green lines are $1/3$ \hkl<1-210> dislocations, blue lines are \hkl<0001> dislocations, magenta lines are \hkl<1-100> dislocations, orange lines are $1/3$\hkl<1-100> dislocations, yellow lines are $1/3$ \hkl<1-213> dislocations and red lines are dislocations with type ''other'' as identified by DXA analysis.}
\end{figure}

\begin{figure}[ht!]
    \centering
    \includegraphics[width=0.2\linewidth]{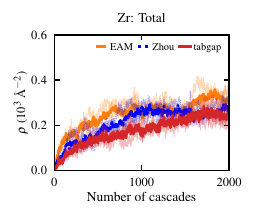}
    \includegraphics[width=0.2\linewidth]{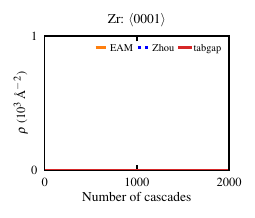}
    \includegraphics[width=0.2\linewidth]{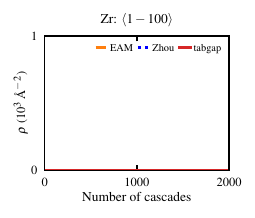}
    \includegraphics[width=0.2\linewidth]{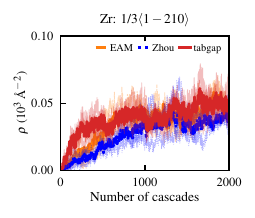}
    \includegraphics[width=0.2\linewidth]{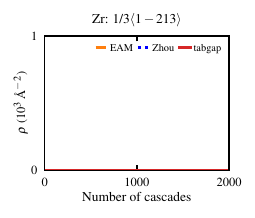}
    \includegraphics[width=0.2\linewidth]{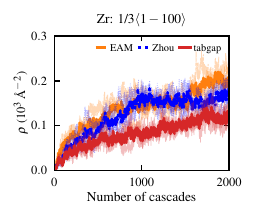}
    \includegraphics[width=0.2\linewidth]{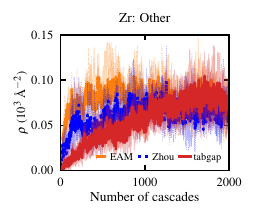}
    \caption{Dislocation density obtained from DXA analysis. Faint curves are individual runs and bold is the average.}
\end{figure}

\clearpage

\section{Overlapping cascades: Niobium $\text{Nb}~~Z=41$}

\begin{table}[ht!]
    \centering
    \begin{tabular}{c|c|c}
        \toprule
        Potential & Ref. & \hspace{5cm} notes \hspace{5cm} \\
        \midrule
        Nb-tabGAP & \cite{byggmastar2026nine} & \\
        \bottomrule
    \end{tabular}
    \caption{Details of potentials}
\end{table}
\vspace{-0.5cm}
\begin{figure}[ht!]
    \centering
    \includegraphics[width=0.35\linewidth]{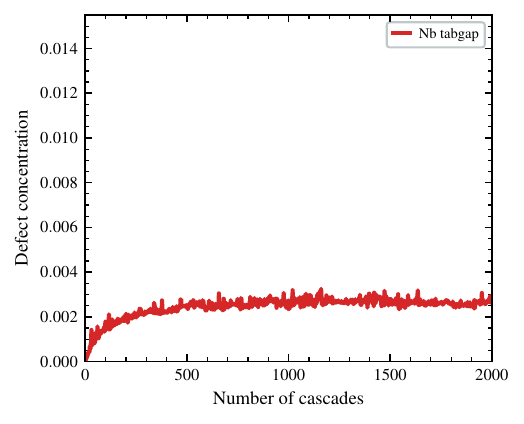}
    \includegraphics[width=0.35\linewidth]{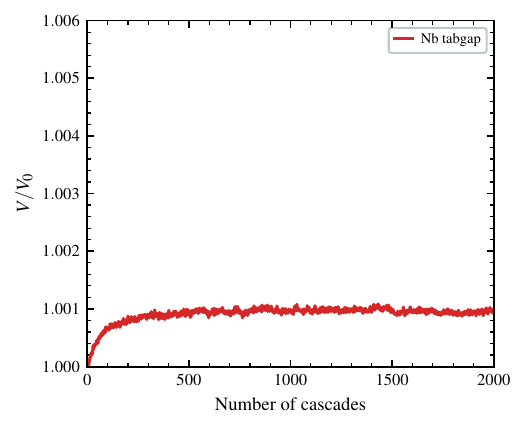}
    \caption{Left is the defect  concentration (interstitials and vacancies) as a function of number of cascades. Right is the volume change as a function of number of cascades.}
\end{figure}

\vspace{-1cm}

\begin{figure}[ht!]
    \centering
\begin{subfigure}{0.29\textwidth}
    \includegraphics[width=1.1\textwidth]{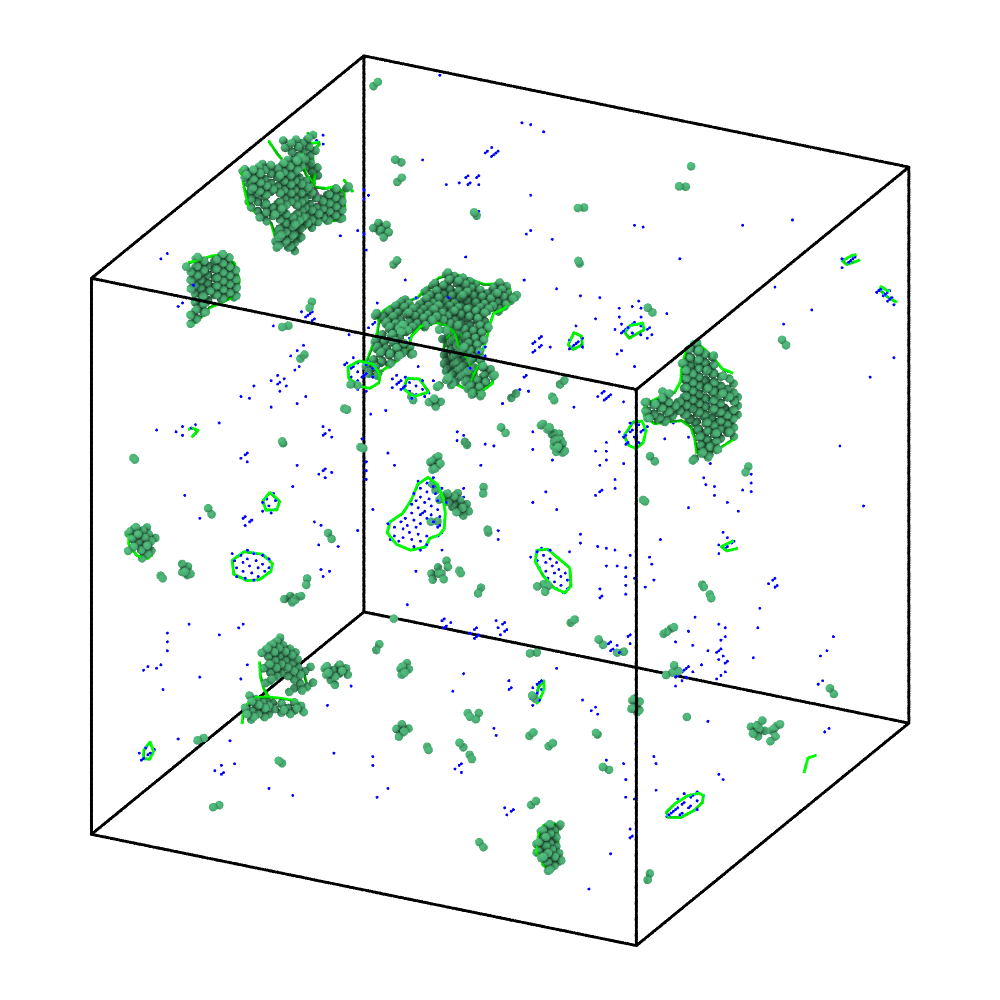}
      \caption{tabGAP}
\end{subfigure}
    \caption{Final configuration after 2000 overlapping collision cascades. small blue spheres represent vacancies, colored spheres represent interstitial atoms, green lines represent 1/2 \hkl<111> dislocations, and pink lines \hkl<100> dislocations.}
\end{figure}

\begin{figure}[ht!]
    \centering
    \includegraphics[width=0.2\linewidth]{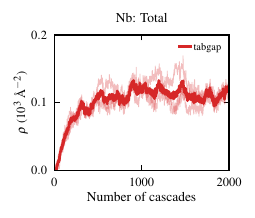}
    \includegraphics[width=0.2\linewidth]{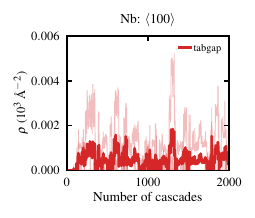}
    \includegraphics[width=0.2\linewidth]{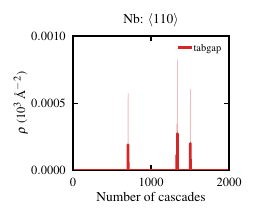}
    \includegraphics[width=0.2\linewidth]{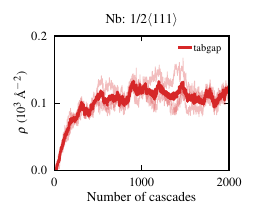}
    \includegraphics[width=0.2\linewidth]{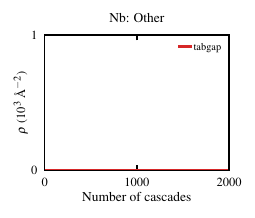}
    \caption{Dislocation density obtained from DXA analysis. Faint curves are individual runs and bold is the average.}
\end{figure}

\clearpage

\section{Overlapping cascades: Molybdenum $\text{Mo}~~Z=42$}

\begin{table}[ht!]
    \centering
    \begin{tabular}{c|c|c}
        \toprule
        Potential & Ref. & \hspace{5cm} notes \hspace{5cm} \\
        \midrule
        Mo-tabGAP & \cite{byggmastar2026nine}  &  \\
        \midrule
        Mo-Zhou & \cite{zhou_misfit-energy-increasing_2004} & \\
        \bottomrule
    \end{tabular}
    \caption{Details of potentials}
\end{table}
\vspace{-0.5cm}
\begin{figure}[ht!]
    \centering
    \includegraphics[width=0.35\linewidth]{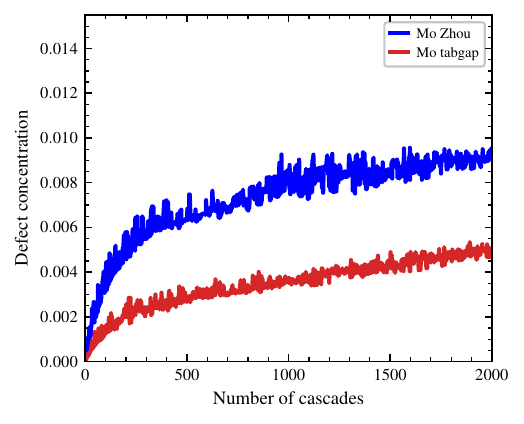}
    \includegraphics[width=0.35\linewidth]{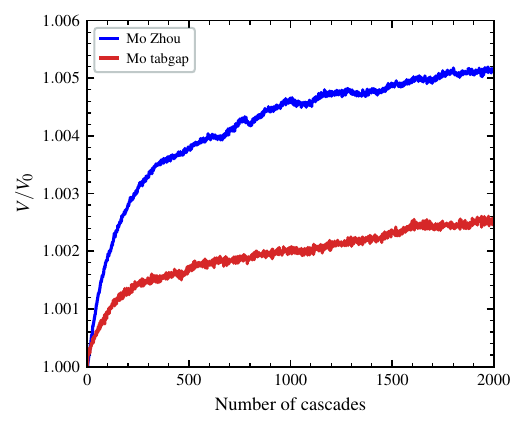}
    \caption{Left is the defect  concentration (interstitials and vacancies) as a function of number of cascades. Right is the volume change as a function of number of cascades.}
\end{figure}

\vspace{-1cm}

\begin{figure}[ht!]
    \centering
\begin{subfigure}{0.29\textwidth}
    \includegraphics[width=1.1\textwidth]{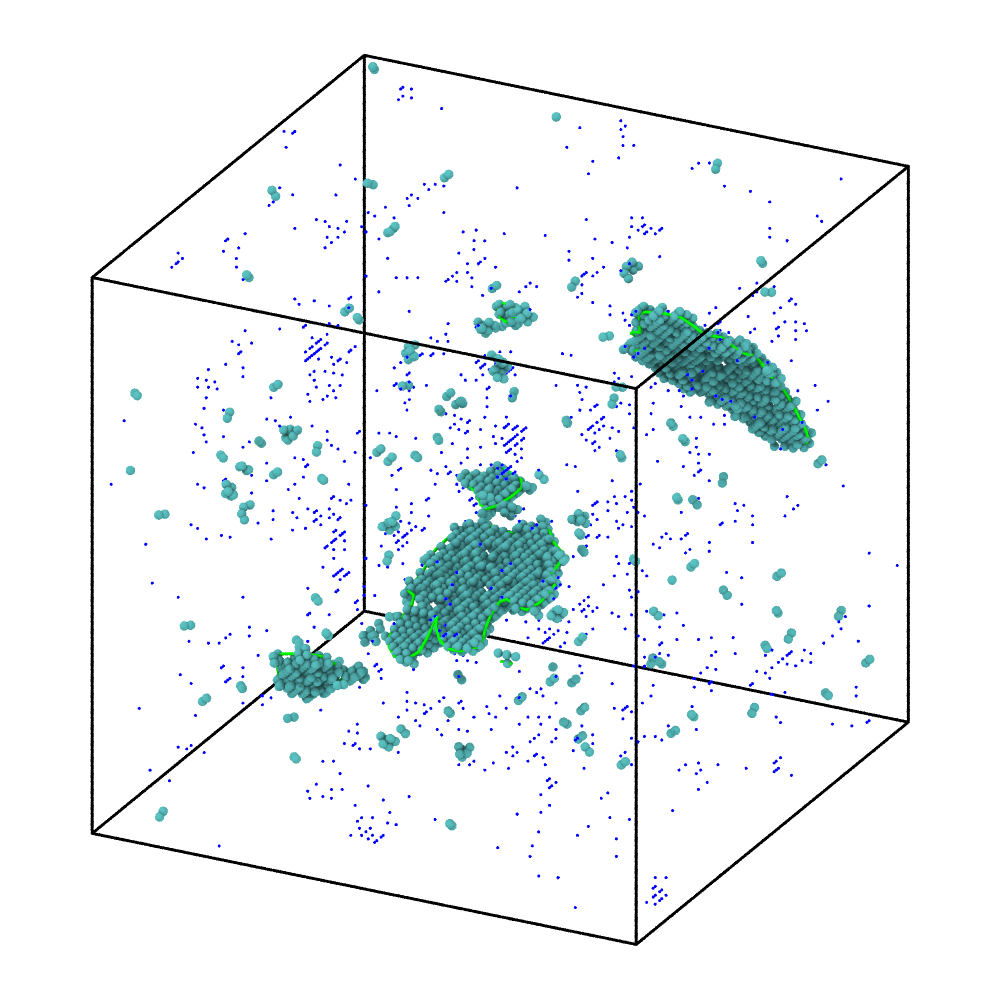}
      \caption{tabGAP}
\end{subfigure}
\begin{subfigure}{0.29\textwidth}
    \includegraphics[width=1.1\textwidth]{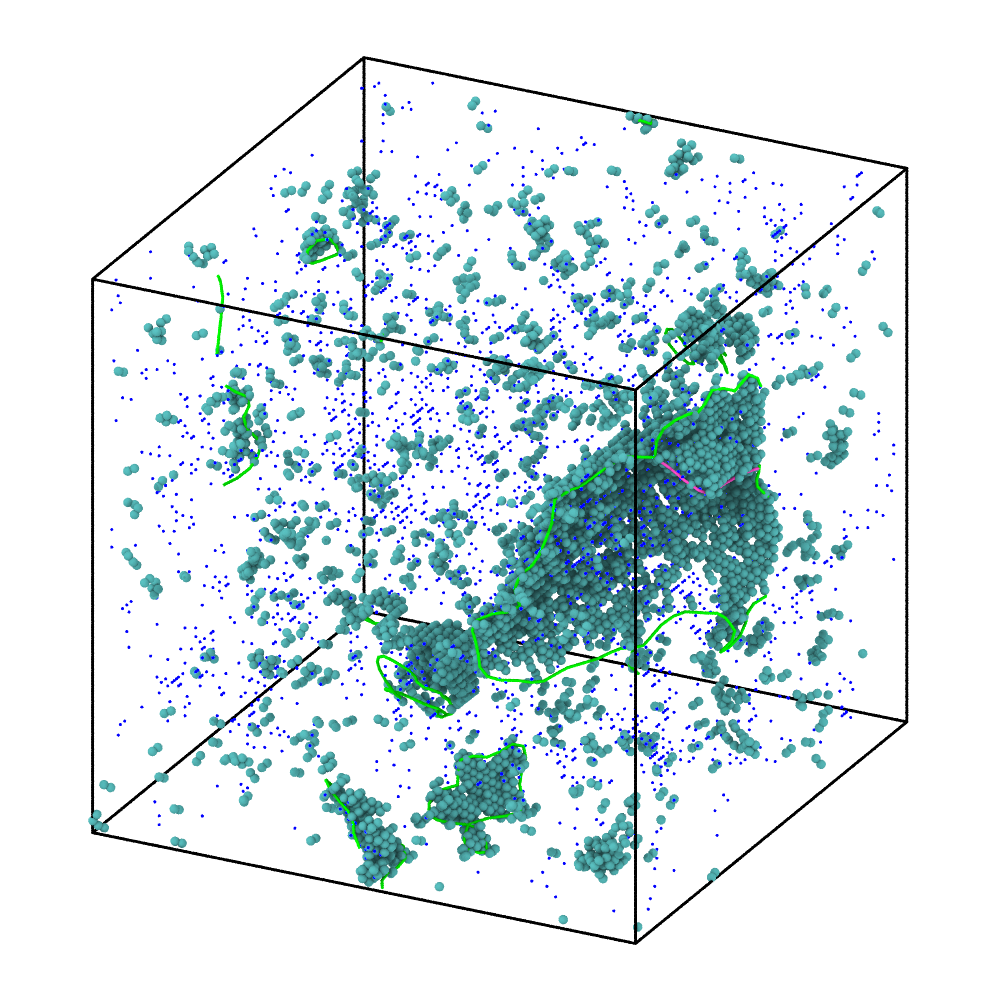}
    \caption{Modified Zhou EAM}
\end{subfigure}
    \caption{Final configuration after 2000 overlapping collision cascades. small blue spheres represent vacancies, colored spheres represent interstitial atoms, green lines represent 1/2 \hkl<111> dislocations, and pink lines \hkl<100> dislocations.}
\end{figure}

\begin{figure}[ht!]
    \centering
    \includegraphics[width=0.2\linewidth]{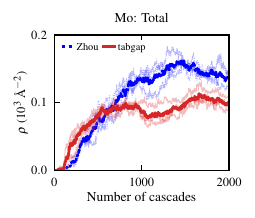}
    \includegraphics[width=0.2\linewidth]{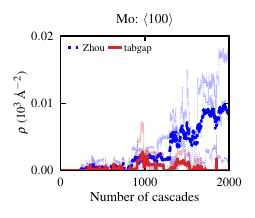}
    \includegraphics[width=0.2\linewidth]{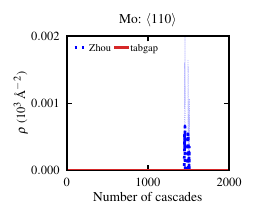}
    \includegraphics[width=0.2\linewidth]{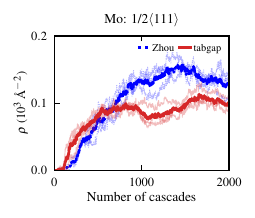}
    \includegraphics[width=0.2\linewidth]{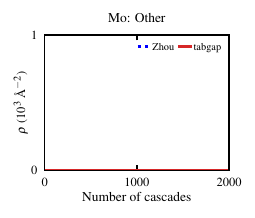}
    \caption{Dislocation density obtained from DXA analysis. Faint curves are individual runs and bold is the average.}
\end{figure}

\clearpage

\section{Overlapping cascades: Rhodium $\text{Rh}~~Z=45$}

\begin{table}[ht!]
    \centering
    \begin{tabular}{c|c|c}
        \toprule
        Potential & Ref. & \hspace{5cm} notes \hspace{5cm} \\
        \midrule
        EAM & \cite{sheng_highly_2011} & Rh-EAM ''hardened'' by J. Byggmästar \\
        \bottomrule
    \end{tabular}
    \caption{Details of potentials}
\end{table}
\vspace{-0.5cm}
\begin{figure}[ht!]
    \centering
    \includegraphics[width=0.35\linewidth]{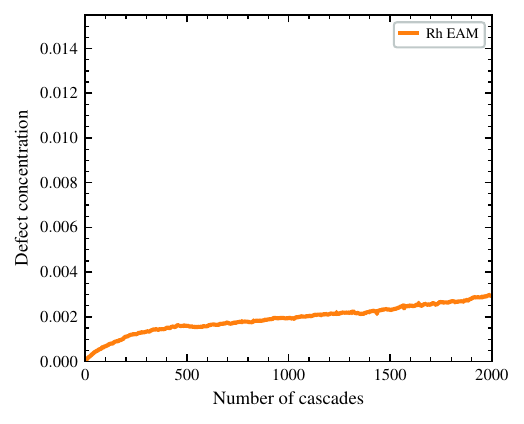}
    \includegraphics[width=0.35\linewidth]{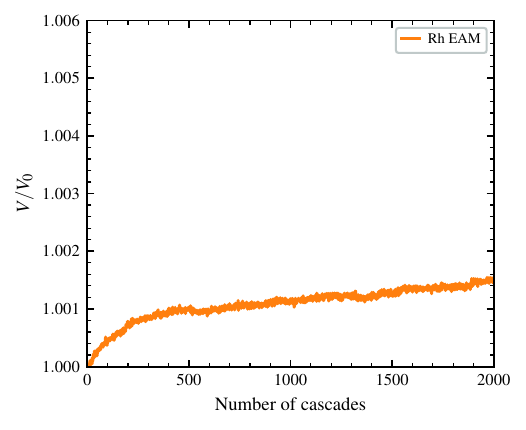}
    \caption{Left is the defect  concentration (interstitials and vacancies) as a function of number of cascades. Right is the volume change as a function of number of cascades.}
\end{figure}

\vspace{-1cm}

\begin{figure}[ht!]
    \centering
\begin{subfigure}{0.29\textwidth}
    \includegraphics[width=1.1\textwidth]{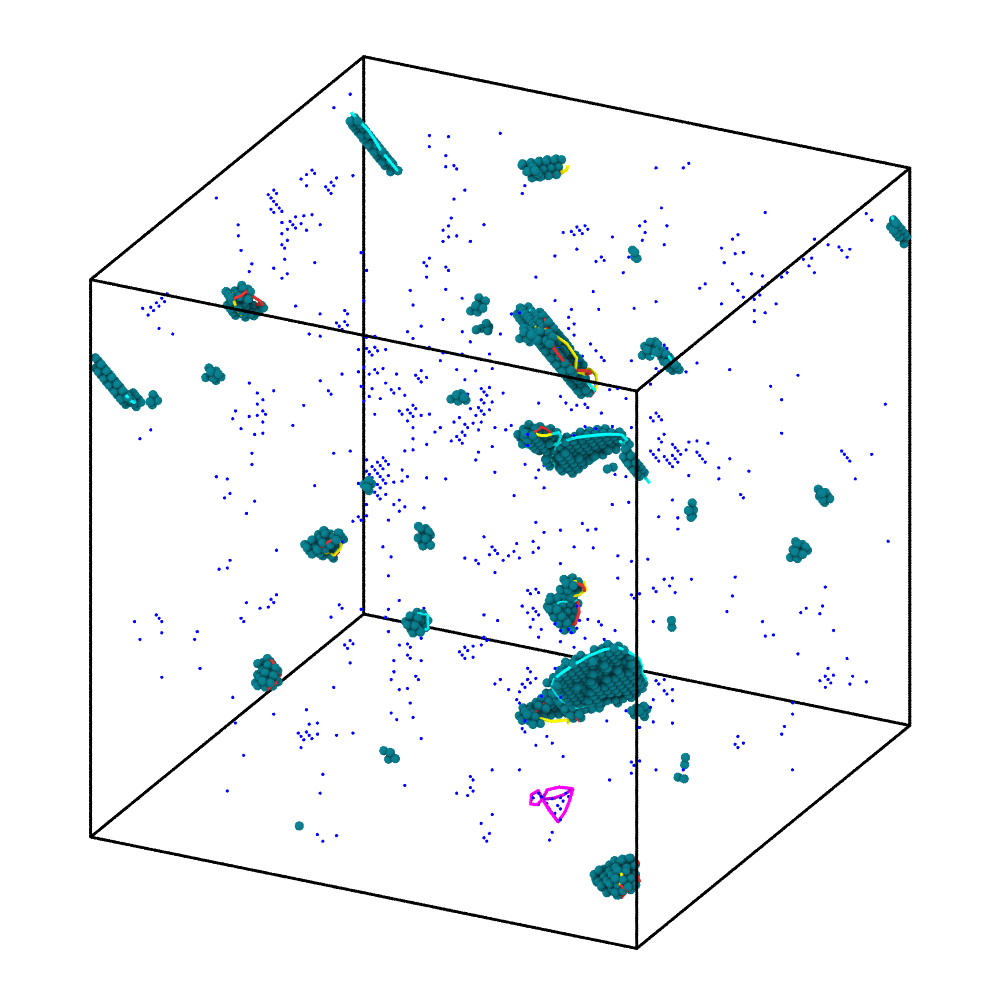}
    \caption{EAM}
\end{subfigure}
    \caption{Final configuration after 2000 overlapping collision cascades. blue spheres represent vacancies, colored spheres represent interstitial atoms, green lines represent Shockley dislocations, and cyan lines represent Frank dislocations and pink lines represent stair-rod dislocations.}
\end{figure}

\begin{figure}[ht!]
    \centering
    \includegraphics[width=0.2\linewidth]{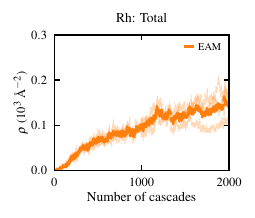}
    \includegraphics[width=0.2\linewidth]{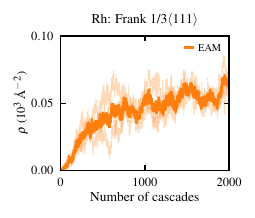}
    \includegraphics[width=0.2\linewidth]{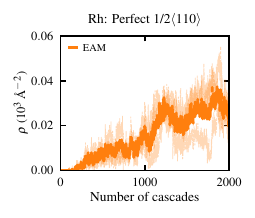}
    \includegraphics[width=0.2\linewidth]{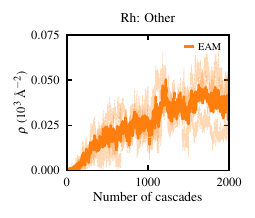}
    \includegraphics[width=0.2\linewidth]{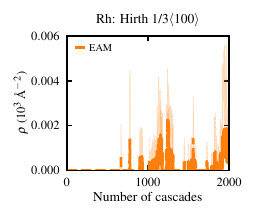}
    \includegraphics[width=0.2\linewidth]{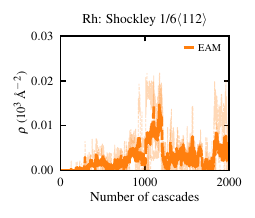}
    \includegraphics[width=0.2\linewidth]{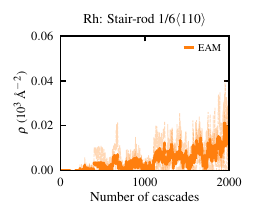}
    \caption{Dislocation density obtained from DXA analysis. Faint curves are individual runs and bold is the average.}
\end{figure}

\clearpage

\section{Overlapping cascades: Palladium $\text{Pd}~~Z=46$}

\begin{table}[ht!]
    \centering
    \begin{tabular}{c|c|c}
        \toprule
        Potential & Ref. & \hspace{5cm} notes \hspace{5cm} \\
        \midrule
        Pd-EAM & \cite{foiles1985application,nordlund1998role} & Pd-EAM  ''hardened'' by K. Nordlund~\cite{nordlund1998role}\\
        \midrule
        Pd-Zhou & \cite{zhou_misfit-energy-increasing_2004} & \\
        \bottomrule
    \end{tabular}
    \caption{Details of potentials}
\end{table}
\vspace{-0.5cm}
\begin{figure}[ht!]
    \centering
    \includegraphics[width=0.35\linewidth]{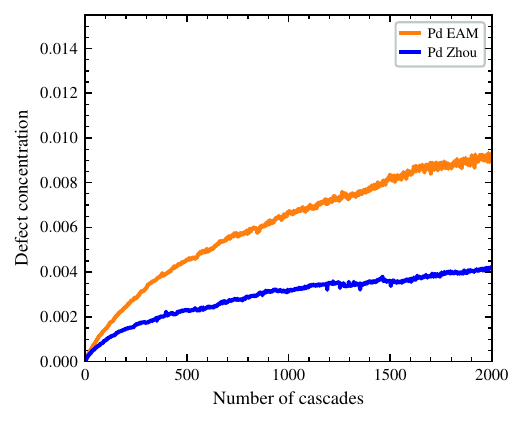}
    \includegraphics[width=0.35\linewidth]{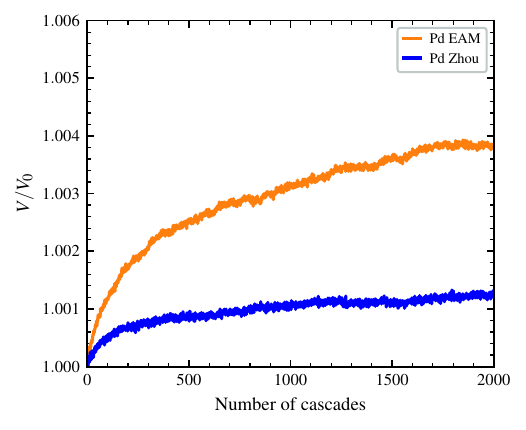}
    \caption{Left is the defect  concentration (interstitials and vacancies) as a function of number of cascades. Right is the volume change as a function of number of cascades.}
\end{figure}

\vspace{-1cm}

\begin{figure}[ht!]
    \centering
\begin{subfigure}{0.29\textwidth}
    \includegraphics[width=1.1\textwidth]{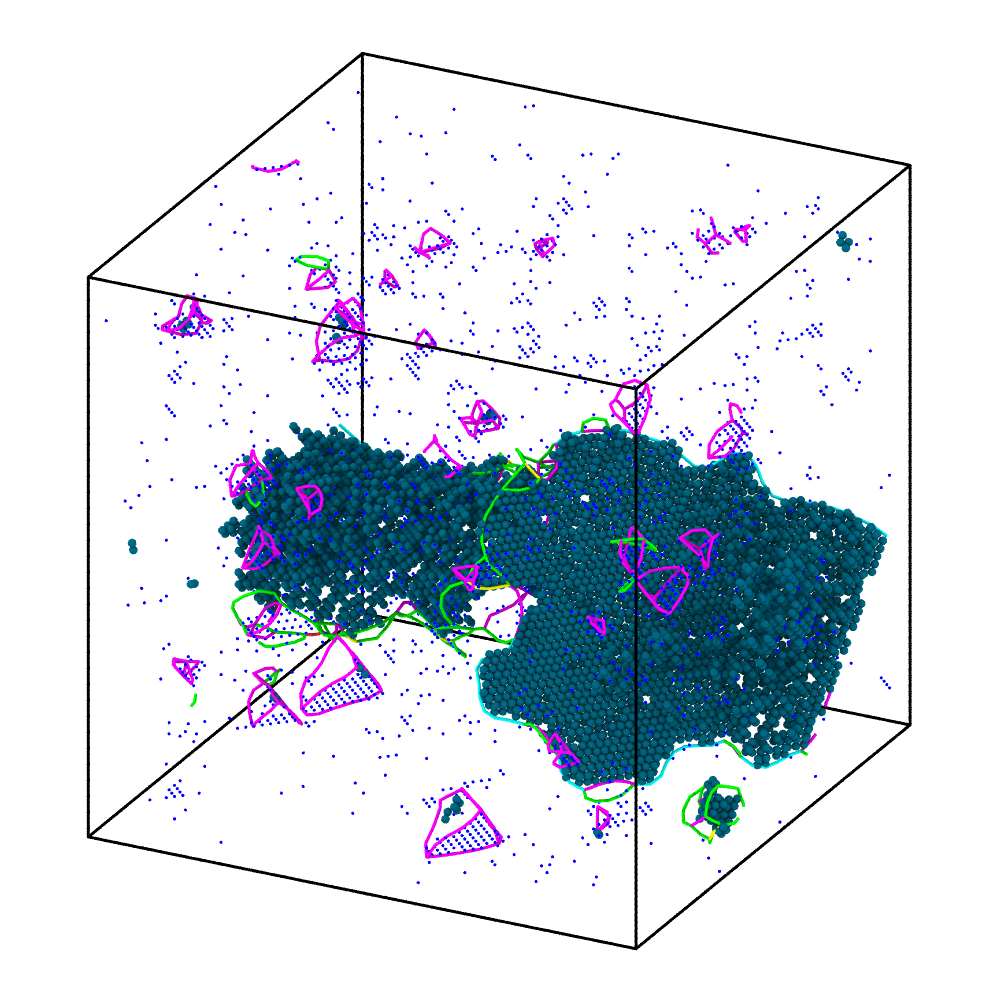}
    \caption{EAM}
\end{subfigure}
\begin{subfigure}{0.29\textwidth}
    \includegraphics[width=1.1\textwidth]{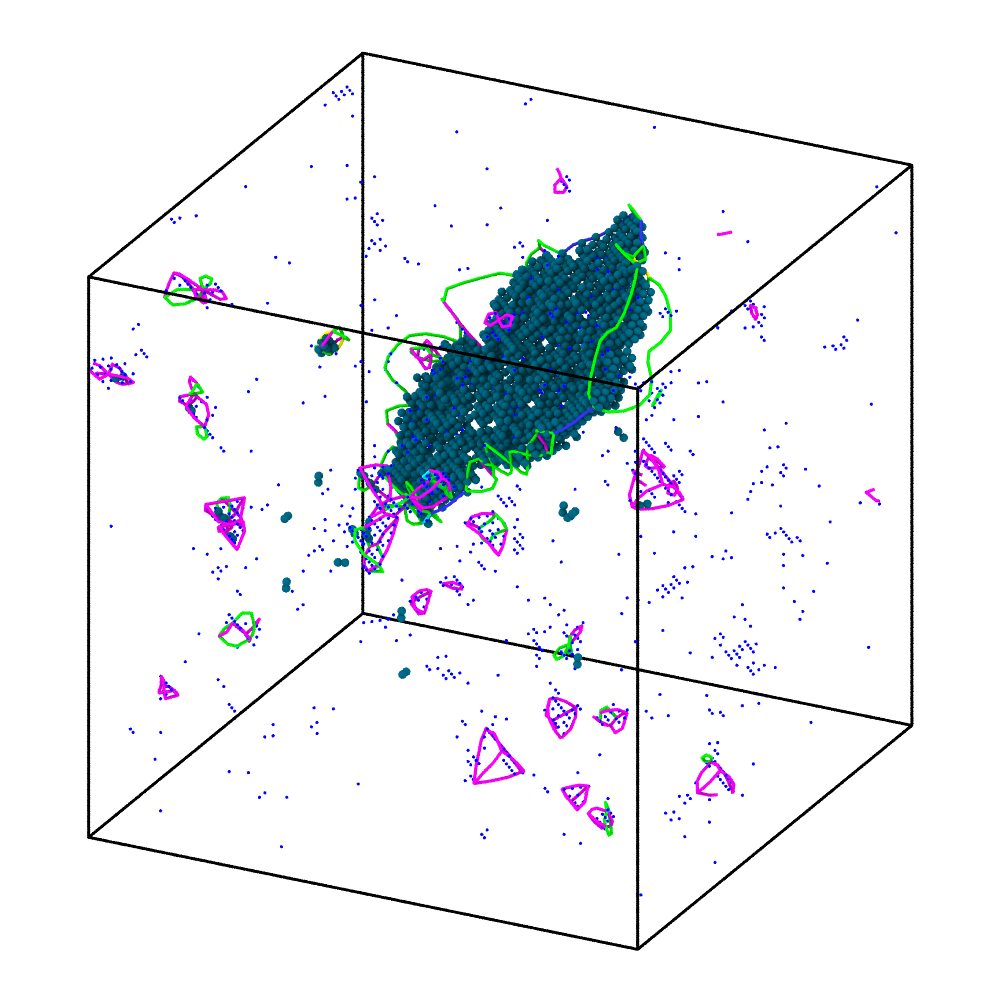}
    \caption{Modified Zhou EAM}
\end{subfigure}
    \caption{Final configuration after 2000 overlapping collision cascades. blue spheres represent vacancies, colored spheres represent interstitial atoms, green lines represent Shockley dislocations, and cyan lines represent Frank dislocations and pink lines represent stair-rod dislocations.}
\end{figure}

\begin{figure}[ht!]
    \centering
    \includegraphics[width=0.2\linewidth]{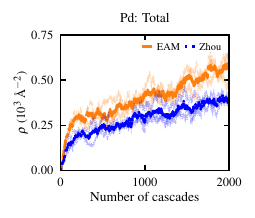}
    \includegraphics[width=0.2\linewidth]{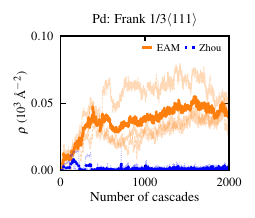}
    \includegraphics[width=0.2\linewidth]{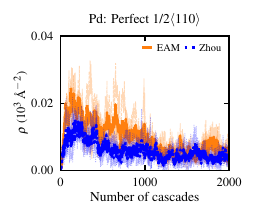}
    \includegraphics[width=0.2\linewidth]{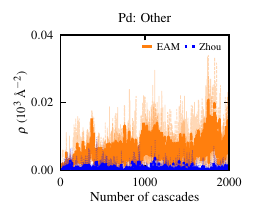}
    \includegraphics[width=0.2\linewidth]{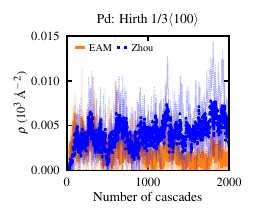}
    \includegraphics[width=0.2\linewidth]{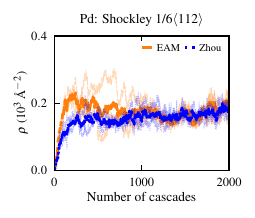}
    \includegraphics[width=0.2\linewidth]{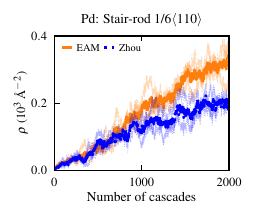}
    \caption{Dislocation density obtained from DXA analysis. Faint curves are individual runs and bold is the average.}
\end{figure}

\clearpage

\section{Overlapping cascades: Silver $\text{Ag}~~Z=47$}

\begin{table}[ht!]
    \centering
    \begin{tabular}{c|c|c}
        \toprule
        Potential & Ref. & \hspace{5cm} notes \hspace{5cm} \\
        \midrule
        Ag-EAM & \cite{foiles_embedded-atom-method_1986,colla2001ion}  & Ag-EAM ''hardened'' by K. Nordlund~\cite{colla2001ion}\\
        \midrule
        Ag-Zhou & \cite{zhou_misfit-energy-increasing_2004} & \\
        \bottomrule
    \end{tabular}
    \caption{Details of potentials}
\end{table}
\vspace{-0.5cm}
\begin{figure}[ht!]
    \centering
    \includegraphics[width=0.35\linewidth]{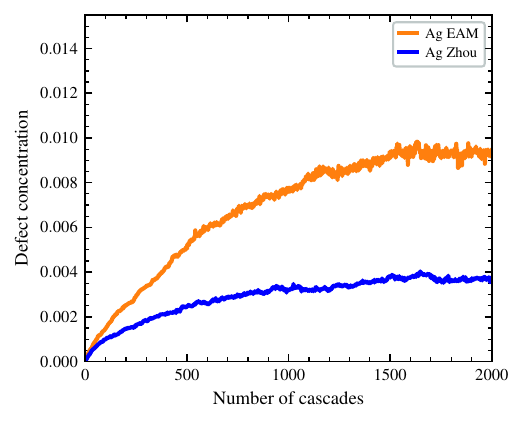}
    \includegraphics[width=0.35\linewidth]{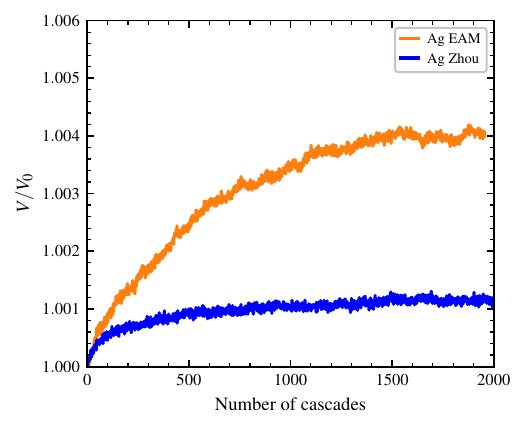}
    \caption{Left is the defect  concentration (interstitials and vacancies) as a function of number of cascades. Right is the volume change as a function of number of cascades.}
\end{figure}

\vspace{-1cm}

\begin{figure}[ht!]
    \centering
\begin{subfigure}{0.29\textwidth}
    \includegraphics[width=1.1\textwidth]{figures/frames/frame-Ag-EAM-1-run-2-2000.png}
    \caption{EAM}
\end{subfigure}
\begin{subfigure}{0.29\textwidth}
    \includegraphics[width=1.1\textwidth]{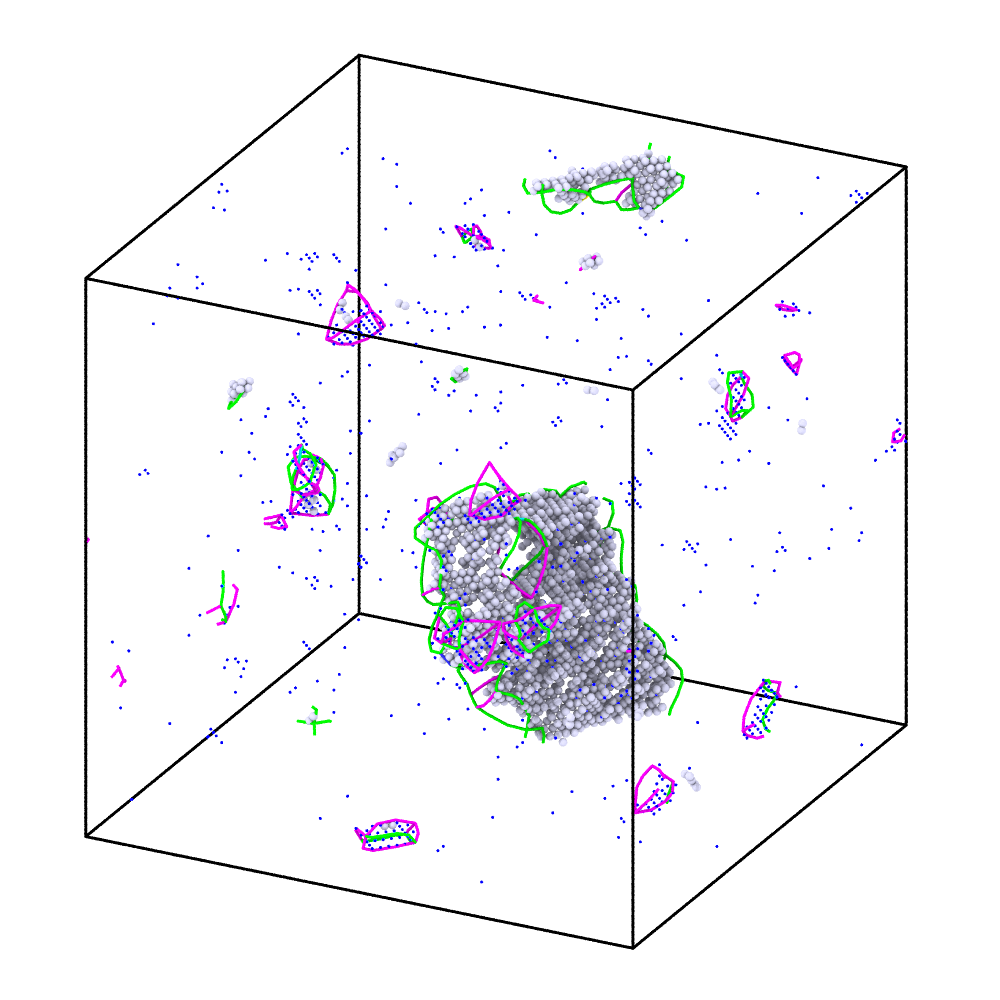}
    \caption{Modified Zhou EAM}
\end{subfigure}
    \caption{Final configuration after 2000 overlapping collision cascades. blue spheres represent vacancies, colored spheres represent interstitial atoms, green lines represent Shockley dislocations, and cyan lines represent Frank dislocations and pink lines represent stair-rod dislocations.}
\end{figure}

\begin{figure}[ht!]
    \centering
    \includegraphics[width=0.2\linewidth]{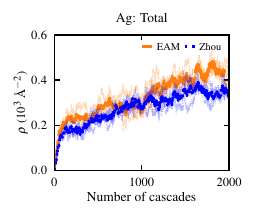}
    \includegraphics[width=0.2\linewidth]{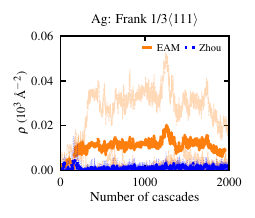}
    \includegraphics[width=0.2\linewidth]{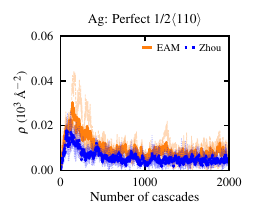}
    \includegraphics[width=0.2\linewidth]{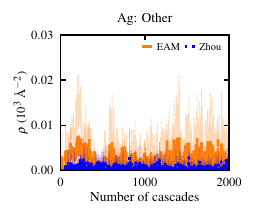}
    \includegraphics[width=0.2\linewidth]{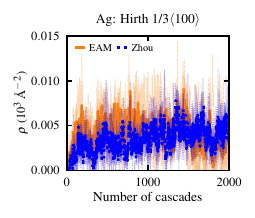}
    \includegraphics[width=0.2\linewidth]{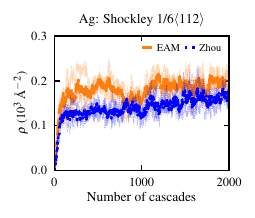}
    \includegraphics[width=0.2\linewidth]{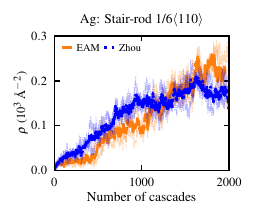}
    \caption{Dislocation density obtained from DXA analysis. Faint curves are individual runs and bold is the average.}
\end{figure}

\clearpage

\section{Overlapping cascades: Hafnium $\text{Hf}~~Z=72$}

\begin{table}[ht!]
    \centering
    \begin{tabular}{c|c|c}
        \toprule
        Potential & Ref. & \hspace{5cm} notes \hspace{5cm} \\
        \midrule
        Hf-tabGAP & \cite{byggmastar2026nine} &  \\
        \bottomrule
    \end{tabular}
    \caption{Details of potentials}
\end{table}
\vspace{-0.5cm}
\begin{figure}[ht!]
    \centering
    \includegraphics[width=0.35\linewidth]{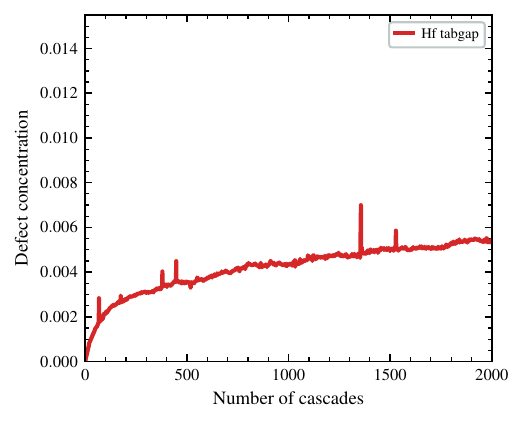}
    \includegraphics[width=0.35\linewidth]{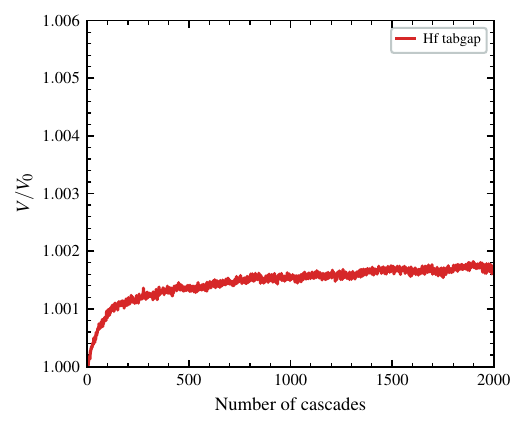}
    \caption{Left is the defect  concentration (interstitials and vacancies) as a function of number of cascades. Right is the volume change as a function of number of cascades.}
\end{figure}

\vspace{-1cm}

\begin{figure}[ht!]
    \centering
\begin{subfigure}{0.29\textwidth}
    \includegraphics[width=1.1\textwidth]{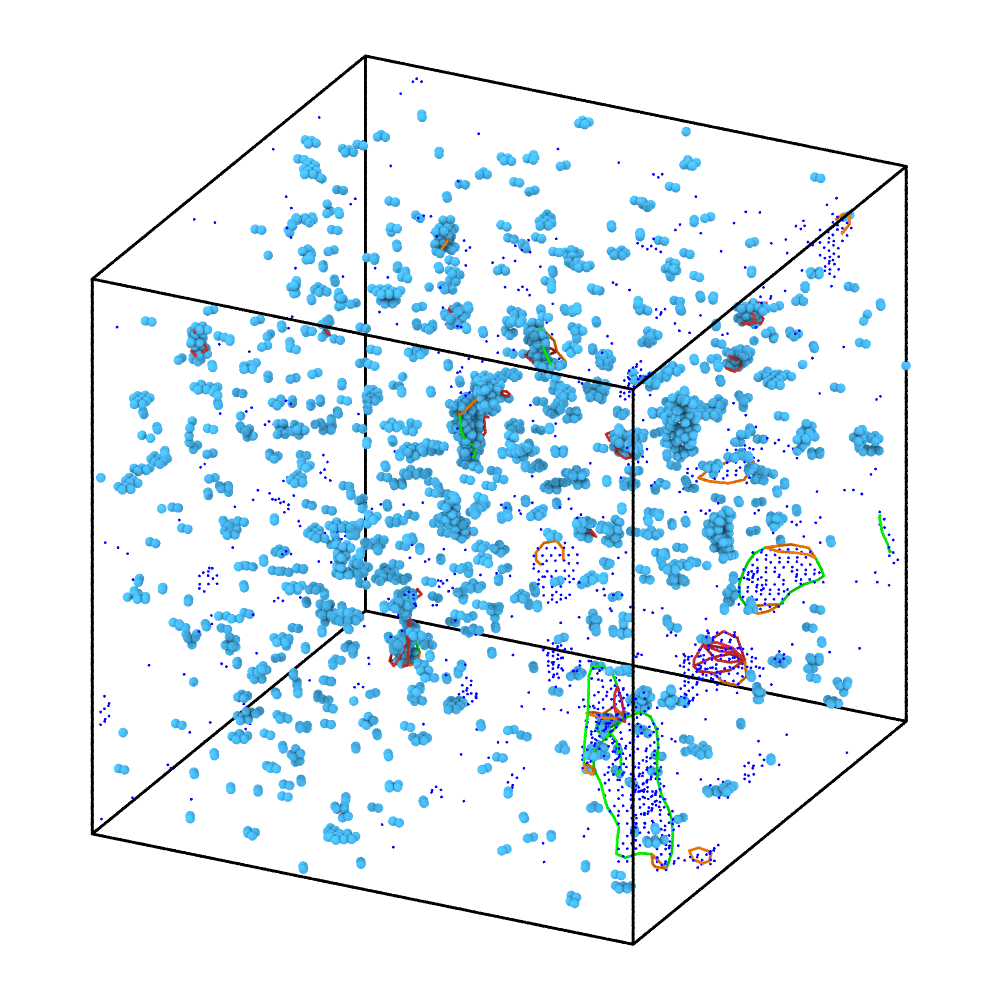}
      \caption{tabGAP}
\end{subfigure}
    \caption{Final configuration after 2000 overlapping collision cascades. small blue spheres represent vacancies, colored spheres represent interstitial atoms, green lines are $1/3$ \hkl<1-210> dislocations, blue lines are \hkl<0001> dislocations, magenta lines are \hkl<1-100> dislocations, orange lines are $1/3$\hkl<1-100> dislocations, yellow lines are $1/3$ \hkl<1-213> dislocations and red lines are dislocations with type ''other'' as identified by DXA analysis.}
\end{figure}

\begin{figure}[ht!]
    \centering
    \includegraphics[width=0.2\linewidth]{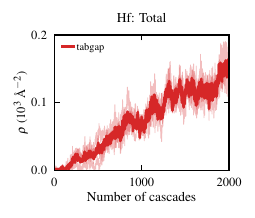}
    \includegraphics[width=0.2\linewidth]{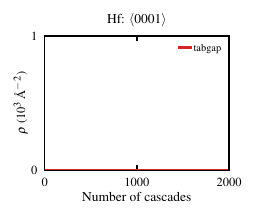}
    \includegraphics[width=0.2\linewidth]{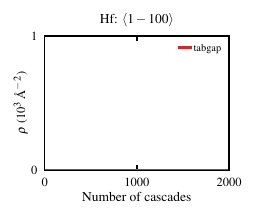}
    \includegraphics[width=0.2\linewidth]{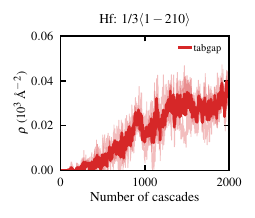}
    \includegraphics[width=0.2\linewidth]{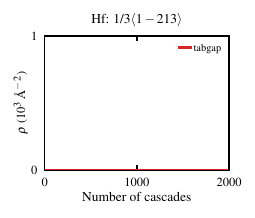}
    \includegraphics[width=0.2\linewidth]{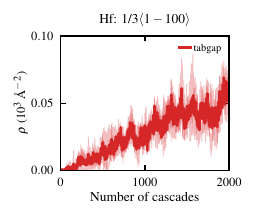}
    \includegraphics[width=0.2\linewidth]{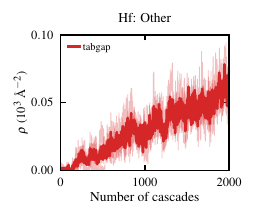}
    \caption{Dislocation density obtained from DXA analysis. Faint curves are individual runs and bold is the average.}
\end{figure}

\clearpage

\section{Overlapping cascades: Tantalum $\text{Ta}~~Z=73$}

\begin{table}[ht!]
    \centering
    \begin{tabular}{c|c|c}
        \toprule
        Potential & Ref. & \hspace{5cm} notes \hspace{5cm} \\
        \midrule
        Ta-tabGAP & \cite{byggmastar2026nine} &  \\
        \midrule
        Ta-Zhou & \cite{zhou_misfit-energy-increasing_2004} & \\
        \bottomrule
    \end{tabular}
    \caption{Details of potentials}
\end{table}
\vspace{-0.5cm}
\begin{figure}[ht!]
    \centering
    \includegraphics[width=0.35\linewidth]{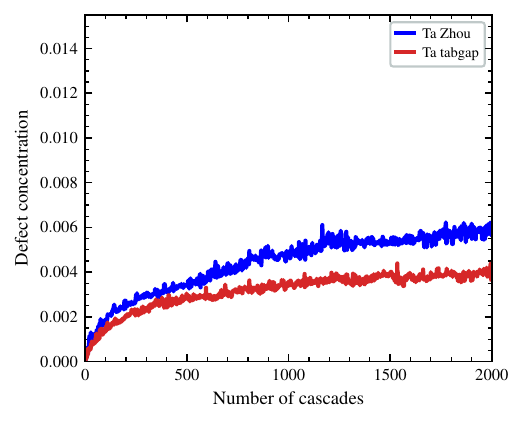}
    \includegraphics[width=0.35\linewidth]{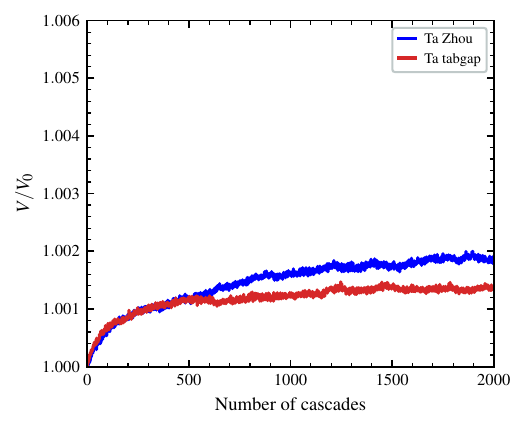}
    \caption{Left is the defect  concentration (interstitials and vacancies) as a function of number of cascades. Right is the volume change as a function of number of cascades.}
\end{figure}

\vspace{-1cm}

\begin{figure}[ht!]
    \centering
\begin{subfigure}{0.29\textwidth}
    \includegraphics[width=1.1\textwidth]{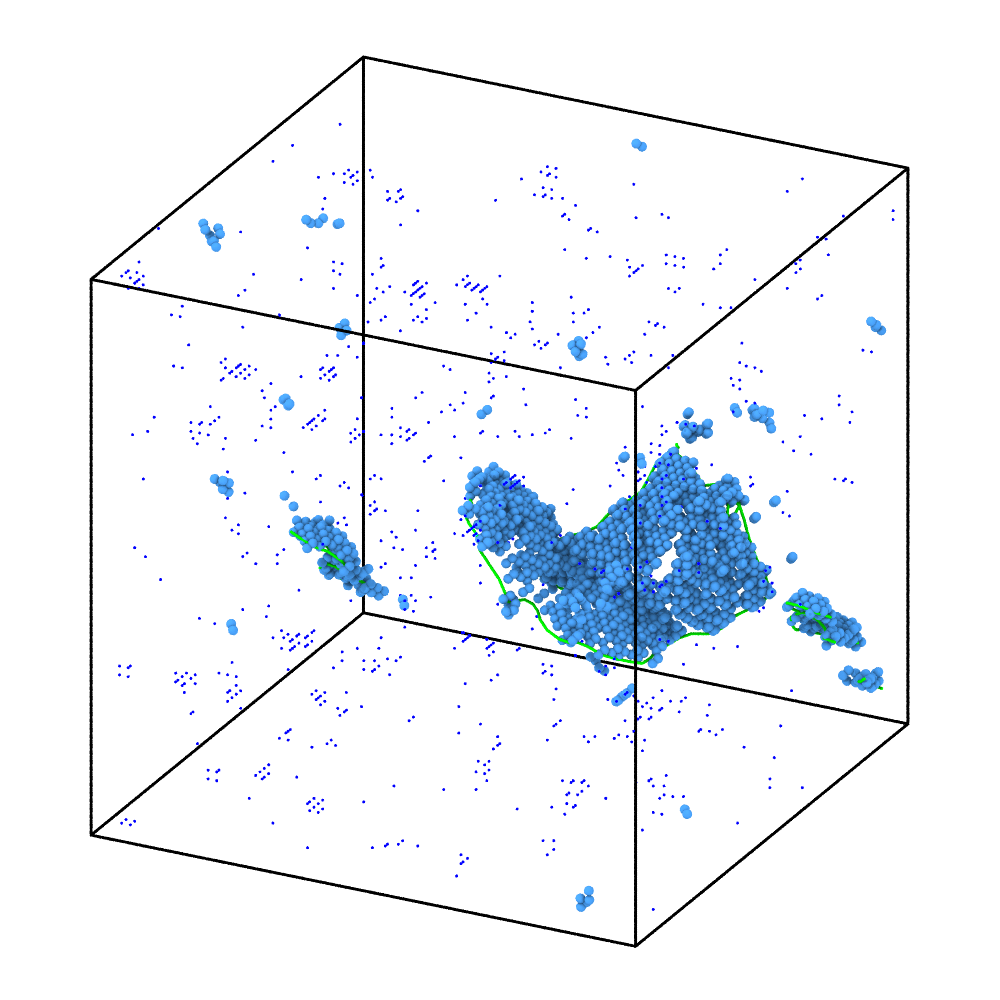}
      \caption{tabGAP}
\end{subfigure}
\begin{subfigure}{0.29\textwidth}
    \includegraphics[width=1.1\textwidth]{figures/frames/frame-Ta-Zhou-run-3-2000.png}
    \caption{Modified Zhou EAM}
\end{subfigure}
    \caption{Final configuration after 2000 overlapping collision cascades. small blue spheres represent vacancies, colored spheres represent interstitial atoms, green lines represent 1/2 \hkl<111> dislocations, and pink lines \hkl<100> dislocations.}
\end{figure}

\begin{figure}[ht!]
    \centering
    \includegraphics[width=0.2\linewidth]{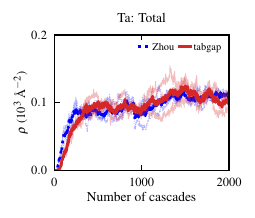}
    \includegraphics[width=0.2\linewidth]{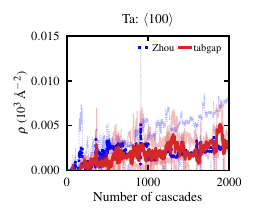}
    \includegraphics[width=0.2\linewidth]{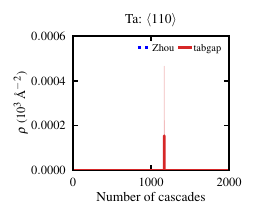}
    \includegraphics[width=0.2\linewidth]{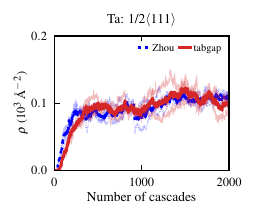}
    \includegraphics[width=0.2\linewidth]{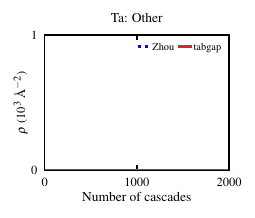}
    \caption{Dislocation density obtained from DXA analysis. Faint curves are individual runs and bold is the average.}
\end{figure}

\clearpage

\section{Overlapping cascades: Tungsten $\text{W}~~Z=74$}

\begin{table}[ht!]
    \centering
    \begin{tabular}{c|c|c}
        \toprule
        Potential & Ref. & \hspace{5cm} notes \hspace{5cm} \\
        \midrule
        W-tabGAP & \cite{byggmastar_simple_2022} & \\
        \midrule
        W-EAM & \cite{mason2023empirical} & \\
        \midrule
        W-Zhou & \cite{zhou_misfit-energy-increasing_2004} & \\
        \bottomrule
    \end{tabular}
    \caption{Details of potentials}
\end{table}
\vspace{-0.5cm}
\begin{figure}[ht!]
    \centering
    \includegraphics[width=0.35\linewidth]{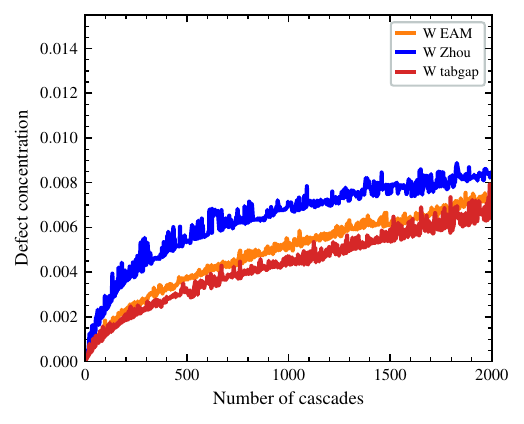}
    \includegraphics[width=0.35\linewidth]{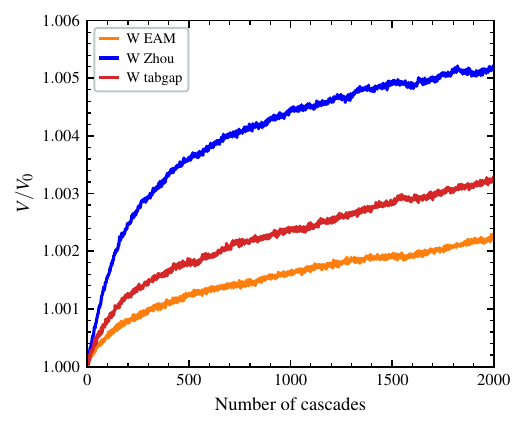}
    \caption{Left is the defect  concentration (interstitials and vacancies) as a function of number of cascades. Right is the volume change as a function of number of cascades.}
\end{figure}

\vspace{-1cm}

\begin{figure}[ht!]
    \centering
\begin{subfigure}{0.29\textwidth}
    \includegraphics[width=1.1\textwidth]{figures/frames/frame-W-tabgap-run-1-2000.png}
      \caption{tabGAP}
\end{subfigure}
\begin{subfigure}{0.29\textwidth}
    \includegraphics[width=1.1\textwidth]{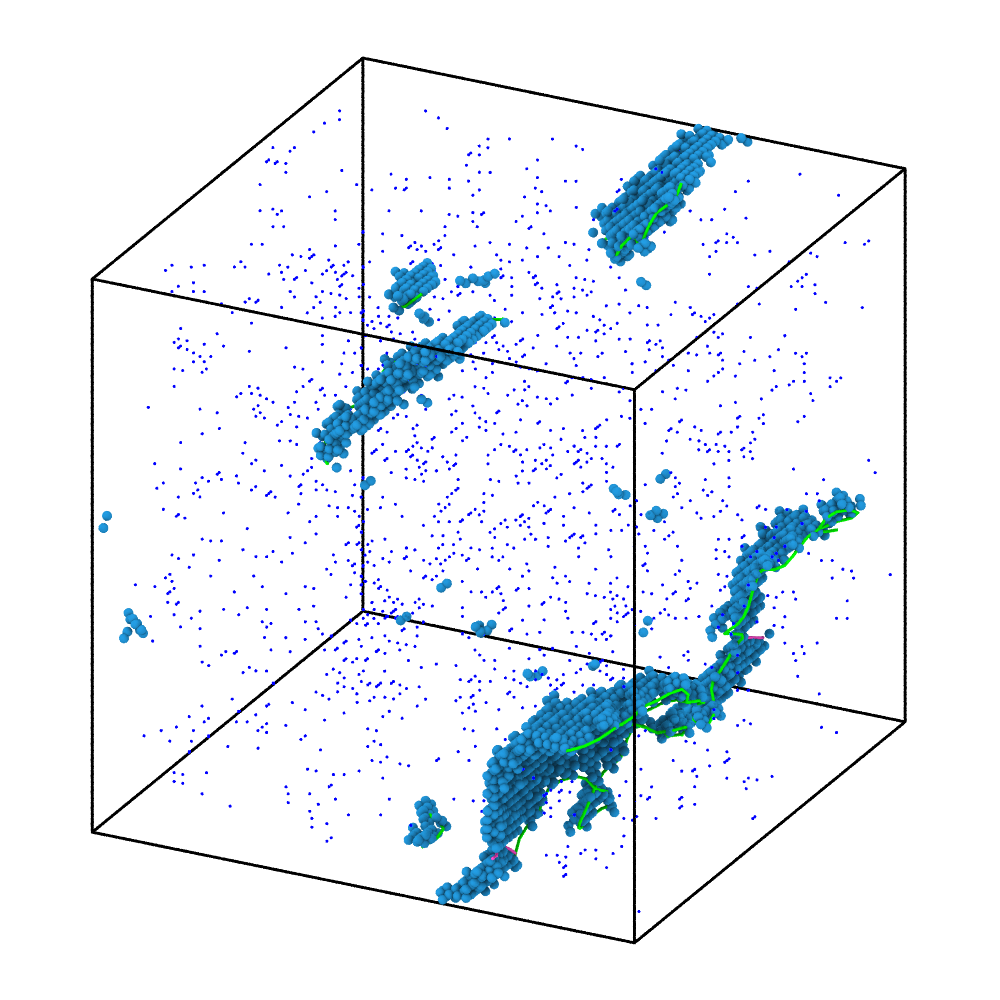}
    \caption{EAM}
\end{subfigure}
\begin{subfigure}{0.29\textwidth}
    \includegraphics[width=1.1\textwidth]{figures/frames/frame-W-Zhou-run-1-2000.png}
    \caption{Modified Zhou EAM}
\end{subfigure}
    \caption{Final configuration after 2000 overlapping collision cascades. small blue spheres represent vacancies, colored spheres represent interstitial atoms, green lines represent 1/2 \hkl<111> dislocations, and pink lines \hkl<100> dislocations.}
\end{figure}

\begin{figure}[ht!]
    \centering
    \includegraphics[width=0.2\linewidth]{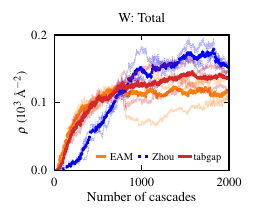}
    \includegraphics[width=0.2\linewidth]{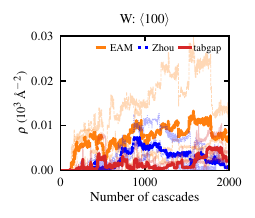}
    \includegraphics[width=0.2\linewidth]{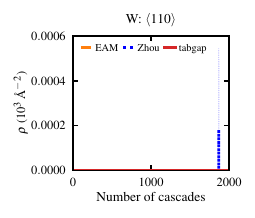}
    \includegraphics[width=0.2\linewidth]{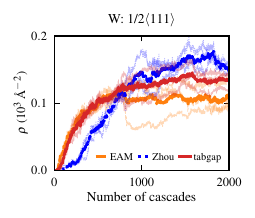}
    \includegraphics[width=0.2\linewidth]{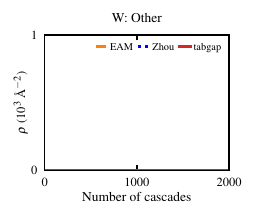}
    \caption{Dislocation density obtained from DXA analysis. Faint curves are individual runs and bold is the average.}
\end{figure}

\clearpage

\section{Overlapping cascades: Platinum $\text{Pt}~~Z=78$}

\begin{table}[ht!]
    \centering
    \begin{tabular}{c|c|c}
        \toprule
        Potential & Ref. & \hspace{5cm} notes \hspace{5cm} \\
        \midrule
        Pt-EAM & \cite{foiles1985application,nordlund1998role} & Pt-EAM ''hardened'' by K. Nordlund~\cite{nordlund1998role}\\
        \midrule
        Pt-Zhou & \cite{zhou_misfit-energy-increasing_2004} & \\
        \bottomrule
    \end{tabular}
    \caption{Details of potentials}
\end{table}
\vspace{-0.5cm}
\begin{figure}[ht!]
    \centering
    \includegraphics[width=0.35\linewidth]{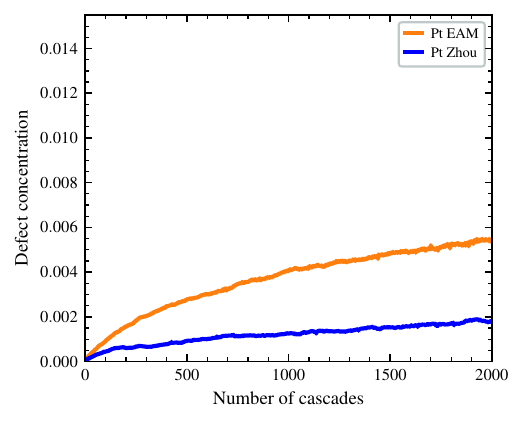}
    \includegraphics[width=0.35\linewidth]{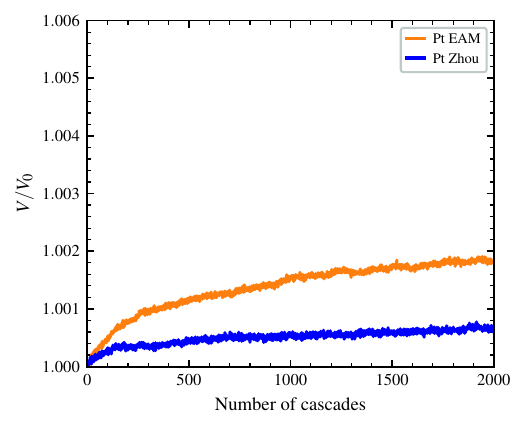}
    \caption{Left is the defect  concentration (interstitials and vacancies) as a function of number of cascades. Right is the volume change as a function of number of cascades.}
\end{figure}

\vspace{-1cm}

\begin{figure}[ht!]
    \centering
\begin{subfigure}{0.29\textwidth}
    \includegraphics[width=1.1\textwidth]{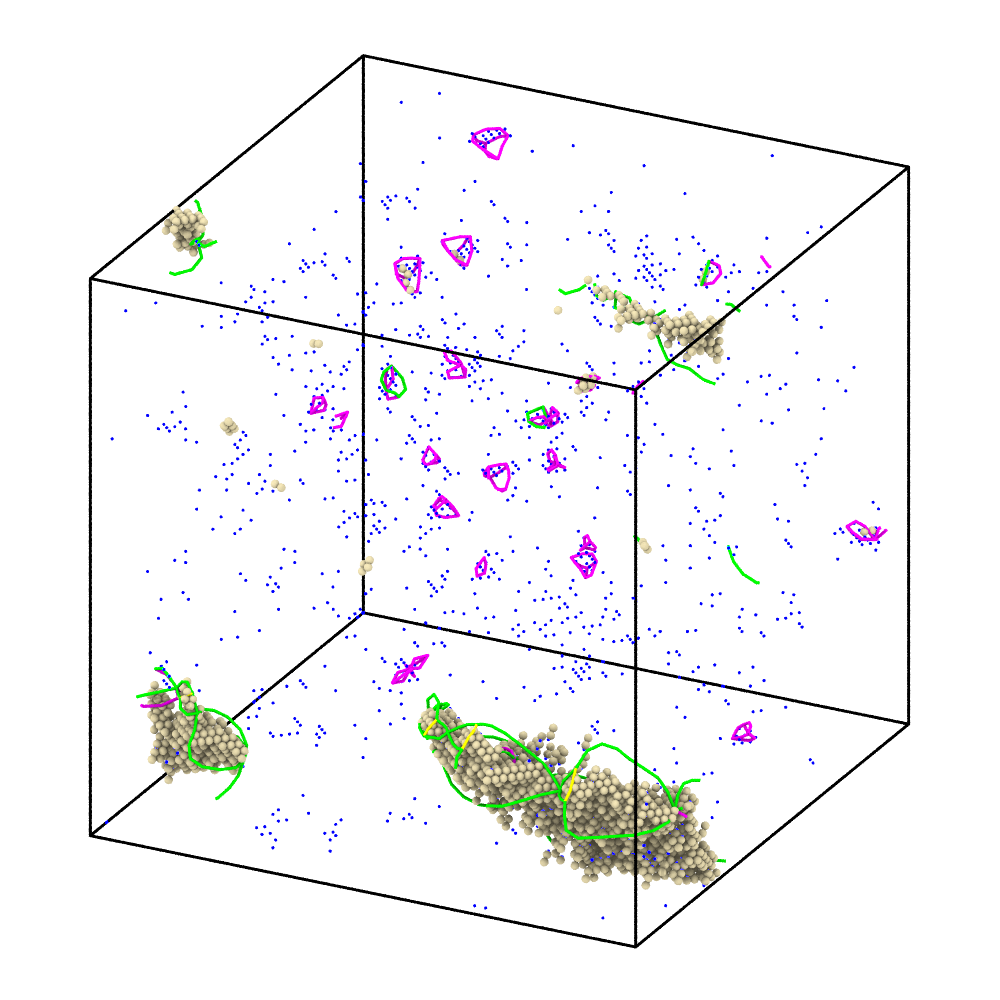}
    \caption{EAM}
\end{subfigure}
\begin{subfigure}{0.29\textwidth}
    \includegraphics[width=1.1\textwidth]{figures/frames/frame-Pt-Zhou-run-2-2000.png}
    \caption{Modified Zhou EAM}
\end{subfigure}
    \caption{Final configuration after 2000 overlapping collision cascades. blue spheres represent vacancies, colored spheres represent interstitial atoms, green lines represent Shockley dislocations, and cyan lines represent Frank dislocations and pink lines represent stair-rod dislocations.}
\end{figure}

\begin{figure}[ht!]
    \centering
    \includegraphics[width=0.2\linewidth]{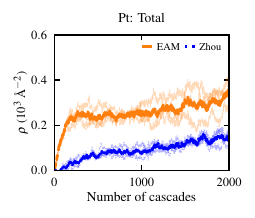}
    \includegraphics[width=0.2\linewidth]{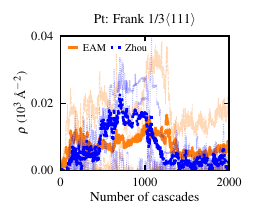}
    \includegraphics[width=0.2\linewidth]{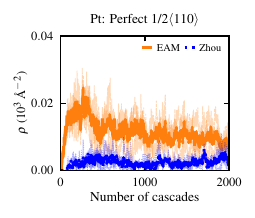}
    \includegraphics[width=0.2\linewidth]{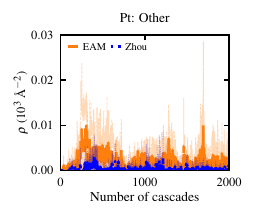}
    \includegraphics[width=0.2\linewidth]{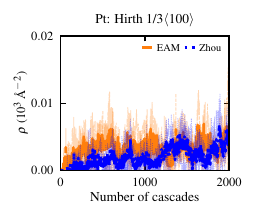}
    \includegraphics[width=0.2\linewidth]{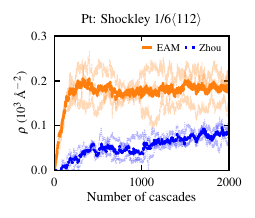}
    \includegraphics[width=0.2\linewidth]{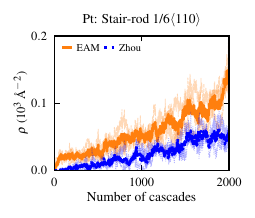}
    \caption{Dislocation density obtained from DXA analysis. Faint curves are individual runs and bold is the average.}
\end{figure}

\clearpage

\section{Overlapping cascades: Gold $\text{Au}~~Z=79$}

\begin{table}[ht!]
    \centering
    \begin{tabular}{c|c|c}
        \toprule
        Potential & Ref. & \hspace{5cm} notes \hspace{5cm} \\
        \midrule
        Au-EAM & \cite{kelchner1994construction,henriksson2005fragmentation} & Au-EAM ''hardened'' By K.O.E. Henrikson~\cite{henriksson2005fragmentation}\\
        \midrule
        Au-Zhou & \cite{zhou_misfit-energy-increasing_2004} & \\
        \bottomrule
    \end{tabular}
    \caption{Details of potentials}
\end{table}
\vspace{-0.5cm}
\begin{figure}[ht!]
    \centering
    \includegraphics[width=0.35\linewidth]{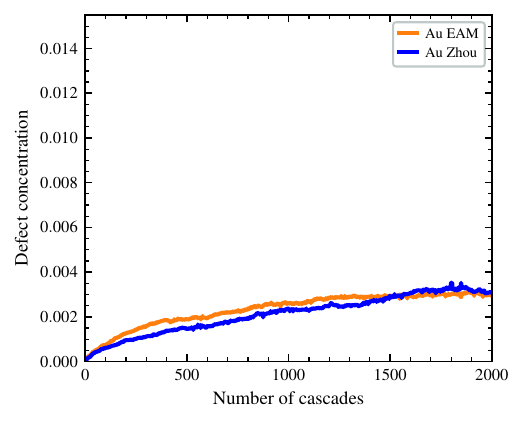}
    \includegraphics[width=0.35\linewidth]{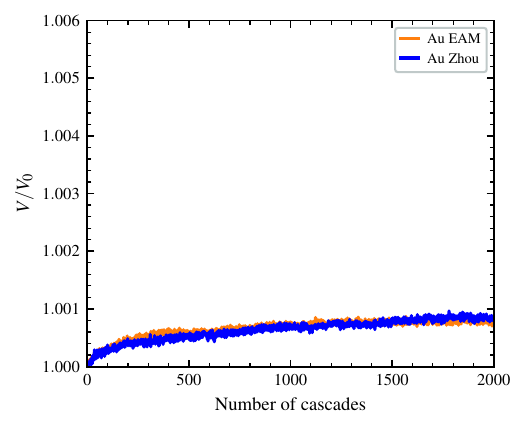}
    \caption{Left is the defect  concentration (interstitials and vacancies) as a function of number of cascades. Right is the volume change as a function of number of cascades.}
\end{figure}

\vspace{-1cm}

\begin{figure}[ht!]
    \centering
\begin{subfigure}{0.29\textwidth}
    \includegraphics[width=1.1\textwidth]{figures/frames/frame-Au-EAM-1-run-1-2000.png}
    \caption{EAM}
\end{subfigure}
\begin{subfigure}{0.29\textwidth}
    \includegraphics[width=1.1\textwidth]{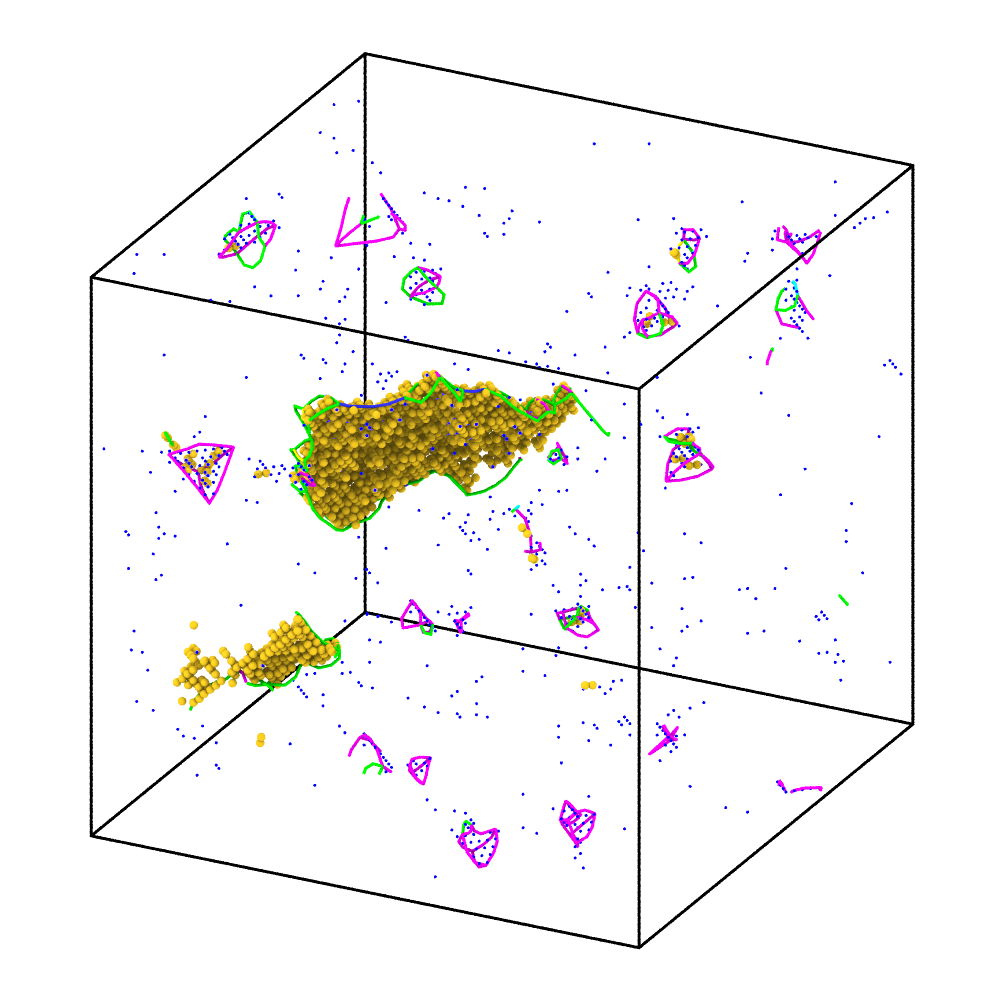}
    \caption{Modified Zhou EAM}
\end{subfigure}
    \caption{Final configuration after 2000 overlapping collision cascades. blue spheres represent vacancies, colored spheres represent interstitial atoms, green lines represent Shockley dislocations, and cyan lines represent Frank dislocations and pink lines represent stair-rod dislocations.}
\end{figure}

\begin{figure}[ht!]
    \centering
    \includegraphics[width=0.2\linewidth]{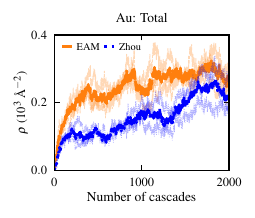}
    \includegraphics[width=0.2\linewidth]{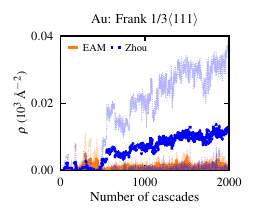}
    \includegraphics[width=0.2\linewidth]{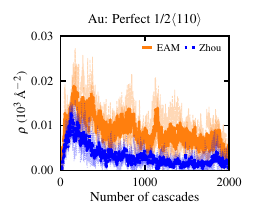}
    \includegraphics[width=0.2\linewidth]{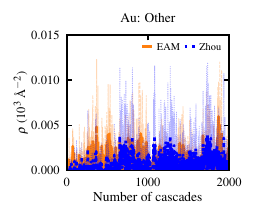}
    \includegraphics[width=0.2\linewidth]{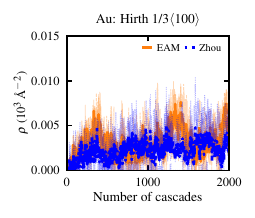}
    \includegraphics[width=0.2\linewidth]{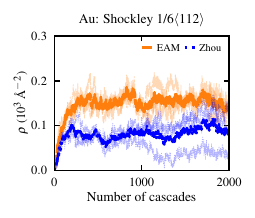}
    \includegraphics[width=0.2\linewidth]{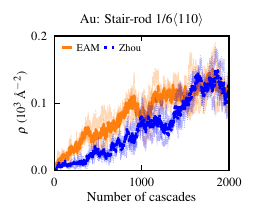}
    \caption{Dislocation density obtained from DXA analysis. Faint curves are individual runs and bold is the average.}
\end{figure}

\clearpage

\section{Overlapping cascades: Lead  $\text{Pb}~~Z=82$}

\begin{table}[ht!]
    \centering
    \begin{tabular}{c|c|c}
        \toprule
        Potential & Ref. & \hspace{5cm} notes \hspace{5cm} \\
        \midrule
        Pb-Zhou & \cite{zhou_misfit-energy-increasing_2004} & \\
        \bottomrule
    \end{tabular}
    \caption{Details of potentials}
\end{table}
\vspace{-0.5cm}
\begin{figure}[ht!]
    \centering
    \includegraphics[width=0.35\linewidth]{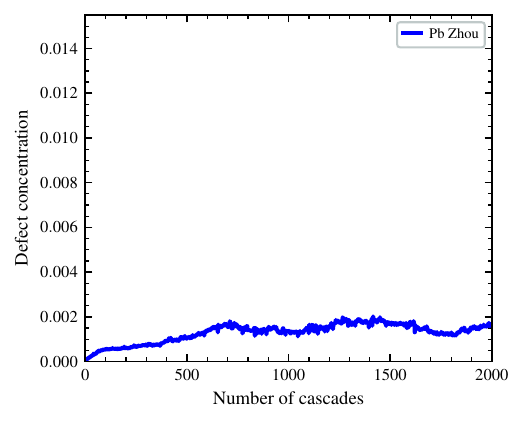}
    \includegraphics[width=0.35\linewidth]{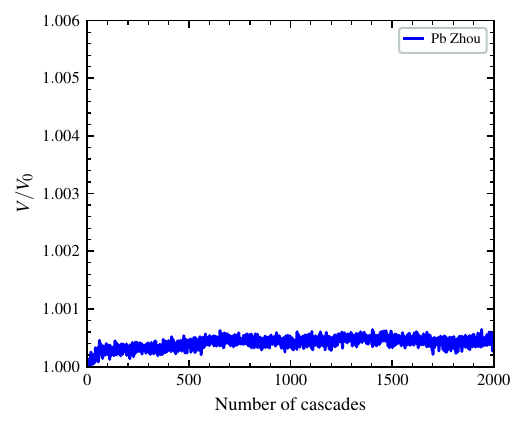}
    \caption{Left is the defect  concentration (interstitials and vacancies) as a function of number of cascades. Right is the volume change as a function of number of cascades.}
\end{figure}

\vspace{-1cm}

\begin{figure}[ht!]
    \centering
\begin{subfigure}{0.29\textwidth}
    \includegraphics[width=1.1\textwidth]{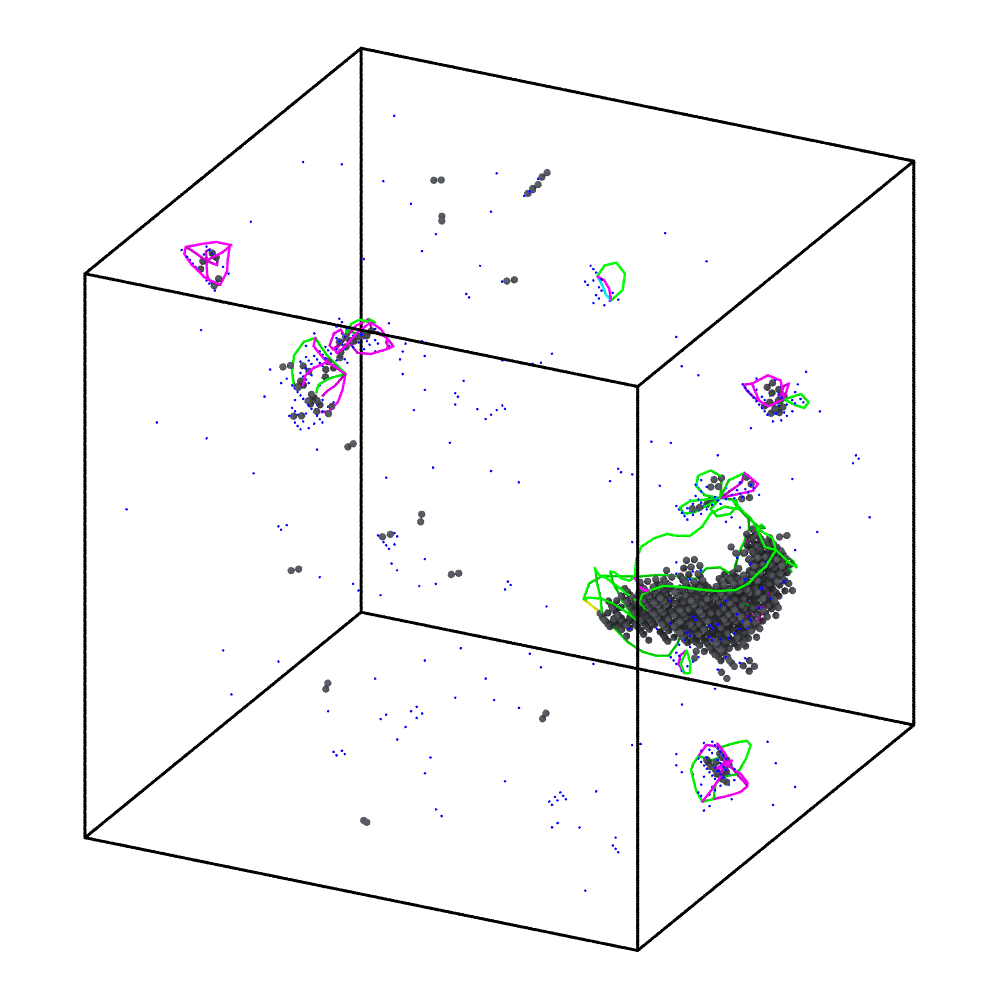}
    \caption{Modified Zhou EAM}
\end{subfigure}
    \caption{Final configuration after 2000 overlapping collision cascades. blue spheres represent vacancies, colored spheres represent interstitial atoms, green lines represent Shockley dislocations, and cyan lines represent Frank dislocations and pink lines represent stair-rod dislocations.}
\end{figure}

\begin{figure}[ht!]
    \centering
    \includegraphics[width=0.2\linewidth]{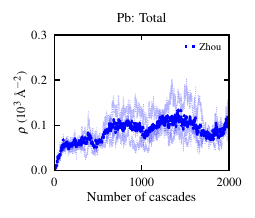}
    \includegraphics[width=0.2\linewidth]{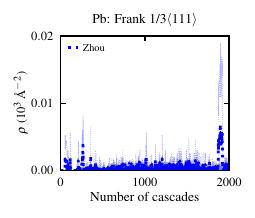}
    \includegraphics[width=0.2\linewidth]{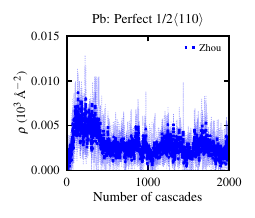}
    \includegraphics[width=0.2\linewidth]{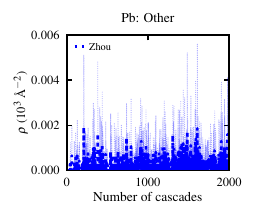}
    \includegraphics[width=0.2\linewidth]{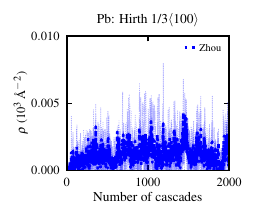}
    \includegraphics[width=0.2\linewidth]{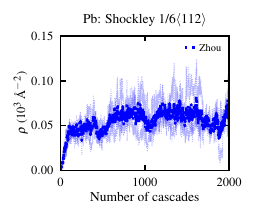}
    \includegraphics[width=0.2\linewidth]{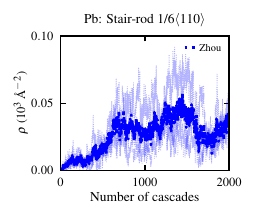}
    \caption{Dislocation density obtained from DXA analysis. Faint curves are individual runs and bold is the average.}
\end{figure}

\bibliography{mybib}